\documentclass{eptcs}

\newif\ifisdraft
\isdraftfalse

\usepackage{underscore}           

\usepackage[utf8]{inputenc}
\usepackage{amsthm}
\usepackage{amssymb}
\usepackage{amsmath}
\usepackage{float}
\usepackage{graphicx}
\usepackage{xcolor}
\usepackage{listings}
\usepackage{hyperref}
\usepackage{glossaries}
\usepackage{xspace}
\usepackage{tabularx}
\usepackage{adjustbox}
\usepackage{listings}
\usepackage{titlesec}
\usepackage{wrapfig}
\usepackage[all]{xy}
\usepackage{isabelle,isabellesym,latexsym}
\usepackage{comment}

\isabellestyle{it}
\renewcommand{\setisabellecontext}[1]{\markright{THEORY~``#1''}}

\usepackage[font=small,skip=7pt]{caption}

\usepackage[colorinlistoftodos,prependcaption]{todonotes}

\usepackage{tikz}
\usetikzlibrary{shapes,arrows}

\ifisdraft
\newcommand{\redcomment}[1]{\textcolor{red}{\textbf{#1}} }
\newcommand{\sidenote}[1]{\todo[linecolor=orange,backgroundcolor=yellow!25,bordercolor=orange]{#1}}
\newcommand{\camtodo}[1]{\todo[linecolor=blue,backgroundcolor=blue!25,bordercolor=blue]{#1}}
\newcommand{\camdone}[1]{\todo[linecolor=green,backgroundcolor=green!25,bordercolor=green]{#1}}
\newcommand{\camready}[1]{\textcolor{green}{#1}}
\else
\newcommand{\sidenote}[1]{}
\newcommand{\redcomment}[1]{ }
\newcommand{\camtodo}[1]{ }
\newcommand{\camdone}[1]{ }
\newcommand{\camready}[1]{#1}
\fi

\titlespacing{\paragraph}{ 0pt}{0.0\baselineskip}{1em}%

\providecommand{\event}{MARS 2017} 

\title{Formalizing Memory Accesses 
and Interrupts}
\def\titlerunning{Formalizing Memory Accesses 
    and Interrupts}

\author{Reto Achermann \quad Lukas Humbel \quad David Cock \quad Timothy Roscoe
    \institute{Systems Group, Department of Computer Science, ETH Zurich\\}
}
\def\authorrunning{R. Achermann, L. Humbel, D. Cock and T. Roscoe}

\newcommand{\etal}{\emph{et al.}\xspace}

\newcommand{\isempty}[3]{%
    \if\relax\detokenize{#1}\relax
    #2%
    \else
    #3%
    \fi}

\usepackage{mdframed}

\newcommand{\N}{\mathbb{N}}

\newcommand{\syntaxfont}[1]{\textbf{#1}}

\newcommand{\Is}{\ \syntaxfont{is}}
\newcommand{\Are}{\ \syntaxfont{are}}
\newcommand{\Accept}{\ \syntaxfont{accept}\ }
\newcommand{\Map}{\ \syntaxfont{map}\ }
\newcommand{\Over}{\ \syntaxfont{over}\ }
\newcommand{\To}{\ \syntaxfont{to}\ }
\newcommand{\At}{\ \syntaxfont{at}\ }

\newcommand{\vieweq}[1]{\mathop{\ \sim_{#1}\ }}

\newcommand{\Acc}{\textrm{accept}}
\newcommand{\Trans}{\textrm{translate}}
\newcommand{\Split}{\textrm{split}}
\newcommand{\SplitC}{\textrm{split}_\textrm{C}}
\newcommand{\Parse}{\textrm{parse}}

\newcommand{\addr}[1]{\texttt{#1}}
\newcommand{\traddr}[2]{\addr{#1}_{#2}}
\newcommand{\rgblk}[2]{\addr{#1}-\addr{#2}}
\newcommand{\alblk}[2]{#1/#2}
\newcommand{\trblk}[3]{\traddr{#1}{#2}/#3}
\newcommand{\spc}[1]{\textit{#1}}
\newcommand{\sbspc}[2]{\spc{#1}_\spc{#2}}
\newcommand{\pspc}[1]{\sbspc{P}{#1}}
\newcommand{\vspc}[1]{\sbspc{V}{#1}}

\newcommand{\ei}{{\it i)}\xspace}
\newcommand{\eii}{{\it ii)}\xspace}

\newcommand{\prolog}[1]{{\lstinline!#1!}}
\begin{document}

\setlength{\abovedisplayskip}{4pt plus 1pt minus 2pt}
\setlength{\belowdisplayskip}{4pt plus 1pt minus 2pt}
\setlength{\abovedisplayshortskip}{4pt plus 1pt minus 2pt}
\setlength{\belowdisplayshortskip}{4pt plus 1pt minus 2pt}

\maketitle

\ifisdraft
\listoftodos
\else
\fi

\begin{abstract}

The hardware/software boundary in modern heterogeneous multicore
computers is increasingly complex, and diverse across different
platforms.
A single memory access by a core or DMA engine traverses
multiple hardware translation and caching steps, and the destination
memory cell or register often appears at different physical addresses
for different cores.
Interrupts pass through a complex topology of interrupt controllers and
remappers before delivery to one or more cores, each with specific
constraints on their configurations.
System software must not only correctly understand the
specific hardware at hand, but also configure it appropriately at
runtime.
We propose a formal model of address spaces and resources in a system that 
allows us to express and verify invariants of the system's runtime 
configuration, and illustrate (and motivate) it with several real
platforms we have encountered in the process of OS implementation.

\end{abstract}

\section{Introduction}
\label{sec:introduction}

We present a formal model for the interpretation of memory accesses
(loads, stores, DMA operations, etc.) and interrupts of a modern
computer system which captures the relevant features of
contemporary hardware (described below).  The model gives an
unambiguous interpretation of memory accesses and interrupts, its
applications include a foundation for system software verification,
identifying problematic hardware designs, and generating
correct-by-construction OS code.

A naive view of memory addressing (often repeated in OS textbooks) is
as follows: a processor issues a load or store to a virtual address,
which is translated in the MMU by a given page table to a page fault
or a physical address, which in turn corresponds to a memory cell,
device register, or bus fault.
Similarly, interrupts are asserted by a device, translated by a
Programmable Interrupt Controller (PIC) into a local ``vector number''
indexed by the processor into a jump table of handlers.

This view is both plain wrong, and unsuitable as a basis for
verification (or, indeed, well-written software).  Modern 
systems, from mobile phone Systems-on-Chip to large
servers, are complex networks of cores, memory, devices, and
translation units.  Multiple caches in this network 
interpose on memory addresses.  Virtualization support creates
additional layers of address and interrupt translation. 

Moreover, a real system has many physical address spaces.  Memory
accesses are routed between them, often involving transforming the
address value itself from one space to another.  Answering a question
like ``Do two virtual addresses in different processes on different
cores refer to the same DRAM cell?'' requires knowledge of the
contents of TLBs, page tables, and caches on the cores \emph{and},
crucially, a representation of the system interconnect, translation
units, and topology.
Determining which core runs what code when a device raises
an interrupt similarly requires a comprehensive description of the
many stages of interrupt routing in the system.


We know of no prior formal model capturing the complexity of address-
and interrupt routing in modern hardware.  Even existing
\emph{informal} attempts to capture the increasingly diverse range of
hardware to facilitate OS software development, fail to adequately
cover the software/hardware interface -- our starting point for this work
was the need for a practical domain-specific
language with clear semantics for use in the Barrelfish research OS.


To exemplify the challenge we address, consider the Texas Instruments
OMAP4460 Multimedia SoC, a good example because it is representative
of the varied class of mobile processors, has been used in products
like the Amazon Kindle Fire 7'' and PandaBoard ES, and has some of the
best publicly available documentation.  The software manual for this
chip~\cite{ti:2011:omaptrm} has 5820 pages.

This chip has numerous processors, including two ARM Cortex
A9 cores (typically running Linux or Android), two Cortex M3's,
two DSPs, a GPU, and an ARM968 for power management, along with other
DMA-capable devices which can issue loads and stores.  Each of these, 
along with RAM and other peripherals, is attached
to one of several ``interconnects'', corresponding to physical address
spaces.  The OMAP has four main interconnects and numerous
smaller ones\footnote{For the curious reader, the relevant diagram is
  on page 157 of the manual~\cite{ti:2011:omaptrm}.}.  These
are themselves interconnected: apertures in one
map to address ranges in another.   These translations
are also subject to programmable access control checks (used, for
example, in phones to sequester the baseband radio stack on the
DSPs). 

Different cores must thus issue different \emph{physical}
addresses (after MMU translation) for the same
resource.  For example, the \texttt{GPTIMER5} timer
device has (at least) three physical addresses depending on
the accessing core: an A9 uses \texttt{0x40138000},
a DSP uses \texttt{0x01D38000} and a DMA-capable device
on the \texttt{L3} interconnect uses \texttt{0x49038000}.

Consider also the M3 cores, which use two address translation
levels: a shared cache with L1 MMU means both cores use
the same virtual address space at all times.  The output of
this MMU is fed into an address splitter and forwarded to
local ROM or RAM, or translated by another, ``L2'' MMU providing
a 1.5GB window starting at \texttt{0x0} in the \texttt{L3}
interconnect.  The M3's thus never see main system RAM at
the same physical address as the A9 cores.  
Figure~\ref{fig:system:omap44xx} shows a simplified view of 
addressing on the OMAP.

An interrupt raised by a device on the OMAP4460 can be routed to a
single designated core or one dynamically selected.  Cores themselves
can also send interrupts between themselves.  Figure
\ref{fig:system:omap44xxint} shows a subset of the interrupt topology
on the chip. The A9 cores each have private timer interrupts.  Devices
like the SDMA engine can generate four different interrupts, 
of which two can be sent to any core, and two cannot be sent
to the DSP.  Interrupts appear with different numbers in the M3 and A9
cores.  The A9 cores can initiate interrupts among themselves but
not target other cores. Interrupts from devices like the
on-chip \texttt{GPTIMER5} cannot be routed to the M3s. M3 page
faults interrupt an A9 core.

Similar complexity exists in almost all modern systems,
including PCs (as we discuss below).  Sophisticated CAD
systems, market volumes, and Moore's law have led to a great profusion in 
different platforms.

Reasoning about software running on this system clearly can neither
rely on unique nor unambiguous physical addresses, even ignoring the
effect of caches and ``conventional'' MMUs.  Software is also
constrained in terms of which interrupts can be received by which
threads.  It is not possible to make strong formal statements about
the semantics of software running on the OMAP4460 or any other
SoC without a formal description which captures the complexity
of this addressing network.
Moreover, correct software operation requires the various levels of
address translation and access control to be programmed accordingly.
Verifying system software on such a range of platforms is simply not
feasible without a clear specification of how the hardware handles
interrupts and memory references.

Our contributions in this paper are as follows. We make the case for a
formal representation of hardware to help systems programmers
understand the hardware at hand, and present a model and its syntax
(\autoref{sec:model}) to express and reason about interrupt routing,
address spaces and their interactions. We express the hardware
configuration of \camready{three different systems in subsections 
~\ref{sec:realsystems:omap44xx} to~\ref{sec:realsystems:server}}
 and give reduction results and algorithms
in~\autoref{sec:reductions}. \camready{We compare our work with the literature 
in~\autoref{sec:relatedwork} followed by a presentation of}
a roadmap of future work in
\autoref{sec:roadmap} and conclude in \autoref{sec:conclusion}.
\camdone{p. 2: Our contributions ...
    You could announce that Section 3 will present three different architectures
    in subsections 3.1 to 3.3. Section 5 is not presented.}

\section{Model and Syntax}
\label{sec:model}

\camready{In this section, we give a brief description of our model and refer 
to ~\autoref{isabelle:decodingnet} for a full description.
We model the system's handling of emitted addresses by a \emph{decoding
net}: a directed graph where each node represents a hardware component.
Addresses and interrupts vectors are natural numbers. We use the term address
and interrupt vector interchangeably in this paper. Decoding of an address 
starts at a particular node. Thus, a \emph{name} is an address 
qualified by the node at which it is decoded. We therefore define a name as 
a tuple of \ei a node identifier (\textit{nodeid}) and \eii an address 
(\textit{addr}).  Nodes are
labeled with natural numbers. A decoding net is then an assignment of nodes to 
identifiers. }
\begin{align*}
\textit{nodeid} = \mathbb{N}
& \quad\quad \textit{addr} = \mathbb{N} \\
    \textit{name} = (\textit{nodeid},\textit{addr})
    & \quad\quad \textit{net} : \textit{nodeid} \rightarrow \textit{node}
\end{align*}
\camready{ 
Hardware components either \emph{accept} addresses (e.g.\ RAM or device
registers), they \emph{translate} addresses and pass them on (e.g.\ MMU or
lookup tables), or both (e.g.\ caches). A node is completely defined by two
properties: a set of accepted addresses and a set of names it decodes an input
address to. Formally:} 
\camdone{For clarity, the "node"
    type should be defined explicitly, rather than implicitly using "accept"
    and "translate". All in one, this section should help understanding the
    annexes, but unfortunately, it makes the job harder.}
\camdone{
    p. 3: the semantics of node, name, accept, etc. is not easy to grasp, as
    everything is a natural number (CPU numbers, addresses, interrupt vectors,
    etc.). Having different type names would make it more readable (it seems
    that such names are introduced later in Annex A). }
\begin{align*}
\textit{node} =\ &\text{accept} : \{\textit{addr}\} \\
                 &\text{translate} : \textit{addr} \rightarrow
                                     \{\textit{name}\}
\end{align*}
\camready{The result of $accept$ is the set of addresses accepted by a node
without forwarding; The result of $translate$ is the set of translated names
for each input address.}
The same model can also be used to represent interrupt delivery where a node 
forwards interrupts (e.g.\ interrupt controllers) or accepts them (e.g.\ CPUs).
All definitions are formalized in Isabelle/HOL, and all results we present are 
proven in the accompanying Isabelle theories. \camready{We want to emphasize 
that our model captures the \emph{static} state of the system. Dynamic aspects 
such as     concurrency and caching can be modeled on top of the decoding net, 
something we plan to tackle in the future.}
\camdone{Modern architectures are intrinsically concurrent. However, the model
    presented in the paper does not seem to have anything really specific to
    concurrency. Perhaps, this point should be discussed.}

The model allows a node to
both accept and translate an address, and to translate an
address to multiple outputs --- real memory hardware doesn't, but interrupt
controllers can and it is useful while normalizing nodes.
Every decoding net defines a \emph{decode relation} and an
\emph{accepted-names-set} (names accepted anywhere in the net):
\begin{align*}
((n^\prime,a^\prime),(n,a)) \in \text{decode}(\textit{net})
    \ &\Leftrightarrow\ 
(n^\prime,a^\prime) \in \text{translate}(\textit{net}(n),a) \\
    \text{accepted}(\textit{net}) &=
\{(n,a) : a \in \text{accept}(\textit{net}(n))\}
\end{align*}

From these, we define the \emph{resolution function}, which maps an input name
(an address presented to a particular node) to a set of \emph{resolved names}:
the nodes at which the input address \emph{could} end up being accepted,
together with the translated input address to that node.
\begin{displaymath}
\text{resolve}(\textit{net},n) =
    \{n\}\ \cap\ \text{accepted}(\textit net)\ \ \cup\ \ 
    \bigcup (n^\prime,n) \in \text{decode}(\textit{net}).\ 
        \text{resolve}(\textit{net},n^\prime)
\end{displaymath}

This is the input name if and only if the start node accepts the 
start address, and then
all names reachable recursively via the decode relation.  The termination of
this recursion depends on the structure of the net, specifically the presence
of loops.  We present a necessary and sufficient condition for termination in
\autoref{sec:reductions}.

\camready{
Nets are expressed in the following concrete syntax (EBNF, terminals are
bold).  This corresponds to the abstract syntax of \autoref{sec:isasyntax}.
}
\begin{align*}
\textit{net}_s &\mathop{=} 
    \Big\{\N\ \textbf{is}\ \textit{node}_s\ \Big|\ 
     \N\textbf{..}\N\ \textbf{are}\ \textit{node}_s\Big\} \\
\textit{node}_s &\mathop{=}
    \Big[\textbf{accept}\ \textbf{[}\ \Big\{\textit{block}_s\Big\}\ \textbf{]}
    \Big]\ 
    \Big[\textbf{map}\  \textbf{[}\ \Big\{\textit{map}_s\Big\}\ \textbf{]} \Big]\ 
    \Big[\textbf{over}\ \N \Big] \\
\textit{map}_s &\mathop{:=}
     \textit{block}_s\ \To \N\ \Big[\At \mathbb{N}\Big]\ 
        \Big\{\textbf{,}\ \N\ \Big[\At \mathbb{N}\Big]\Big\}\\
\textit{block}_s &\mathop{:=} \N\textbf{--}\N\
\end{align*}
\camdone{p. 3: the meta-language used to define the "concrete syntax" (isn't 
rather
    an abstract syntax?) is all but clear. One can guess that "(..)?" means 
    optional. But what is the signification of "[...]"? Usually, "[...]" means
    optional, but here it seems that [A, B, C] means either A or B or C. The 
    authors should forget about this custom meta-language and use standard BNF 
    instead. Or, at least, they should define precisely their metalanguage. }
Here all translations are specified by mapping a contiguous block of input
addresses (specified by range $base-limit$) to some output
node, with all address values shifted to a new block base address.  Nodes are
specified by a finite set of accepting and mapping blocks and may be
initialized using an \emph{overlay} ($\textbf{over}$). If so, the node maps
all input addresses 1--1 to the specified \emph{overlay node}, unless they are
captured by an accept or map block.  Nodes are assigned identifiers with
$\textbf{is}$ or $\textbf{are}$ (for repeated nodes). One can use the 
identifier instead of the assigned number when referring to the node. The map 
specification allows us to declare multiple destinations, which is necessary 
to describe interrupt systems.

In the following section, we demonstrate that this syntax, together with our
model, concisely represents the address decoding of real hardware.  In
\autoref{sec:reductions} we further show that our model supports reasoning
about address translation, for example by showing that networks can be reduced
to a normal form, and that this reduction can be expressed by a simple
algorithm on the concrete representation of the network i.e. that the
representation \emph{refines} the model.

\section{Modeling Real Systems}
\label{sec:realsystems}

We now express real systems (all of which we use for Barrelfish
development) using the syntax from the previous section.  \camready{We
present the OMAP4460 SoC and two different x86 systems here. In addition, 
we show fully detailed models of those systems (\ref{isabelle:omap}to 
\ref{isabelle:server}) and two additional systems (cluster 
system~\ref{isabelle:scc} and the Intel SCC~\ref{isabelle:cluster}) in the 
appendix.}
We focus our modelling on 
software-visible hardware features. We represent a block of 
addresses in the form $\trblk{p}{z}{b}$ with a (hex) prefix $p$ 
followed by $z$ zeros and the block size is $2^b$  e.g.\ 
\texttt{0x20000000-0x20000fff} is represented as 
$\trblk{20000}{3}{12}$ with a block size is 12-bits (4kB).

\camdone{There are two more architectures in annexes B.4 and B.5: the article
    should at least mention them, even if they are not detailed in pages
    1-12.}
\camdone{make sure to mention the appendix clearly}

%
%

\begin{figure}[ht]
    \centering
    \includegraphics{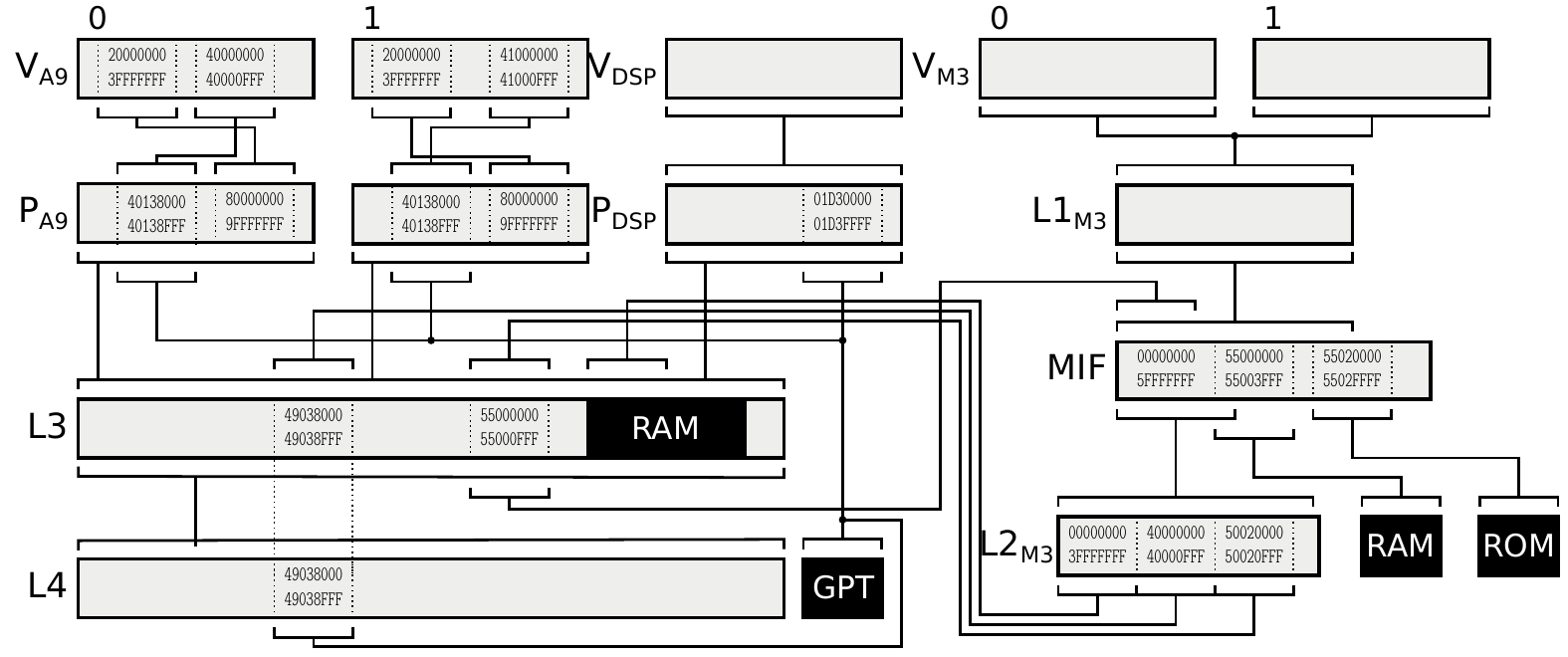}
    \vspace{-2.5em}
    \begin{align*}
    \vspc{A9:0}\Is
    &\Map [\trblk{20000}{3}{12}\To\pspc{A9:0}\At\traddr{80000}{3}]
    &\vspc{A9:1}\Is
    &\Map [\trblk{20000}{3}{12}\To\pspc{A9:1}\At\traddr{80000}{3}]\\
    \pspc{A9:0},\pspc{A9:1}\Are
    &\Map [\trblk{40138}{3}{12}\To\spc{GPT}\At\addr{0}] 
    \Over\spc{L3}
    &\vspc{DSP}\Is
    &\Over\pspc{DSP} \\
    \pspc{DSP}\Is
    &\Map [\trblk{1d3e}{3}{12}\To\spc{GPT}\At\addr{0}]
    \Over\spc{L3} 
    &\sbspc{L2}{M3}\Is
    &\Map [\traddr{0}{30}\To\spc{L3}\At\traddr{80000}{3}] \\     
    \vspc{M3},\vspc{M3}\Are
    &\Over\sbspc{L1}{M3}
    & \sbspc{L1}{M3}\Is
    &\Map [\traddr{0}{28}\To\spc{MIF}]\\
    \sbspc{RAM}{M3}\Is
    &\Accept [\trblk{55020}{3}{16}]
    &\spc{L4}\Is
    &\Map [\trblk{49038}{3}{12}\To\spc{GPT}\At\addr{0}] \\    
    \sbspc{ROM}{M3}\Is
    &\Accept [\trblk{55000}{3}{14}]
    &\spc{GPT}\Is
    &\Accept [\alblk{\addr{0}}{12}]
    \end{align*}
    \vspace{-3.0em}
    \begin{align*}    
    \spc{MIF}\Is
    &\Map [\rgblk{0}{5fffffff}\To\sbspc{L2}{M3},
    \trblk{55000}{3}{14}\To\sbspc{RAM}{M3},
    \trblk{55020}{3}{16}\To\sbspc{ROM}{M3}] \quad \\
    \spc{L3}\Is
    &\Map [\trblk{49000}{3}{24}\To\spc{L4}\At\traddr{40100}{3},
    \trblk{55000}{3}{12}\To\spc{MIF}]
    \Accept [\trblk{80000}{3}{30}]
    \end{align*}
    \vspace{-2em}
    \caption{Addressing block diagram and model description of the Texas 
    Instrument OMAP 
    44xx SoC. }
    \label{fig:system:omap44xx}
\end{figure}

\subsection{A Mobile Device SoC: the OMAP4460}
\label{sec:realsystems:omap44xx}

\begin{figure}[!ht]
    \centering 
    \includegraphics{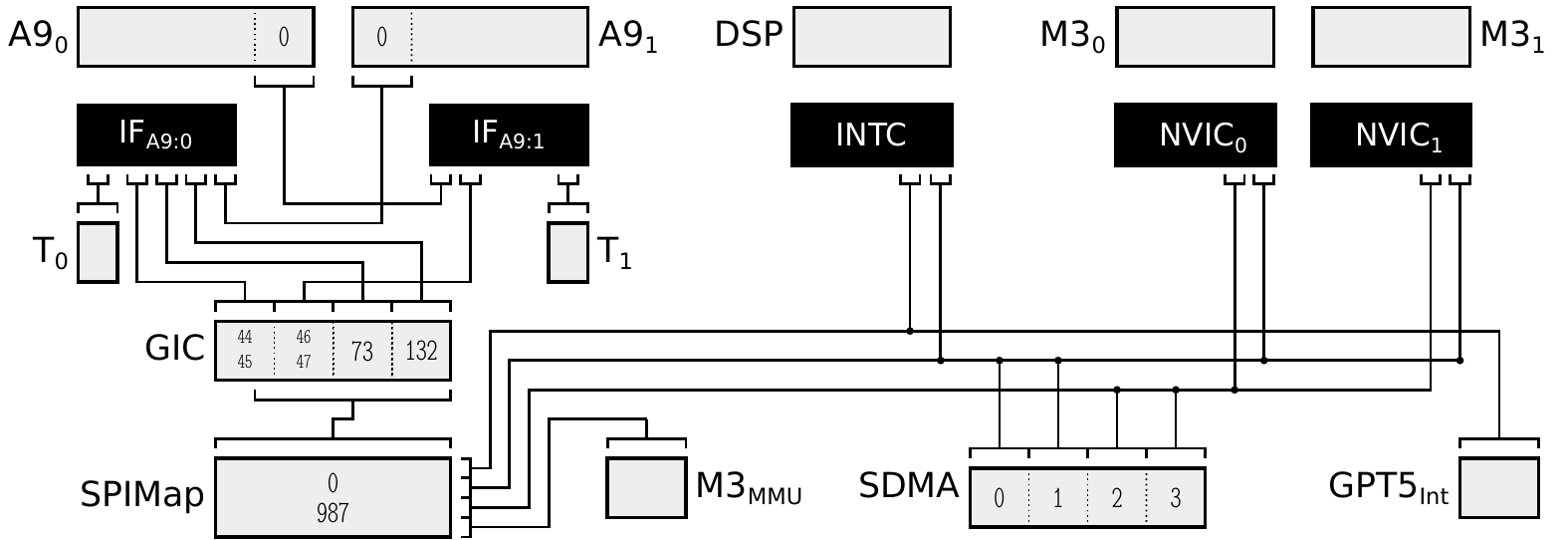}
    \vspace{-2em}
    \begin{align*}
    \spc{SDMA}\Is
    & \Map [\addr{0} \To \spc{SPIMap} \At \addr{12}
    \To \spc{INTC} \At \addr{18}
    \To \sbspc{NVIC}{0} \At \addr{18}
    \To \sbspc{NVIC}{1} \At \addr{18}, \\&
    \addr{1} \To \spc{SPIMap} \At \addr{13}
    \To \spc{INTC} \At \addr{19}
    \To \sbspc{NVIC}{M3:0} \At \addr{19}
    \To \sbspc{NVIC}{M3:1} \At \addr{19},\\&
    \addr{2} \To \spc{SPIMap} \At \addr{14}
    \To \sbspc{NVIC}{0} \At \addr{20}
    \To \sbspc{NVIC}{1} \At \addr{20},\\&
    \addr{3} \To \spc{SPIMap} \At \addr{15}
    \To \sbspc{NVIC}{0} \At \addr{21}
    \To \sbspc{NVIC}{1} \At \addr{21}
    ] \\
    \sbspc{GPT5}{Int}\Is
    & \Map [\addr{0} \To \spc{SPIMap} \At \addr{41} 
    \To \spc{INTC} \At \addr{41} ] \\
    \spc{GIC}\Is
    & \Map [\rgblk{44}{45}\To\sbspc{IF}{A9:0}\At\addr{44},
            \rgblk{46}{47}\To\sbspc{IF}{A9:1}\At\addr{46},
        \ldots] 
    \end{align*}
    \vspace{-3em}
    \begin{align*}
        \sbspc{A9}{0}\Is 
        & \Map [\addr{0} \To \sbspc{IF}{A9:1} \At \addr{0}] &
        \sbspc{A9}{1}\Is 
        & \Map [\addr{0} \To \sbspc{IF}{A9:0} \At \addr{0}] \\
        \sbspc{T}{0}\Is
        & \Map [\addr{0} \To \sbspc{IF}{A9:0} \At \addr{29}] &
        \sbspc{T}{1}\Is
        & \Map [\addr{0} \To \sbspc{IF}{A9:1} \At \addr{29}] \\
        \sbspc{M3}{MMU}\Is
        & \Map [\addr{0} \To \spc{SPIMap} \At \addr{100}] &
        \spc{SPIMap}\Is
        & \Map [\rgblk{0}{987}\To\spc{GIC}\At\addr{32}] \\
        \spc{INTC}, \sbspc{NVIC}{*} \Are
        &\Accept [ ] &
        \sbspc{IF}{*} \Are
        & \Accept [\rgblk{0}{1020}] 
    \end{align*}
    \vspace{-0.8cm}
    \caption{Overview and model of the interrupt system of the Texas Instrument 
    OMAP 44xx SoC}
    \label{fig:system:omap44xxint}
\end{figure}

We introduced the OMAP4460~\cite{ti:2011:omaptrm} in~\autoref{sec:introduction}. 
Each core (A9, M3 and DSP) has a distinct view of the system: some 
resources are core-private while others appear at different addresses,
etc.  We show examples of addressing and interrupt models 
(Figures~\ref{fig:system:omap44xx} and~\ref{fig:system:omap44xxint})
and refer
to section~\ref{isabelle:omap} for the full model.

\paragraph{Accessing DRAM from the A9 and M3 cores:}
From the A9 core's virtual address space, $\vspc{A9:0}$, the address 
\texttt{0x20000000} is translated and forwarded to ($\pspc{A9:0}$, 
\texttt{0x8000000}) which overlays the $\spc{L3}$ interconnect address space. 
$\spc{L3}$ accepts the input and decoding terminates. Addresses from the 
$\vspc{DSP}$ are handled similarly. The M3 cores share the translation tables 
and hence accesses from $\vspc{M3:*}$ are overlaid onto $\sbspc{L1}{M3}$ which 
maps and forwards addresses to the $\spc{MIF}$, an address splitter. 
In our case, $\sbspc{L1}{M3}$ outputs address \texttt{0x0} which is forwarded 
by the $\spc{MIF}$ to the $\sbspc{L2}{M3}$ and further translated to 
\texttt{0x80000000} and forwarded to and accepted by $\spc{L3}$.

\paragraph{Accessing the \texttt{GPTIMER5} device:}
The \texttt{GPTIMER5} has multiple 
names that resolve to the $\spc{GPT}$ node:  $(\pspc{DSP}, 
\texttt{0x01d3e000})$, 
$(\pspc{A9}, \texttt{0x40138000})$, and $(\pspc{M3},
\texttt{0x30038000})$, along with $(\spc{L3}, \texttt{0x49038000})$
that means that $\pspc{A9}$ and $\pspc{DSP}$ can also access
$\spc{GPT}$ as they overlay $\spc{L3}$.

\paragraph{Almost a loop:}
We can construct a configuration where an access goes through the same 
address space twice, but we can show that resolution still works as the two 
decoding steps have different addresses. We start at the $(\spc{MIF}, 
\texttt{0x50020000})$ and the decoding steps go through $\sbspc{L2}{M3}$, 
$\spc{L3}$, $\spc{L4}$ and end up in $(\spc{MIF}, \texttt{0x55000000})$. This 
is the same node where we started suggesting a loop in the decoding, however 
the address is \emph{different}. The 
$\spc{MIF}$ now maps the request to $\sbspc{ROM}{M3}$


\paragraph{SDMA triggers interrupt 2:}
The SDMA device can generate four different interrupts. We
model this as as a node that maps consecutive vectors starting from
zero.  Once it triggers interrupt 2 in the \spc{SDMA} node, the
interrupt will be forwarded to multiple controllers with
\textit{different vectors}. Specifically, it will be multicast towards
$(\spc{SPIMap}, \texttt{12})$ and $(\sbspc{NVIC}{*}, \texttt{18})$ of
the M3 subsystem. Since the $\spc{NVIC}$s are configured to ignore
(mask) the interrupt, the nodes will neither accept nor map
these interrupts.  $\spc{SPIMap}$ translates the vector and forwards
the signal to $(\spc{GIC}, \texttt{44})$ and is finally accepted
$(\sbspc{IF}{A9:0}, \texttt{44})$

%
%
\subsection{A Desktop PC}
\label{sec:realsystems:desktop}
\begin{figure}[ht]
    \centering 
    \includegraphics[width=\textwidth]{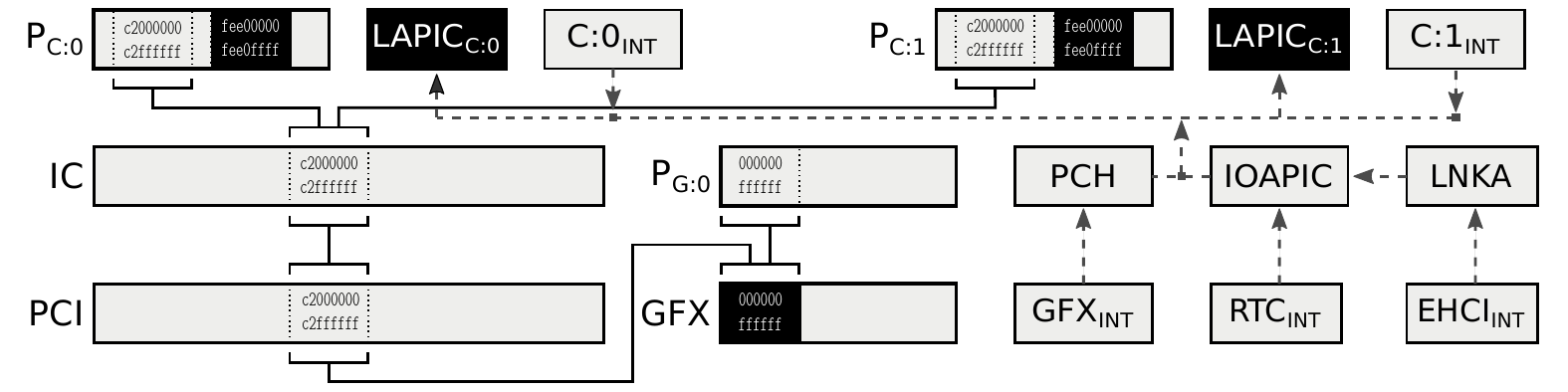}
\begin{align*}
        \pspc{C:0}\Is
        & \Map [\trblk{0xc2000}{3}{24}\To\spc{IC}] 
        \Accept [\alblk{\addr{0xfee00000}}{13}]\\
        \pspc{C:1}\Is
        & \Map [\trblk{0xc2000}{3}{24}\To\spc{IC}] 
        \Accept [\alblk{\addr{0xfee00000}}{13}] \\
        \spc{IOAPIC}\Is
        & \Map [\addr{4} \To \sbspc{LAPIC}{C:0} \At \addr{48}, \addr{8}\To 
        \sbspc{LAPIC}{C:0} \At \addr{40}]\\
        \spc{PCH}\Is
        & \Map [\addr{fee002b800000029}\To\sbspc{LAPIC}{C:0}\At 
        125] \\
        \sbspc{GFX}{INT}\Is
        & \Map [\addr{0} \To\spc{PCH}\At\addr{fee002b800000029}]\\
        \spc{IOAPIC}\Is
        &\Map [\addr{4} \To \sbspc{LAPIC}{C:0} \At \addr{48},  \addr{8}\To 
        \sbspc{LAPIC}{C:0} \At \addr{40}]        
        \end{align*}
        \vspace{-3em}
        \begin{align*}
        \spc{IC}\Is
        &  \Map [\trblk{c2000}{3}{24}\To\spc{PCI}]
        &\spc{PCI}\Is
        &\Map [\trblk{c2000}{3}{24}\To\spc{GFX} \At \addr{0}]\\
        \sbspc{P}{G:0}\Is 
        & \Map [\alblk{0}{24}\To\spc{GFX}] 
        &\spc{GFX}\Is
        &\Accept [\alblk{\addr{0}}{24}] \\
        \sbspc{C:0}{INT}\Is
        & \Map [\addr{251}\To\sbspc{LAPIC}{C:1}\At\addr{251}] \quad \quad
        &\sbspc{C:1}{INT}\Is 
        &\Map [\addr{251}\To\sbspc{LAPIC}{C:0}\At\addr{251}]\\
        \sbspc{EHCI}{INT}\Is
        & \Map [\addr{0} \To\spc{LNKA}]
        \quad
        &\spc{LNKA}\Is
        &\Map [\addr{0} \To\spc{IOAPIC}\At\addr{4}]\\
        \sbspc{RTC}{INT}\Is
        & \Map [\addr{0} \To\spc{IOAPIC}\At\addr{8}] &
        \sbspc{LAPIC}{C:*} \Are
        & \Accept [\addr{32}-\addr{255}] \\
    \end{align*}
    \vspace{-1.5cm}
    \caption{Schematic overview and model representation of the desktop PC}
    \label{fig:system:desktop}
\end{figure}

Our example desktop machine has a quad-core processor, 32GB of 
main
memory and several I/O devices.  We focus on two aspects of the memory
model: a resource can respond to multiple addresses, and different
resources respond to the same address.  We further highlight the
diversity of interrupt paths and vector formats.  The relevant nodes
(with 2 cores) are shown in~\autoref{fig:system:desktop}; the
full model is in ~\autoref{isabelle:desktop}.

\paragraph{Address homonyms:}
Consider cores $\pspc{C:*}$ in~\autoref{fig:system:desktop}.  If they
are using the same MMU page table one might think -- erroneously -- that
they access the same view of physical memory.  In fact, core-physical
address \texttt{0xfee00000} (the local APIC address) on each core is
accepted locally by both $\pspc{C:0}$ and  $\pspc{C:1}$.  Each core's
MMU sees a different physical address space. 

\paragraph{Address synonyms:}
Conversely, a single resource, the GDDR region of $\spc{GFX}$, appears
at multiple addresses depending on the starting node. For instance
$(\sbspc{GFX}{Core:0}, \texttt{0x0})$ will decode to $(\spc{GFX},
\texttt{0x0})$.  The same resource can be reached via a different
address: $(\pspc{C:0}, \texttt{0xc2000000})\rightarrow (\spc{IC},
\texttt{0xc2000000})\rightarrow (\spc{PCI},
\texttt{0xc2000000})\rightarrow (\spc{GFX}, \texttt{0x0})$.  Physical
addresses are not unique identifiers for the resource.

\paragraph{Message-signalled interrupts: } The GPU issues message
signalled interrupts (MSI), memory writes to a platform-specific
address range with a data word. We start with $(\sbspc{GFX}{INT},
\texttt{0})$ which is translated to a memory write to $(\spc{PCH},
\texttt{0xfee002b800000029})$. We model this MSI as the 
concatenation of
the address and data word since $\spc{PCH}$ is able to distinguish
both. The PCH transforms these memory writes back to a regular
interrupt message, here forwarded to $(\sbspc{LAPIC}{C:0},\texttt{125})$ which accepts the signal.

\paragraph{Legacy interrupt path:} The EHCI USB controller raises a legacy
interrupt, which appears at $\sbspc{EHCI}{INT}$. It propagates 
to the PCI link device $\spc{LNKA}$ with vector zero, which redirects 
it to the $\spc{IOAPIC}$ with vector 4.  The $\spc{IOAPIC}$ is configured to forward 
interrupt number 4 to $\sbspc{LAPIC}{C:0}$, with vector 48.
Alternatively, the RTC device follows a similar path to the EHCI
but it is directly connected to the IOAPIC.  The decoding steps are
$(\sbspc{RTC}{INT},0)\rightarrow (\spc{IOAPIC},8) \rightarrow (\sbspc{LAPIC}{C:0},40)$.

\paragraph{Vector sharing:} The ARM platform (as
in~\autoref{sec:realsystems:omap44xx}) separates interrupts generated
by peripheral devices and inter-processor interrupts initiated by
software.  The Intel x86 platform in contrast does not, and so for
each core we split inbound ($\sbspc{LAPIC}{C:*}$) and outbound
($\sbspc{C:*}{INT}$) interrupts into separate nodes 
(see~\autoref{sec:reductions} for proof). In our example, 
$\sbspc{C:1}{INT}$ triggers an interrupt which
decodes to the same destination as the $\sbspc{EHCI}{INT}$. We start
with $(\sbspc{C:1}{INT}, \texttt{1})$ which is directly forwarded to
the accepting node $(\sbspc{LAPIC}{C:0}, \texttt{48})$.  Sharing may
be unavoidable: an MSI device can issue up to 2048 different
interrupts while an x86 core can only distinguish 256 vectors.

%
%
%
%

\subsection{A Heterogeneous x86 Server}
\label{sec:realsystems:server}
Servers, particularly for high-performance computing, are more complex
than commodity desktops.  Our example has 2 $\times$ 10-core sockets,
each with its own DRAM controller, PCIe root complex, and IOMMU.  Both
PCIe buses have a Xeon Phi co-processor with 57 cores and 6GB
GDDR RAM.   The host cores support virtualization, including
nested paging.  This hardware is discussed in more detail in
~\cite{Gerber:2015:YPP:2831090.2831106}).

\begin{figure}[!ht]
    \centering 
    \includegraphics[width=\textwidth]{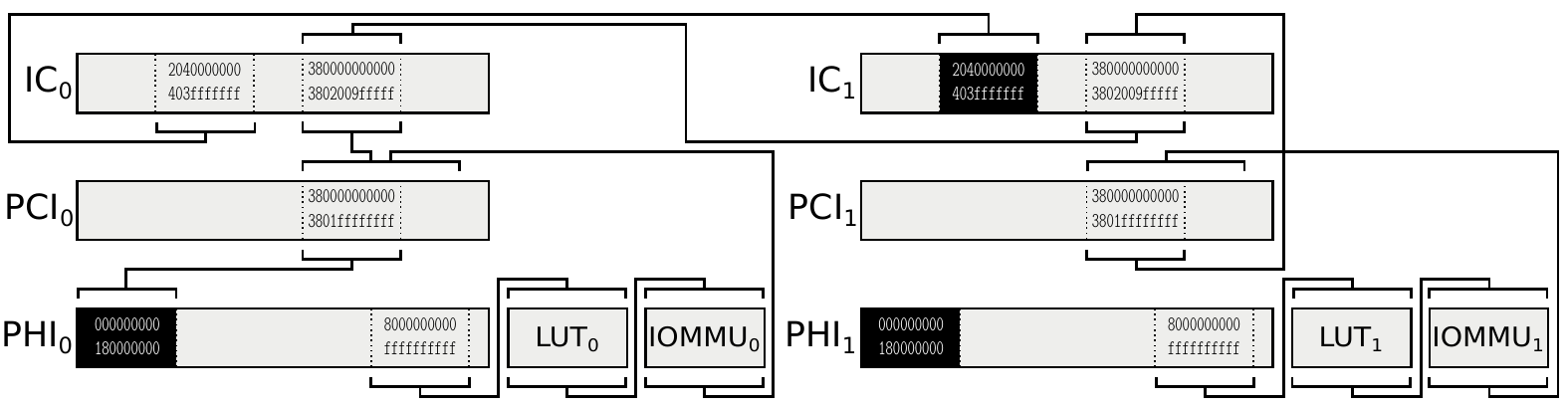}
    \begin{align*}
    \textit{IC}_\textit{0}\ \Is\ &\Map\ [
    \texttt{2040000}_3/37\ \To\ \textit{IC}_\textit{1},\ 
    \texttt{380}_{9}-\texttt{3802009fffff}\ \To\ \textit{PCI}_\textit{0}] \\
    \textit{IC}_\textit{1}\ \Is\ &\Accept\ [\texttt{2040000}_3/37]\ 
    \Map\ [\texttt{380}_{9}-\texttt{3802009fffff}\ \To\ \textit{IC}_\textit{0}] 
    \\
    \textit{PCI}_\textit{0}\ \Is\ &\Map\ [\texttt{380}_{9}/34 \To\ 
    \textit{Phi}_\textit{0}\ \At\ \texttt{0}]\\
    \textit{PCI}_\textit{1}\ \Is\ &\Map\ 
    [\texttt{380}_{9}-\texttt{3802009fffff}\ 
    \To\ \textit{IC}_\textit{1}\ \At\ \texttt{380}_{9}/43] \\
    \textit{LUT}_\textit{0}\ \Is\ & \Map\ [\texttt{c00000}_{3}/35 \To\ 
    \textit{IOMMU}_\textit{0}\ \At\ \texttt{800}_3]\\
    \textit{LUT}_\textit{1}\ \Is\ & \Map\ [\texttt{800000}_{3}/35 \To\ 
    \textit{IOMMU}_\textit{1}\ \At\ \texttt{600}_3]\\
    %
    \textit{IOMMU}_\textit{0}\ \Is\ & \Map\ [\texttt{800}_3/21\ \To\ 
    \textit{PCI}_\textit{0}\ \At\ \texttt{380}_{9}/43]\\
    \textit{IOMMU}_\textit{1}\ \Is\ & \Map\ [\texttt{600}_3/21\ \To\ 
    \textit{PCI}_\textit{1}\ \At\ \texttt{380}_{9}/43] \\
    \textit{PHI}_\textit{0}\ \Is\ & \Accept\ [\texttt{0}/34]\ 
    \Map\ [\texttt{8000000}_{3}/39\ \To\ \textit{LUT}_\textit{0}\ \At\ 
    \texttt{0}]\\
    \textit{PHI}_\textit{1}\ \Is\ & \Accept\ [\texttt{0}/34]\ 
    \Map\ [\texttt{8000000}_{3}/39\ \To\ \textit{LUT}_\textit{1}\ \At\ 
    \texttt{0}]\\
    \end{align*}
    \vspace{-1.5cm}
    \caption{Schematic overview of a heterogeneous server with Xeon Phi 
        co-processors}
    \label{fig:system:server}
\end{figure}

The features we highlight (and make the server different from the
desktop) are shown in~\autoref{fig:system:server}: the NUMA
memory topology, PCI devices accessing each other's memory through the
IOMMU~\cite{intel:2016:vtd}, and aliasing in the  
same address space. We refer to ~\autoref{isabelle:server} for a full 
model.

\paragraph{NUMA topology:}
The NUMA nodes are abstracted by $\sbspc{IC}{0}$ and 
$\sbspc{IC}{1}$. We model the behavior as follows: starting from 
$(\sbspc{IC}{0}, \texttt{0x2040000000})$ decodes to a forward to 
$\sbspc{IC}{1}$ with the same address and is accepted (remote access). Local 
accesses start from  $(\sbspc{IC}{1}, \texttt{0x2040000000})$ and are directly 
accepted by $\sbspc{IC}{1}$.

\paragraph{Aliasing:}
The Xeon Phi can be configured such that its local GDDR region is
aliased through the system memory interface.  In our example, emitted
addresses $(\sbspc{PHI}{0}, \texttt{0x0})$ are directly accepted by
$\sbspc{PHI}{0}$ whereas $(\sbspc{PHI}{0}, \texttt{0x8c00000000})$
triggers the translation chain: $ (\sbspc{LUT}{0}, \texttt{c00000000})
\rightarrow (\sbspc{IOMMU}{0}, \\ \texttt{0x800000}) \rightarrow
(\sbspc{PCI}{0}, \texttt{0x380000000000}) \rightarrow (\sbspc{PHI}{0},
\texttt{0x0}) $. Here, input address $(\sbspc{PHI}{0},
\texttt{0x8c00000000})$ decodes to $(\sbspc{PHI}{0}, \texttt{0x0})$,
i.e. there are multiple addresses for this RAM cell: the LUT and
IOMMU allow multiple mappings, although  notably with different
coherency and latency characteristics.

\paragraph{PCI-PCI access:}
The local resources of $\sbspc{PHI}{0}$ 
can be accessed from  $\sbspc{Phi}{1}$ through the system memory 
region: emitted address $(\sbspc{Phi}{1}, 
\texttt{0x8800000000})$ is forwarded and mapped through 
$\sbspc{LUT}{1}$ and $\sbspc{IOMMU}{1}$ to the interconnect 
$(\sbspc{IC}{1}, \texttt{0x380000000000})$. This address is in the range 
of the other socket, is forwarded there, and eventually accepted by 
$(\sbspc{PHI}{0}, \texttt{0x0})$.

\paragraph{Interrupts:}  The main difference from the system
in~\autoref{sec:realsystems:desktop} is the IOMMU's  
interrupt controller on the path between devices and host cores. The Xeon Phis
can raise regular PCIe interrupts, but also have their own local
interrupt subsystem resembling another x86 system.  These two subsystems 
are isolated: an interrupt on the Xeon Phi cannot be directly 
forwarded to the host or vice versa, another example of the limited
reachability we referred to in~\autoref{sec:realsystems:omap44xx}.

%
%

%
%

\section{Reductions and Algorithms}
\label{sec:reductions}

The purpose of this model is to accurately represent the complex structure of
address resolution and interrupt routing in a format that can be easily 
generated and manipulated at runtime. We now demonstrate that the model also 
has nice logical properties: it is amenable to formal analysis and the 
verification of routines that interpret or manipulate the concrete syntax.
All results presented here are verified, and presented with a reference to the
proof in the accompanying Isabelle/HOL source.
\camdone{Section 4 gives applications regarding verification and runtime
    features. The write-up of this section is less clear than the rest of
    the paper, perhaps this could be improved.
    }

\subsection{Termination}

The underlying model, as introduced in \autoref{sec:model}, is strictly
graphical---it is defined entirely by the $\textrm{decode}$ relation and the
$\textrm{accept}$ set.  This permitted us to directly specify the hardware
behavior, including decoding loops, but says nothing directly about an agent's
view of the system.  Which resources are visible at such and such an address,
in such and such an address space?  The $\textrm{resolve}()$ function provides
the link, giving a mapping from local names (addresses relative to a
viewpoint) to global names (nodes that may accept the address, and the local
address at which they accept it).

HOL is a logic of total functions, and we can thus express the mapping from
(address space, address) to (a set of) resources as a function if, and only
if, the decoding process terminates.  This occurs when every path in the
decode relation beginning at the input address is finite.  We express this as
the existence of some well-formed \emph{ranking function} $f : \textit{name}
\rightarrow \mathbb{N}$, that decreases on every step of the decode relation
(\autoref{sec:isawfnet}):
\begin{equation*}
\text{wf\_rank}(f,n,\textit{net})\ \longleftrightarrow\ 
    \forall x,y.\ 
        ((y,x) \in \text{decode}(net)) \wedge
         (x,n) \in \text{decode}(net)^* \longrightarrow f(y) < f(x)
\end{equation*}
Termination follows as $f$ is bounded below by $0$ (here
$\text{dom}(\textit{net},n)$ means that resolution terminates from $n$ i.e.
the arguments are in the \emph{domain} of the function
(\autoref{sec:isaterm}):
\begin{equation*}
\exists f.\ \textrm{wf\_rank}(f,n,net) \longleftrightarrow
    \textrm{dom}(net,n)
\end{equation*}
The duality between the graphical and operational views of an address space is
fundamental: the result of any well-defined resolution is the set of
names reachable via the decode relation (i.e. lie in the image of the
reflexive, transitive closure of the relation), that refer to actual resources
(i.e.\ are in the accept set, \autoref{sec:isares}):
\begin{equation*}
\textrm{dom}(net,n) \longrightarrow
\textrm{resolve}(net,n) = \textrm{accepted}(net) \cap
    (\text{decode}(net)^{-1})^*(\{n\})
\end{equation*}
This result lets us freely substitute one view of the system for another.

\subsection{Normalization and Refinement}

To use this model, we need efficient algorithms in the OS to manipulate it
e.g.\ calculate the set of visible resources from a processor. This
information is implicit in the graphical model, but not easily accessible.  We
might, for example, wish to produce a flattened representation that preserves
the view from each processor, while making such a query efficient.  One way to
achieve this is to split all nodes into nodes that only accept addresses
(\emph{resources}), and ones that only map to other nodes (\emph{address
spaces}).  Then merging, or \emph{flattening} the mapping nodes gives us the
desired result.

We would like to verify such algorithms.  We show here that we can define and
verify \emph{equivalence-preserving transformations} on the semantic model,
together with a notion of \emph{refinement}.  In the remainder of this
section, we demonstrate equivalence-preservation and refinement for the first
step: splitting, with reference to the accompanying sources. Further results
regarding flattening are also provided in the Isabelle sources for the
interested reader. 




Two nets are view-equivalent, written $(f,\textit{net}) \vieweq{S}
(g,\textit{net}^\prime)$ if all observers in $S$ have the same view (i.e.
the results of $\textrm{resolve}()$ are the same), modulo some renaming ($f$ and
$g$) of the accepting nodes.  Let $c$ be greater than the label of any extant
node.  The split net is then defined as (\autoref{sec:isasplit}):
\begin{align*}
\Acc(\Split(net),nd) &= \emptyset \\
\Acc(\Split(net),nd + c) &= \text{accept}(net,nd) \\
\Trans(\Split(net),nd,a) &=
    \{(nd+c,a) : a \in \Acc(net,nd)\}\ \mathop{\cup}\ \Trans(net,nd,a) \\
\Trans(\Split(net),nd+c,a) &= \emptyset
\end{align*}
This new net is view-equivalent to the original, with names that were accepted
at $nd$ now accepted at $nd + c$, and no node both accepting and translating
addresses (\autoref{sec:isasplit}):
\begin{equation}\label{eq:spliteq}
(nd \mapsto nd + c,\textit{net}) \vieweq{S}
(\emptyset,\textrm{split}(\textit{net}))
\end{equation}
Splitting on the concrete representation is a simple syntactic operation.
Each node is replaced as follows (\autoref{sec:isasplitC}):
\begin{equation*}
nd \Is \Accept A \Map M \Over O
\mapsto \left[ nd + c \Is \Accept A,\ 
nd \Is \Map M(nd \mapsto nd') \Over O \right]
\end{equation*}
Refinement is, as usual, expressed as the commutativity of the operations
(here $\Split()$ and $\SplitC()$) with the state relation (here $\Parse()$,
which constructs a net from its syntactic representation,
\autoref{sec:isasplitC}):
\begin{equation}\label{eq:splitrefine}
\Split(\Parse(s)) = \Parse(\SplitC(s))
\end{equation}
Combining \autoref{eq:spliteq} with \autoref{eq:splitrefine} we have the
desired result, that the concrete implementation preserves the equivalence of
the nets constructed by parsing (\autoref{sec:isasplitC}):
\begin{equation}\label{eq:fulleq}
(nd \mapsto nd + c,\Parse(s)) \vieweq{S} (\emptyset,\Parse(\SplitC(s)))
\end{equation}
Together with the equivalent result for flattening, we can verify that the
physical address spaces that we read directly from the transformed model are
exactly those that we would have found by (expensively) traversing the
original hardware-derived model for all addresses.

\section{Related Work}
\label{sec:relatedwork}
The difficulty of writing OS code in C for increasingly complex
hardware has led to several non-formal mechanisms for the OS to
discover, and configure, the hardware platform at runtime.

System firmware provides software with configuration information through 
ACPI~\cite{ACPI} tables and, more recently, UEFI~\cite{uefi}. Both 
provide somewhat abstracted information about the NUMA affinity 
of
memory controllers and other devices, information required to 
write fast memory intensive applications. Hardware connection
standards like PCI Express provide a measure of device enumeration
(including address discovery and interrupt routing requirements),
giving a hierarchy of the PCI bridges and connected devices.
Schupbach \etal~\cite{Schupbach:2011:DLA:1950365.1950382} applied a declarative 
approach to the configuration of the memory windows of PCI bridges.
Similarly, USB devices are discovered by hierarchical enumeration
of hubs.  Processors provide cache hierarchy data, for example using
the x86 \texttt{cpuid} instruction.

A more comprehensive description of a hardware platform is attempted
by Device Trees~\cite{devicetree}, which describes a binary file
format designed for bootloaders to find hardware at startup and is
now used extensively in the Linux kernel to handle non-discoverable
devices.  While a Device Tree captures some information about, for example, the
addresses of devices as seen from a single core (the root of the
tree), it has no well-defined semantics for interpreting the
data. Moreover, it is not well-suited for heterogenous systems where a 
single hierarchy is a poor match for hardware, and does not capture
caches, TLBs, or the view of the system from DMA-capable devices. 

We are not aware of any work on \textit{formalizing} the interrupt subsystem.
Commodity operating systems often only support a specific
mode of interrupt delivery. Linux for instance always distributes
all interrupts to all CPUs~\cite{bovet2005understanding}.
Similarly, FreeBSD refers to interrupts using the ACPI enumeration resulting in 
clashes when naming MSI sources. The system assumes that MSIs directly reach
the CPUs, which is no longer true since the introduction of the I/OMMU 
\cite{baldwin2008pci}.



%

Alglave \etal~\cite{Alglave:2010:FWM:2144310.2144342} applied the technique of
litmus testing to define allowable execution traces in the presence of memory
operation reordering (e.g.~write buffering or speculation), and test them
against real hardware.  Further work of the
authors~\cite{Alglave:2012:FHW:2385125.2385127, Flur:2016:MAA:2837614.2837615}
develops this into a semantic model of weak-memory systems (such as IBM Power
and ARM), in particular taking advantage of a close relationship with ARM to
ensure the faithfulness of their models to (the intended behavior of)
production silicon.  Our work is complementary: we provide a means to specify
and reason about the connectivity of address spaces in a system, on top of
which a rigorous model of weak memory would fully define the behavior of the
memory system.

Earlier work in hardware verification, such as that of
Velev~\cite{Velev:2001:AAM:646485.691776} or
Ganai~\cite{Ganai:2005:VEM:1048925.1049281}, considered the influence of
memory system microarchitecture on the verification of instruction-set
semantics.  The widespread adoption of out-of-order execution and weak memory
models has greatly increased the complexity of the problem, which as mentioned
is now being tackled.  Our focus is rather on the explosion of complexity in
the physical interconnection of devices, an area that is not yet well studied.

On the programming-language side, there has long been
interest~\cite{Zakkak:2016:DCM:2972206.2972212,
Sarkar:2012:SCP:2254064.2254102} in the interaction of language-specified memory
models (particularly that of Java, and now C11/C++11) and that provided by the
hardware.  Again, these models do not describe the low-level interconnection
of hardware, which has been relegated to an `OS problem', where it is solved
(badly) with tools such as device trees and ACPI.

\section{Roadmap}
\label{sec:roadmap}

While our model is a useful first step, we are extending it
considerably as we apply it to engineering Barrelfish.  We list our future
directions here.


We used the same model of interrupts and memory because both resemble
network forwarding, but in reality they are deeply connected in
hardware: message-signaled interrupts and inter-processor interrupts
are initiated by writing to memory addresses, and virtualization
hardware can translate interrupts into memory writes.  Unifying memory
and interrupts is a natural next step.


We also do not distinguish reads, writes, and other types of
transaction in our addressing model.  In practice, different request
types may traverse different paths in the interconnect and/or be accepted
by different components: examples include read-only memory protection,
and some PCIe DMA controllers which can only copy data in one
direction for certain address ranges. 


A further step is extending our static model to capture dynamic state,
starting with caches.  Caches are a challenge because they may or may
not respond to an address depending on their contents, they may
themselves emit addresses to other caches and resources, they may also
perform address translation, and they can be bypassed by non-cacheable
reads and writes.  Our current model expresses the set of resources
that may respond to a request, but not which resource actually 
responds.  Furthermore, the future behavior of the cache changes in
response to requests.
\camdone{Directory-based cache coherence protocols are essential to 
    scalability.
    Does your model support them, and how? Are there particular cases of
    address translation units? How is their dynamic nature reflected?}
Coherence protocols like MOESI also allow a line to be
fetched from another cache instead of main memory.  Not all cores
or devices might participate in the coherence protocol, but others
might be able to directly write to remote caches.  In both cases,
caches can be inconsistent with main memory -- something our model
cannot yet handle.


The dynamic state of a system also includes the configuration of
translation units.  Unfortunately, such units (lookup tables,
interrupt controllers, etc.) have varying constraints on their
configuration.  Some interrupt controllers perform 
fixed translation, some allow selective masking, some can remap blocks
of interrupt requests, and others can arbitrarily translate vectors. 
Memory translation can be achieved with page tables or lookup tables
of varying, fixed-size pages.  The addresses that can be
mapped between address spaces can be limited in range or domain. 

Our goal is, given a topology, these constraints, and a desired
end-to-end view, to \textit{synthesize} a correct configuration for
the translation units in the system.  Since many feasible
configurations can exist, we would also define an optimization goal
such as minimizing the required space for page tables or interrupt
mapping tables in limited-resource hardware like IOMMUs.


Synthesizing a valid configuration is insufficient for correct
operation, however.  The system requirements are dynamic: devices are
hotplugged, and brought up and down by power management functions.
Threads are migrated between cores by schedulers, etc.  This means
that the \emph{transition} between correct configurations, achieved by
reprogramming individual controllers or management units, must be
achieved without violating security or correctness guarantees, 
by creating a sequence of correct intermediate states and/or
determining which tasks must be paused while reconfiguration occurs.  
A simple example is reconfiguring a set of memory translation units in
sequence such that at no point does a process have unauthorized access
to an area of main memory, and that caches are consistent at all
points. 


Finally, while we capture the semantics of memory access and
interrupts, the performance of such operations also depends on the
platform configuration.  By annotating mapping functions with
performance characteristics, our model might be used to generate
hardware optimized data structures and messaging protocols as in
~\cite{Kaestle:2016:MAB:3026877.3026881}, or minimized interrupt
latency by choosing an appropriate delivery
mechanism~\cite{intelmsilatency}.

\section{Conclusion}
\label{sec:conclusion}
Contemporary hardware exposes a memory system and interrupt system structure
more complicated than usually assumed.  We have seen three examples of current
systems that violate common assumptions such as that a physical address uniquely
identifies a resource or that interrupts can be directed to all CPUs.  This 
implies that there is no --- or had never been a --- single physical address 
space and interrupts can not be directed to all cores anymore. 

We presented a formal model to express the interactions and topologies of 
address spaces and interrupts. Our model is capable of capturing the 
characteristics of a broad range of current systems and we show view 
equivalence preserving transformations that can be used to convert a complex 
system model into a flattened representation.

%
%

\bibliographystyle{eptcs}
\bibliography{compressed}

%
%
\newpage
\appendix
\section{Decoding Net Model}
\label{isabelle:decodingnet}
\camdone{Annexes,
    Hexadecimal constants are sometimes written in lowercase (which gives 
    questionable results in italics), and sometimes in uppercase, 
    The latter is better. Both should be unified and, if possible, a fixed-width
    font should be used.}

This appendix presents the formal development (in Isabelle/HOL) of the
decoding net model outlined in \autoref{sec:model}, together with proofs of
some key results.  The text here is generated directly from the Isabelle
sources, with some sections excluded for space and not being (in our
subjective opinion) particularly interesting.  The proofs may therefore
occasionally refer to definitions and lemmas which are not stated here---in
any such case, the full, machine-checked proof is available in the theory
files associated with this paper.

\begin{isabellebody}%
\setisabellecontext{Model}%
\isadelimtheory
\endisadelimtheory
\isatagtheory
\endisatagtheory
{\isafoldtheory}%
\isadelimtheory
\endisadelimtheory
\label{sec:isamodel}
\begin{isamarkuptext}%
First, we nail down some types.  For ease in getting started, we're using natural numbers for
  addresses.  It should be possible to use the same definitions to handle finite-length words
  without much modification.%
\end{isamarkuptext}\isamarkuptrue%
\isacommand{type{\isacharunderscore}synonym}\isamarkupfalse%
\ nodeid\ {\isacharequal}\ nat\isanewline
\isacommand{type{\isacharunderscore}synonym}\isamarkupfalse%
\ addr\ {\isacharequal}\ nat%
\begin{isamarkuptext}%
A name is a qualified address: which is defined with respect to some context, in this
  case the node at which decoding begins.%
\end{isamarkuptext}\isamarkuptrue%
\isacommand{type{\isacharunderscore}synonym}\isamarkupfalse%
\ name\ {\isacharequal}\ {\isachardoublequoteopen}nodeid\ {\isasymtimes}\ addr{\isachardoublequoteclose}%
\begin{isamarkuptext}%
A node can accept an input address, in which case resolution terminates here, or it can
  translate it into an input address for another node, or both.  We allow the sets of accepted
  and translated addresses to overlap to model nondeterministic behaviour e.g. a cache, which has a
  well-defined translation for every input address, but could potentially respond to any request
  with a locally-cached value.  In general, we're interested in the set of (node, address) pairs
  at which a given input address might be accepted.%
\end{isamarkuptext}\isamarkuptrue%
\isacommand{record}\isamarkupfalse%
\ node\ {\isacharequal}\isanewline
\ \ accept\ {\isacharcolon}{\isacharcolon}\ {\isachardoublequoteopen}addr\ set{\isachardoublequoteclose}\isanewline
\ \ translate\ {\isacharcolon}{\isacharcolon}\ {\isachardoublequoteopen}addr\ {\isasymRightarrow}\ name\ set{\isachardoublequoteclose}\isanewline
\isanewline
\ \ %
\isadelimproof
\endisadelimproof
\isatagproof
\endisatagproof
{\isafoldproof}%
\isadelimproof
\endisadelimproof
\isamarkupsubsection{Address Decoding Nets.%
}
\isamarkuptrue%
\label{sec:nets}
\begin{isamarkuptext}%
A decode net is an assignment of nodes to identifiers.%
\end{isamarkuptext}\isamarkuptrue%
\isacommand{type{\isacharunderscore}synonym}\isamarkupfalse%
\ net\ {\isacharequal}\ {\isachardoublequoteopen}nodeid\ {\isasymRightarrow}\ node{\isachardoublequoteclose}%
\begin{isamarkuptext}%
One step of address decoding, mapping an input name (node, address) to an output name (or
  nothing).%
\end{isamarkuptext}\isamarkuptrue%
\isacommand{definition}\isamarkupfalse%
\ decode{\isacharunderscore}step\ {\isacharcolon}{\isacharcolon}\ {\isachardoublequoteopen}net\ {\isasymRightarrow}\ name\ {\isasymRightarrow}\ name\ set{\isachardoublequoteclose}\isanewline
\ \ \isakeyword{where}\isanewline
\ \ \ \ {\isachardoublequoteopen}decode{\isacharunderscore}step\ net\ name\ {\isacharequal}\ translate\ {\isacharparenleft}net\ {\isacharparenleft}fst\ name{\isacharparenright}{\isacharparenright}\ {\isacharparenleft}snd\ name{\isacharparenright}{\isachardoublequoteclose}%
\begin{isamarkuptext}%
The decode relation is, in general, a directed graph.  If it's actually a DAG, then all
  addresses can be decoded in a well-defined manner.%
\end{isamarkuptext}\isamarkuptrue%
\isacommand{definition}\isamarkupfalse%
\ decodes{\isacharunderscore}to\ {\isacharcolon}{\isacharcolon}\ {\isachardoublequoteopen}net\ {\isasymRightarrow}\ name\ rel{\isachardoublequoteclose}\isanewline
\ \ \isakeyword{where}\isanewline
\ \ \ \ {\isachardoublequoteopen}decodes{\isacharunderscore}to\ net\ {\isacharequal}\ {\isacharbraceleft}\ {\isacharparenleft}n{\isacharprime}{\isacharcomma}\ n{\isacharparenright}{\isachardot}\ n{\isacharprime}\ {\isasymin}\ decode{\isacharunderscore}step\ net\ n\ {\isacharbraceright}{\isachardoublequoteclose}%
\begin{isamarkuptext}%
The set of names that can be accepted anywhere in this decoding net i.e. the union of
  the accept sets of all nodes.%
\end{isamarkuptext}\isamarkuptrue%
\isacommand{definition}\isamarkupfalse%
\ accepted{\isacharunderscore}names\ {\isacharcolon}{\isacharcolon}\ {\isachardoublequoteopen}net\ {\isasymRightarrow}\ name\ set{\isachardoublequoteclose}\isanewline
\ \ \isakeyword{where}\ {\isachardoublequoteopen}accepted{\isacharunderscore}names\ net\ {\isacharequal}\ {\isacharbraceleft}{\isacharparenleft}nd{\isacharcomma}a{\isacharparenright}\ {\isacharbar}nd\ a{\isachardot}\ a\ {\isasymin}\ accept\ {\isacharparenleft}net\ nd{\isacharparenright}{\isacharbraceright}{\isachardoublequoteclose}\isanewline
\isadelimtheory
\endisadelimtheory
\isatagtheory
\endisatagtheory
{\isafoldtheory}%
\isadelimtheory
\endisadelimtheory
\end{isabellebody}%

%
\begin{isabellebody}%
\setisabellecontext{Resolution}%
\isadelimtheory
\endisadelimtheory
\isatagtheory
\endisatagtheory
{\isafoldtheory}%
\isadelimtheory
\endisadelimtheory
\isamarkupsubsection{Resolution%
}
\isamarkuptrue%
\label{sec:isares}
\begin{isamarkuptext}%
To resolve an input name, start with the name itself if the net accepts it, and recurse on
  all names reachable via the \isa{decodes{\isacharunderscore}to} relation%
\end{isamarkuptext}\isamarkuptrue%
\isacommand{function}\isamarkupfalse%
\ {\isacharparenleft}domintros{\isacharparenright}\ resolve\ {\isacharcolon}{\isacharcolon}\ {\isachardoublequoteopen}net\ {\isasymRightarrow}\ name\ {\isasymRightarrow}\ name\ set{\isachardoublequoteclose}\isanewline
\ \ \isakeyword{where}\ {\isachardoublequoteopen}resolve\ net\ n\ {\isacharequal}\isanewline
\ \ \ {\isacharparenleft}{\isacharbraceleft}n{\isacharbraceright}\ {\isasyminter}\ accepted{\isacharunderscore}names\ net{\isacharparenright}\ {\isasymunion}\isanewline
\ \ \ {\isacharparenleft}{\isasymUnion}n{\isacharprime}{\isachardot}\ if\ {\isacharparenleft}n{\isacharprime}{\isacharcomma}n{\isacharparenright}\ {\isasymin}\ decodes{\isacharunderscore}to\ net\ then\ resolve\ net\ n{\isacharprime}\ else\ {\isacharbraceleft}{\isacharbraceright}{\isacharparenright}{\isachardoublequoteclose}\isanewline
\isadelimproof
\ \ %
\endisadelimproof
\isatagproof
\isacommand{by}\isamarkupfalse%
{\isacharparenleft}pat{\isacharunderscore}completeness{\isacharcomma}\ auto{\isacharparenright}\isanewline
\ \ \ \ %
\endisatagproof
{\isafoldproof}%
\isadelimproof
\endisadelimproof
\isadelimproof
\endisadelimproof
\isatagproof
\endisatagproof
{\isafoldproof}%
\isadelimproof
\endisadelimproof
\isadelimproof
\endisadelimproof
\isatagproof
\endisatagproof
{\isafoldproof}%
\isadelimproof
\endisadelimproof
\isadelimproof
\endisadelimproof
\isatagproof
\endisatagproof
{\isafoldproof}%
\isadelimproof
\endisadelimproof
\begin{isamarkuptext}%
The defining relation for \isa{resolve} is simply \isa{decodes{\isacharunderscore}to}:%
\end{isamarkuptext}\isamarkuptrue%
\isacommand{lemma}\isamarkupfalse%
\ resolve{\isacharunderscore}rel{\isacharunderscore}decodes{\isacharunderscore}to{\isacharcolon}\isanewline
\ \ {\isachardoublequoteopen}resolve{\isacharunderscore}rel\ x\ y\ {\isasymlongleftrightarrow}\ {\isacharparenleft}fst\ x\ {\isacharequal}\ fst\ y{\isacharparenright}\ {\isasymand}\ {\isacharparenleft}snd\ x{\isacharcomma}\ snd\ y{\isacharparenright}\ {\isasymin}\ decodes{\isacharunderscore}to\ {\isacharparenleft}fst\ x{\isacharparenright}{\isachardoublequoteclose}\isanewline
\isadelimproof
\ \ %
\endisadelimproof
\isatagproof
\isacommand{by}\isamarkupfalse%
{\isacharparenleft}cases\ x{\isacharcomma}\ cases\ y{\isacharcomma}\ auto\ elim{\isacharcolon}resolve{\isacharunderscore}rel{\isachardot}cases\ intro{\isacharcolon}resolve{\isacharunderscore}rel{\isachardot}intros{\isacharparenright}\isanewline
\ \ \ \ %
\endisatagproof
{\isafoldproof}%
\isadelimproof
\endisadelimproof
\isadelimproof
\endisadelimproof
\isatagproof
\endisatagproof
{\isafoldproof}%
\isadelimproof
\endisadelimproof
\isadelimproof
\endisadelimproof
\isatagproof
\endisatagproof
{\isafoldproof}%
\isadelimproof
\endisadelimproof
\isadelimproof
\endisadelimproof
\isatagproof
\endisatagproof
{\isafoldproof}%
\isadelimproof
\endisadelimproof
\isadelimproof
\endisadelimproof
\isatagproof
\endisatagproof
{\isafoldproof}%
\isadelimproof
\endisadelimproof
\isadelimproof
\endisadelimproof
\isatagproof
\endisatagproof
{\isafoldproof}%
\isadelimproof
\endisadelimproof
\isadelimproof
\endisadelimproof
\isatagproof
\endisatagproof
{\isafoldproof}%
\isadelimproof
\endisadelimproof
\begin{isamarkuptext}%
We can express resolution in an equivalent non-recursive fashion, as the image of the
  closure of the decoding relation:%
\end{isamarkuptext}\isamarkuptrue%
\isacommand{lemma}\isamarkupfalse%
\ resolve{\isacharunderscore}eval{\isacharcolon}\isanewline
\ \ \isakeyword{assumes}\ dom{\isacharcolon}\ {\isachardoublequoteopen}resolve{\isacharunderscore}dom\ {\isacharparenleft}net{\isacharcomma}\ n{\isacharparenright}{\isachardoublequoteclose}\isanewline
\ \ \isakeyword{shows}\ {\isachardoublequoteopen}resolve\ net\ n\ {\isacharequal}\ accepted{\isacharunderscore}names\ net\ {\isasyminter}\ {\isacharparenleft}{\isacharparenleft}decodes{\isacharunderscore}to\ net{\isacharparenright}{\isasyminverse}{\isacharparenright}\isactrlsup {\isacharasterisk}\ {\isacharbackquote}{\isacharbackquote}\ {\isacharbraceleft}n{\isacharbraceright}{\isachardoublequoteclose}\isanewline
\ \ \ \ %
\isadelimproof
\endisadelimproof
\isatagproof
\endisatagproof
{\isafoldproof}%
\isadelimproof
\endisadelimproof
\isamarkupsubsection{Well-Formed Decoding Nets%
}
\isamarkuptrue%
\label{sec:isawfnet}
\begin{isamarkuptext}%
The most general condition for the decoding of a given name to be well-defined is that the
  decoding process terminates i.e. that all paths of decoding steps (elements of the decodes-to
  relation) are finite (we eventually reach a node that either accepts its input address, or
  faults).

  A well-formed rank function \isa{f} assigns a natural number to every name, such that if some
  name \isa{n} decodes to \isa{n{\isacharprime}}, \isa{f\ n{\isacharprime}\ {\isacharless}\ f\ n}.  From this, it is trivial to show
  that decoding terminates.  Note that it is only necessary for the ranking to be well-formed for
  the name that we're resolving: it may not be possible to assign a consistent ranking to all names,
  that is well-formed for all starting points, although in well-designed systems it probably should
  be.%
\end{isamarkuptext}\isamarkuptrue%
\isacommand{definition}\isamarkupfalse%
\ wf{\isacharunderscore}rank\ {\isacharcolon}{\isacharcolon}\ {\isachardoublequoteopen}{\isacharparenleft}name\ {\isasymRightarrow}\ nat{\isacharparenright}\ {\isasymRightarrow}\ name\ {\isasymRightarrow}\ net\ {\isasymRightarrow}\ bool{\isachardoublequoteclose}\isanewline
\ \ \isakeyword{where}\isanewline
\ \ \ \ {\isachardoublequoteopen}wf{\isacharunderscore}rank\ f\ n\ net\ {\isasymlongleftrightarrow}\isanewline
\ \ \ \ \ \ {\isacharparenleft}{\isasymforall}x\ y{\isachardot}\ {\isacharparenleft}x{\isacharcomma}n{\isacharparenright}\ {\isasymin}\ rtrancl\ {\isacharparenleft}decodes{\isacharunderscore}to\ net{\isacharparenright}\ {\isasymand}\ {\isacharparenleft}y{\isacharcomma}x{\isacharparenright}\ {\isasymin}\ decodes{\isacharunderscore}to\ net\ {\isasymlongrightarrow}\ f\ y\ {\isacharless}\ f\ x{\isacharparenright}{\isachardoublequoteclose}\isanewline
\ \ \ \ %
\isadelimproof
\endisadelimproof
\isatagproof
\endisatagproof
{\isafoldproof}%
\isadelimproof
\endisadelimproof
\isadelimproof
\endisadelimproof
\isatagproof
\endisatagproof
{\isafoldproof}%
\isadelimproof
\endisadelimproof
\isadelimproof
\endisadelimproof
\isatagproof
\endisatagproof
{\isafoldproof}%
\isadelimproof
\endisadelimproof
\isadelimproof
\endisadelimproof
\isatagproof
\endisatagproof
{\isafoldproof}%
\isadelimproof
\endisadelimproof
\isadelimproof
\endisadelimproof
\isatagproof
\endisatagproof
{\isafoldproof}%
\isadelimproof
\endisadelimproof
\begin{isamarkuptext}%
We use our well-formedness predicate to insist that all node both accept and translate
  a finite set of addresses.  While this isn't strictly necessary for a lot of the theory, it's
  essential for termination.%
\end{isamarkuptext}\isamarkuptrue%
\isacommand{definition}\isamarkupfalse%
\ wf{\isacharunderscore}net\ {\isacharcolon}{\isacharcolon}\ {\isachardoublequoteopen}net\ {\isasymRightarrow}\ bool{\isachardoublequoteclose}\isanewline
\ \ \isakeyword{where}\ {\isachardoublequoteopen}wf{\isacharunderscore}net\ net\ {\isasymlongleftrightarrow}\isanewline
\ \ \ \ {\isacharparenleft}{\isasymforall}nd{\isachardot}\ finite\ {\isacharparenleft}accept\ {\isacharparenleft}net\ nd{\isacharparenright}{\isacharparenright}{\isacharparenright}\ {\isasymand}\isanewline
\ \ \ \ {\isacharparenleft}{\isasymforall}n{\isachardot}\ resolve{\isacharunderscore}dom\ {\isacharparenleft}net{\isacharcomma}n{\isacharparenright}\ {\isasymlongrightarrow}\ finite\ {\isacharparenleft}{\isacharparenleft}decodes{\isacharunderscore}to\ net{\isacharparenright}{\isasyminverse}\ {\isacharbackquote}{\isacharbackquote}\ {\isacharbraceleft}n{\isacharbraceright}{\isacharparenright}{\isacharparenright}{\isachardoublequoteclose}\isanewline
\ \ \ \ %
\isadelimproof
\endisadelimproof
\isatagproof
\endisatagproof
{\isafoldproof}%
\isadelimproof
\endisadelimproof
\isadelimproof
\endisadelimproof
\isatagproof
\endisatagproof
{\isafoldproof}%
\isadelimproof
\endisadelimproof
\isadelimproof
\endisadelimproof
\isatagproof
\endisatagproof
{\isafoldproof}%
\isadelimproof
\endisadelimproof
\isamarkupsubsection{Termination%
}
\isamarkuptrue%
\label{sec:isaterm}
\begin{isamarkuptext}%
If we can supply a ranking function that is well-formed for all names reachable from the
  name we wish to decode, then the decoding function is well-defined here (this name lies in its
  \emph{domain}).%
\end{isamarkuptext}\isamarkuptrue%
\isacommand{lemma}\isamarkupfalse%
\ wf{\isacharunderscore}resolve{\isacharunderscore}dom{\isacharcolon}\isanewline
\ \ \isakeyword{fixes}\ f\ {\isacharcolon}{\isacharcolon}\ {\isachardoublequoteopen}name\ {\isasymRightarrow}\ nat{\isachardoublequoteclose}\ \isakeyword{and}\ n\ {\isacharcolon}{\isacharcolon}\ name\ \isakeyword{and}\ net\ {\isacharcolon}{\isacharcolon}\ net\isanewline
\ \ \isakeyword{assumes}\ wf{\isacharunderscore}at{\isacharcolon}\ {\isachardoublequoteopen}wf{\isacharunderscore}rank\ f\ n\ net{\isachardoublequoteclose}\isanewline
\ \ \isakeyword{shows}\ {\isachardoublequoteopen}resolve{\isacharunderscore}dom\ {\isacharparenleft}net{\isacharcomma}n{\isacharparenright}{\isachardoublequoteclose}\isanewline
\isadelimproof
\endisadelimproof
\isatagproof
\isacommand{proof}\isamarkupfalse%
\ {\isacharminus}\isanewline
\ \ \isacommand{{\isacharbraceleft}}\isamarkupfalse%
\begin{isamarkuptext}%
We argue by (strong) induction on the rank of the name, but we need to carry the
      assumption of reachability into the induction hypothesis (as otherwise we can't appeal to a
      well-formed ranking.  We then trivially discard this assumption as \isa{n} is reachable from
      itself, by definition.%
\end{isamarkuptext}\isamarkuptrue%
\ \ \ \ \isacommand{fix}\isamarkupfalse%
\ a\isanewline
\ \ \ \ \isacommand{assume}\isamarkupfalse%
\ {\isachardoublequoteopen}{\isacharparenleft}a{\isacharcomma}n{\isacharparenright}\ {\isasymin}\ {\isacharparenleft}decodes{\isacharunderscore}to\ net{\isacharparenright}\isactrlsup {\isacharasterisk}{\isachardoublequoteclose}\isanewline
\ \ \ \ \isacommand{hence}\isamarkupfalse%
\ {\isachardoublequoteopen}resolve{\isacharunderscore}dom\ {\isacharparenleft}net{\isacharcomma}a{\isacharparenright}{\isachardoublequoteclose}\isanewline
\ \ \ \ \isacommand{proof}\isamarkupfalse%
{\isacharparenleft}induct\ {\isachardoublequoteopen}f\ a{\isachardoublequoteclose}\ arbitrary{\isacharcolon}a\ rule{\isacharcolon}nat{\isacharunderscore}less{\isacharunderscore}induct{\isacharparenright}\isanewline
\ \ \ \ \ \ \isacommand{fix}\isamarkupfalse%
\ b%
\begin{isamarkuptext}%
Assume the current node is reachable, and all reachable nodes of lesser rank lie in
        the domain of \isa{resolve}.%
\end{isamarkuptext}\isamarkuptrue%
\ \ \ \ \ \ \isacommand{assume}\isamarkupfalse%
\ reachable{\isacharcolon}\ {\isachardoublequoteopen}{\isacharparenleft}b{\isacharcomma}n{\isacharparenright}\ {\isasymin}\ {\isacharparenleft}decodes{\isacharunderscore}to\ net{\isacharparenright}\isactrlsup {\isacharasterisk}{\isachardoublequoteclose}\isanewline
\ \ \ \ \ \ \ \ \ \isakeyword{and}\ IH{\isacharcolon}\ {\isachardoublequoteopen}{\isasymforall}m{\isacharless}f\ b{\isachardot}\ {\isasymforall}x{\isachardot}\ m\ {\isacharequal}\ f\ x\ {\isasymlongrightarrow}\ {\isacharparenleft}x{\isacharcomma}n{\isacharparenright}\ {\isasymin}\ {\isacharparenleft}decodes{\isacharunderscore}to\ net{\isacharparenright}\isactrlsup {\isacharasterisk}\ {\isasymlongrightarrow}\ resolve{\isacharunderscore}dom\ {\isacharparenleft}net{\isacharcomma}\ x{\isacharparenright}{\isachardoublequoteclose}%
\begin{isamarkuptext}%
We show that the arguments of any recursive call to \isa{resolve} must lie in
        the domain, as new node is both reachable, and has strictly lesser rank, thanks to well-
        formedness.%
\end{isamarkuptext}\isamarkuptrue%
\ \ \ \ \ \ \isacommand{show}\isamarkupfalse%
\ {\isachardoublequoteopen}resolve{\isacharunderscore}dom\ {\isacharparenleft}net{\isacharcomma}\ b{\isacharparenright}{\isachardoublequoteclose}\isanewline
\ \ \ \ \ \ \isacommand{proof}\isamarkupfalse%
{\isacharparenleft}rule\ resolve{\isacharunderscore}domI{\isacharparenright}\isanewline
\ \ \ \ \ \ \ \ \isacommand{fix}\isamarkupfalse%
\ a%
\begin{isamarkuptext}%
Assume that there \emph{is} a translation/decoding step.  We don't need to show
          anything for the terminating case, as there's no recursive call.%
\end{isamarkuptext}\isamarkuptrue%
\ \ \ \ \ \ \ \ \isacommand{assume}\isamarkupfalse%
\ step{\isacharcolon}\ {\isachardoublequoteopen}{\isacharparenleft}a{\isacharcomma}b{\isacharparenright}\ {\isasymin}\ decodes{\isacharunderscore}to\ net{\isachardoublequoteclose}%
\begin{isamarkuptext}%
The two names lie in the decoding relation, and the new name is also reachable
          from \isa{n}.%
\end{isamarkuptext}\isamarkuptrue%
\ \ \ \ \ \ \ \ \isacommand{from}\isamarkupfalse%
\ step\ reachable\ \isacommand{have}\isamarkupfalse%
\ reachable{\isacharunderscore}yz{\isacharcolon}\ {\isachardoublequoteopen}{\isacharparenleft}a{\isacharcomma}n{\isacharparenright}\ {\isasymin}\ {\isacharparenleft}decodes{\isacharunderscore}to\ net{\isacharparenright}\isactrlsup {\isacharasterisk}{\isachardoublequoteclose}\ \isacommand{by}\isamarkupfalse%
{\isacharparenleft}simp{\isacharparenright}%
\begin{isamarkuptext}%
From the (assumed) reachability of \isa{b}, we can appeal to well-formedness to
          show that the rank decreases.%
\end{isamarkuptext}\isamarkuptrue%
\ \ \ \ \ \ \ \ \isacommand{from}\isamarkupfalse%
\ wf{\isacharunderscore}at\ reachable\ step\ \isacommand{have}\isamarkupfalse%
\ {\isachardoublequoteopen}f\ a\ {\isacharless}\ f\ b{\isachardoublequoteclose}\isanewline
\ \ \ \ \ \ \ \ \ \ \isacommand{unfolding}\isamarkupfalse%
\ wf{\isacharunderscore}rank{\isacharunderscore}def\ \isacommand{by}\isamarkupfalse%
{\isacharparenleft}blast{\isacharparenright}%
\begin{isamarkuptext}%
Thus with the reachability of the new name, we have the result by appealing to the
          induction hypothesis.%
\end{isamarkuptext}\isamarkuptrue%
\ \ \ \ \ \ \ \ \isacommand{with}\isamarkupfalse%
\ reachable{\isacharunderscore}yz\ IH\ \isacommand{show}\isamarkupfalse%
\ {\isachardoublequoteopen}resolve{\isacharunderscore}dom\ {\isacharparenleft}net{\isacharcomma}\ a{\isacharparenright}{\isachardoublequoteclose}\ \isacommand{by}\isamarkupfalse%
{\isacharparenleft}blast{\isacharparenright}\isanewline
\ \ \ \ \ \ \isacommand{qed}\isamarkupfalse%
\isanewline
\ \ \ \ \isacommand{qed}\isamarkupfalse%
\isanewline
\ \ \isacommand{{\isacharbraceright}}\isamarkupfalse%
\begin{isamarkuptext}%
Finally, we discharge the reachability assumption.%
\end{isamarkuptext}\isamarkuptrue%
\ \ \isacommand{thus}\isamarkupfalse%
\ {\isacharquery}thesis\ \isacommand{by}\isamarkupfalse%
{\isacharparenleft}auto{\isacharparenright}\isanewline
\isacommand{qed}\isamarkupfalse%
\endisatagproof
{\isafoldproof}%
\isadelimproof
\endisadelimproof
\begin{isamarkuptext}%
This is the converse of the previous lemma, for decoding nets that with finite branching:
  If a single decoding step maps a name to a finite number of new names, then there must exist a
  well-formed ranking for each resolvable name.%
\end{isamarkuptext}\isamarkuptrue%
\isacommand{lemma}\isamarkupfalse%
\ mkrank{\isacharcolon}\isanewline
\ \ \isakeyword{fixes}\ \ n\ {\isacharcolon}{\isacharcolon}\ name\ \isakeyword{and}\ net\ {\isacharcolon}{\isacharcolon}\ net\isanewline
\ \ \isakeyword{assumes}\ branching{\isacharcolon}\ {\isachardoublequoteopen}{\isasymAnd}n{\isachardot}\ resolve{\isacharunderscore}dom\ {\isacharparenleft}net{\isacharcomma}n{\isacharparenright}\ {\isasymLongrightarrow}\ finite\ {\isacharparenleft}{\isacharparenleft}decodes{\isacharunderscore}to\ net{\isacharparenright}{\isasyminverse}\ {\isacharbackquote}{\isacharbackquote}\ {\isacharbraceleft}n{\isacharbraceright}{\isacharparenright}{\isachardoublequoteclose}\isanewline
\ \ \isakeyword{shows}\ {\isachardoublequoteopen}resolve{\isacharunderscore}dom\ {\isacharparenleft}net{\isacharcomma}n{\isacharparenright}\ {\isasymLongrightarrow}\ {\isasymexists}f{\isachardot}\ wf{\isacharunderscore}rank\ f\ n\ net{\isachardoublequoteclose}\isanewline
\isadelimproof
\endisadelimproof
\isatagproof
\isacommand{proof}\isamarkupfalse%
{\isacharparenleft}induction\ {\isachardoublequoteopen}{\isacharparenleft}net{\isacharcomma}n{\isacharparenright}{\isachardoublequoteclose}\ arbitrary{\isacharcolon}n\ rule{\isacharcolon}accp{\isachardot}induct{\isacharparenright}\isanewline
\ \ \isacommand{fix}\isamarkupfalse%
\ n%
\begin{isamarkuptext}%
Assume that there exists a well-formed ranking for every direct descendent.%
\end{isamarkuptext}\isamarkuptrue%
\ \ \isacommand{assume}\isamarkupfalse%
\ IH{\isacharcolon}\ {\isachardoublequoteopen}{\isasymAnd}n{\isacharprime}{\isachardot}\ resolve{\isacharunderscore}rel\ {\isacharparenleft}net{\isacharcomma}\ n{\isacharprime}{\isacharparenright}\ {\isacharparenleft}net{\isacharcomma}\ n{\isacharparenright}\ {\isasymLongrightarrow}\ {\isasymexists}f{\isachardot}\ wf{\isacharunderscore}rank\ f\ n{\isacharprime}\ net{\isachardoublequoteclose}\isanewline
\ \ \ \ \ \isakeyword{and}\ dom{\isacharcolon}\ {\isachardoublequoteopen}{\isasymAnd}y{\isachardot}\ resolve{\isacharunderscore}rel\ y\ {\isacharparenleft}net{\isacharcomma}\ n{\isacharparenright}\ {\isasymLongrightarrow}\ resolve{\isacharunderscore}dom\ y{\isachardoublequoteclose}\isanewline
\ \ \ \ \ \isanewline
\ \ \isacommand{have}\isamarkupfalse%
\ rd{\isacharcolon}\ {\isachardoublequoteopen}resolve{\isacharunderscore}dom\ {\isacharparenleft}net{\isacharcomma}n{\isacharparenright}{\isachardoublequoteclose}\ \isacommand{by}\isamarkupfalse%
{\isacharparenleft}blast\ intro{\isacharcolon}accp{\isachardot}intros\ dom{\isacharparenright}%
\begin{isamarkuptext}%
By appealing to the axiom of choice (although as we're finite we could do without),
    construct \isa{g} which, for every ancestor \isa{n{\isacharprime}} of \isa{n}, gives a ranking function
    that is well-formed at \isa{n{\isacharprime}}.%
\end{isamarkuptext}\isamarkuptrue%
\ \ \isacommand{from}\isamarkupfalse%
\ IH\isanewline
\ \ \isacommand{have}\isamarkupfalse%
\ {\isachardoublequoteopen}{\isasymforall}n{\isacharprime}{\isasymin}\ {\isacharparenleft}decodes{\isacharunderscore}to\ net{\isacharparenright}{\isasyminverse}\ {\isacharbackquote}{\isacharbackquote}\ {\isacharbraceleft}n{\isacharbraceright}{\isachardot}\ {\isasymexists}f{\isachardot}\ wf{\isacharunderscore}rank\ f\ n{\isacharprime}\ net{\isachardoublequoteclose}\isanewline
\ \ \ \ \isacommand{by}\isamarkupfalse%
{\isacharparenleft}blast\ intro{\isacharcolon}resolve{\isacharunderscore}rel{\isachardot}intros{\isacharparenright}\isanewline
\ \ \isacommand{hence}\isamarkupfalse%
\ {\isachardoublequoteopen}{\isasymexists}g{\isachardot}\ {\isasymforall}n{\isacharprime}{\isasymin}\ {\isacharparenleft}decodes{\isacharunderscore}to\ net{\isacharparenright}{\isasyminverse}\ {\isacharbackquote}{\isacharbackquote}\ {\isacharbraceleft}n{\isacharbraceright}{\isachardot}\ wf{\isacharunderscore}rank\ {\isacharparenleft}g\ n{\isacharprime}{\isacharparenright}\ n{\isacharprime}\ net{\isachardoublequoteclose}\isanewline
\ \ \ \ \isacommand{by}\isamarkupfalse%
{\isacharparenleft}rule\ bchoice{\isacharparenright}\isanewline
\ \ \isacommand{then}\isamarkupfalse%
\ \isacommand{obtain}\isamarkupfalse%
\ g\ \isakeyword{where}\ wf{\isacharunderscore}g{\isacharcolon}\ {\isachardoublequoteopen}{\isasymforall}n{\isacharprime}\ {\isasymin}\ {\isacharparenleft}decodes{\isacharunderscore}to\ net{\isacharparenright}{\isasyminverse}\ {\isacharbackquote}{\isacharbackquote}\ {\isacharbraceleft}n{\isacharbraceright}{\isachardot}\ wf{\isacharunderscore}rank\ {\isacharparenleft}g\ n{\isacharprime}{\isacharparenright}\ n{\isacharprime}\ net{\isachardoublequoteclose}\isanewline
\ \ \ \ \isacommand{by}\isamarkupfalse%
{\isacharparenleft}blast\ dest{\isacharcolon}bchoice{\isacharparenright}%
\begin{isamarkuptext}%
For any node \isa{n{\isacharprime}}, this is the set of its ancestors that are direct descendents
    of \isa{n} i.e. the set of nodes that any path from \isa{n} to \isa{n{\isacharprime}} \emph{must}
    pass through.  This set is finite.%
\end{isamarkuptext}\isamarkuptrue%
\ \ \isacommand{let}\isamarkupfalse%
\ {\isachardoublequoteopen}{\isacharquery}ancs\ n{\isacharprime}{\isachardoublequoteclose}\ {\isacharequal}\ {\isachardoublequoteopen}{\isacharbraceleft}n{\isacharprime}{\isacharprime}{\isachardot}\ {\isacharparenleft}n{\isacharprime}{\isacharprime}{\isacharcomma}n{\isacharparenright}\ {\isasymin}\ decodes{\isacharunderscore}to\ net\ {\isasymand}\ {\isacharparenleft}n{\isacharprime}{\isacharcomma}n{\isacharprime}{\isacharprime}{\isacharparenright}\ {\isasymin}\ {\isacharparenleft}decodes{\isacharunderscore}to\ net{\isacharparenright}\isactrlsup {\isacharasterisk}{\isacharbraceright}{\isachardoublequoteclose}\isanewline
\ \ \isacommand{have}\isamarkupfalse%
\ {\isachardoublequoteopen}{\isasymAnd}x{\isachardot}\ {\isacharquery}ancs\ x\ {\isasymsubseteq}\ {\isacharparenleft}decodes{\isacharunderscore}to\ net{\isacharparenright}{\isasyminverse}\ {\isacharbackquote}{\isacharbackquote}\ {\isacharbraceleft}n{\isacharbraceright}{\isachardoublequoteclose}\ \isacommand{by}\isamarkupfalse%
{\isacharparenleft}auto{\isacharparenright}\isanewline
\ \ \isacommand{with}\isamarkupfalse%
\ branching\ rd\ \isacommand{have}\isamarkupfalse%
\ finite{\isacharunderscore}ancs{\isacharcolon}\ {\isachardoublequoteopen}{\isasymAnd}x{\isachardot}\ finite\ {\isacharparenleft}{\isacharquery}ancs\ x{\isacharparenright}{\isachardoublequoteclose}\ \isacommand{by}\isamarkupfalse%
{\isacharparenleft}blast\ dest{\isacharcolon}finite{\isacharunderscore}subset{\isacharparenright}%
\begin{isamarkuptext}%
From \isa{g}, construct \isa{g{\isacharprime}}, by taking, for each \isa{n{\isacharprime}}, the least rank
    assigned it by any of the well-formed rankings associated with its ancestors.  This new
    ranking is still well-formed for all of the direct descendents of \isa{n}.%
\end{isamarkuptext}\isamarkuptrue%
\ \ \isacommand{let}\isamarkupfalse%
\ {\isacharquery}g{\isacharprime}\ {\isacharequal}\ {\isachardoublequoteopen}{\isasymlambda}n{\isacharprime}{\isachardot}\ Min\ {\isacharparenleft}{\isacharparenleft}{\isasymlambda}n{\isacharprime}{\isacharprime}{\isachardot}\ g\ n{\isacharprime}{\isacharprime}\ n{\isacharprime}{\isacharparenright}\ {\isacharbackquote}\ {\isacharquery}ancs\ n{\isacharprime}{\isacharparenright}{\isachardoublequoteclose}\isanewline
\ \ \isacommand{have}\isamarkupfalse%
\ wf{\isacharunderscore}g{\isacharprime}{\isacharcolon}\ {\isachardoublequoteopen}{\isasymforall}n{\isacharprime}\ {\isasymin}\ {\isacharparenleft}decodes{\isacharunderscore}to\ net{\isacharparenright}{\isasyminverse}\ {\isacharbackquote}{\isacharbackquote}\ {\isacharbraceleft}n{\isacharbraceright}{\isachardot}\ wf{\isacharunderscore}rank\ {\isacharquery}g{\isacharprime}\ n{\isacharprime}\ net{\isachardoublequoteclose}\isanewline
\ \ \isacommand{proof}\isamarkupfalse%
{\isacharparenleft}intro\ ballI\ wf{\isacharunderscore}rankI{\isacharparenright}\isanewline
\ \ \ \ \isacommand{fix}\isamarkupfalse%
\ w\ x\ y%
\begin{isamarkuptext}%
Assume \isa{y} and \isa{x} are reachable, and in the decode relation.  They
      therefore have the same set of ancestors.%
\end{isamarkuptext}\isamarkuptrue%
\ \ \ \ \isacommand{assume}\isamarkupfalse%
\ {\isachardoublequoteopen}w\ {\isasymin}\ {\isacharparenleft}decodes{\isacharunderscore}to\ net{\isacharparenright}{\isasyminverse}\ {\isacharbackquote}{\isacharbackquote}\ {\isacharbraceleft}n{\isacharbraceright}{\isachardoublequoteclose}\isanewline
\ \ \ \ \isacommand{hence}\isamarkupfalse%
\ wn{\isacharcolon}\ {\isachardoublequoteopen}{\isacharparenleft}w{\isacharcomma}n{\isacharparenright}\ {\isasymin}\ decodes{\isacharunderscore}to\ net{\isachardoublequoteclose}\ \isacommand{by}\isamarkupfalse%
{\isacharparenleft}simp{\isacharparenright}\isanewline
\ \ \ \ \isacommand{assume}\isamarkupfalse%
\ xw{\isacharcolon}\ {\isachardoublequoteopen}{\isacharparenleft}x{\isacharcomma}w{\isacharparenright}\ {\isasymin}\ {\isacharparenleft}decodes{\isacharunderscore}to\ net{\isacharparenright}\isactrlsup {\isacharasterisk}{\isachardoublequoteclose}\isanewline
\ \ \ \ \ \ \ \isakeyword{and}\ yx{\isacharcolon}\ {\isachardoublequoteopen}{\isacharparenleft}y{\isacharcomma}x{\isacharparenright}\ {\isasymin}\ decodes{\isacharunderscore}to\ net{\isachardoublequoteclose}%
\begin{isamarkuptext}%
We show that for any ancestor of \isa{x}, the rank assigned \isa{x} is greater
      than the new rank we've constructed for \isa{y}, and thus so is the minimum over these
      i.e. \isa{g{\isacharprime}}.%
\end{isamarkuptext}\isamarkuptrue%
\ \ \ \ \isacommand{show}\isamarkupfalse%
\ {\isachardoublequoteopen}{\isacharquery}g{\isacharprime}\ y\ {\isacharless}\ {\isacharquery}g{\isacharprime}\ x{\isachardoublequoteclose}\isanewline
\ \ \ \ \isacommand{proof}\isamarkupfalse%
{\isacharparenleft}intro\ iffD{\isadigit{2}}{\isacharbrackleft}OF\ Min{\isacharunderscore}gr{\isacharunderscore}iff{\isacharbrackright}{\isacharparenright}%
\begin{isamarkuptext}%
The ancestors are finite, and there is at least one.%
\end{isamarkuptext}\isamarkuptrue%
\ \ \ \ \ \ \isacommand{from}\isamarkupfalse%
\ finite{\isacharunderscore}ancs\ \isacommand{show}\isamarkupfalse%
\ {\isachardoublequoteopen}finite\ {\isacharparenleft}{\isacharparenleft}{\isasymlambda}n{\isacharprime}{\isacharprime}{\isachardot}\ g\ n{\isacharprime}{\isacharprime}\ x{\isacharparenright}\ {\isacharbackquote}\ {\isacharquery}ancs\ x{\isacharparenright}{\isachardoublequoteclose}\ \isacommand{by}\isamarkupfalse%
{\isacharparenleft}auto{\isacharparenright}\isanewline
\ \ \ \ \ \ \isacommand{from}\isamarkupfalse%
\ wn\ xw\ \isacommand{show}\isamarkupfalse%
\ {\isachardoublequoteopen}{\isacharparenleft}{\isasymlambda}n{\isacharprime}{\isacharprime}{\isachardot}\ g\ n{\isacharprime}{\isacharprime}\ x{\isacharparenright}\ {\isacharbackquote}\ {\isacharquery}ancs\ x\ {\isasymnoteq}\ {\isacharbraceleft}{\isacharbraceright}{\isachardoublequoteclose}\ \isacommand{by}\isamarkupfalse%
{\isacharparenleft}blast{\isacharparenright}\isanewline
\isanewline
\ \ \ \ \ \ \isacommand{show}\isamarkupfalse%
\ {\isachardoublequoteopen}{\isasymforall}a{\isasymin}{\isacharparenleft}{\isasymlambda}n{\isacharprime}{\isacharprime}{\isachardot}\ g\ n{\isacharprime}{\isacharprime}\ x{\isacharparenright}\ {\isacharbackquote}\ {\isacharquery}ancs\ x{\isachardot}\ Min\ {\isacharparenleft}{\isacharparenleft}{\isasymlambda}n{\isacharprime}{\isacharprime}{\isachardot}\ g\ n{\isacharprime}{\isacharprime}\ y{\isacharparenright}\ {\isacharbackquote}\ {\isacharquery}ancs\ y{\isacharparenright}\ {\isacharless}\ a{\isachardoublequoteclose}\isanewline
\ \ \ \ \ \ \isacommand{proof}\isamarkupfalse%
\isanewline
\ \ \ \ \ \ \ \ \isacommand{fix}\isamarkupfalse%
\ a\isanewline
\ \ \ \ \ \ \ \ \isacommand{assume}\isamarkupfalse%
\ {\isachardoublequoteopen}a\ {\isasymin}\ {\isacharparenleft}{\isasymlambda}n{\isacharprime}{\isacharprime}{\isachardot}\ g\ n{\isacharprime}{\isacharprime}\ x{\isacharparenright}\ {\isacharbackquote}\ {\isacharquery}ancs\ x{\isachardoublequoteclose}\isanewline
\ \ \ \ \ \ \ \ \isacommand{then}\isamarkupfalse%
\ \isacommand{obtain}\isamarkupfalse%
\ n{\isacharprime}\ \isakeyword{where}\ step{\isacharcolon}\ {\isachardoublequoteopen}n{\isacharprime}\ {\isasymin}\ {\isacharparenleft}decodes{\isacharunderscore}to\ net{\isacharparenright}{\isasyminverse}\ {\isacharbackquote}{\isacharbackquote}\ {\isacharbraceleft}n{\isacharbraceright}{\isachardoublequoteclose}\isanewline
\ \ \ \ \ \ \ \ \ \ \ \ \ \ \ \ \ \ \ \ \ \ \ \ \ \isakeyword{and}\ anc{\isacharcolon}\ \ {\isachardoublequoteopen}{\isacharparenleft}x{\isacharcomma}n{\isacharprime}{\isacharparenright}\ {\isasymin}\ {\isacharparenleft}decodes{\isacharunderscore}to\ net{\isacharparenright}\isactrlsup {\isacharasterisk}{\isachardoublequoteclose}\isanewline
\ \ \ \ \ \ \ \ \ \ \ \ \ \ \ \ \ \ \ \ \ \ \ \ \ \isakeyword{and}\ aeq{\isacharcolon}\ {\isachardoublequoteopen}a\ {\isacharequal}\ g\ n{\isacharprime}\ x{\isachardoublequoteclose}\ \isacommand{by}\isamarkupfalse%
{\isacharparenleft}blast{\isacharparenright}%
\begin{isamarkuptext}%
As any ancestor \isa{n{\isacharprime}} of \isa{x} is also an ancestor of \isa{y}, we can
          appeal to well-formedness to show that \isa{g\ n{\isacharprime}\ y\ {\isacharless}\ g\ n{\isacharprime}\ x}.  It thus follows that
          the minimum is also lower.%
\end{isamarkuptext}\isamarkuptrue%
\ \ \ \ \ \ \ \ \isacommand{from}\isamarkupfalse%
\ anc\ yx\ \isacommand{have}\isamarkupfalse%
\ {\isachardoublequoteopen}{\isacharparenleft}y{\isacharcomma}n{\isacharprime}{\isacharparenright}\ {\isasymin}\ {\isacharparenleft}decodes{\isacharunderscore}to\ net{\isacharparenright}\isactrlsup {\isacharasterisk}{\isachardoublequoteclose}\ \isacommand{by}\isamarkupfalse%
{\isacharparenleft}auto{\isacharparenright}\isanewline
\ \ \ \ \ \ \ \ \isacommand{with}\isamarkupfalse%
\ step\ \isacommand{have}\isamarkupfalse%
\ {\isachardoublequoteopen}g\ n{\isacharprime}\ y\ {\isasymin}\ {\isacharparenleft}{\isasymlambda}n{\isacharprime}{\isacharprime}{\isachardot}\ g\ n{\isacharprime}{\isacharprime}\ y{\isacharparenright}\ {\isacharbackquote}\ {\isacharquery}ancs\ y{\isachardoublequoteclose}\ \isacommand{by}\isamarkupfalse%
{\isacharparenleft}blast{\isacharparenright}\isanewline
\ \ \ \ \ \ \ \ \isacommand{moreover}\isamarkupfalse%
\ \isacommand{from}\isamarkupfalse%
\ finite{\isacharunderscore}ancs\ \isacommand{have}\isamarkupfalse%
\ {\isachardoublequoteopen}finite\ {\isacharparenleft}{\isacharparenleft}{\isasymlambda}n{\isacharprime}{\isacharprime}{\isachardot}\ g\ n{\isacharprime}{\isacharprime}\ y{\isacharparenright}\ {\isacharbackquote}\ {\isacharquery}ancs\ y{\isacharparenright}{\isachardoublequoteclose}\ \isacommand{by}\isamarkupfalse%
{\isacharparenleft}auto{\isacharparenright}\isanewline
\ \ \ \ \ \ \ \ \isacommand{ultimately}\isamarkupfalse%
\ \isacommand{have}\isamarkupfalse%
\ {\isachardoublequoteopen}Min\ {\isacharparenleft}{\isacharparenleft}{\isasymlambda}n{\isacharprime}{\isacharprime}{\isachardot}\ g\ n{\isacharprime}{\isacharprime}\ y{\isacharparenright}\ {\isacharbackquote}\ {\isacharquery}ancs\ y{\isacharparenright}\ {\isasymle}\ g\ n{\isacharprime}\ y{\isachardoublequoteclose}\ \isacommand{by}\isamarkupfalse%
{\isacharparenleft}auto{\isacharparenright}\isanewline
\ \ \ \ \ \ \ \ \isacommand{also}\isamarkupfalse%
\ \isacommand{{\isacharbraceleft}}\isamarkupfalse%
\isanewline
\ \ \ \ \ \ \ \ \ \ \isacommand{from}\isamarkupfalse%
\ step\ wf{\isacharunderscore}g\ \isacommand{have}\isamarkupfalse%
\ {\isachardoublequoteopen}wf{\isacharunderscore}rank\ {\isacharparenleft}g\ n{\isacharprime}{\isacharparenright}\ n{\isacharprime}\ net{\isachardoublequoteclose}\ \isacommand{by}\isamarkupfalse%
{\isacharparenleft}blast{\isacharparenright}\isanewline
\ \ \ \ \ \ \ \ \ \ \isacommand{with}\isamarkupfalse%
\ yx\ anc\ \isacommand{have}\isamarkupfalse%
\ {\isachardoublequoteopen}g\ n{\isacharprime}\ y\ {\isacharless}\ g\ n{\isacharprime}\ x{\isachardoublequoteclose}\ \isacommand{unfolding}\isamarkupfalse%
\ wf{\isacharunderscore}rank{\isacharunderscore}def\ \isacommand{by}\isamarkupfalse%
{\isacharparenleft}blast{\isacharparenright}\isanewline
\ \ \ \ \ \ \ \ \isacommand{{\isacharbraceright}}\isamarkupfalse%
\isanewline
\ \ \ \ \ \ \ \ \isacommand{finally}\isamarkupfalse%
\ \isacommand{show}\isamarkupfalse%
\ {\isachardoublequoteopen}Min\ {\isacharparenleft}{\isacharparenleft}{\isasymlambda}n{\isacharprime}{\isacharprime}{\isachardot}\ g\ n{\isacharprime}{\isacharprime}\ y{\isacharparenright}\ {\isacharbackquote}\ {\isacharquery}ancs\ y{\isacharparenright}\ {\isacharless}\ a{\isachardoublequoteclose}\ \isacommand{by}\isamarkupfalse%
{\isacharparenleft}simp\ add{\isacharcolon}aeq{\isacharparenright}\isanewline
\ \ \ \ \ \ \isacommand{qed}\isamarkupfalse%
\isanewline
\ \ \ \ \isacommand{qed}\isamarkupfalse%
\isanewline
\ \ \isacommand{qed}\isamarkupfalse%
\begin{isamarkuptext}%
We will appeal to the fact that we must be in the accessible portion of the decode
    relation to show that \isa{n} can never appear as a descendent of itself, and can thus be
    assigned a higher rank than all of its descendents.%
\end{isamarkuptext}\isamarkuptrue%
\ \ \isacommand{from}\isamarkupfalse%
\ rd\ \isacommand{have}\isamarkupfalse%
\ nacc{\isacharcolon}\ {\isachardoublequoteopen}n\ {\isasymin}\ Wellfounded{\isachardot}acc\ {\isacharparenleft}decodes{\isacharunderscore}to\ net{\isacharparenright}{\isachardoublequoteclose}\isanewline
\ \ \ \ \isacommand{by}\isamarkupfalse%
{\isacharparenleft}auto\ dest{\isacharcolon}resolve{\isacharunderscore}dom{\isacharunderscore}decodes{\isacharunderscore}to{\isacharparenright}%
\begin{isamarkuptext}%
Finally, we construct our ranking function by assigning to \isa{n} a rank greater than
    any of its descendents.  Doing so relies on there being only finitely many of these.%
\end{isamarkuptext}\isamarkuptrue%
\ \ \isacommand{let}\isamarkupfalse%
\ {\isacharquery}max\ {\isacharequal}\ {\isachardoublequoteopen}Max\ {\isacharparenleft}{\isacharquery}g{\isacharprime}\ {\isacharbackquote}\ {\isacharparenleft}decodes{\isacharunderscore}to\ net{\isacharparenright}{\isasyminverse}\ {\isacharbackquote}{\isacharbackquote}\ {\isacharbraceleft}n{\isacharbraceright}{\isacharparenright}{\isachardoublequoteclose}\isanewline
\ \ \isacommand{let}\isamarkupfalse%
\ {\isacharquery}h\ {\isacharequal}\ {\isachardoublequoteopen}{\isacharquery}g{\isacharprime}{\isacharparenleft}n\ {\isacharcolon}{\isacharequal}\ Suc\ {\isacharquery}max{\isacharparenright}{\isachardoublequoteclose}\isanewline
\ \ \isacommand{have}\isamarkupfalse%
\ {\isachardoublequoteopen}wf{\isacharunderscore}rank\ {\isacharquery}h\ n\ net{\isachardoublequoteclose}%
\begin{isamarkuptext}%
We can show this be appealing to the well-formedness for each descendents that we just
      proved, and the fact that we assigned a greater rank to \isa{n}.%
\end{isamarkuptext}\isamarkuptrue%
\ \ \isacommand{proof}\isamarkupfalse%
{\isacharparenleft}rule\ wf{\isacharunderscore}rank{\isacharunderscore}step{\isacharparenright}\isanewline
\ \ \ \ \isacommand{fix}\isamarkupfalse%
\ y\isanewline
\ \ \ \ \isacommand{assume}\isamarkupfalse%
\ yn{\isacharcolon}\ {\isachardoublequoteopen}{\isacharparenleft}y{\isacharcomma}n{\isacharparenright}\ {\isasymin}\ decodes{\isacharunderscore}to\ net{\isachardoublequoteclose}%
\begin{isamarkuptext}%
We must show that the ranking is still well-defined for all descendents, even though
      we've changed the rank assigned to \isa{n}.  This is where we need to know that \isa{n}
      never appears as its own descendent.%
\end{isamarkuptext}\isamarkuptrue%
\ \ \ \ \isacommand{show}\isamarkupfalse%
\ {\isachardoublequoteopen}wf{\isacharunderscore}rank\ {\isacharquery}h\ y\ net{\isachardoublequoteclose}\isanewline
\ \ \ \ \isacommand{proof}\isamarkupfalse%
{\isacharparenleft}rule\ wf{\isacharunderscore}rankI{\isacharparenright}\isanewline
\ \ \ \ \ \ \isacommand{fix}\isamarkupfalse%
\ x\ z\isanewline
\ \ \ \ \ \ \isacommand{assume}\isamarkupfalse%
\ xy{\isacharcolon}\ {\isachardoublequoteopen}{\isacharparenleft}x{\isacharcomma}y{\isacharparenright}\ {\isasymin}\ {\isacharparenleft}decodes{\isacharunderscore}to\ net{\isacharparenright}\isactrlsup {\isacharasterisk}{\isachardoublequoteclose}\isanewline
\ \ \ \ \ \ \ \ \ \isakeyword{and}\ zx{\isacharcolon}\ {\isachardoublequoteopen}{\isacharparenleft}z{\isacharcomma}x{\isacharparenright}\ {\isasymin}\ decodes{\isacharunderscore}to\ net{\isachardoublequoteclose}%
\begin{isamarkuptext}%
Appeal to the absence of loops in the accessible portion.%
\end{isamarkuptext}\isamarkuptrue%
\ \ \ \ \ \ \isacommand{from}\isamarkupfalse%
\ xy\ yn\ \isacommand{have}\isamarkupfalse%
\ xreach{\isacharcolon}\ {\isachardoublequoteopen}{\isacharparenleft}x{\isacharcomma}n{\isacharparenright}\ {\isasymin}\ {\isacharparenleft}decodes{\isacharunderscore}to\ net{\isacharparenright}\isactrlsup {\isacharplus}{\isachardoublequoteclose}\ \isacommand{by}\isamarkupfalse%
{\isacharparenleft}auto{\isacharparenright}\isanewline
\ \ \ \ \ \ \isacommand{with}\isamarkupfalse%
\ nacc\ \isacommand{have}\isamarkupfalse%
\ x{\isacharunderscore}ne{\isacharunderscore}n{\isacharcolon}\ {\isachardoublequoteopen}x\ {\isasymnoteq}\ n{\isachardoublequoteclose}\ \isacommand{by}\isamarkupfalse%
{\isacharparenleft}auto\ dest{\isacharcolon}no{\isacharunderscore}loops{\isacharunderscore}acc{\isacharparenright}\isanewline
\isanewline
\ \ \ \ \ \ \isacommand{from}\isamarkupfalse%
\ zx\ xy\ yn\ \isacommand{have}\isamarkupfalse%
\ {\isachardoublequoteopen}{\isacharparenleft}z{\isacharcomma}n{\isacharparenright}\ {\isasymin}\ {\isacharparenleft}decodes{\isacharunderscore}to\ net{\isacharparenright}\isactrlsup {\isacharplus}{\isachardoublequoteclose}\ \isacommand{by}\isamarkupfalse%
{\isacharparenleft}auto{\isacharparenright}\isanewline
\ \ \ \ \ \ \isacommand{with}\isamarkupfalse%
\ nacc\ \isacommand{have}\isamarkupfalse%
\ z{\isacharunderscore}ne{\isacharunderscore}n{\isacharcolon}\ {\isachardoublequoteopen}z\ {\isasymnoteq}\ n{\isachardoublequoteclose}\ \isacommand{by}\isamarkupfalse%
{\isacharparenleft}auto\ dest{\isacharcolon}no{\isacharunderscore}loops{\isacharunderscore}acc{\isacharparenright}\isanewline
\ \ \ \ \ \ \isanewline
\ \ \ \ \ \ \isacommand{from}\isamarkupfalse%
\ yn\ \isacommand{have}\isamarkupfalse%
\ {\isachardoublequoteopen}y{\isasymin}{\isacharparenleft}decodes{\isacharunderscore}to\ net{\isacharparenright}{\isasyminverse}\ {\isacharbackquote}{\isacharbackquote}\ {\isacharbraceleft}n{\isacharbraceright}{\isachardoublequoteclose}\ \isacommand{by}\isamarkupfalse%
{\isacharparenleft}simp{\isacharparenright}\isanewline
\ \ \ \ \ \ \isacommand{with}\isamarkupfalse%
\ wf{\isacharunderscore}g{\isacharprime}\ \isacommand{have}\isamarkupfalse%
\ wf{\isacharunderscore}g{\isacharunderscore}y{\isacharcolon}\ {\isachardoublequoteopen}wf{\isacharunderscore}rank\ {\isacharquery}g{\isacharprime}\ y\ net{\isachardoublequoteclose}\ \isacommand{by}\isamarkupfalse%
{\isacharparenleft}simp{\isacharparenright}%
\begin{isamarkuptext}%
Appeal to well-formedness.%
\end{isamarkuptext}\isamarkuptrue%
\ \ \ \ \ \ \isacommand{from}\isamarkupfalse%
\ wf{\isacharunderscore}g{\isacharunderscore}y\ zx\ xy\ \isacommand{have}\isamarkupfalse%
\ {\isachardoublequoteopen}{\isacharquery}g{\isacharprime}\ z\ {\isacharless}\ {\isacharquery}g{\isacharprime}\ x{\isachardoublequoteclose}\ \isacommand{unfolding}\isamarkupfalse%
\ wf{\isacharunderscore}rank{\isacharunderscore}def\ \isacommand{by}\isamarkupfalse%
{\isacharparenleft}blast{\isacharparenright}\isanewline
\ \ \ \ \ \ \isacommand{thus}\isamarkupfalse%
\ {\isachardoublequoteopen}{\isacharquery}h\ z\ {\isacharless}\ {\isacharquery}h\ x{\isachardoublequoteclose}\ \isacommand{by}\isamarkupfalse%
{\isacharparenleft}simp\ add{\isacharcolon}x{\isacharunderscore}ne{\isacharunderscore}n\ z{\isacharunderscore}ne{\isacharunderscore}n{\isacharparenright}\isanewline
\ \ \ \ \isacommand{qed}\isamarkupfalse%
\begin{isamarkuptext}%
Showing that the rank increases is now trivial.%
\end{isamarkuptext}\isamarkuptrue%
\ \ \ \ \isacommand{from}\isamarkupfalse%
\ nacc\ yn\ \isacommand{have}\isamarkupfalse%
\ y{\isacharunderscore}ne{\isacharunderscore}n{\isacharcolon}\ {\isachardoublequoteopen}y\ {\isasymnoteq}\ n{\isachardoublequoteclose}\ \isacommand{by}\isamarkupfalse%
{\isacharparenleft}auto\ dest{\isacharcolon}no{\isacharunderscore}loops{\isacharunderscore}acc{\isacharparenright}\isanewline
\ \ \ \ \isacommand{from}\isamarkupfalse%
\ branching\ rd\ yn\isanewline
\ \ \ \ \isacommand{show}\isamarkupfalse%
\ {\isachardoublequoteopen}{\isacharquery}h\ y\ {\isacharless}\ {\isacharquery}h\ n{\isachardoublequoteclose}\isanewline
\ \ \ \ \ \ \isacommand{by}\isamarkupfalse%
{\isacharparenleft}simp\ add{\isacharcolon}y{\isacharunderscore}ne{\isacharunderscore}n{\isacharcomma}\ intro\ le{\isacharunderscore}imp{\isacharunderscore}less{\isacharunderscore}Suc{\isacharbrackleft}OF\ Max{\isacharunderscore}ge{\isacharbrackright}{\isacharcomma}\ auto{\isacharparenright}\isanewline
\ \ \isacommand{qed}\isamarkupfalse%
\isanewline
\ \ \isacommand{thus}\isamarkupfalse%
\ {\isachardoublequoteopen}{\isasymexists}f{\isachardot}\ wf{\isacharunderscore}rank\ f\ n\ net{\isachardoublequoteclose}\ \isacommand{by}\isamarkupfalse%
{\isacharparenleft}blast{\isacharparenright}\isanewline
\isacommand{qed}\isamarkupfalse%
\endisatagproof
{\isafoldproof}%
\isadelimproof
\endisadelimproof
\isadelimproof
\endisadelimproof
\isatagproof
\endisatagproof
{\isafoldproof}%
\isadelimproof
\endisadelimproof
\isadelimtheory
\endisadelimtheory
\isatagtheory
\endisatagtheory
{\isafoldtheory}%
\isadelimtheory
\endisadelimtheory
\end{isabellebody}%

%
\begin{isabellebody}%
\setisabellecontext{Equivalence}%
\isadelimtheory
\endisadelimtheory
\isatagtheory
\endisatagtheory
{\isafoldtheory}%
\isadelimtheory
\endisadelimtheory
\isamarkupsubsection{View Equivalence%
}
\isamarkuptrue%
\label{sec:vieweq}
\begin{isamarkuptext}%
A view is a function that encodes the result of all address resolutions beginning at a
  given node.%
\end{isamarkuptext}\isamarkuptrue%
\isacommand{type{\isacharunderscore}synonym}\isamarkupfalse%
\ view\ {\isacharequal}\ {\isachardoublequoteopen}addr\ {\isasymRightarrow}\ name\ set{\isachardoublequoteclose}\isanewline
\isanewline
\isacommand{definition}\isamarkupfalse%
\ view{\isacharunderscore}from\ {\isacharcolon}{\isacharcolon}\ {\isachardoublequoteopen}nodeid\ {\isasymRightarrow}\ net\ {\isasymRightarrow}\ view{\isachardoublequoteclose}\isanewline
\ \ \isakeyword{where}\ {\isachardoublequoteopen}view{\isacharunderscore}from\ node\ net\ {\isacharequal}\ {\isacharparenleft}{\isasymlambda}addr{\isachardot}\ resolve\ net\ {\isacharparenleft}node{\isacharcomma}\ addr{\isacharparenright}{\isacharparenright}{\isachardoublequoteclose}%
\begin{isamarkuptext}%
A remapping is a renaming of nodes, leaving addresses intact.%
\end{isamarkuptext}\isamarkuptrue%
\isacommand{type{\isacharunderscore}synonym}\isamarkupfalse%
\ remap\ {\isacharequal}\ {\isachardoublequoteopen}nodeid\ {\isasymRightarrow}\ nodeid{\isachardoublequoteclose}\isanewline
\isanewline
\isacommand{definition}\isamarkupfalse%
\ rename\ {\isacharcolon}{\isacharcolon}\ {\isachardoublequoteopen}remap\ {\isasymRightarrow}\ name\ {\isasymRightarrow}\ name{\isachardoublequoteclose}\isanewline
\ \ \isakeyword{where}\ {\isachardoublequoteopen}rename\ m\ n\ {\isacharequal}\ {\isacharparenleft}m\ {\isacharparenleft}fst\ n{\isacharparenright}{\isacharcomma}\ snd\ n{\isacharparenright}{\isachardoublequoteclose}\isanewline
\ \ \ \ %
\isadelimproof
\endisadelimproof
\isatagproof
\endisatagproof
{\isafoldproof}%
\isadelimproof
\isanewline
\endisadelimproof
\isacommand{primrec}\isamarkupfalse%
\ rename{\isacharunderscore}list\ {\isacharcolon}{\isacharcolon}\ {\isachardoublequoteopen}{\isacharparenleft}nodeid\ {\isasymRightarrow}\ nodeid{\isacharparenright}\ {\isasymRightarrow}\ nodeid\ list\ {\isasymRightarrow}\ {\isacharparenleft}nodeid\ {\isasymRightarrow}\ nodeid{\isacharparenright}{\isachardoublequoteclose}\isanewline
\ \ \isakeyword{where}\ {\isachardoublequoteopen}rename{\isacharunderscore}list\ f\ {\isacharbrackleft}{\isacharbrackright}\ {\isacharequal}\ id{\isachardoublequoteclose}\ {\isacharbar}\isanewline
\ \ \ \ \ \ \ \ {\isachardoublequoteopen}rename{\isacharunderscore}list\ f\ {\isacharparenleft}nd{\isacharhash}nds{\isacharparenright}\ {\isacharequal}\ {\isacharparenleft}{\isasymlambda}x{\isachardot}\ if\ x\ {\isacharequal}\ nd\ then\ f\ nd\ else\ x{\isacharparenright}\ o\ {\isacharparenleft}rename{\isacharunderscore}list\ f\ nds{\isacharparenright}{\isachardoublequoteclose}%
\begin{isamarkuptext}%
Two nets are view-equivalent for some node, if the two views have the same domain, and
  give the same result for all addresses, modulo a renaming of accepting nodes.%
\end{isamarkuptext}\isamarkuptrue%
\isacommand{definition}\isamarkupfalse%
\ view{\isacharunderscore}eq\ {\isacharcolon}{\isacharcolon}\ {\isachardoublequoteopen}nodeid\ {\isasymRightarrow}\ {\isacharparenleft}remap\ {\isasymtimes}\ net{\isacharparenright}\ {\isasymRightarrow}\ {\isacharparenleft}remap\ {\isasymtimes}\ net{\isacharparenright}\ {\isasymRightarrow}\ bool{\isachardoublequoteclose}\isanewline
\ \ \isakeyword{where}\ {\isachardoublequoteopen}view{\isacharunderscore}eq\ nd\ x\ y\ {\isasymlongleftrightarrow}\isanewline
\ \ \ \ \ \ \ \ \ \ {\isacharparenleft}{\isasymforall}a{\isachardot}\ resolve{\isacharunderscore}dom\ {\isacharparenleft}snd\ x{\isacharcomma}{\isacharparenleft}nd{\isacharcomma}a{\isacharparenright}{\isacharparenright}\ {\isacharequal}\ resolve{\isacharunderscore}dom\ {\isacharparenleft}snd\ y{\isacharcomma}{\isacharparenleft}nd{\isacharcomma}a{\isacharparenright}{\isacharparenright}{\isacharparenright}\ {\isasymand}\isanewline
\ \ \ \ \ \ \ \ \ \ {\isacharparenleft}{\isasymforall}a{\isachardot}\ resolve{\isacharunderscore}dom\ {\isacharparenleft}snd\ x{\isacharcomma}{\isacharparenleft}nd{\isacharcomma}a{\isacharparenright}{\isacharparenright}\ {\isasymlongrightarrow}\isanewline
\ \ \ \ \ \ \ \ \ \ \ \ rename\ {\isacharparenleft}fst\ x{\isacharparenright}\ {\isacharbackquote}\ view{\isacharunderscore}from\ nd\ {\isacharparenleft}snd\ x{\isacharparenright}\ a\ {\isacharequal}\isanewline
\ \ \ \ \ \ \ \ \ \ \ \ rename\ {\isacharparenleft}fst\ y{\isacharparenright}\ {\isacharbackquote}\ view{\isacharunderscore}from\ nd\ {\isacharparenleft}snd\ y{\isacharparenright}\ a{\isacharparenright}{\isachardoublequoteclose}\isanewline
\ \ \ \ \isanewline
\isacommand{definition}\isamarkupfalse%
\ view{\isacharunderscore}eq{\isacharunderscore}on\ {\isacharcolon}{\isacharcolon}\ {\isachardoublequoteopen}nodeid\ set\ {\isasymRightarrow}\ {\isacharparenleft}remap\ {\isasymtimes}\ net{\isacharparenright}\ {\isasymRightarrow}\ {\isacharparenleft}remap\ {\isasymtimes}\ net{\isacharparenright}\ {\isasymRightarrow}\ bool{\isachardoublequoteclose}\isanewline
\ \ \isakeyword{where}\ {\isachardoublequoteopen}view{\isacharunderscore}eq{\isacharunderscore}on\ S\ x\ y\ {\isasymlongleftrightarrow}\ {\isacharparenleft}{\isasymforall}nd{\isasymin}S{\isachardot}\ view{\isacharunderscore}eq\ nd\ x\ y{\isacharparenright}{\isachardoublequoteclose}\isanewline
\ \ \ \ %
\begin{isamarkuptext}%
Two nodes are equivalent (for a given net) if they have the same view.%
\end{isamarkuptext}\isamarkuptrue%
\isacommand{definition}\isamarkupfalse%
\ node{\isacharunderscore}eq\ {\isacharcolon}{\isacharcolon}\ {\isachardoublequoteopen}net\ {\isasymRightarrow}\ nodeid\ {\isasymRightarrow}\ nodeid\ {\isasymRightarrow}\ bool{\isachardoublequoteclose}\isanewline
\ \ \isakeyword{where}\ {\isachardoublequoteopen}node{\isacharunderscore}eq\ net\ nd\ nd{\isacharprime}\ {\isasymlongleftrightarrow}\ \isanewline
\ \ \ \ \ \ \ \ \ \ {\isacharparenleft}{\isasymforall}a{\isachardot}\ resolve{\isacharunderscore}dom\ {\isacharparenleft}net{\isacharcomma}{\isacharparenleft}nd{\isacharcomma}a{\isacharparenright}{\isacharparenright}\ {\isacharequal}\ resolve{\isacharunderscore}dom\ {\isacharparenleft}net{\isacharcomma}{\isacharparenleft}nd{\isacharprime}{\isacharcomma}a{\isacharparenright}{\isacharparenright}{\isacharparenright}\ {\isasymand}\isanewline
\ \ \ \ \ \ \ \ \ \ {\isacharparenleft}{\isasymforall}a{\isachardot}\ resolve{\isacharunderscore}dom\ {\isacharparenleft}net{\isacharcomma}{\isacharparenleft}nd{\isacharcomma}a{\isacharparenright}{\isacharparenright}\ {\isasymlongrightarrow}\ view{\isacharunderscore}from\ nd\ net\ a\ {\isacharequal}\ view{\isacharunderscore}from\ nd{\isacharprime}\ net\ a{\isacharparenright}{\isachardoublequoteclose}\isanewline
\ \ \ \ %
\isadelimproof
\endisadelimproof
\isatagproof
\endisatagproof
{\isafoldproof}%
\isadelimproof
\endisadelimproof
\isadelimproof
\endisadelimproof
\isatagproof
\endisatagproof
{\isafoldproof}%
\isadelimproof
\endisadelimproof
\isadelimproof
\endisadelimproof
\isatagproof
\endisatagproof
{\isafoldproof}%
\isadelimproof
\endisadelimproof
\isadelimproof
\endisadelimproof
\isatagproof
\endisatagproof
{\isafoldproof}%
\isadelimproof
\endisadelimproof
\isadelimproof
\endisadelimproof
\isatagproof
\endisatagproof
{\isafoldproof}%
\isadelimproof
\endisadelimproof
\isadelimproof
\endisadelimproof
\isatagproof
\endisatagproof
{\isafoldproof}%
\isadelimproof
\endisadelimproof
\isadelimproof
\endisadelimproof
\isatagproof
\endisatagproof
{\isafoldproof}%
\isadelimproof
\endisadelimproof
\isadelimproof
\endisadelimproof
\isatagproof
\endisatagproof
{\isafoldproof}%
\isadelimproof
\endisadelimproof
\isadelimproof
\endisadelimproof
\isatagproof
\endisatagproof
{\isafoldproof}%
\isadelimproof
\endisadelimproof
\begin{isamarkuptext}%
Both view-equivalence and node-equivalence are proper equivalence relations.%
\end{isamarkuptext}\isamarkuptrue%
\isacommand{lemma}\isamarkupfalse%
\ equivp{\isacharunderscore}view{\isacharunderscore}eq{\isacharcolon}\isanewline
\ \ {\isachardoublequoteopen}{\isasymAnd}nd{\isachardot}\ equivp\ {\isacharparenleft}view{\isacharunderscore}eq\ nd{\isacharparenright}{\isachardoublequoteclose}\isanewline
\ \ \ \ %
\isadelimproof
\endisadelimproof
\isatagproof
\endisatagproof
{\isafoldproof}%
\isadelimproof
\isanewline
\endisadelimproof
\isacommand{lemma}\isamarkupfalse%
\ equivp{\isacharunderscore}view{\isacharunderscore}eq{\isacharunderscore}on{\isacharcolon}\isanewline
\ \ \isakeyword{fixes}\ S\ {\isacharcolon}{\isacharcolon}\ {\isachardoublequoteopen}nodeid\ set{\isachardoublequoteclose}\isanewline
\ \ \isakeyword{shows}\ {\isachardoublequoteopen}equivp\ {\isacharparenleft}view{\isacharunderscore}eq{\isacharunderscore}on\ S{\isacharparenright}{\isachardoublequoteclose}\isanewline
\ \ \ \ %
\isadelimproof
\endisadelimproof
\isatagproof
\endisatagproof
{\isafoldproof}%
\isadelimproof
\endisadelimproof
\isadelimproof
\endisadelimproof
\isatagproof
\endisatagproof
{\isafoldproof}%
\isadelimproof
\endisadelimproof
\isadelimproof
\endisadelimproof
\isatagproof
\endisatagproof
{\isafoldproof}%
\isadelimproof
\endisadelimproof
\isadelimproof
\endisadelimproof
\isatagproof
\endisatagproof
{\isafoldproof}%
\isadelimproof
\endisadelimproof
\isadelimproof
\endisadelimproof
\isatagproof
\endisatagproof
{\isafoldproof}%
\isadelimproof
\endisadelimproof
\isadelimproof
\endisadelimproof
\isatagproof
\endisatagproof
{\isafoldproof}%
\isadelimproof
\endisadelimproof
\isadelimproof
\endisadelimproof
\isatagproof
\endisatagproof
{\isafoldproof}%
\isadelimproof
\endisadelimproof
\isadelimproof
\endisadelimproof
\isatagproof
\endisatagproof
{\isafoldproof}%
\isadelimproof
\endisadelimproof
\begin{isamarkuptext}%
Both equivalence relations preserve resolution.%
\end{isamarkuptext}\isamarkuptrue%
\isacommand{lemma}\isamarkupfalse%
\ node{\isacharunderscore}eq{\isacharunderscore}resolve{\isacharcolon}\isanewline
\ \ \isakeyword{fixes}\ nd\ nd{\isacharprime}\ {\isacharcolon}{\isacharcolon}\ nodeid\ \isakeyword{and}\ net\ {\isacharcolon}{\isacharcolon}\ net\ \isakeyword{and}\ a\ {\isacharcolon}{\isacharcolon}\ addr\isanewline
\ \ \isakeyword{shows}\ {\isachardoublequoteopen}node{\isacharunderscore}eq\ net\ nd\ nd{\isacharprime}\ {\isasymLongrightarrow}\ resolve{\isacharunderscore}dom\ {\isacharparenleft}net{\isacharcomma}nd{\isacharcomma}a{\isacharparenright}\ {\isasymLongrightarrow}\ resolve\ net\ {\isacharparenleft}nd{\isacharcomma}a{\isacharparenright}\ {\isacharequal}\ resolve\ net\ {\isacharparenleft}nd{\isacharprime}{\isacharcomma}a{\isacharparenright}{\isachardoublequoteclose}\isanewline
\ \ \ \ %
\isadelimproof
\endisadelimproof
\isatagproof
\endisatagproof
{\isafoldproof}%
\isadelimproof
\isanewline
\endisadelimproof
\isacommand{lemma}\isamarkupfalse%
\ view{\isacharunderscore}eq{\isacharunderscore}resolve{\isacharcolon}\isanewline
\ \ \isakeyword{fixes}\ nd\ {\isacharcolon}{\isacharcolon}\ nodeid\ \isakeyword{and}\ x\ y\ {\isacharcolon}{\isacharcolon}\ {\isachardoublequoteopen}remap\ {\isasymtimes}\ net{\isachardoublequoteclose}\ \isakeyword{and}\ a\ {\isacharcolon}{\isacharcolon}\ addr\isanewline
\ \ \isakeyword{shows}\ {\isachardoublequoteopen}view{\isacharunderscore}eq\ nd\ x\ y\ {\isasymLongrightarrow}\ resolve{\isacharunderscore}dom\ {\isacharparenleft}snd\ x{\isacharcomma}nd{\isacharcomma}a{\isacharparenright}\ {\isasymLongrightarrow}\isanewline
\ \ \ \ \ \ \ \ \ rename\ {\isacharparenleft}fst\ x{\isacharparenright}\ {\isacharbackquote}\ resolve\ {\isacharparenleft}snd\ x{\isacharparenright}\ {\isacharparenleft}nd{\isacharcomma}a{\isacharparenright}\ {\isacharequal}\ rename\ {\isacharparenleft}fst\ y{\isacharparenright}\ {\isacharbackquote}\ resolve\ {\isacharparenleft}snd\ y{\isacharparenright}\ {\isacharparenleft}nd{\isacharcomma}a{\isacharparenright}{\isachardoublequoteclose}\isanewline
\ \ \ \ %
\isadelimproof
\endisadelimproof
\isatagproof
\endisatagproof
{\isafoldproof}%
\isadelimproof
\endisadelimproof
\isadelimproof
\endisadelimproof
\isatagproof
\endisatagproof
{\isafoldproof}%
\isadelimproof
\endisadelimproof
\begin{isamarkuptext}%
View-equivalence is preserved by any further node renaming.%
\end{isamarkuptext}\isamarkuptrue%
\isadelimproof
\endisadelimproof
\isatagproof
\endisatagproof
{\isafoldproof}%
\isadelimproof
\endisadelimproof
\isacommand{lemma}\isamarkupfalse%
\ view{\isacharunderscore}eq{\isacharunderscore}comp{\isacharcolon}\isanewline
\ \ {\isachardoublequoteopen}view{\isacharunderscore}eq\ nd\ {\isacharparenleft}f{\isacharcomma}net{\isacharparenright}\ {\isacharparenleft}g{\isacharcomma}net{\isacharprime}{\isacharparenright}\ {\isasymLongrightarrow}\ view{\isacharunderscore}eq\ nd\ {\isacharparenleft}h\ o\ f{\isacharcomma}net{\isacharparenright}\ {\isacharparenleft}h\ o\ g{\isacharcomma}net{\isacharprime}{\isacharparenright}{\isachardoublequoteclose}\isanewline
\ \ \ \ %
\isadelimproof
\endisadelimproof
\isatagproof
\endisatagproof
{\isafoldproof}%
\isadelimproof
\endisadelimproof
\isadelimproof
\endisadelimproof
\isatagproof
\endisatagproof
{\isafoldproof}%
\isadelimproof
\endisadelimproof
\isadelimproof
\endisadelimproof
\isatagproof
\endisatagproof
{\isafoldproof}%
\isadelimproof
\endisadelimproof
\begin{isamarkuptext}%
For transformations that add nodes, we need to know that the new node has no descendents
  or ancestors.%
\end{isamarkuptext}\isamarkuptrue%
\isacommand{definition}\isamarkupfalse%
\ fresh{\isacharunderscore}node\ {\isacharcolon}{\isacharcolon}\ {\isachardoublequoteopen}net\ {\isasymRightarrow}\ nodeid\ {\isasymRightarrow}\ bool{\isachardoublequoteclose}\isanewline
\ \ \isakeyword{where}\ {\isachardoublequoteopen}fresh{\isacharunderscore}node\ net\ nd\ {\isasymlongleftrightarrow}\isanewline
\ \ \ \ \ \ \ \ \ \ {\isacharparenleft}{\isasymforall}a{\isachardot}\ translate\ {\isacharparenleft}net\ nd{\isacharparenright}\ a\ {\isacharequal}\ {\isacharbraceleft}{\isacharbraceright}{\isacharparenright}\ {\isasymand}\isanewline
\ \ \ \ \ \ \ \ \ \ {\isacharparenleft}{\isasymforall}x\ y{\isachardot}\ {\isacharparenleft}x{\isacharcomma}y{\isacharparenright}\ {\isasymin}\ decodes{\isacharunderscore}to\ net\ {\isasymlongrightarrow}\ fst\ x\ {\isasymnoteq}\ nd{\isacharparenright}\ {\isasymand}\isanewline
\ \ \ \ \ \ \ \ \ \ accept\ {\isacharparenleft}net\ nd{\isacharparenright}\ {\isacharequal}\ {\isacharbraceleft}{\isacharbraceright}{\isachardoublequoteclose}\isanewline
\isadelimproof
\endisadelimproof
\isatagproof
\endisatagproof
{\isafoldproof}%
\isadelimproof
\endisadelimproof
\isadelimproof
\endisadelimproof
\isatagproof
\endisatagproof
{\isafoldproof}%
\isadelimproof
\endisadelimproof
\isadelimproof
\endisadelimproof
\isatagproof
\endisatagproof
{\isafoldproof}%
\isadelimproof
\endisadelimproof
\isadelimtheory
\endisadelimtheory
\isatagtheory
\endisatagtheory
{\isafoldtheory}%
\isadelimtheory
\endisadelimtheory
\end{isabellebody}%

%
\begin{isabellebody}%
\setisabellecontext{AbstractOps}%
\isadelimtheory
\endisadelimtheory
\isatagtheory
\endisatagtheory
{\isafoldtheory}%
\isadelimtheory
\endisadelimtheory
\isamarkupsubsection{Equivalence-Preserving Transformations%
}
\isamarkuptrue%
\isamarkupsubsubsection{Splitting Nodes%
}
\isamarkuptrue%
\label{sec:isasplit}
\begin{isamarkuptext}%
The acceptor accepts all addresses accepted by the original node, but translates none.%
\end{isamarkuptext}\isamarkuptrue%
\isacommand{definition}\isamarkupfalse%
\ acceptor{\isacharunderscore}node\ {\isacharcolon}{\isacharcolon}\ {\isachardoublequoteopen}node\ {\isasymRightarrow}\ node{\isachardoublequoteclose}\isanewline
\ \ \isakeyword{where}\ {\isachardoublequoteopen}acceptor{\isacharunderscore}node\ node\ {\isacharequal}\ node\ {\isasymlparr}\ translate\ {\isacharcolon}{\isacharequal}\ {\isasymlambda}{\isacharunderscore}{\isachardot}\ {\isacharbraceleft}{\isacharbraceright}\ {\isasymrparr}{\isachardoublequoteclose}%
\begin{isamarkuptext}%
Forward all addresses to the acceptor node, maintaining existing translations.%
\end{isamarkuptext}\isamarkuptrue%
\isacommand{definition}\isamarkupfalse%
\ redirector{\isacharunderscore}node\ {\isacharcolon}{\isacharcolon}\ {\isachardoublequoteopen}nodeid\ {\isasymRightarrow}\ node\ {\isasymRightarrow}\ node{\isachardoublequoteclose}\isanewline
\ \ \isakeyword{where}\ {\isachardoublequoteopen}redirector{\isacharunderscore}node\ nd{\isacharprime}\ node\ {\isacharequal}\ {\isasymlparr}\ accept\ {\isacharequal}\ {\isacharbraceleft}{\isacharbraceright}{\isacharcomma}\isanewline
\ \ \ \ \ \ \ \ \ \ translate\ {\isacharequal}\ {\isacharparenleft}{\isasymlambda}a{\isachardot}\ if\ a\ {\isasymin}\ accept\ node\ then\ insert\ {\isacharparenleft}nd{\isacharprime}{\isacharcomma}a{\isacharparenright}\ {\isacharparenleft}translate\ node\ a{\isacharparenright}\ else\ translate\ node\ a{\isacharparenright}\ {\isasymrparr}{\isachardoublequoteclose}%
\begin{isamarkuptext}%
Split a node into an acceptor, that accepts all addresses accepted by the original node, and
  a redirector, which forwards all addresses that it would have accepted to the acceptor.%
\end{isamarkuptext}\isamarkuptrue%
\isacommand{definition}\isamarkupfalse%
\ split{\isacharunderscore}node\ {\isacharcolon}{\isacharcolon}\ {\isachardoublequoteopen}nodeid\ {\isasymRightarrow}\ nodeid\ {\isasymRightarrow}\ net\ {\isasymRightarrow}\ net{\isachardoublequoteclose}\isanewline
\ \ \isakeyword{where}\ {\isachardoublequoteopen}split{\isacharunderscore}node\ nd\ nd{\isacharprime}\ net\ {\isacharequal}\isanewline
\ \ \ \ net{\isacharparenleft}nd\ {\isacharcolon}{\isacharequal}\ redirector{\isacharunderscore}node\ nd{\isacharprime}\ {\isacharparenleft}net\ nd{\isacharparenright}{\isacharcomma}\ nd{\isacharprime}\ {\isacharcolon}{\isacharequal}\ acceptor{\isacharunderscore}node\ {\isacharparenleft}net\ nd{\isacharparenright}{\isacharparenright}{\isachardoublequoteclose}%
\begin{isamarkuptext}%
We can represent the effect of node splitting by its action on the set of accepted names,
  and on the decoding relation.  Recall from \autoref{sec:isares} that this is sufficient to
  fully define the result of resolution.%
\end{isamarkuptext}\isamarkuptrue%
\begin{isamarkuptext}%
Splitting only adds the new decode edges:%
\end{isamarkuptext}\isamarkuptrue%
\isacommand{lemma}\isamarkupfalse%
\ split{\isacharunderscore}decode{\isacharcolon}\isanewline
\ \ {\isachardoublequoteopen}nd\ {\isasymnoteq}\ nd{\isacharprime}\ {\isasymLongrightarrow}\ {\isacharparenleft}{\isasymAnd}a{\isachardot}\ translate\ {\isacharparenleft}net\ nd{\isacharprime}{\isacharparenright}\ a\ {\isacharequal}\ {\isacharbraceleft}{\isacharbraceright}{\isacharparenright}\ {\isasymLongrightarrow}\isanewline
\ \ \ decodes{\isacharunderscore}to\ {\isacharparenleft}split{\isacharunderscore}node\ nd\ nd{\isacharprime}\ net{\isacharparenright}\ {\isacharequal}\isanewline
\ \ \ decodes{\isacharunderscore}to\ net\ {\isasymunion}\ {\isacharparenleft}{\isasymlambda}a{\isachardot}\ {\isacharparenleft}Pair\ nd{\isacharprime}\ a{\isacharcomma}\ Pair\ nd\ a{\isacharparenright}{\isacharparenright}\ {\isacharbackquote}\ accept\ {\isacharparenleft}net\ nd{\isacharparenright}{\isachardoublequoteclose}%
\isadelimproof
\endisadelimproof
\isatagproof
\endisatagproof
{\isafoldproof}%
\isadelimproof
\endisadelimproof
\begin{isamarkuptext}%
Splitting only touches the named nodes:%
\end{isamarkuptext}\isamarkuptrue%
\isacommand{lemma}\isamarkupfalse%
\ fresh{\isacharunderscore}split{\isacharunderscore}node{\isacharcolon}\isanewline
\ \ \isakeyword{assumes}\ fresh{\isacharcolon}\ {\isachardoublequoteopen}fresh{\isacharunderscore}node\ net\ x{\isachardoublequoteclose}\isanewline
\ \ \ \ \ \ \isakeyword{and}\ neq{\isacharcolon}\ {\isachardoublequoteopen}x\ {\isasymnoteq}\ nd{\isachardoublequoteclose}\ {\isachardoublequoteopen}x\ {\isasymnoteq}\ nd{\isacharprime}{\isachardoublequoteclose}\isanewline
\ \ \ \ \isakeyword{shows}\ {\isachardoublequoteopen}fresh{\isacharunderscore}node\ {\isacharparenleft}split{\isacharunderscore}node\ nd\ nd{\isacharprime}\ net{\isacharparenright}\ x{\isachardoublequoteclose}%
\isadelimproof
\endisadelimproof
\isatagproof
\endisatagproof
{\isafoldproof}%
\isadelimproof
\endisadelimproof
\begin{isamarkuptext}%
Splitting neither adds nor removes accepted names, it simply renames those accepted by the
  original node:%
\end{isamarkuptext}\isamarkuptrue%
\isacommand{lemma}\isamarkupfalse%
\ split{\isacharunderscore}accepted{\isacharcolon}\isanewline
\ \ \isakeyword{assumes}\ empty{\isacharcolon}\ {\isachardoublequoteopen}accept\ {\isacharparenleft}net\ nd{\isacharprime}{\isacharparenright}\ {\isacharequal}\ {\isacharbraceleft}{\isacharbraceright}{\isachardoublequoteclose}\isanewline
\ \ \ \ \isakeyword{shows}\ {\isachardoublequoteopen}accepted{\isacharunderscore}names\ {\isacharparenleft}split{\isacharunderscore}node\ nd\ nd{\isacharprime}\ net{\isacharparenright}\ {\isacharequal}\ rename\ {\isacharparenleft}id{\isacharparenleft}nd\ {\isacharcolon}{\isacharequal}\ nd{\isacharprime}{\isacharparenright}{\isacharparenright}\ {\isacharbackquote}\ accepted{\isacharunderscore}names\ net{\isachardoublequoteclose}%
\isadelimproof
\endisadelimproof
\isatagproof
\endisatagproof
{\isafoldproof}%
\isadelimproof
\endisadelimproof
\begin{isamarkuptext}%
Splitting a node has no effect on the termination of \isa{resolve}:%
\end{isamarkuptext}\isamarkuptrue%
\isacommand{lemma}\isamarkupfalse%
\ split{\isacharunderscore}node{\isacharunderscore}domeq{\isacharcolon}\isanewline
\ \ \isakeyword{fixes}\ S\ {\isacharcolon}{\isacharcolon}\ {\isachardoublequoteopen}nodeid\ set{\isachardoublequoteclose}\ \isakeyword{and}\ nd\ nd{\isacharprime}\ {\isacharcolon}{\isacharcolon}\ nodeid\ \isakeyword{and}\ net\ {\isacharcolon}{\isacharcolon}\ net\ \isakeyword{and}\ n{\isacharcolon}{\isacharcolon}name\isanewline
\ \ \isakeyword{assumes}\ neq{\isacharcolon}\ {\isachardoublequoteopen}nd\ {\isasymnoteq}\ nd{\isacharprime}{\isachardoublequoteclose}\isanewline
\ \ \ \ \ \ \isakeyword{and}\ wf{\isacharunderscore}net{\isacharcolon}\ {\isachardoublequoteopen}wf{\isacharunderscore}net\ net{\isachardoublequoteclose}\isanewline
\ \ \ \ \ \ \isakeyword{and}\ fresh{\isacharcolon}\ {\isachardoublequoteopen}fresh{\isacharunderscore}node\ net\ nd{\isacharprime}{\isachardoublequoteclose}\isanewline
\ \ \ \ \isakeyword{shows}\ {\isachardoublequoteopen}resolve{\isacharunderscore}dom\ {\isacharparenleft}net{\isacharcomma}n{\isacharparenright}\ {\isacharequal}\ resolve{\isacharunderscore}dom\ {\isacharparenleft}split{\isacharunderscore}node\ nd\ nd{\isacharprime}\ net{\isacharcomma}n{\isacharparenright}{\isachardoublequoteclose}%
\isadelimproof
\endisadelimproof
\isatagproof
\endisatagproof
{\isafoldproof}%
\isadelimproof
\endisadelimproof
\begin{isamarkuptext}%
The effect of splitting a node is just to rename anything that was accepted by the split
  node.%
\end{isamarkuptext}\isamarkuptrue%
\isacommand{lemma}\isamarkupfalse%
\ split{\isacharunderscore}node{\isacharunderscore}resolveeq{\isacharcolon}\isanewline
\ \ \isakeyword{fixes}\ S\ {\isacharcolon}{\isacharcolon}\ {\isachardoublequoteopen}nodeid\ set{\isachardoublequoteclose}\ \isakeyword{and}\ nd\ nd{\isacharprime}\ {\isacharcolon}{\isacharcolon}\ nodeid\ \isakeyword{and}\ net\ {\isacharcolon}{\isacharcolon}\ net\ \isakeyword{and}\ n{\isacharcolon}{\isacharcolon}name\isanewline
\ \ \isakeyword{assumes}\ neq{\isacharcolon}\ {\isachardoublequoteopen}nd\ {\isasymnoteq}\ nd{\isacharprime}{\isachardoublequoteclose}\isanewline
\ \ \ \ \ \ \isakeyword{and}\ wf{\isacharunderscore}net{\isacharcolon}\ {\isachardoublequoteopen}wf{\isacharunderscore}net\ net{\isachardoublequoteclose}\isanewline
\ \ \ \ \ \ \isakeyword{and}\ fresh{\isacharcolon}\ {\isachardoublequoteopen}fresh{\isacharunderscore}node\ net\ nd{\isacharprime}{\isachardoublequoteclose}\isanewline
\ \ \ \ \ \ \isakeyword{and}\ dom{\isacharcolon}\ {\isachardoublequoteopen}resolve{\isacharunderscore}dom\ {\isacharparenleft}net{\isacharcomma}n{\isacharparenright}{\isachardoublequoteclose}\isanewline
\ \ \ \ \isakeyword{shows}\ {\isachardoublequoteopen}fst\ n\ {\isasymnoteq}\ nd{\isacharprime}\ {\isasymLongrightarrow}\isanewline
\ \ \ \ \ \ \ \ \ \ \ rename\ {\isacharparenleft}id{\isacharparenleft}nd\ {\isacharcolon}{\isacharequal}\ nd{\isacharprime}{\isacharparenright}{\isacharparenright}\ {\isacharbackquote}\ resolve\ net\ n\ {\isacharequal}\isanewline
\ \ \ \ \ \ \ \ \ \ \ rename\ id\ {\isacharbackquote}\ resolve\ {\isacharparenleft}split{\isacharunderscore}node\ nd\ nd{\isacharprime}\ net{\isacharparenright}\ n{\isachardoublequoteclose}%
\isadelimproof
\endisadelimproof
\isatagproof
\endisatagproof
{\isafoldproof}%
\isadelimproof
\endisadelimproof
\begin{isamarkuptext}%
From these two lemmas, we have view-equivalence under splitting.%
\end{isamarkuptext}\isamarkuptrue%
\isacommand{lemma}\isamarkupfalse%
\ split{\isacharunderscore}node{\isacharunderscore}eq{\isacharcolon}\isanewline
\ \ \isakeyword{fixes}\ S\ {\isacharcolon}{\isacharcolon}\ {\isachardoublequoteopen}nodeid\ set{\isachardoublequoteclose}\ \isakeyword{and}\ nd\ nd{\isacharprime}\ {\isacharcolon}{\isacharcolon}\ nodeid\ \isakeyword{and}\ net\ {\isacharcolon}{\isacharcolon}\ net\isanewline
\ \ \isakeyword{assumes}\ neq{\isacharcolon}\ {\isachardoublequoteopen}nd\ {\isasymnoteq}\ nd{\isacharprime}{\isachardoublequoteclose}\isanewline
\ \ \ \ \ \ \isakeyword{and}\ wf{\isacharunderscore}net{\isacharcolon}\ {\isachardoublequoteopen}wf{\isacharunderscore}net\ net{\isachardoublequoteclose}\isanewline
\ \ \ \ \ \ \isakeyword{and}\ fresh{\isacharcolon}\ {\isachardoublequoteopen}fresh{\isacharunderscore}node\ net\ nd{\isacharprime}{\isachardoublequoteclose}\isanewline
\ \ \ \ \ \ \isakeyword{and}\ notin{\isacharcolon}\ {\isachardoublequoteopen}nd{\isacharprime}\ {\isasymnotin}\ S{\isachardoublequoteclose}\isanewline
\ \ \ \ \isakeyword{shows}\ {\isachardoublequoteopen}view{\isacharunderscore}eq{\isacharunderscore}on\ S\ {\isacharparenleft}id{\isacharparenleft}nd\ {\isacharcolon}{\isacharequal}\ nd{\isacharprime}{\isacharparenright}{\isacharcomma}\ net{\isacharparenright}\ {\isacharparenleft}id{\isacharcomma}\ split{\isacharunderscore}node\ nd\ nd{\isacharprime}\ net{\isacharparenright}{\isachardoublequoteclose}%
\isadelimproof
\endisadelimproof
\isatagproof
\endisatagproof
{\isafoldproof}%
\isadelimproof
\endisadelimproof
\isadelimproof
\endisadelimproof
\isatagproof
\endisatagproof
{\isafoldproof}%
\isadelimproof
\endisadelimproof
\begin{isamarkuptext}%
Since a single split preserves equivalence, so does splitting a finite list of nodes
  (\autoref{eq:spliteq}):%
\end{isamarkuptext}\isamarkuptrue%
\isacommand{primrec}\isamarkupfalse%
\ split{\isacharunderscore}all\ {\isacharcolon}{\isacharcolon}\ {\isachardoublequoteopen}nodeid\ list\ {\isasymRightarrow}\ {\isacharparenleft}nodeid\ {\isasymRightarrow}\ nodeid{\isacharparenright}\ {\isasymRightarrow}\ net\ {\isasymRightarrow}\ net{\isachardoublequoteclose}\isanewline
\ \ \isakeyword{where}\ {\isachardoublequoteopen}split{\isacharunderscore}all\ {\isacharbrackleft}{\isacharbrackright}\ {\isacharunderscore}\ net\ {\isacharequal}\ net{\isachardoublequoteclose}\ {\isacharbar}\isanewline
\ \ \ \ \ \ \ \ {\isachardoublequoteopen}split{\isacharunderscore}all\ {\isacharparenleft}nd{\isacharhash}nds{\isacharparenright}\ f\ net\ {\isacharequal}\ split{\isacharunderscore}node\ nd\ {\isacharparenleft}f\ nd{\isacharparenright}\ {\isacharparenleft}split{\isacharunderscore}all\ nds\ f\ net{\isacharparenright}{\isachardoublequoteclose}\isanewline
\isadelimproof
\endisadelimproof
\isatagproof
\endisatagproof
{\isafoldproof}%
\isadelimproof
\endisadelimproof
\isadelimproof
\endisadelimproof
\isatagproof
\endisatagproof
{\isafoldproof}%
\isadelimproof
\endisadelimproof
\isadelimproof
\endisadelimproof
\isatagproof
\endisatagproof
{\isafoldproof}%
\isadelimproof
\isanewline
\endisadelimproof
\isacommand{lemma}\isamarkupfalse%
\ view{\isacharunderscore}eq{\isacharunderscore}split{\isacharunderscore}all{\isacharcolon}\isanewline
\ \ \isakeyword{assumes}\ distinct{\isacharcolon}\ {\isachardoublequoteopen}distinct\ nds{\isachardoublequoteclose}\isanewline
\ \ \ \ \ \ \isakeyword{and}\ nonds{\isacharcolon}\ {\isachardoublequoteopen}{\isasymAnd}nd{\isachardot}\ f\ nd\ {\isasymnotin}\ set\ nds{\isachardoublequoteclose}\isanewline
\ \ \ \ \ \ \isakeyword{and}\ fresh{\isacharcolon}\ {\isachardoublequoteopen}{\isasymAnd}nd{\isachardot}\ fresh{\isacharunderscore}node\ net\ {\isacharparenleft}f\ nd{\isacharparenright}{\isachardoublequoteclose}\isanewline
\ \ \ \ \ \ \isakeyword{and}\ inj{\isacharcolon}\ {\isachardoublequoteopen}inj{\isacharunderscore}on\ f\ {\isacharparenleft}set\ nds{\isacharparenright}{\isachardoublequoteclose}\isanewline
\ \ \ \ \ \ \isakeyword{and}\ wf{\isacharcolon}\ {\isachardoublequoteopen}wf{\isacharunderscore}net\ net{\isachardoublequoteclose}\isanewline
\ \ \ \ \ \ \isakeyword{and}\ noS{\isacharcolon}\ {\isachardoublequoteopen}{\isasymAnd}nd{\isachardot}\ f\ nd\ {\isasymnotin}\ S{\isachardoublequoteclose}\isanewline
\ \ \isakeyword{shows}\ {\isachardoublequoteopen}view{\isacharunderscore}eq{\isacharunderscore}on\ S\ {\isacharparenleft}rename{\isacharunderscore}list\ f\ nds\ {\isacharcomma}\ net{\isacharparenright}\ {\isacharparenleft}id{\isacharcomma}\ split{\isacharunderscore}all\ nds\ f\ net{\isacharparenright}{\isachardoublequoteclose}%
\isadelimproof
\endisadelimproof
\isatagproof
\endisatagproof
{\isafoldproof}%
\isadelimproof
\endisadelimproof
\isadelimproof
\endisadelimproof
\isatagproof
\endisatagproof
{\isafoldproof}%
\isadelimproof
\endisadelimproof
\isadelimproof
\endisadelimproof
\isatagproof
\endisatagproof
{\isafoldproof}%
\isadelimproof
\endisadelimproof
\isadelimproof
\endisadelimproof
\isatagproof
\endisatagproof
{\isafoldproof}%
\isadelimproof
\endisadelimproof
\isadelimproof
\endisadelimproof
\isatagproof
\endisatagproof
{\isafoldproof}%
\isadelimproof
\endisadelimproof
\isadelimproof
\endisadelimproof
\isatagproof
\endisatagproof
{\isafoldproof}%
\isadelimproof
\endisadelimproof
\isadelimproof
\endisadelimproof
\isatagproof
\endisatagproof
{\isafoldproof}%
\isadelimproof
\endisadelimproof
\isadelimproof
\endisadelimproof
\isatagproof
\endisatagproof
{\isafoldproof}%
\isadelimproof
\endisadelimproof
\isadelimproof
\endisadelimproof
\isatagproof
\endisatagproof
{\isafoldproof}%
\isadelimproof
\endisadelimproof
\isadelimproof
\endisadelimproof
\isatagproof
\endisatagproof
{\isafoldproof}%
\isadelimproof
\endisadelimproof
\isadelimproof
\endisadelimproof
\isatagproof
\endisatagproof
{\isafoldproof}%
\isadelimproof
\endisadelimproof
\isadelimproof
\endisadelimproof
\isatagproof
\endisatagproof
{\isafoldproof}%
\isadelimproof
\endisadelimproof
\isadelimproof
\endisadelimproof
\isatagproof
\endisatagproof
{\isafoldproof}%
\isadelimproof
\endisadelimproof
\isadelimproof
\endisadelimproof
\isatagproof
\endisatagproof
{\isafoldproof}%
\isadelimproof
\endisadelimproof
\isadelimtheory
\endisadelimtheory
\isatagtheory
\endisatagtheory
{\isafoldtheory}%
\isadelimtheory
\endisadelimtheory
\end{isabellebody}%

%
\begin{isabellebody}%
\setisabellecontext{Syntax}%
\isadelimtheory
\endisadelimtheory
\isatagtheory
\endisatagtheory
{\isafoldtheory}%
\isadelimtheory
\endisadelimtheory
\isamarkupsubsection{Abstract Syntax for Nets%
}
\isamarkuptrue%
\label{sec:isasyntax}
\begin{isamarkuptext}%
This is the abstract syntax, corresponding to the concrete sytax introduced in
  \autoref{sec:model}.  We do not yet have a parser, and thus models are constructed by hand.%
\end{isamarkuptext}\isamarkuptrue%
\begin{isamarkuptext}%
A contiguous block of addresses, expressed as a base-limit pair:%
\end{isamarkuptext}\isamarkuptrue%
\isacommand{type{\isacharunderscore}synonym}\isamarkupfalse%
\ block{\isacharunderscore}spec\ {\isacharequal}\ {\isachardoublequoteopen}addr\ {\isasymtimes}\ addr{\isachardoublequoteclose}%
\begin{isamarkuptext}%
For each syntax item (nonterminal), we have a translation function into the abstract
  semantic model.  Together these define the parse() function of \autoref{sec:reductions}.%
\end{isamarkuptext}\isamarkuptrue%
\isacommand{definition}\isamarkupfalse%
\ mk{\isacharunderscore}block\ {\isacharcolon}{\isacharcolon}\ {\isachardoublequoteopen}block{\isacharunderscore}spec\ {\isasymRightarrow}\ addr\ set{\isachardoublequoteclose}\isanewline
\ \ \isakeyword{where}\ {\isachardoublequoteopen}mk{\isacharunderscore}block\ s\ {\isacharequal}\ {\isacharbraceleft}a{\isachardot}\ fst\ s\ {\isasymle}\ a\ {\isasymand}\ a\ {\isasymle}\ snd\ s{\isacharbraceright}{\isachardoublequoteclose}%
\begin{isamarkuptext}%
A single block mapping that maps the specified source block to the given destination
  node, beginning at the given base address:%
\end{isamarkuptext}\isamarkuptrue%
\isacommand{record}\isamarkupfalse%
\ map{\isacharunderscore}spec\ {\isacharequal}\isanewline
\ \ src{\isacharunderscore}block\ {\isacharcolon}{\isacharcolon}\ block{\isacharunderscore}spec\isanewline
\ \ dest{\isacharunderscore}node\ {\isacharcolon}{\isacharcolon}\ nodeid\isanewline
\ \ dest{\isacharunderscore}base\ {\isacharcolon}{\isacharcolon}\ addr%
\begin{isamarkuptext}%
Map a block without changing its base address:%
\end{isamarkuptext}\isamarkuptrue%
\isacommand{definition}\isamarkupfalse%
\ direct{\isacharunderscore}map\ {\isacharcolon}{\isacharcolon}\ {\isachardoublequoteopen}block{\isacharunderscore}spec\ {\isasymRightarrow}\ nodeid\ {\isasymRightarrow}\ map{\isacharunderscore}spec{\isachardoublequoteclose}\isanewline
\ \ \isakeyword{where}\ {\isachardoublequoteopen}direct{\isacharunderscore}map\ block\ node\ {\isacharequal}\ {\isasymlparr}\ src{\isacharunderscore}block\ {\isacharequal}\ block{\isacharcomma}\ dest{\isacharunderscore}node\ {\isacharequal}\ node{\isacharcomma}\ dest{\isacharunderscore}base\ {\isacharequal}\ fst\ block\ {\isasymrparr}{\isachardoublequoteclose}\isanewline
\isanewline
\isacommand{definition}\isamarkupfalse%
\ block{\isacharunderscore}map\ {\isacharcolon}{\isacharcolon}\ {\isachardoublequoteopen}block{\isacharunderscore}spec\ {\isasymRightarrow}\ nodeid\ {\isasymRightarrow}\ addr\ {\isasymRightarrow}\ map{\isacharunderscore}spec{\isachardoublequoteclose}\isanewline
\ \ \isakeyword{where}\ {\isachardoublequoteopen}block{\isacharunderscore}map\ block\ node\ base\ {\isacharequal}\ {\isasymlparr}\ src{\isacharunderscore}block\ {\isacharequal}\ block{\isacharcomma}\ dest{\isacharunderscore}node\ {\isacharequal}\ node{\isacharcomma}\ dest{\isacharunderscore}base\ {\isacharequal}\ base\ {\isasymrparr}{\isachardoublequoteclose}\isanewline
\ \ \ \ \isanewline
\isacommand{definition}\isamarkupfalse%
\ one{\isacharunderscore}map\ {\isacharcolon}{\isacharcolon}\ {\isachardoublequoteopen}addr\ {\isasymRightarrow}\ nodeid\ {\isasymRightarrow}\ addr\ {\isasymRightarrow}\ map{\isacharunderscore}spec{\isachardoublequoteclose}\isanewline
\ \ \isakeyword{where}\ {\isachardoublequoteopen}one{\isacharunderscore}map\ src\ node\ base\ {\isacharequal}\ {\isasymlparr}\ src{\isacharunderscore}block\ {\isacharequal}\ {\isacharparenleft}src{\isacharcomma}src{\isacharparenright}{\isacharcomma}\ dest{\isacharunderscore}node\ {\isacharequal}\ node{\isacharcomma}\ dest{\isacharunderscore}base\ {\isacharequal}\ base\ {\isasymrparr}{\isachardoublequoteclose}\isanewline
\isanewline
\isacommand{definition}\isamarkupfalse%
\ mk{\isacharunderscore}map\ {\isacharcolon}{\isacharcolon}\ {\isachardoublequoteopen}map{\isacharunderscore}spec\ {\isasymRightarrow}\ addr\ {\isasymRightarrow}\ name\ set{\isachardoublequoteclose}\isanewline
\ \ \isakeyword{where}\ {\isachardoublequoteopen}mk{\isacharunderscore}map\ s\ {\isacharequal}\isanewline
\ \ \ \ {\isacharparenleft}{\isasymlambda}a{\isachardot}\ if\ a\ {\isasymin}\ mk{\isacharunderscore}block\ {\isacharparenleft}src{\isacharunderscore}block\ s{\isacharparenright}\isanewline
\ \ \ \ \ \ then\ {\isacharbraceleft}{\isacharparenleft}dest{\isacharunderscore}node\ s{\isacharcomma}\ dest{\isacharunderscore}base\ s\ {\isacharplus}\ {\isacharparenleft}a\ {\isacharminus}\ fst\ {\isacharparenleft}src{\isacharunderscore}block\ s{\isacharparenright}{\isacharparenright}{\isacharparenright}{\isacharbraceright}\isanewline
\ \ \ \ \ \ else\ {\isacharbraceleft}{\isacharbraceright}{\isacharparenright}{\isachardoublequoteclose}%
\begin{isamarkuptext}%
A finitely-specified decoding node, with a list of blocks to accept locally, and a
  list of those to translate:%
\end{isamarkuptext}\isamarkuptrue%
\isacommand{record}\isamarkupfalse%
\ node{\isacharunderscore}spec\ {\isacharequal}\isanewline
\ \ acc{\isacharunderscore}blocks\ {\isacharcolon}{\isacharcolon}\ {\isachardoublequoteopen}block{\isacharunderscore}spec\ list{\isachardoublequoteclose}\isanewline
\ \ map{\isacharunderscore}blocks\ {\isacharcolon}{\isacharcolon}\ {\isachardoublequoteopen}map{\isacharunderscore}spec\ list{\isachardoublequoteclose}\isanewline
\ \ overlay\ \ \ \ {\isacharcolon}{\isacharcolon}\ {\isachardoublequoteopen}nodeid\ option{\isachardoublequoteclose}\isanewline
\isanewline
\isacommand{definition}\isamarkupfalse%
\ empty{\isacharunderscore}spec\ {\isacharcolon}{\isacharcolon}\ {\isachardoublequoteopen}node{\isacharunderscore}spec{\isachardoublequoteclose}\isanewline
\ \ \isakeyword{where}\ {\isachardoublequoteopen}empty{\isacharunderscore}spec\ {\isacharequal}\ {\isasymlparr}\ acc{\isacharunderscore}blocks\ {\isacharequal}\ {\isacharbrackleft}{\isacharbrackright}{\isacharcomma}\ map{\isacharunderscore}blocks\ {\isacharequal}\ {\isacharbrackleft}{\isacharbrackright}{\isacharcomma}\ overlay\ {\isacharequal}\ None\ {\isasymrparr}{\isachardoublequoteclose}%
\isadelimproof
\endisadelimproof
\isatagproof
\endisatagproof
{\isafoldproof}%
\isadelimproof
\endisadelimproof
\begin{isamarkuptext}%
If an overlay node is specified, initialise the map by forwarding all addresses to that
  node:%
\end{isamarkuptext}\isamarkuptrue%
\isacommand{definition}\isamarkupfalse%
\ mk{\isacharunderscore}overlay\ {\isacharcolon}{\isacharcolon}\ {\isachardoublequoteopen}nodeid\ option\ {\isasymRightarrow}\ node{\isachardoublequoteclose}\isanewline
\ \ \isakeyword{where}\ {\isachardoublequoteopen}mk{\isacharunderscore}overlay\ ov\ {\isacharequal}\ {\isasymlparr}\isanewline
\ \ \ \ \ \ \ \ \ \ accept\ {\isacharequal}\ {\isacharbraceleft}{\isacharbraceright}{\isacharcomma}\isanewline
\ \ \ \ \ \ \ \ \ \ translate\ {\isacharequal}\ {\isacharparenleft}case\ ov\ of\ None\ {\isasymRightarrow}\ {\isasymlambda}{\isacharunderscore}{\isachardot}\ {\isacharbraceleft}{\isacharbraceright}\ {\isacharbar}\ Some\ n\ {\isasymRightarrow}\ {\isacharparenleft}{\isasymlambda}a{\isachardot}\ {\isacharbraceleft}{\isacharparenleft}n{\isacharcomma}a{\isacharparenright}{\isacharbraceright}{\isacharparenright}{\isacharparenright}\ {\isasymrparr}{\isachardoublequoteclose}\isanewline
\isadelimproof
\endisadelimproof
\isatagproof
\endisatagproof
{\isafoldproof}%
\isadelimproof
\isanewline
\endisadelimproof
\isacommand{primrec}\isamarkupfalse%
\ add{\isacharunderscore}blocks\ {\isacharcolon}{\isacharcolon}\ {\isachardoublequoteopen}block{\isacharunderscore}spec\ list\ {\isasymRightarrow}\ node\ {\isasymRightarrow}\ node{\isachardoublequoteclose}\isanewline
\ \ \isakeyword{where}\ {\isachardoublequoteopen}add{\isacharunderscore}blocks\ {\isacharbrackleft}{\isacharbrackright}\ node\ {\isacharequal}\ node{\isachardoublequoteclose}\ {\isacharbar}\isanewline
\ \ \ \ \ \ \ \ {\isachardoublequoteopen}add{\isacharunderscore}blocks\ {\isacharparenleft}s{\isacharhash}ss{\isacharparenright}\ node\ {\isacharequal}\ accept{\isacharunderscore}update\ {\isacharparenleft}op\ {\isasymunion}\ {\isacharparenleft}mk{\isacharunderscore}block\ s{\isacharparenright}{\isacharparenright}\ {\isacharparenleft}add{\isacharunderscore}blocks\ ss\ node{\isacharparenright}{\isachardoublequoteclose}\isanewline
\isadelimproof
\endisadelimproof
\isatagproof
\endisatagproof
{\isafoldproof}%
\isadelimproof
\endisadelimproof
\isadelimproof
\endisadelimproof
\isatagproof
\endisatagproof
{\isafoldproof}%
\isadelimproof
\isanewline
\endisadelimproof
\isacommand{primrec}\isamarkupfalse%
\ add{\isacharunderscore}maps\ {\isacharcolon}{\isacharcolon}\ {\isachardoublequoteopen}map{\isacharunderscore}spec\ list\ {\isasymRightarrow}\ node\ {\isasymRightarrow}\ node{\isachardoublequoteclose}\isanewline
\ \ \isakeyword{where}\ {\isachardoublequoteopen}add{\isacharunderscore}maps\ {\isacharbrackleft}{\isacharbrackright}\ node\ {\isacharequal}\ node{\isachardoublequoteclose}\ {\isacharbar}\isanewline
\ \ \ \ \ \ \ \ {\isachardoublequoteopen}add{\isacharunderscore}maps\ {\isacharparenleft}s{\isacharhash}ss{\isacharparenright}\ node\ {\isacharequal}\ translate{\isacharunderscore}update\ {\isacharparenleft}{\isasymlambda}t\ a{\isachardot}\ mk{\isacharunderscore}map\ s\ a\ {\isasymunion}\ t\ a{\isacharparenright}\ {\isacharparenleft}add{\isacharunderscore}maps\ ss\ node{\isacharparenright}{\isachardoublequoteclose}\isanewline
\isadelimproof
\endisadelimproof
\isatagproof
\endisatagproof
{\isafoldproof}%
\isadelimproof
\endisadelimproof
\isadelimproof
\endisadelimproof
\isatagproof
\endisatagproof
{\isafoldproof}%
\isadelimproof
\isanewline
\endisadelimproof
\isacommand{definition}\isamarkupfalse%
\ mk{\isacharunderscore}node\ {\isacharcolon}{\isacharcolon}\ {\isachardoublequoteopen}node{\isacharunderscore}spec\ {\isasymRightarrow}\ node{\isachardoublequoteclose}\isanewline
\ \ \isakeyword{where}\ {\isachardoublequoteopen}mk{\isacharunderscore}node\ s\ {\isacharequal}\ add{\isacharunderscore}maps\ {\isacharparenleft}map{\isacharunderscore}blocks\ s{\isacharparenright}\ {\isacharparenleft}add{\isacharunderscore}blocks\ {\isacharparenleft}acc{\isacharunderscore}blocks\ s{\isacharparenright}\ {\isacharparenleft}mk{\isacharunderscore}overlay\ {\isacharparenleft}overlay\ s{\isacharparenright}{\isacharparenright}{\isacharparenright}{\isachardoublequoteclose}\isanewline
\isadelimproof
\endisadelimproof
\isatagproof
\endisatagproof
{\isafoldproof}%
\isadelimproof
\endisadelimproof
\isadelimproof
\endisadelimproof
\isatagproof
\endisatagproof
{\isafoldproof}%
\isadelimproof
\endisadelimproof
\isadelimproof
\endisadelimproof
\isatagproof
\endisatagproof
{\isafoldproof}%
\isadelimproof
\isanewline
\endisadelimproof
\isacommand{type{\isacharunderscore}synonym}\isamarkupfalse%
\ net{\isacharunderscore}spec\ {\isacharequal}\ {\isachardoublequoteopen}{\isacharparenleft}nodeid\ {\isasymtimes}\ node{\isacharunderscore}spec{\isacharparenright}\ list{\isachardoublequoteclose}\isanewline
\isanewline
\isacommand{definition}\isamarkupfalse%
\ {\isachardoublequoteopen}empty{\isacharunderscore}net\ {\isacharequal}\ {\isacharparenleft}{\isasymlambda}{\isacharunderscore}{\isachardot}\ empty{\isacharunderscore}node{\isacharparenright}{\isachardoublequoteclose}\isanewline
\isanewline
\isacommand{primrec}\isamarkupfalse%
\ repeat{\isacharunderscore}node\ {\isacharcolon}{\isacharcolon}\ {\isachardoublequoteopen}node{\isacharunderscore}spec\ {\isasymRightarrow}\ nodeid\ {\isasymRightarrow}\ nat\ {\isasymRightarrow}\ net{\isacharunderscore}spec{\isachardoublequoteclose}\isanewline
\ \ \isakeyword{where}\ {\isachardoublequoteopen}repeat{\isacharunderscore}node\ node\ base\ {\isadigit{0}}\ {\isacharequal}\ {\isacharbrackleft}{\isacharbrackright}{\isachardoublequoteclose}\ {\isacharbar}\isanewline
\ \ \ \ \ \ \ \ {\isachardoublequoteopen}repeat{\isacharunderscore}node\ node\ base\ {\isacharparenleft}Suc\ n{\isacharparenright}\ {\isacharequal}\ {\isacharparenleft}base{\isacharcomma}\ node{\isacharparenright}\ {\isacharhash}\ repeat{\isacharunderscore}node\ node\ {\isacharparenleft}Suc\ base{\isacharparenright}\ n{\isachardoublequoteclose}\isanewline
\isanewline
\isacommand{primrec}\isamarkupfalse%
\ mk{\isacharunderscore}net\ {\isacharcolon}{\isacharcolon}\ {\isachardoublequoteopen}net{\isacharunderscore}spec\ {\isasymRightarrow}\ net{\isachardoublequoteclose}\isanewline
\ \ \isakeyword{where}\ {\isachardoublequoteopen}mk{\isacharunderscore}net\ {\isacharbrackleft}{\isacharbrackright}\ {\isacharequal}\ empty{\isacharunderscore}net{\isachardoublequoteclose}\ {\isacharbar}\isanewline
\ \ \ \ \ \ \ \ {\isachardoublequoteopen}mk{\isacharunderscore}net\ {\isacharparenleft}s{\isacharhash}ss{\isacharparenright}\ {\isacharequal}\ {\isacharparenleft}mk{\isacharunderscore}net\ ss{\isacharparenright}{\isacharparenleft}fst\ s\ {\isacharcolon}{\isacharequal}\ mk{\isacharunderscore}node\ {\isacharparenleft}snd\ s{\isacharparenright}{\isacharparenright}{\isachardoublequoteclose}%
\isadelimproof
\endisadelimproof
\isatagproof
\endisatagproof
{\isafoldproof}%
\isadelimproof
\endisadelimproof
\isadelimproof
\endisadelimproof
\isatagproof
\endisatagproof
{\isafoldproof}%
\isadelimproof
\endisadelimproof
\isadelimproof
\endisadelimproof
\isatagproof
\endisatagproof
{\isafoldproof}%
\isadelimproof
\endisadelimproof
\isadelimproof
\endisadelimproof
\isatagproof
\endisatagproof
{\isafoldproof}%
\isadelimproof
\endisadelimproof
\isadelimproof
\endisadelimproof
\isatagproof
\endisatagproof
{\isafoldproof}%
\isadelimproof
\endisadelimproof
\isadelimproof
\endisadelimproof
\isatagproof
\endisatagproof
{\isafoldproof}%
\isadelimproof
\endisadelimproof
\isadelimproof
\endisadelimproof
\isatagproof
\endisatagproof
{\isafoldproof}%
\isadelimproof
\endisadelimproof
\isadelimproof
\endisadelimproof
\isatagproof
\endisatagproof
{\isafoldproof}%
\isadelimproof
\endisadelimproof
\isadelimproof
\endisadelimproof
\isatagproof
\endisatagproof
{\isafoldproof}%
\isadelimproof
\endisadelimproof
\begin{isamarkuptext}%
Nets built from abstract syntax are correct by construction:%
\end{isamarkuptext}\isamarkuptrue%
\isacommand{lemma}\isamarkupfalse%
\ wf{\isacharunderscore}mk{\isacharunderscore}net{\isacharcolon}\isanewline
\ \ {\isachardoublequoteopen}wf{\isacharunderscore}net\ {\isacharparenleft}mk{\isacharunderscore}net\ ss{\isacharparenright}{\isachardoublequoteclose}%
\isadelimproof
\endisadelimproof
\isatagproof
\endisatagproof
{\isafoldproof}%
\isadelimproof
\endisadelimproof
\isamarkupsubsubsection{Finding Fresh Nodes%
}
\isamarkuptrue%
\begin{isamarkuptext}%
These functions are guaranteed to return a node that's unused in the supplied
  specification.%
\end{isamarkuptext}\isamarkuptrue%
\isacommand{definition}\isamarkupfalse%
\ ff{\isacharunderscore}overlay\ {\isacharcolon}{\isacharcolon}\ {\isachardoublequoteopen}nodeid\ option\ {\isasymRightarrow}\ nodeid{\isachardoublequoteclose}\isanewline
\ \ \isakeyword{where}\ {\isachardoublequoteopen}ff{\isacharunderscore}overlay\ s\ {\isacharequal}\ {\isacharparenleft}case\ s\ of\ Some\ nd\ {\isasymRightarrow}\ Suc\ nd\ {\isacharbar}\ None\ {\isasymRightarrow}\ {\isadigit{0}}{\isacharparenright}{\isachardoublequoteclose}\isanewline
\isadelimproof
\endisadelimproof
\isatagproof
\endisatagproof
{\isafoldproof}%
\isadelimproof
\isanewline
\endisadelimproof
\isacommand{primrec}\isamarkupfalse%
\ ff{\isacharunderscore}map\ {\isacharcolon}{\isacharcolon}\ {\isachardoublequoteopen}map{\isacharunderscore}spec\ list\ {\isasymRightarrow}\ nodeid{\isachardoublequoteclose}\isanewline
\ \ \isakeyword{where}\ {\isachardoublequoteopen}ff{\isacharunderscore}map\ {\isacharbrackleft}{\isacharbrackright}\ {\isacharequal}\ {\isadigit{0}}{\isachardoublequoteclose}\ {\isacharbar}\isanewline
\ \ \ \ \ \ \ \ {\isachardoublequoteopen}ff{\isacharunderscore}map\ {\isacharparenleft}s{\isacharhash}ss{\isacharparenright}\ {\isacharequal}\ max\ {\isacharparenleft}Suc\ {\isacharparenleft}dest{\isacharunderscore}node\ s{\isacharparenright}{\isacharparenright}\ {\isacharparenleft}ff{\isacharunderscore}map\ ss{\isacharparenright}{\isachardoublequoteclose}\isanewline
\isadelimproof
\endisadelimproof
\isatagproof
\endisatagproof
{\isafoldproof}%
\isadelimproof
\endisadelimproof
\isadelimproof
\endisadelimproof
\isatagproof
\endisatagproof
{\isafoldproof}%
\isadelimproof
\isanewline
\endisadelimproof
\isacommand{definition}\isamarkupfalse%
\ ff{\isacharunderscore}node\ {\isacharcolon}{\isacharcolon}\ {\isachardoublequoteopen}node{\isacharunderscore}spec\ {\isasymRightarrow}\ nodeid{\isachardoublequoteclose}\isanewline
\ \ \isakeyword{where}\ {\isachardoublequoteopen}ff{\isacharunderscore}node\ s\ {\isacharequal}\ max\ {\isacharparenleft}ff{\isacharunderscore}overlay\ {\isacharparenleft}overlay\ s{\isacharparenright}{\isacharparenright}\ {\isacharparenleft}ff{\isacharunderscore}map\ {\isacharparenleft}map{\isacharunderscore}blocks\ s{\isacharparenright}{\isacharparenright}{\isachardoublequoteclose}\isanewline
\isadelimproof
\endisadelimproof
\isatagproof
\endisatagproof
{\isafoldproof}%
\isadelimproof
\isanewline
\endisadelimproof
\isacommand{primrec}\isamarkupfalse%
\ ff{\isacharunderscore}net\ {\isacharcolon}{\isacharcolon}\ {\isachardoublequoteopen}net{\isacharunderscore}spec\ {\isasymRightarrow}\ nodeid{\isachardoublequoteclose}\isanewline
\ \ \isakeyword{where}\ {\isachardoublequoteopen}ff{\isacharunderscore}net\ {\isacharbrackleft}{\isacharbrackright}\ {\isacharequal}\ {\isadigit{0}}{\isachardoublequoteclose}\ {\isacharbar}\isanewline
\ \ \ \ \ \ \ \ {\isachardoublequoteopen}ff{\isacharunderscore}net\ {\isacharparenleft}s{\isacharhash}ss{\isacharparenright}\ {\isacharequal}\ Max\ {\isacharbraceleft}ff{\isacharunderscore}node\ {\isacharparenleft}snd\ s{\isacharparenright}{\isacharcomma}\ ff{\isacharunderscore}net\ ss{\isacharcomma}\ Suc\ {\isacharparenleft}fst\ s{\isacharparenright}{\isacharbraceright}{\isachardoublequoteclose}\isanewline
\isadelimproof
\endisadelimproof
\isatagproof
\endisatagproof
{\isafoldproof}%
\isadelimproof
\endisadelimproof
\isadelimproof
\endisadelimproof
\isatagproof
\endisatagproof
{\isafoldproof}%
\isadelimproof
\endisadelimproof
\isadelimproof
\endisadelimproof
\isatagproof
\endisatagproof
{\isafoldproof}%
\isadelimproof
\endisadelimproof
\isadelimproof
\endisadelimproof
\isatagproof
\endisatagproof
{\isafoldproof}%
\isadelimproof
\endisadelimproof
\isadelimproof
\endisadelimproof
\isatagproof
\endisatagproof
{\isafoldproof}%
\isadelimproof
\endisadelimproof
\isadelimtheory
\endisadelimtheory
\isatagtheory
\endisatagtheory
{\isafoldtheory}%
\isadelimtheory
\endisadelimtheory
\end{isabellebody}%

%
\begin{isabellebody}%
\setisabellecontext{ConcreteOps}%
\isadelimtheory
\endisadelimtheory
\isatagtheory
\endisatagtheory
{\isafoldtheory}%
\isadelimtheory
\endisadelimtheory
\isamarkupsubsection{Transformations on Abstract Syntax%
}
\isamarkuptrue%
\label{sec:isasplitC}
\begin{isamarkuptext}%
These are the equivalents, in abstract syntax, of the abstract acceptor
        and translator nodes.  Somewhat confusingly, we will now refer to the
        \emph{abstract} syntax as the \emph{concrete} model (relative to the
        abstract directed graph model).%
\end{isamarkuptext}\isamarkuptrue%
\isacommand{primrec}\isamarkupfalse%
\ remap{\isacharunderscore}all\ {\isacharcolon}{\isacharcolon}\ {\isachardoublequoteopen}nodeid\ {\isasymRightarrow}\ block{\isacharunderscore}spec\ list\ {\isasymRightarrow}\ map{\isacharunderscore}spec\ list{\isachardoublequoteclose}\isanewline
\ \ \isakeyword{where}\ {\isachardoublequoteopen}remap{\isacharunderscore}all\ {\isacharunderscore}\ {\isacharbrackleft}{\isacharbrackright}\ {\isacharequal}\ {\isacharbrackleft}{\isacharbrackright}{\isachardoublequoteclose}\ {\isacharbar}\isanewline
\ \ \ \ \ \ \ \ {\isachardoublequoteopen}remap{\isacharunderscore}all\ nd\ {\isacharparenleft}b{\isacharhash}bs{\isacharparenright}\ {\isacharequal}\ direct{\isacharunderscore}map\ b\ nd\ {\isacharhash}\ remap{\isacharunderscore}all\ nd\ bs{\isachardoublequoteclose}\isanewline
\isanewline
\isacommand{definition}\isamarkupfalse%
\ redirector{\isacharunderscore}node{\isacharunderscore}C\ {\isacharcolon}{\isacharcolon}\ {\isachardoublequoteopen}nodeid\ {\isasymRightarrow}\ node{\isacharunderscore}spec\ {\isasymRightarrow}\ node{\isacharunderscore}spec{\isachardoublequoteclose}\isanewline
\ \ \isakeyword{where}\ {\isachardoublequoteopen}redirector{\isacharunderscore}node{\isacharunderscore}C\ nd{\isacharprime}\ ns\ {\isacharequal}\isanewline
\ \ \ \ \ \ \ \ \ \ empty{\isacharunderscore}spec\ {\isasymlparr}\ map{\isacharunderscore}blocks\ {\isacharcolon}{\isacharequal}\ remap{\isacharunderscore}all\ nd{\isacharprime}\ {\isacharparenleft}acc{\isacharunderscore}blocks\ ns{\isacharparenright}\ {\isacharat}\ map{\isacharunderscore}blocks\ ns{\isacharcomma}\isanewline
\ \ \ \ \ \ \ \ \ \ \ \ \ \ \ \ \ \ \ \ \ \ \ overlay\ {\isacharcolon}{\isacharequal}\ overlay\ ns\ {\isasymrparr}{\isachardoublequoteclose}\isanewline
\isanewline
\isacommand{definition}\isamarkupfalse%
\ acceptor{\isacharunderscore}node{\isacharunderscore}C\ {\isacharcolon}{\isacharcolon}\ {\isachardoublequoteopen}node{\isacharunderscore}spec\ {\isasymRightarrow}\ node{\isacharunderscore}spec{\isachardoublequoteclose}\isanewline
\ \ \isakeyword{where}\ {\isachardoublequoteopen}acceptor{\isacharunderscore}node{\isacharunderscore}C\ ns\ {\isacharequal}\ empty{\isacharunderscore}spec\ {\isasymlparr}\ acc{\isacharunderscore}blocks\ {\isacharcolon}{\isacharequal}\ acc{\isacharunderscore}blocks\ ns\ {\isasymrparr}{\isachardoublequoteclose}\isanewline
\isadelimproof
\endisadelimproof
\isatagproof
\endisatagproof
{\isafoldproof}%
\isadelimproof
\endisadelimproof
\isadelimproof
\endisadelimproof
\isatagproof
\endisatagproof
{\isafoldproof}%
\isadelimproof
\endisadelimproof
\isadelimproof
\endisadelimproof
\isatagproof
\endisatagproof
{\isafoldproof}%
\isadelimproof
\endisadelimproof
\isadelimproof
\endisadelimproof
\isatagproof
\endisatagproof
{\isafoldproof}%
\isadelimproof
\endisadelimproof
\begin{isamarkuptext}%
The concrete redirector node refines the abstract:%
\end{isamarkuptext}\isamarkuptrue%
\isacommand{lemma}\isamarkupfalse%
\ redirector{\isacharunderscore}rel{\isacharcolon}\isanewline
\ \ {\isachardoublequoteopen}mk{\isacharunderscore}node\ {\isacharparenleft}redirector{\isacharunderscore}node{\isacharunderscore}C\ nd{\isacharprime}\ ns{\isacharparenright}\ {\isacharequal}\ redirector{\isacharunderscore}node\ nd{\isacharprime}\ {\isacharparenleft}mk{\isacharunderscore}node\ ns{\isacharparenright}{\isachardoublequoteclose}%
\isadelimproof
\endisadelimproof
\isatagproof
\endisatagproof
{\isafoldproof}%
\isadelimproof
\endisadelimproof
\begin{isamarkuptext}%
The concrete acceptor node refines the abstract:%
\end{isamarkuptext}\isamarkuptrue%
\isacommand{lemma}\isamarkupfalse%
\ acceptor{\isacharunderscore}rel{\isacharcolon}\isanewline
\ \ {\isachardoublequoteopen}mk{\isacharunderscore}node\ {\isacharparenleft}acceptor{\isacharunderscore}node{\isacharunderscore}C\ ns{\isacharparenright}\ {\isacharequal}\ acceptor{\isacharunderscore}node\ {\isacharparenleft}mk{\isacharunderscore}node\ ns{\isacharparenright}{\isachardoublequoteclose}%
\isadelimproof
\endisadelimproof
\isatagproof
\endisatagproof
{\isafoldproof}%
\isadelimproof
\endisadelimproof
\isacommand{primrec}\isamarkupfalse%
\ split{\isacharunderscore}all{\isacharunderscore}C\ {\isacharcolon}{\isacharcolon}\ {\isachardoublequoteopen}nodeid\ {\isasymRightarrow}\ net{\isacharunderscore}spec\ {\isasymRightarrow}\ net{\isacharunderscore}spec{\isachardoublequoteclose}\isanewline
\ \ \isakeyword{where}\ {\isachardoublequoteopen}split{\isacharunderscore}all{\isacharunderscore}C\ {\isacharunderscore}\ {\isacharbrackleft}{\isacharbrackright}\ {\isacharequal}\ {\isacharbrackleft}{\isacharbrackright}{\isachardoublequoteclose}\ {\isacharbar}\isanewline
\ \ \ \ \ \ \ \ {\isachardoublequoteopen}split{\isacharunderscore}all{\isacharunderscore}C\ off\ {\isacharparenleft}s{\isacharhash}ss{\isacharparenright}\ {\isacharequal}\isanewline
\ \ \ \ \ \ \ \ \ \ {\isacharbrackleft}{\isacharparenleft}off\ {\isacharplus}\ fst\ s{\isacharcomma}\ acceptor{\isacharunderscore}node{\isacharunderscore}C\ {\isacharparenleft}snd\ s{\isacharparenright}{\isacharparenright}{\isacharcomma}\isanewline
\ \ \ \ \ \ \ \ \ \ \ {\isacharparenleft}fst\ s{\isacharcomma}\ redirector{\isacharunderscore}node{\isacharunderscore}C\ {\isacharparenleft}off\ {\isacharplus}\ fst\ s{\isacharparenright}\ {\isacharparenleft}snd\ s{\isacharparenright}{\isacharparenright}{\isacharbrackright}\ {\isacharat}\ split{\isacharunderscore}all{\isacharunderscore}C\ off\ ss{\isachardoublequoteclose}%
\begin{isamarkuptext}%
Thus (by induction), splitting all nodes in the concrete spec is equivalent to doing so
  in the abstract (\autoref{eq:splitrefine}):%
\end{isamarkuptext}\isamarkuptrue%
\isacommand{lemma}\isamarkupfalse%
\ split{\isacharunderscore}all{\isacharunderscore}rel{\isacharcolon}\isanewline
\ \ \isakeyword{assumes}\ {\isachardoublequoteopen}distinct\ {\isacharparenleft}map\ fst\ ss{\isacharparenright}{\isachardoublequoteclose}\isanewline
\ \ \ \ \ \ \isakeyword{and}\ \ {\isachardoublequoteopen}{\isasymAnd}nd{\isachardot}\ nd\ {\isasymin}\ set\ {\isacharparenleft}map\ fst\ ss{\isacharparenright}\ {\isasymLongrightarrow}\ nd\ {\isacharless}\ off{\isachardoublequoteclose}\isanewline
\ \ \ \ \isakeyword{shows}\ {\isachardoublequoteopen}mk{\isacharunderscore}net\ {\isacharparenleft}split{\isacharunderscore}all{\isacharunderscore}C\ off\ ss{\isacharparenright}\ {\isacharequal}\ split{\isacharunderscore}all\ {\isacharparenleft}map\ fst\ ss{\isacharparenright}\ {\isacharparenleft}op\ {\isacharplus}\ off{\isacharparenright}\ {\isacharparenleft}mk{\isacharunderscore}net\ ss{\isacharparenright}{\isachardoublequoteclose}%
\isadelimproof
\endisadelimproof
\isatagproof
\endisatagproof
{\isafoldproof}%
\isadelimproof
\endisadelimproof
\begin{isamarkuptext}%
Finally, we have refinement between the concrete and abstract split operations on whole
  nets (\autoref{eq:fulleq}):%
\end{isamarkuptext}\isamarkuptrue%
\isacommand{definition}\isamarkupfalse%
\ split{\isacharunderscore}net{\isacharunderscore}C\ {\isacharcolon}{\isacharcolon}\ {\isachardoublequoteopen}net{\isacharunderscore}spec\ {\isasymRightarrow}\ net{\isacharunderscore}spec{\isachardoublequoteclose}\isanewline
\ \ \isakeyword{where}\ {\isachardoublequoteopen}split{\isacharunderscore}net{\isacharunderscore}C\ s\ {\isacharequal}\ split{\isacharunderscore}all{\isacharunderscore}C\ {\isacharparenleft}ff{\isacharunderscore}net\ s{\isacharparenright}\ s{\isachardoublequoteclose}\isanewline
\isanewline
\isacommand{lemma}\isamarkupfalse%
\ split{\isacharunderscore}net{\isacharunderscore}equiv{\isacharcolon}\isanewline
\ \ \isakeyword{assumes}\ distinct{\isacharcolon}\ {\isachardoublequoteopen}distinct\ {\isacharparenleft}map\ fst\ s{\isacharparenright}{\isachardoublequoteclose}\isanewline
\ \ \isakeyword{shows}\ {\isachardoublequoteopen}view{\isacharunderscore}eq{\isacharunderscore}on\ {\isacharbraceleft}{\isadigit{0}}{\isachardot}{\isachardot}{\isacharless}ff{\isacharunderscore}net\ s{\isacharbraceright}\ {\isacharparenleft}rename{\isacharunderscore}list\ {\isacharparenleft}op\ {\isacharplus}\ {\isacharparenleft}ff{\isacharunderscore}net\ s{\isacharparenright}{\isacharparenright}\ {\isacharparenleft}map\ fst\ s{\isacharparenright}{\isacharcomma}\ mk{\isacharunderscore}net\ s{\isacharparenright}\ {\isacharparenleft}id{\isacharcomma}\ mk{\isacharunderscore}net\ {\isacharparenleft}split{\isacharunderscore}net{\isacharunderscore}C\ s{\isacharparenright}{\isacharparenright}{\isachardoublequoteclose}\isanewline
\isadelimproof
\endisadelimproof
\isatagproof
\isacommand{proof}\isamarkupfalse%
\ {\isacharminus}%
\begin{isamarkuptext}%
The abstract split preserves view equivalence for any node that was defined in the
    original spec:%
\end{isamarkuptext}\isamarkuptrue%
\ \ \isacommand{have}\isamarkupfalse%
\ {\isachardoublequoteopen}view{\isacharunderscore}eq{\isacharunderscore}on\ {\isacharbraceleft}{\isadigit{0}}{\isachardot}{\isachardot}{\isacharless}ff{\isacharunderscore}net\ s{\isacharbraceright}\ {\isacharparenleft}rename{\isacharunderscore}list\ {\isacharparenleft}op\ {\isacharplus}\ {\isacharparenleft}ff{\isacharunderscore}net\ s{\isacharparenright}{\isacharparenright}\ {\isacharparenleft}map\ fst\ s{\isacharparenright}{\isacharcomma}\ mk{\isacharunderscore}net\ s{\isacharparenright}\isanewline
\ \ \ \ \ \ \ \ \ \ \ \ \ \ \ \ \ \ \ \ \ \ \ \ \ \ \ \ \ \ \ \ \ \ {\isacharparenleft}id{\isacharcomma}\ split{\isacharunderscore}all\ {\isacharparenleft}map\ fst\ s{\isacharparenright}\ {\isacharparenleft}op\ {\isacharplus}\ {\isacharparenleft}ff{\isacharunderscore}net\ s{\isacharparenright}{\isacharparenright}\ {\isacharparenleft}mk{\isacharunderscore}net\ s{\isacharparenright}{\isacharparenright}{\isachardoublequoteclose}\isanewline
\ \ \ \ \isacommand{using}\isamarkupfalse%
\ ff{\isacharunderscore}net{\isacharunderscore}idbound\ distinct\ wf{\isacharunderscore}mk{\isacharunderscore}net\isanewline
\ \ \ \ \isacommand{by}\isamarkupfalse%
{\isacharparenleft}intro\ view{\isacharunderscore}eq{\isacharunderscore}split{\isacharunderscore}all{\isacharcomma}\ auto\ intro{\isacharcolon}ff{\isacharunderscore}net{\isacharunderscore}fresh\ view{\isacharunderscore}eq{\isacharunderscore}split{\isacharunderscore}all{\isacharparenright}%
\begin{isamarkuptext}%
The concrete split refines the abstract split:%
\end{isamarkuptext}\isamarkuptrue%
\ \ \isacommand{moreover}\isamarkupfalse%
\ \isacommand{have}\isamarkupfalse%
\ {\isachardoublequoteopen}mk{\isacharunderscore}net\ {\isacharparenleft}split{\isacharunderscore}all{\isacharunderscore}C\ {\isacharparenleft}ff{\isacharunderscore}net\ s{\isacharparenright}\ s{\isacharparenright}\ {\isacharequal}\ split{\isacharunderscore}all\ {\isacharparenleft}map\ fst\ s{\isacharparenright}\ {\isacharparenleft}op\ {\isacharplus}\ {\isacharparenleft}ff{\isacharunderscore}net\ s{\isacharparenright}{\isacharparenright}\ {\isacharparenleft}mk{\isacharunderscore}net\ s{\isacharparenright}{\isachardoublequoteclose}\isanewline
\ \ \ \ \isacommand{by}\isamarkupfalse%
{\isacharparenleft}auto\ intro{\isacharcolon}split{\isacharunderscore}all{\isacharunderscore}rel\ distinct\ ff{\isacharunderscore}net{\isacharunderscore}idbound{\isacharparenright}%
\begin{isamarkuptext}%
Thus we have view equivalence for the concrete operation, too.%
\end{isamarkuptext}\isamarkuptrue%
\ \ \isacommand{ultimately}\isamarkupfalse%
\ \isacommand{show}\isamarkupfalse%
\ {\isacharquery}thesis\ \isacommand{by}\isamarkupfalse%
{\isacharparenleft}simp\ add{\isacharcolon}split{\isacharunderscore}net{\isacharunderscore}C{\isacharunderscore}def{\isacharparenright}\isanewline
\isacommand{qed}\isamarkupfalse%
\isanewline
\endisatagproof
{\isafoldproof}%
\isadelimproof
\endisadelimproof
\isadelimtheory
\endisadelimtheory
\isatagtheory
\endisatagtheory
{\isafoldtheory}%
\isadelimtheory
\endisadelimtheory
\end{isabellebody}%

\section{System Models}
\label{isabelle:system:models}
\camdone{p. 24: Annex B is very abrupt.
    An introduction to Annex B would be welcome, to explain the forthcoming
    contents of this section.
    For instance, section 3.1 should mention that the model will be fully
    detailed in annex B.1, and annex B.1 should mention that the model has
    been presented in section 3.1.
    Same remark for B.2 and B.3, which are likely related to 3.2 and 3.3
    although the section titles are not exactly the same.}

\camready{
This appendix gives the full definitions of the models
in~\autoref{sec:realsystems}, in the abstract syntax of
\autoref{sec:isasyntax}: ~\autoref{isabelle:omap} corresponds to the OMAP4460
SoC of ~\autoref{sec:realsystems:omap44xx}, ~\autoref{isabelle:desktop}
corresponds to the desktop system of ~\autoref{sec:realsystems:desktop} and
~\autoref{isabelle:server} corresponds to the server of
~\autoref{sec:realsystems:server}. In additon, we present two supplementary
systems to show the applicability of our model to clusters
(~\autoref{isabelle:cluster}) and exotic hardware such as the Intel Single
Chip Cloud Computer (SCC, ~\autoref{sec:system:scc}).
}

\isabellestyle{tt}

\subsection{A mobile device SoC: the OMAP4460}
\label{isabelle:omap}
\camready{This is the full model representation of the Texas Instruments 
OMAP4460 SoC that we already introduced in~\autoref{sec:realsystems:omap44xx}.}

\begin{isabellebody}%
\setisabellecontext{OMAP{\isadigit{4}}{\isadigit{4}}xx}%
\isadelimtheory
\endisadelimtheory
\isatagtheory
\endisatagtheory
{\isafoldtheory}%
\isadelimtheory
\endisadelimtheory
\isamarkupsubsubsection{Address spaces%
}
\isamarkuptrue%
\begin{isamarkuptext}%
RAM%
\end{isamarkuptext}\isamarkuptrue%
\isacommand{definition}\isamarkupfalse%
\ {\isachardoublequoteopen}ram\ {\isacharequal}\ {\isacharparenleft}{\isadigit{0}}x{\isadigit{8}}{\isadigit{0}}{\isadigit{0}}{\isadigit{0}}{\isadigit{0}}{\isadigit{0}}{\isadigit{0}}{\isadigit{0}}{\isacharcomma}\ {\isadigit{0}}xBFFFFFFF{\isacharparenright}{\isachardoublequoteclose}\isanewline
\isacommand{definition}\isamarkupfalse%
\ {\isachardoublequoteopen}node{\isacharunderscore}{\isadigit{1}}{\isacharunderscore}ram\ {\isacharequal}\ empty{\isacharunderscore}spec\ {\isasymlparr}\isanewline
\ \ acc{\isacharunderscore}blocks\ {\isacharcolon}{\isacharequal}\ {\isacharbrackleft}ram{\isacharbrackright}{\isacharcomma}\isanewline
\ \ map{\isacharunderscore}blocks\ {\isacharcolon}{\isacharequal}\ {\isacharbrackleft}{\isacharbrackright}\isanewline
{\isasymrparr}{\isachardoublequoteclose}%
\begin{isamarkuptext}%
General-Purpose Timer 5%
\end{isamarkuptext}\isamarkuptrue%
\isacommand{definition}\isamarkupfalse%
\ {\isachardoublequoteopen}gptimer{\isadigit{5}}\ {\isacharequal}\ {\isacharparenleft}{\isadigit{0}}x{\isadigit{0}}{\isacharcomma}\ {\isadigit{0}}x{\isadigit{1}}{\isadigit{0}}{\isadigit{0}}{\isadigit{0}}{\isacharparenright}{\isachardoublequoteclose}\isanewline
\isacommand{definition}\isamarkupfalse%
\ {\isachardoublequoteopen}node{\isacharunderscore}{\isadigit{2}}{\isacharunderscore}gptimer{\isadigit{5}}\ {\isacharequal}\ empty{\isacharunderscore}spec\ {\isasymlparr}\isanewline
\ \ acc{\isacharunderscore}blocks\ {\isacharcolon}{\isacharequal}\ {\isacharbrackleft}gptimer{\isadigit{5}}{\isacharbrackright}{\isacharcomma}\isanewline
\ \ map{\isacharunderscore}blocks\ {\isacharcolon}{\isacharequal}\ {\isacharbrackleft}{\isacharbrackright}\isanewline
{\isasymrparr}{\isachardoublequoteclose}%
\begin{isamarkuptext}%
The L3 Interconnect%
\end{isamarkuptext}\isamarkuptrue%
\isacommand{definition}\isamarkupfalse%
\ {\isachardoublequoteopen}l{\isadigit{3}}{\isacharunderscore}boot\ {\isacharequal}\ {\isacharparenleft}{\isadigit{0}}x{\isadigit{0}}{\isadigit{0}}{\isadigit{0}}{\isadigit{0}}{\isadigit{0}}{\isadigit{0}}{\isadigit{0}}{\isadigit{0}}{\isacharcomma}\ {\isadigit{0}}x{\isadigit{4}}{\isadigit{0}}{\isadigit{0}}{\isadigit{0}}{\isadigit{0}}{\isadigit{0}}{\isadigit{0}}{\isadigit{0}}{\isacharparenright}{\isachardoublequoteclose}\isanewline
\isacommand{definition}\isamarkupfalse%
\ {\isachardoublequoteopen}l{\isadigit{3}}{\isacharunderscore}l{\isadigit{4}}\ {\isacharequal}\ {\isacharparenleft}{\isadigit{0}}x{\isadigit{4}}{\isadigit{9}}{\isadigit{0}}{\isadigit{0}}{\isadigit{0}}{\isadigit{0}}{\isadigit{0}}{\isadigit{0}}{\isacharcomma}\ {\isadigit{0}}x{\isadigit{4}}{\isadigit{9}}FFFFFF{\isacharparenright}{\isachardoublequoteclose}\isanewline
\isacommand{definition}\isamarkupfalse%
\ {\isachardoublequoteopen}l{\isadigit{3}}{\isacharunderscore}sdma\ {\isacharequal}\ {\isacharparenleft}{\isadigit{0}}x{\isadigit{4}}A{\isadigit{0}}{\isadigit{5}}{\isadigit{6}}{\isadigit{0}}{\isadigit{0}}{\isadigit{0}}{\isacharcomma}\ {\isadigit{0}}x{\isadigit{4}}A{\isadigit{0}}{\isadigit{5}}{\isadigit{6}}FFF{\isacharparenright}{\isachardoublequoteclose}\isanewline
\isacommand{definition}\isamarkupfalse%
\ {\isachardoublequoteopen}l{\isadigit{3}}{\isacharunderscore}ram\ {\isacharequal}\ {\isacharparenleft}{\isadigit{0}}x{\isadigit{8}}{\isadigit{0}}{\isadigit{0}}{\isadigit{0}}{\isadigit{0}}{\isadigit{0}}{\isadigit{0}}{\isacharcomma}\ {\isadigit{0}}xBFFFFFFF{\isacharparenright}{\isachardoublequoteclose}\isanewline
\isanewline
\isacommand{definition}\isamarkupfalse%
\ {\isachardoublequoteopen}node{\isacharunderscore}{\isadigit{3}}{\isacharunderscore}l{\isadigit{3}}{\isacharunderscore}interconnect\ {\isacharequal}\ empty{\isacharunderscore}spec\ {\isasymlparr}\isanewline
\ \ acc{\isacharunderscore}blocks\ {\isacharcolon}{\isacharequal}\ {\isacharbrackleft}l{\isadigit{3}}{\isacharunderscore}boot{\isacharbrackright}{\isacharcomma}\isanewline
\ \ map{\isacharunderscore}blocks\ {\isacharcolon}{\isacharequal}\ {\isacharbrackleft}direct{\isacharunderscore}map\ l{\isadigit{3}}{\isacharunderscore}ram\ {\isadigit{1}}{\isacharcomma}\ direct{\isacharunderscore}map\ l{\isadigit{3}}{\isacharunderscore}l{\isadigit{4}}\ {\isadigit{2}}{\isacharcomma}\ direct{\isacharunderscore}map\ l{\isadigit{3}}{\isacharunderscore}sdma\ {\isadigit{3}}\ {\isacharbrackright}\isanewline
{\isasymrparr}{\isachardoublequoteclose}%
\begin{isamarkuptext}%
The L4 Interconnect%
\end{isamarkuptext}\isamarkuptrue%
\isacommand{definition}\isamarkupfalse%
\ {\isachardoublequoteopen}l{\isadigit{4}}{\isacharunderscore}gptimer{\isadigit{5}}{\isacharunderscore}a{\isadigit{9}}{\isacharunderscore}module\ {\isacharequal}\ {\isacharparenleft}{\isadigit{0}}x{\isadigit{4}}{\isadigit{0}}{\isadigit{1}}{\isadigit{3}}{\isadigit{8}}{\isadigit{0}}{\isadigit{0}}{\isadigit{0}}{\isacharcomma}\ {\isadigit{0}}x{\isadigit{4}}{\isadigit{0}}{\isadigit{1}}{\isadigit{3}}{\isadigit{8}}FFF{\isacharparenright}{\isachardoublequoteclose}\isanewline
\isacommand{definition}\isamarkupfalse%
\ {\isachardoublequoteopen}l{\isadigit{4}}{\isacharunderscore}gptimer{\isadigit{5}}{\isacharunderscore}a{\isadigit{9}}{\isacharunderscore}l{\isadigit{4}}\ {\isacharequal}\ {\isacharparenleft}{\isadigit{0}}x{\isadigit{4}}{\isadigit{0}}{\isadigit{1}}{\isadigit{3}}{\isadigit{9}}{\isadigit{0}}{\isadigit{0}}{\isadigit{0}}{\isacharcomma}\ {\isadigit{0}}x{\isadigit{4}}{\isadigit{0}}{\isadigit{1}}{\isadigit{3}}{\isadigit{9}}FFF{\isacharparenright}{\isachardoublequoteclose}\isanewline
\isacommand{definition}\isamarkupfalse%
\ {\isachardoublequoteopen}l{\isadigit{4}}{\isacharunderscore}gptimer{\isadigit{5}}{\isacharunderscore}l{\isadigit{3}}{\isacharunderscore}module\ {\isacharequal}\ {\isacharparenleft}{\isadigit{0}}x{\isadigit{4}}{\isadigit{9}}{\isadigit{0}}{\isadigit{3}}{\isadigit{8}}{\isadigit{0}}{\isadigit{0}}{\isadigit{0}}{\isacharcomma}\ {\isadigit{0}}x{\isadigit{4}}{\isadigit{9}}{\isadigit{0}}{\isadigit{3}}{\isadigit{8}}FFF{\isacharparenright}{\isachardoublequoteclose}\isanewline
\isacommand{definition}\isamarkupfalse%
\ {\isachardoublequoteopen}l{\isadigit{4}}{\isacharunderscore}gptimer{\isadigit{5}}{\isacharunderscore}l{\isadigit{3}}{\isacharunderscore}l{\isadigit{4}}\ {\isacharequal}\ {\isacharparenleft}{\isadigit{0}}x{\isadigit{4}}{\isadigit{9}}{\isadigit{0}}{\isadigit{3}}{\isadigit{9}}{\isadigit{0}}{\isadigit{0}}{\isadigit{0}}{\isacharcomma}\ {\isadigit{0}}x{\isadigit{4}}{\isadigit{9}}{\isadigit{0}}{\isadigit{3}}{\isadigit{9}}FFF{\isacharparenright}{\isachardoublequoteclose}\isanewline
\isanewline
\isacommand{definition}\isamarkupfalse%
\ {\isachardoublequoteopen}l{\isadigit{4}}{\isacharunderscore}gptimer{\isadigit{5}}{\isacharunderscore}dsp{\isacharunderscore}module\ {\isacharequal}\ {\isacharparenleft}{\isadigit{0}}x{\isadigit{0}}{\isadigit{1}}D{\isadigit{3}}{\isadigit{8}}{\isadigit{0}}{\isadigit{0}}{\isadigit{0}}{\isacharcomma}\ {\isadigit{0}}x{\isadigit{0}}{\isadigit{1}}D{\isadigit{3}}{\isadigit{8}}FFF{\isacharparenright}{\isachardoublequoteclose}\isanewline
\isacommand{definition}\isamarkupfalse%
\ {\isachardoublequoteopen}l{\isadigit{4}}{\isacharunderscore}gptimer{\isadigit{5}}{\isacharunderscore}dsp{\isacharunderscore}l{\isadigit{4}}\ {\isacharequal}\ {\isacharparenleft}{\isadigit{0}}x{\isadigit{0}}{\isadigit{1}}D{\isadigit{3}}{\isadigit{9}}{\isadigit{0}}{\isadigit{0}}{\isadigit{0}}{\isacharcomma}\ {\isadigit{0}}x{\isadigit{0}}{\isadigit{1}}D{\isadigit{3}}{\isadigit{9}}FFF{\isacharparenright}{\isachardoublequoteclose}\isanewline
\isanewline
\ \ \isanewline
\isacommand{definition}\isamarkupfalse%
\ {\isachardoublequoteopen}l{\isadigit{4}}{\isacharunderscore}sdma{\isacharunderscore}module\ {\isacharequal}\ {\isacharparenleft}{\isadigit{0}}x{\isadigit{4}}A{\isadigit{0}}{\isadigit{5}}{\isadigit{6}}{\isadigit{0}}{\isadigit{0}}{\isadigit{0}}{\isacharcomma}\ {\isadigit{0}}x{\isadigit{4}}A{\isadigit{0}}{\isadigit{5}}{\isadigit{6}}FFFF{\isacharparenright}{\isachardoublequoteclose}\isanewline
\isacommand{definition}\isamarkupfalse%
\ {\isachardoublequoteopen}l{\isadigit{4}}{\isacharunderscore}sdma{\isacharunderscore}l{\isadigit{4}}\ {\isacharequal}\ {\isacharparenleft}{\isadigit{0}}x{\isadigit{4}}A{\isadigit{0}}{\isadigit{5}}{\isadigit{7}}{\isadigit{0}}{\isadigit{0}}{\isadigit{0}}{\isacharcomma}\ {\isadigit{0}}x{\isadigit{4}}A{\isadigit{0}}{\isadigit{5}}{\isadigit{7}}FFF{\isacharparenright}{\isachardoublequoteclose}\isanewline
\isanewline
\isacommand{definition}\isamarkupfalse%
\ {\isachardoublequoteopen}node{\isacharunderscore}{\isadigit{4}}{\isacharunderscore}l{\isadigit{4}}{\isacharunderscore}interconnect\ {\isacharequal}\ empty{\isacharunderscore}spec\ {\isasymlparr}\isanewline
\ \ acc{\isacharunderscore}blocks\ {\isacharcolon}{\isacharequal}\ {\isacharbrackleft}l{\isadigit{3}}{\isacharunderscore}boot{\isacharbrackright}{\isacharcomma}\isanewline
\ \ map{\isacharunderscore}blocks\ {\isacharcolon}{\isacharequal}\ {\isacharbrackleft}{\isacharbrackright}\isanewline
{\isasymrparr}{\isachardoublequoteclose}%
\begin{isamarkuptext}%
The DSP Subsystem%
\end{isamarkuptext}\isamarkuptrue%
\isacommand{definition}\isamarkupfalse%
\ {\isachardoublequoteopen}dspvirt{\isacharunderscore}gptimer\ {\isacharequal}\ {\isacharparenleft}{\isadigit{0}}x{\isadigit{1}}{\isadigit{0}}{\isadigit{0}}{\isadigit{0}}{\isadigit{0}}{\isadigit{0}}{\isadigit{0}}{\isacharcomma}\ {\isadigit{0}}x{\isadigit{1}}{\isadigit{0}}{\isadigit{0}}{\isadigit{0}}{\isadigit{0}}FFF{\isacharparenright}{\isachardoublequoteclose}\isanewline
\isacommand{definition}\isamarkupfalse%
\ {\isachardoublequoteopen}dspvirt{\isacharunderscore}ram\ {\isacharequal}\ {\isacharparenleft}{\isadigit{0}}x{\isadigit{2}}{\isadigit{0}}{\isadigit{0}}{\isadigit{0}}{\isadigit{0}}{\isadigit{0}}{\isadigit{0}}{\isacharcomma}\ {\isadigit{0}}x{\isadigit{5}}FFFFFF{\isacharparenright}{\isachardoublequoteclose}\isanewline
\isacommand{definition}\isamarkupfalse%
\ {\isachardoublequoteopen}node{\isacharunderscore}{\isadigit{5}}{\isacharunderscore}dspvirt\ {\isacharequal}\ empty{\isacharunderscore}spec\ {\isasymlparr}\isanewline
\ \ acc{\isacharunderscore}blocks\ {\isacharcolon}{\isacharequal}\ {\isacharbrackleft}{\isacharbrackright}{\isacharcomma}\isanewline
\ \ map{\isacharunderscore}blocks\ {\isacharcolon}{\isacharequal}\ {\isacharbrackleft}block{\isacharunderscore}map\ dspvirt{\isacharunderscore}gptimer\ {\isadigit{6}}\ {\isadigit{0}}x{\isadigit{0}}{\isadigit{1}}D{\isadigit{3}}{\isadigit{8}}{\isadigit{0}}{\isadigit{0}}{\isadigit{0}}{\isacharcomma}\ \isanewline
\ \ \ \ \ \ \ \ \ \ \ \ \ \ \ \ \ block{\isacharunderscore}map\ dspvirt{\isacharunderscore}ram\ {\isadigit{6}}\ {\isadigit{0}}x{\isadigit{8}}{\isadigit{0}}{\isadigit{0}}{\isadigit{0}}{\isadigit{0}}{\isadigit{0}}{\isadigit{0}}{\isacharbrackright}\isanewline
{\isasymrparr}{\isachardoublequoteclose}\isanewline
\isanewline
\isacommand{definition}\isamarkupfalse%
\ {\isachardoublequoteopen}dspphys{\isacharunderscore}ram\ {\isacharequal}\ \ {\isacharparenleft}{\isadigit{0}}x{\isadigit{8}}{\isadigit{0}}{\isadigit{0}}{\isadigit{0}}{\isadigit{0}}{\isadigit{0}}{\isadigit{0}}{\isadigit{0}}{\isacharcomma}\ {\isadigit{0}}xBFFFFFFF{\isacharparenright}{\isachardoublequoteclose}\isanewline
\isacommand{definition}\isamarkupfalse%
\ {\isachardoublequoteopen}dspphys{\isacharunderscore}gptimer{\isadigit{5}}\ {\isacharequal}\ {\isacharparenleft}{\isadigit{0}}x{\isadigit{0}}{\isadigit{1}}D{\isadigit{3}}{\isadigit{8}}{\isadigit{0}}{\isadigit{0}}{\isadigit{0}}{\isacharcomma}\ {\isadigit{0}}x{\isadigit{0}}{\isadigit{1}}D{\isadigit{3}}{\isadigit{8}}FFF{\isacharparenright}{\isachardoublequoteclose}\isanewline
\isacommand{definition}\isamarkupfalse%
\ {\isachardoublequoteopen}node{\isacharunderscore}{\isadigit{6}}{\isacharunderscore}dspphys\ {\isacharequal}\ empty{\isacharunderscore}spec\ {\isasymlparr}\isanewline
\ \ acc{\isacharunderscore}blocks\ {\isacharcolon}{\isacharequal}\ {\isacharbrackleft}{\isacharbrackright}{\isacharcomma}\isanewline
\ \ map{\isacharunderscore}blocks\ {\isacharcolon}{\isacharequal}\ {\isacharbrackleft}direct{\isacharunderscore}map\ dspphys{\isacharunderscore}ram\ {\isadigit{3}}{\isacharcomma}\ block{\isacharunderscore}map\ dspphys{\isacharunderscore}gptimer{\isadigit{5}}\ {\isadigit{2}}\ {\isadigit{0}}{\isacharbrackright}\isanewline
{\isasymrparr}{\isachardoublequoteclose}%
\begin{isamarkuptext}%
The SDMA Module%
\end{isamarkuptext}\isamarkuptrue%
\isacommand{definition}\isamarkupfalse%
\ {\isachardoublequoteopen}sdma\ {\isacharequal}\ {\isacharparenleft}{\isadigit{0}}x{\isadigit{4}}A{\isadigit{0}}{\isadigit{5}}{\isadigit{6}}{\isadigit{0}}{\isadigit{0}}{\isadigit{0}}{\isacharcomma}\ {\isadigit{0}}x{\isadigit{4}}A{\isadigit{0}}{\isadigit{5}}{\isadigit{6}}FFF{\isacharparenright}{\isachardoublequoteclose}\isanewline
\isacommand{definition}\isamarkupfalse%
\ {\isachardoublequoteopen}node{\isacharunderscore}{\isadigit{7}}{\isacharunderscore}sdma\ {\isacharequal}\ empty{\isacharunderscore}spec\ {\isasymlparr}\isanewline
\ \ acc{\isacharunderscore}blocks\ {\isacharcolon}{\isacharequal}\ {\isacharbrackleft}sdma{\isacharbrackright}{\isacharcomma}\isanewline
\ \ map{\isacharunderscore}blocks\ {\isacharcolon}{\isacharequal}\ {\isacharbrackleft}direct{\isacharunderscore}map\ ram\ {\isadigit{1}}{\isacharbrackright}\isanewline
{\isasymrparr}{\isachardoublequoteclose}%
\begin{isamarkuptext}%
The Cortex A9 MPU Subsystem%
\end{isamarkuptext}\isamarkuptrue%
\isacommand{definition}\isamarkupfalse%
\ {\isachardoublequoteopen}a{\isadigit{9}}{\isacharunderscore}{\isadigit{0}}{\isacharunderscore}virt{\isacharunderscore}ram\ {\isacharequal}\ \ {\isacharparenleft}{\isadigit{0}}x{\isadigit{0}}{\isadigit{0}}{\isadigit{0}}{\isadigit{0}}{\isadigit{0}}{\isadigit{0}}{\isadigit{0}}{\isadigit{0}}{\isacharcomma}\ {\isadigit{0}}x{\isadigit{3}}FFFFFFF{\isacharparenright}{\isachardoublequoteclose}\isanewline
\isacommand{definition}\isamarkupfalse%
\ {\isachardoublequoteopen}a{\isadigit{9}}{\isacharunderscore}{\isadigit{0}}{\isacharunderscore}virt{\isacharunderscore}gptimer{\isacharunderscore}priv\ {\isacharequal}\ {\isacharparenleft}{\isadigit{0}}x{\isadigit{6}}{\isadigit{0}}{\isadigit{0}}{\isadigit{0}}{\isadigit{0}}{\isadigit{0}}{\isadigit{0}}{\isadigit{0}}{\isacharcomma}\ {\isadigit{0}}x{\isadigit{6}}{\isadigit{0}}{\isadigit{0}}{\isadigit{0}}{\isadigit{0}}FFF{\isacharparenright}{\isachardoublequoteclose}\isanewline
\isacommand{definition}\isamarkupfalse%
\ {\isachardoublequoteopen}a{\isadigit{9}}{\isacharunderscore}{\isadigit{0}}{\isacharunderscore}virt{\isacharunderscore}gptimer\ {\isacharequal}\ {\isacharparenleft}{\isadigit{0}}x{\isadigit{6}}{\isadigit{0}}{\isadigit{0}}{\isadigit{0}}{\isadigit{1}}{\isadigit{0}}{\isadigit{0}}{\isadigit{0}}{\isacharcomma}\ {\isadigit{0}}x{\isadigit{6}}{\isadigit{0}}{\isadigit{0}}{\isadigit{0}}{\isadigit{1}}FFF{\isacharparenright}{\isachardoublequoteclose}\isanewline
\isacommand{definition}\isamarkupfalse%
\ {\isachardoublequoteopen}a{\isadigit{9}}{\isacharunderscore}{\isadigit{0}}{\isacharunderscore}virt{\isacharunderscore}sdma\ {\isacharequal}\ {\isacharparenleft}{\isadigit{0}}x{\isadigit{6}}{\isadigit{0}}{\isadigit{0}}{\isadigit{0}}{\isadigit{2}}{\isadigit{0}}{\isadigit{0}}{\isadigit{0}}{\isacharcomma}\ {\isadigit{0}}x{\isadigit{6}}{\isadigit{0}}{\isadigit{0}}{\isadigit{0}}{\isadigit{2}}FFF{\isacharparenright}{\isachardoublequoteclose}\isanewline
\isanewline
\isacommand{definition}\isamarkupfalse%
\ {\isachardoublequoteopen}node{\isacharunderscore}{\isadigit{8}}{\isacharunderscore}a{\isadigit{9}}virt{\isacharunderscore}{\isadigit{0}}\ {\isacharequal}\ \ empty{\isacharunderscore}spec\ {\isasymlparr}\isanewline
\ \ acc{\isacharunderscore}blocks\ {\isacharcolon}{\isacharequal}\ {\isacharbrackleft}{\isacharbrackright}{\isacharcomma}\isanewline
\ \ map{\isacharunderscore}blocks\ {\isacharcolon}{\isacharequal}\ {\isacharbrackleft}block{\isacharunderscore}map\ a{\isadigit{9}}{\isacharunderscore}{\isadigit{0}}{\isacharunderscore}virt{\isacharunderscore}ram\ {\isadigit{9}}\ {\isadigit{0}}x{\isadigit{8}}{\isadigit{0}}{\isadigit{0}}{\isadigit{0}}{\isadigit{0}}{\isadigit{0}}{\isadigit{0}}{\isacharcomma}\ \isanewline
\ \ \ \ \ \ \ \ \ \ \ \ \ \ \ \ \ block{\isacharunderscore}map\ a{\isadigit{9}}{\isacharunderscore}{\isadigit{0}}{\isacharunderscore}virt{\isacharunderscore}gptimer{\isacharunderscore}priv\ {\isadigit{9}}\ {\isadigit{0}}x{\isadigit{4}}{\isadigit{0}}{\isadigit{1}}{\isadigit{3}}{\isadigit{8}}{\isadigit{0}}{\isadigit{0}}{\isadigit{0}}{\isacharcomma}\ \isanewline
\ \ \ \ \ \ \ \ \ \ \ \ \ \ \ \ \ block{\isacharunderscore}map\ a{\isadigit{9}}{\isacharunderscore}{\isadigit{0}}{\isacharunderscore}virt{\isacharunderscore}gptimer\ {\isadigit{9}}\ {\isadigit{0}}x{\isadigit{4}}{\isadigit{9}}{\isadigit{0}}{\isadigit{3}}{\isadigit{8}}{\isadigit{0}}{\isadigit{0}}{\isadigit{0}}{\isacharcomma}\ \isanewline
\ \ \ \ \ \ \ \ \ \ \ \ \ \ \ \ \ block{\isacharunderscore}map\ a{\isadigit{9}}{\isacharunderscore}{\isadigit{0}}{\isacharunderscore}virt{\isacharunderscore}sdma\ {\isadigit{9}}\ {\isadigit{0}}x{\isadigit{4}}A{\isadigit{0}}{\isadigit{5}}{\isadigit{6}}{\isadigit{0}}{\isadigit{0}}{\isadigit{0}}{\isacharbrackright}\isanewline
{\isasymrparr}{\isachardoublequoteclose}\isanewline
\isanewline
\isacommand{definition}\isamarkupfalse%
\ {\isachardoublequoteopen}a{\isadigit{9}}{\isacharunderscore}{\isadigit{0}}{\isacharunderscore}phys{\isacharunderscore}ram\ {\isacharequal}\ {\isacharparenleft}{\isadigit{0}}x{\isadigit{8}}{\isadigit{0}}{\isadigit{0}}{\isadigit{0}}{\isadigit{0}}{\isadigit{0}}{\isadigit{0}}{\isadigit{0}}{\isadigit{0}}{\isacharcomma}\ {\isadigit{0}}xBFFFFFFF{\isacharparenright}{\isachardoublequoteclose}\isanewline
\isacommand{definition}\isamarkupfalse%
\ {\isachardoublequoteopen}a{\isadigit{9}}{\isacharunderscore}{\isadigit{0}}{\isacharunderscore}phys{\isacharunderscore}gptimer{\isacharunderscore}priv\ {\isacharequal}\ {\isacharparenleft}{\isadigit{0}}x{\isadigit{4}}{\isadigit{0}}{\isadigit{1}}{\isadigit{3}}{\isadigit{8}}{\isadigit{0}}{\isadigit{0}}{\isadigit{0}}{\isacharcomma}\ {\isadigit{0}}x{\isadigit{4}}{\isadigit{0}}{\isadigit{1}}{\isadigit{3}}{\isadigit{8}}FFF{\isacharparenright}{\isachardoublequoteclose}\isanewline
\isacommand{definition}\isamarkupfalse%
\ {\isachardoublequoteopen}a{\isadigit{9}}{\isacharunderscore}{\isadigit{0}}{\isacharunderscore}phys{\isacharunderscore}gptimer\ {\isacharequal}\ {\isacharparenleft}{\isadigit{0}}x{\isadigit{4}}{\isadigit{9}}{\isadigit{0}}{\isadigit{3}}{\isadigit{8}}{\isadigit{0}}{\isadigit{0}}{\isadigit{0}}{\isacharcomma}\ {\isadigit{0}}x{\isadigit{4}}{\isadigit{9}}{\isadigit{0}}{\isadigit{3}}{\isadigit{8}}FFF{\isacharparenright}{\isachardoublequoteclose}\isanewline
\isacommand{definition}\isamarkupfalse%
\ {\isachardoublequoteopen}a{\isadigit{9}}{\isacharunderscore}{\isadigit{0}}{\isacharunderscore}phys{\isacharunderscore}sdma\ {\isacharequal}\ {\isacharparenleft}{\isadigit{0}}x{\isadigit{4}}A{\isadigit{0}}{\isadigit{5}}{\isadigit{6}}{\isadigit{0}}{\isadigit{0}}{\isadigit{0}}{\isacharcomma}\ {\isadigit{0}}x{\isadigit{4}}A{\isadigit{0}}{\isadigit{5}}{\isadigit{6}}FFF{\isacharparenright}{\isachardoublequoteclose}\isanewline
\isacommand{definition}\isamarkupfalse%
\ {\isachardoublequoteopen}node{\isacharunderscore}{\isadigit{9}}{\isacharunderscore}a{\isadigit{9}}phys{\isacharunderscore}{\isadigit{0}}\ {\isacharequal}\ \ empty{\isacharunderscore}spec\ {\isasymlparr}\isanewline
\ \ acc{\isacharunderscore}blocks\ {\isacharcolon}{\isacharequal}\ {\isacharbrackleft}{\isacharbrackright}{\isacharcomma}\isanewline
\ \ map{\isacharunderscore}blocks\ {\isacharcolon}{\isacharequal}\ {\isacharbrackleft}direct{\isacharunderscore}map\ a{\isadigit{9}}{\isacharunderscore}{\isadigit{0}}{\isacharunderscore}phys{\isacharunderscore}ram\ {\isadigit{3}}{\isacharcomma}\ direct{\isacharunderscore}map\ a{\isadigit{9}}{\isacharunderscore}{\isadigit{0}}{\isacharunderscore}phys{\isacharunderscore}gptimer{\isacharunderscore}priv\ {\isadigit{4}}{\isacharcomma}\isanewline
\ \ \ \ \ \ \ \ \ \ \ \ \ \ \ \ \ direct{\isacharunderscore}map\ a{\isadigit{9}}{\isacharunderscore}{\isadigit{0}}{\isacharunderscore}phys{\isacharunderscore}gptimer\ {\isadigit{3}}{\isacharcomma}\ direct{\isacharunderscore}map\ a{\isadigit{9}}{\isacharunderscore}{\isadigit{0}}{\isacharunderscore}phys{\isacharunderscore}sdma\ {\isadigit{3}}{\isacharbrackright}\isanewline
{\isasymrparr}{\isachardoublequoteclose}\isanewline
\isanewline
\isacommand{definition}\isamarkupfalse%
\ {\isachardoublequoteopen}a{\isadigit{9}}{\isacharunderscore}{\isadigit{1}}{\isacharunderscore}virt{\isacharunderscore}ram\ {\isacharequal}\ {\isacharparenleft}{\isadigit{0}}x{\isadigit{1}}{\isadigit{0}}{\isadigit{0}}{\isadigit{0}}{\isadigit{0}}{\isadigit{0}}{\isadigit{0}}{\isadigit{0}}{\isacharcomma}\ {\isadigit{0}}x{\isadigit{4}}FFFFFFF{\isacharparenright}{\isachardoublequoteclose}\isanewline
\isacommand{definition}\isamarkupfalse%
\ {\isachardoublequoteopen}a{\isadigit{9}}{\isacharunderscore}{\isadigit{1}}{\isacharunderscore}virt{\isacharunderscore}gptimer{\isacharunderscore}priv\ {\isacharequal}\ {\isacharparenleft}{\isadigit{0}}x{\isadigit{7}}{\isadigit{0}}{\isadigit{0}}{\isadigit{0}}{\isadigit{0}}{\isadigit{0}}{\isadigit{0}}{\isadigit{0}}{\isacharcomma}\ {\isadigit{0}}x{\isadigit{7}}{\isadigit{0}}{\isadigit{0}}{\isadigit{0}}{\isadigit{0}}FFF{\isacharparenright}{\isachardoublequoteclose}\isanewline
\isacommand{definition}\isamarkupfalse%
\ {\isachardoublequoteopen}a{\isadigit{9}}{\isacharunderscore}{\isadigit{1}}{\isacharunderscore}virt{\isacharunderscore}gptimer\ {\isacharequal}\ {\isacharparenleft}{\isadigit{0}}x{\isadigit{7}}{\isadigit{0}}{\isadigit{0}}{\isadigit{0}}{\isadigit{1}}{\isadigit{0}}{\isadigit{0}}{\isadigit{0}}{\isacharcomma}\ {\isadigit{0}}x{\isadigit{7}}{\isadigit{0}}{\isadigit{0}}{\isadigit{0}}{\isadigit{1}}FFF{\isacharparenright}{\isachardoublequoteclose}\isanewline
\isacommand{definition}\isamarkupfalse%
\ {\isachardoublequoteopen}a{\isadigit{9}}{\isacharunderscore}{\isadigit{1}}{\isacharunderscore}virt{\isacharunderscore}sdma\ {\isacharequal}\ {\isacharparenleft}{\isadigit{0}}x{\isadigit{7}}{\isadigit{0}}{\isadigit{0}}{\isadigit{0}}{\isadigit{1}}{\isadigit{0}}{\isadigit{0}}{\isadigit{0}}{\isacharcomma}\ {\isadigit{0}}x{\isadigit{7}}{\isadigit{0}}{\isadigit{0}}{\isadigit{0}}{\isadigit{1}}FFF{\isacharparenright}{\isachardoublequoteclose}\isanewline
\ \ \isanewline
\isacommand{definition}\isamarkupfalse%
\ {\isachardoublequoteopen}node{\isacharunderscore}{\isadigit{1}}{\isadigit{0}}{\isacharunderscore}a{\isadigit{9}}virt{\isacharunderscore}{\isadigit{1}}\ {\isacharequal}\ empty{\isacharunderscore}spec\ {\isasymlparr}\isanewline
\ \ acc{\isacharunderscore}blocks\ {\isacharcolon}{\isacharequal}\ {\isacharbrackleft}{\isacharbrackright}{\isacharcomma}\isanewline
\ \ map{\isacharunderscore}blocks\ {\isacharcolon}{\isacharequal}\ {\isacharbrackleft}block{\isacharunderscore}map\ a{\isadigit{9}}{\isacharunderscore}{\isadigit{1}}{\isacharunderscore}virt{\isacharunderscore}ram\ {\isadigit{1}}{\isadigit{1}}\ {\isadigit{0}}x{\isadigit{8}}{\isadigit{0}}{\isadigit{0}}{\isadigit{0}}{\isadigit{0}}{\isadigit{0}}{\isadigit{0}}{\isacharcomma}\ \isanewline
\ \ \ \ \ \ \ \ \ \ \ \ \ \ \ \ \ block{\isacharunderscore}map\ a{\isadigit{9}}{\isacharunderscore}{\isadigit{1}}{\isacharunderscore}virt{\isacharunderscore}gptimer{\isacharunderscore}priv\ {\isadigit{1}}{\isadigit{1}}\ {\isadigit{0}}x{\isadigit{4}}{\isadigit{0}}{\isadigit{1}}{\isadigit{3}}{\isadigit{8}}{\isadigit{0}}{\isadigit{0}}{\isadigit{0}}{\isacharcomma}\ \isanewline
\ \ \ \ \ \ \ \ \ \ \ \ \ \ \ \ \ block{\isacharunderscore}map\ a{\isadigit{9}}{\isacharunderscore}{\isadigit{1}}{\isacharunderscore}virt{\isacharunderscore}gptimer\ {\isadigit{1}}{\isadigit{1}}\ {\isadigit{0}}x{\isadigit{4}}{\isadigit{9}}{\isadigit{0}}{\isadigit{3}}{\isadigit{8}}{\isadigit{0}}{\isadigit{0}}{\isadigit{0}}{\isacharcomma}\ \isanewline
\ \ \ \ \ \ \ \ \ \ \ \ \ \ \ \ \ block{\isacharunderscore}map\ a{\isadigit{9}}{\isacharunderscore}{\isadigit{1}}{\isacharunderscore}virt{\isacharunderscore}sdma\ {\isadigit{1}}{\isadigit{1}}\ {\isadigit{0}}x{\isadigit{4}}A{\isadigit{0}}{\isadigit{5}}{\isadigit{6}}{\isadigit{0}}{\isadigit{0}}{\isadigit{0}}{\isacharbrackright}\isanewline
{\isasymrparr}{\isachardoublequoteclose}\isanewline
\isanewline
\isacommand{definition}\isamarkupfalse%
\ {\isachardoublequoteopen}a{\isadigit{9}}{\isacharunderscore}{\isadigit{1}}{\isacharunderscore}phys{\isacharunderscore}ram\ {\isacharequal}\ {\isacharparenleft}{\isadigit{0}}x{\isadigit{8}}{\isadigit{0}}{\isadigit{0}}{\isadigit{0}}{\isadigit{0}}{\isadigit{0}}{\isadigit{0}}{\isadigit{0}}{\isadigit{0}}{\isacharcomma}\ {\isadigit{0}}xBFFFFFFF{\isacharparenright}{\isachardoublequoteclose}\isanewline
\isacommand{definition}\isamarkupfalse%
\ {\isachardoublequoteopen}a{\isadigit{9}}{\isacharunderscore}{\isadigit{1}}{\isacharunderscore}phys{\isacharunderscore}gptimer{\isacharunderscore}priv\ {\isacharequal}\ \ {\isacharparenleft}{\isadigit{0}}x{\isadigit{4}}{\isadigit{0}}{\isadigit{1}}{\isadigit{3}}{\isadigit{8}}{\isadigit{0}}{\isadigit{0}}{\isadigit{0}}{\isacharcomma}\ {\isadigit{0}}x{\isadigit{4}}{\isadigit{0}}{\isadigit{1}}{\isadigit{3}}{\isadigit{8}}FFF{\isacharparenright}{\isachardoublequoteclose}\isanewline
\isacommand{definition}\isamarkupfalse%
\ {\isachardoublequoteopen}a{\isadigit{9}}{\isacharunderscore}{\isadigit{1}}{\isacharunderscore}phys{\isacharunderscore}gptimer\ {\isacharequal}\ {\isacharparenleft}{\isadigit{0}}x{\isadigit{4}}{\isadigit{9}}{\isadigit{0}}{\isadigit{3}}{\isadigit{8}}{\isadigit{0}}{\isadigit{0}}{\isadigit{0}}{\isacharcomma}\ {\isadigit{0}}x{\isadigit{4}}{\isadigit{9}}{\isadigit{0}}{\isadigit{3}}{\isadigit{8}}FFF{\isacharparenright}{\isachardoublequoteclose}\isanewline
\isacommand{definition}\isamarkupfalse%
\ {\isachardoublequoteopen}a{\isadigit{9}}{\isacharunderscore}{\isadigit{1}}{\isacharunderscore}phys{\isacharunderscore}sdma\ {\isacharequal}\ {\isacharparenleft}{\isadigit{0}}x{\isadigit{4}}A{\isadigit{0}}{\isadigit{5}}{\isadigit{6}}{\isadigit{0}}{\isadigit{0}}{\isadigit{0}}{\isacharcomma}\ {\isadigit{0}}x{\isadigit{4}}A{\isadigit{0}}{\isadigit{5}}{\isadigit{6}}FFF{\isacharparenright}{\isachardoublequoteclose}\isanewline
\isacommand{definition}\isamarkupfalse%
\ {\isachardoublequoteopen}node{\isacharunderscore}{\isadigit{1}}{\isadigit{1}}{\isacharunderscore}a{\isadigit{9}}phys{\isacharunderscore}{\isadigit{1}}\ {\isacharequal}\ \ empty{\isacharunderscore}spec\ {\isasymlparr}\isanewline
\ \ acc{\isacharunderscore}blocks\ {\isacharcolon}{\isacharequal}\ {\isacharbrackleft}{\isacharbrackright}{\isacharcomma}\isanewline
\ \ map{\isacharunderscore}blocks\ {\isacharcolon}{\isacharequal}\ {\isacharbrackleft}direct{\isacharunderscore}map\ a{\isadigit{9}}{\isacharunderscore}{\isadigit{1}}{\isacharunderscore}phys{\isacharunderscore}ram\ {\isadigit{3}}{\isacharcomma}\ direct{\isacharunderscore}map\ a{\isadigit{9}}{\isacharunderscore}{\isadigit{1}}{\isacharunderscore}phys{\isacharunderscore}gptimer{\isacharunderscore}priv\ {\isadigit{4}}{\isacharcomma}\isanewline
\ \ \ \ \ \ \ \ \ \ \ \ \ \ \ \ \ direct{\isacharunderscore}map\ a{\isadigit{9}}{\isacharunderscore}{\isadigit{1}}{\isacharunderscore}phys{\isacharunderscore}gptimer\ {\isadigit{3}}{\isacharcomma}\ direct{\isacharunderscore}map\ a{\isadigit{9}}{\isacharunderscore}{\isadigit{1}}{\isacharunderscore}phys{\isacharunderscore}sdma\ {\isadigit{3}}{\isacharbrackright}\isanewline
{\isasymrparr}{\isachardoublequoteclose}%
\begin{isamarkuptext}%
The Cortex M3 Subsystem%
\end{isamarkuptext}\isamarkuptrue%
\isacommand{definition}\isamarkupfalse%
\ {\isachardoublequoteopen}m{\isadigit{3}}{\isacharunderscore}virt{\isacharunderscore}ram{\isacharunderscore}{\isadigit{0}}\ \ {\isacharequal}\ {\isacharparenleft}{\isadigit{0}}x{\isadigit{1}}{\isadigit{0}}{\isadigit{0}}{\isadigit{0}}{\isadigit{0}}{\isadigit{0}}{\isadigit{0}}{\isadigit{0}}{\isacharcomma}\ {\isadigit{0}}x{\isadigit{4}}FFFFFF{\isacharparenright}{\isachardoublequoteclose}\isanewline
\isacommand{definition}\isamarkupfalse%
\ {\isachardoublequoteopen}m{\isadigit{3}}{\isacharunderscore}virt{\isacharunderscore}local{\isacharunderscore}rom{\isacharunderscore}{\isadigit{0}}\ {\isacharequal}\ {\isacharparenleft}{\isadigit{0}}x{\isadigit{5}}{\isadigit{0}}{\isadigit{0}}{\isadigit{0}}{\isadigit{0}}{\isadigit{0}}{\isadigit{0}}{\isadigit{0}}{\isacharcomma}\ {\isadigit{0}}x{\isadigit{5}}{\isadigit{0}}{\isadigit{0}}{\isadigit{0}}{\isadigit{3}}FFF{\isacharparenright}{\isachardoublequoteclose}\isanewline
\isacommand{definition}\isamarkupfalse%
\ {\isachardoublequoteopen}m{\isadigit{3}}{\isacharunderscore}virt{\isacharunderscore}local{\isacharunderscore}ram{\isacharunderscore}{\isadigit{0}}\ {\isacharequal}\ {\isacharparenleft}{\isadigit{0}}x{\isadigit{5}}{\isadigit{0}}{\isadigit{0}}{\isadigit{2}}{\isadigit{0}}{\isadigit{0}}{\isadigit{0}}{\isadigit{0}}{\isacharcomma}\ {\isadigit{0}}x{\isadigit{5}}{\isadigit{0}}{\isadigit{0}}{\isadigit{2}}FFFF{\isacharparenright}{\isachardoublequoteclose}\isanewline
\isacommand{definition}\isamarkupfalse%
\ {\isachardoublequoteopen}node{\isacharunderscore}{\isadigit{1}}{\isadigit{2}}{\isacharunderscore}m{\isadigit{3}}{\isacharunderscore}virt{\isacharunderscore}{\isadigit{0}}\ {\isacharequal}\ empty{\isacharunderscore}spec\ {\isasymlparr}\isanewline
\ \ acc{\isacharunderscore}blocks\ {\isacharcolon}{\isacharequal}\ {\isacharbrackleft}{\isacharbrackright}{\isacharcomma}\isanewline
\ \ map{\isacharunderscore}blocks\ {\isacharcolon}{\isacharequal}\ {\isacharbrackleft}block{\isacharunderscore}map\ m{\isadigit{3}}{\isacharunderscore}virt{\isacharunderscore}ram{\isacharunderscore}{\isadigit{0}}\ {\isadigit{1}}{\isadigit{3}}\ {\isadigit{0}}x{\isadigit{0}}{\isadigit{0}}{\isadigit{0}}{\isadigit{0}}{\isadigit{0}}{\isadigit{0}}{\isadigit{0}}{\isadigit{0}}{\isacharcomma}\ \isanewline
\ \ \ \ \ \ \ \ \ \ \ \ \ \ \ \ \ block{\isacharunderscore}map\ m{\isadigit{3}}{\isacharunderscore}virt{\isacharunderscore}local{\isacharunderscore}rom{\isacharunderscore}{\isadigit{0}}\ {\isadigit{1}}{\isadigit{3}}\ {\isadigit{0}}x{\isadigit{5}}{\isadigit{5}}{\isadigit{0}}{\isadigit{0}}{\isadigit{0}}{\isadigit{0}}{\isadigit{0}}{\isadigit{0}}{\isacharcomma}\isanewline
\ \ \ \ \ \ \ \ \ \ \ \ \ \ \ \ \ block{\isacharunderscore}map\ \ m{\isadigit{3}}{\isacharunderscore}virt{\isacharunderscore}local{\isacharunderscore}ram{\isacharunderscore}{\isadigit{0}}\ {\isadigit{1}}{\isadigit{3}}\ {\isadigit{0}}x{\isadigit{5}}{\isadigit{5}}{\isadigit{0}}{\isadigit{2}}{\isadigit{0}}{\isadigit{0}}{\isadigit{0}}{\isadigit{0}}\ \ {\isacharbrackright}\isanewline
{\isasymrparr}{\isachardoublequoteclose}\ \ \isanewline
\isanewline
\isacommand{definition}\isamarkupfalse%
\ {\isachardoublequoteopen}m{\isadigit{3}}{\isacharunderscore}local{\isacharunderscore}ram\ {\isacharequal}\ {\isacharparenleft}{\isadigit{0}}x{\isadigit{5}}{\isadigit{5}}{\isadigit{0}}{\isadigit{2}}{\isadigit{0}}{\isadigit{0}}{\isadigit{0}}{\isadigit{0}}{\isacharcomma}\ {\isadigit{0}}x{\isadigit{5}}{\isadigit{5}}{\isadigit{0}}{\isadigit{2}}FFFF{\isacharparenright}{\isachardoublequoteclose}\isanewline
\isacommand{definition}\isamarkupfalse%
\ {\isachardoublequoteopen}m{\isadigit{3}}{\isacharunderscore}local{\isacharunderscore}rom\ {\isacharequal}\ {\isacharparenleft}{\isadigit{0}}x{\isadigit{5}}{\isadigit{5}}{\isadigit{0}}{\isadigit{0}}{\isadigit{0}}{\isadigit{0}}{\isadigit{0}}{\isadigit{0}}{\isacharcomma}\ {\isadigit{0}}x{\isadigit{5}}{\isadigit{5}}{\isadigit{0}}{\isadigit{0}}{\isadigit{3}}FFF{\isacharparenright}{\isachardoublequoteclose}\isanewline
\isacommand{definition}\isamarkupfalse%
\ {\isachardoublequoteopen}m{\isadigit{3}}{\isacharunderscore}l{\isadigit{3}}\ {\isacharequal}\ {\isacharparenleft}{\isadigit{0}}x{\isadigit{0}}{\isadigit{0}}{\isadigit{0}}{\isadigit{0}}{\isadigit{0}}{\isadigit{0}}{\isadigit{0}}{\isadigit{0}}{\isacharcomma}\ {\isadigit{0}}x{\isadigit{5}}FFFFFF{\isacharparenright}{\isachardoublequoteclose}\isanewline
\isacommand{definition}\isamarkupfalse%
\ {\isachardoublequoteopen}node{\isacharunderscore}{\isadigit{1}}{\isadigit{3}}{\isacharunderscore}m{\isadigit{3}}{\isacharunderscore}l{\isadigit{2}}{\isacharunderscore}mif\ {\isacharequal}\ empty{\isacharunderscore}spec\ {\isasymlparr}\isanewline
\ \ acc{\isacharunderscore}blocks\ {\isacharcolon}{\isacharequal}\ {\isacharbrackleft}m{\isadigit{3}}{\isacharunderscore}local{\isacharunderscore}ram{\isacharcomma}m{\isadigit{3}}{\isacharunderscore}local{\isacharunderscore}rom{\isacharbrackright}{\isacharcomma}\isanewline
\ \ map{\isacharunderscore}blocks\ {\isacharcolon}{\isacharequal}\ {\isacharbrackleft}direct{\isacharunderscore}map\ m{\isadigit{3}}{\isacharunderscore}l{\isadigit{3}}\ {\isadigit{1}}{\isadigit{4}}{\isacharbrackright}\isanewline
{\isasymrparr}{\isachardoublequoteclose}\isanewline
\isanewline
\isacommand{definition}\isamarkupfalse%
\ {\isachardoublequoteopen}node{\isacharunderscore}{\isadigit{1}}{\isadigit{4}}{\isacharunderscore}m{\isadigit{3}}{\isacharunderscore}phys\ {\isacharequal}\ empty{\isacharunderscore}spec\ {\isasymlparr}\isanewline
\ \ acc{\isacharunderscore}blocks\ {\isacharcolon}{\isacharequal}\ {\isacharbrackleft}{\isacharbrackright}{\isacharcomma}\isanewline
\ \ map{\isacharunderscore}blocks\ {\isacharcolon}{\isacharequal}\ {\isacharbrackleft}block{\isacharunderscore}map\ m{\isadigit{3}}{\isacharunderscore}l{\isadigit{3}}\ {\isadigit{3}}\ {\isadigit{0}}x{\isadigit{8}}{\isadigit{0}}{\isadigit{0}}{\isadigit{0}}{\isadigit{0}}{\isadigit{0}}{\isadigit{0}}\ {\isacharbrackright}\isanewline
{\isasymrparr}{\isachardoublequoteclose}\isanewline
\isanewline
\isacommand{definition}\isamarkupfalse%
\ {\isachardoublequoteopen}sys\ {\isacharequal}\ {\isacharbrackleft}{\isacharparenleft}{\isadigit{1}}{\isacharcomma}node{\isacharunderscore}{\isadigit{1}}{\isacharunderscore}ram{\isacharparenright}{\isacharcomma}\ \isanewline
\ \ \ \ \ \ \ \ \ \ \ \ \ \ \ \ \ \ \ {\isacharparenleft}{\isadigit{2}}{\isacharcomma}\ node{\isacharunderscore}{\isadigit{2}}{\isacharunderscore}gptimer{\isadigit{5}}{\isacharparenright}{\isacharcomma}\ \isanewline
\ \ \ \ \ \ \ \ \ \ \ \ \ \ \ \ \ \ \ {\isacharparenleft}{\isadigit{3}}{\isacharcomma}\ node{\isacharunderscore}{\isadigit{3}}{\isacharunderscore}l{\isadigit{3}}{\isacharunderscore}interconnect{\isacharparenright}{\isacharcomma}\isanewline
\ \ \ \ \ \ \ \ \ \ \ \ \ \ \ \ \ \ \ {\isacharparenleft}{\isadigit{4}}{\isacharcomma}\ node{\isacharunderscore}{\isadigit{4}}{\isacharunderscore}l{\isadigit{4}}{\isacharunderscore}interconnect{\isacharparenright}{\isacharcomma}\isanewline
\ \ \ \ \ \ \ \ \ \ \ \ \ \ \ \ \ \ \ {\isacharparenleft}{\isadigit{5}}{\isacharcomma}\ node{\isacharunderscore}{\isadigit{5}}{\isacharunderscore}dspvirt{\isacharparenright}{\isacharcomma}\ \isanewline
\ \ \ \ \ \ \ \ \ \ \ \ \ \ \ \ \ \ \ {\isacharparenleft}{\isadigit{6}}{\isacharcomma}\ node{\isacharunderscore}{\isadigit{6}}{\isacharunderscore}dspphys{\isacharparenright}{\isacharcomma}\ \isanewline
\ \ \ \ \ \ \ \ \ \ \ \ \ \ \ \ \ \ \ {\isacharparenleft}{\isadigit{7}}{\isacharcomma}\ node{\isacharunderscore}{\isadigit{7}}{\isacharunderscore}sdma{\isacharparenright}{\isacharcomma}\ \isanewline
\ \ \ \ \ \ \ \ \ \ \ \ \ \ \ \ \ \ \ {\isacharparenleft}{\isadigit{8}}{\isacharcomma}node{\isacharunderscore}{\isadigit{8}}{\isacharunderscore}a{\isadigit{9}}virt{\isacharunderscore}{\isadigit{0}}{\isacharparenright}{\isacharcomma}\ \isanewline
\ \ \ \ \ \ \ \ \ \ \ \ \ \ \ \ \ \ \ {\isacharparenleft}{\isadigit{9}}{\isacharcomma}\ node{\isacharunderscore}{\isadigit{9}}{\isacharunderscore}a{\isadigit{9}}phys{\isacharunderscore}{\isadigit{0}}{\isacharparenright}{\isacharcomma}\ \isanewline
\ \ \ \ \ \ \ \ \ \ \ \ \ \ \ \ \ \ \ {\isacharparenleft}{\isadigit{1}}{\isadigit{0}}{\isacharcomma}\ node{\isacharunderscore}{\isadigit{1}}{\isadigit{0}}{\isacharunderscore}a{\isadigit{9}}virt{\isacharunderscore}{\isadigit{1}}{\isacharparenright}{\isacharcomma}\ \isanewline
\ \ \ \ \ \ \ \ \ \ \ \ \ \ \ \ \ \ \ {\isacharparenleft}{\isadigit{1}}{\isadigit{1}}{\isacharcomma}\ node{\isacharunderscore}{\isadigit{1}}{\isadigit{1}}{\isacharunderscore}a{\isadigit{9}}phys{\isacharunderscore}{\isadigit{1}}{\isacharparenright}{\isacharcomma}\ \isanewline
\ \ \ \ \ \ \ \ \ \ \ \ \ \ \ \ \ \ \ {\isacharparenleft}{\isadigit{1}}{\isadigit{2}}{\isacharcomma}node{\isacharunderscore}{\isadigit{1}}{\isadigit{2}}{\isacharunderscore}m{\isadigit{3}}{\isacharunderscore}virt{\isacharunderscore}{\isadigit{0}}\ {\isacharparenright}{\isacharcomma}\ \isanewline
\ \ \ \ \ \ \ \ \ \ \ \ \ \ \ \ \ \ \ {\isacharparenleft}{\isadigit{1}}{\isadigit{3}}{\isacharcomma}node{\isacharunderscore}{\isadigit{1}}{\isadigit{3}}{\isacharunderscore}m{\isadigit{3}}{\isacharunderscore}l{\isadigit{2}}{\isacharunderscore}mif{\isacharparenright}{\isacharcomma}\ \isanewline
\ \ \ \ \ \ \ \ \ \ \ \ \ \ \ \ \ \ \ {\isacharparenleft}{\isadigit{1}}{\isadigit{4}}{\isacharcomma}\ node{\isacharunderscore}{\isadigit{1}}{\isadigit{4}}{\isacharunderscore}m{\isadigit{3}}{\isacharunderscore}phys{\isacharparenright}{\isacharbrackright}{\isachardoublequoteclose}\isanewline
\isadelimtheory
\endisadelimtheory
\isatagtheory
\isacommand{end}\isamarkupfalse%
\endisatagtheory
{\isafoldtheory}%
\isadelimtheory
\endisadelimtheory
\end{isabellebody}%

%
\begin{isabellebody}%
\setisabellecontext{OMAP{\isadigit{4}}{\isadigit{4}}xxInt}%
\isadelimtheory
\endisadelimtheory
\isatagtheory
\endisatagtheory
{\isafoldtheory}%
\isadelimtheory
\endisadelimtheory
\isamarkupsubsubsection{Interrupts%
}
\isamarkuptrue%
\begin{isamarkuptext}%
Note: The DSP core is under NDA. The public datasheet only states 
which interrupts are delivered to the DSP, but not under which vector. Therefore,
the vector numbers have been assumed to be the same as the for the M3 subsystem%
\end{isamarkuptext}\isamarkuptrue%
\begin{isamarkuptext}%
Interrupt domains according to ARM GICv2 specification%
\end{isamarkuptext}\isamarkuptrue%
\isacommand{definition}\isamarkupfalse%
\ {\isachardoublequoteopen}sgi{\isacharunderscore}domain\ {\isacharequal}\ {\isacharparenleft}{\isadigit{0}}{\isacharcomma}{\isadigit{1}}{\isadigit{5}}{\isacharparenright}{\isachardoublequoteclose}\ \isanewline
\isacommand{definition}\isamarkupfalse%
\ {\isachardoublequoteopen}ppi{\isacharunderscore}domain\ {\isacharequal}\ {\isacharparenleft}{\isadigit{1}}{\isadigit{6}}{\isacharcomma}{\isadigit{3}}{\isadigit{1}}{\isacharparenright}{\isachardoublequoteclose}\ \isanewline
\isacommand{definition}\isamarkupfalse%
\ {\isachardoublequoteopen}spi{\isacharunderscore}domain\ {\isacharequal}\ {\isacharparenleft}{\isadigit{3}}{\isadigit{2}}{\isacharcomma}{\isadigit{1}}{\isadigit{0}}{\isadigit{1}}{\isadigit{9}}{\isacharparenright}{\isachardoublequoteclose}\ \isanewline
\isacommand{definition}\isamarkupfalse%
\ {\isachardoublequoteopen}arm{\isacharunderscore}vec{\isacharunderscore}domain\ {\isacharequal}\ {\isacharparenleft}{\isadigit{0}}{\isacharcomma}{\isadigit{1}}{\isadigit{0}}{\isadigit{2}}{\isadigit{0}}{\isacharparenright}{\isachardoublequoteclose}%
\begin{isamarkuptext}%
A9 Core 0. Can create SGI on A9 Core 1.%
\end{isamarkuptext}\isamarkuptrue%
\isacommand{definition}\isamarkupfalse%
\ {\isachardoublequoteopen}node{\isacharunderscore}{\isadigit{0}}{\isacharunderscore}a{\isadigit{9}}{\isacharunderscore}{\isadigit{0}}\ {\isacharequal}\ empty{\isacharunderscore}spec\ {\isasymlparr}\isanewline
\ \ acc{\isacharunderscore}blocks\ {\isacharcolon}{\isacharequal}\ {\isacharbrackleft}{\isacharbrackright}{\isacharcomma}\isanewline
\ \ map{\isacharunderscore}blocks\ {\isacharcolon}{\isacharequal}\ {\isacharbrackleft}one{\isacharunderscore}map\ {\isadigit{0}}\ {\isadigit{1}}\ {\isadigit{0}}{\isacharbrackright}\ \isanewline
{\isasymrparr}{\isachardoublequoteclose}%
\begin{isamarkuptext}%
A9 Core 1. Can create SGI on A9 core 0.%
\end{isamarkuptext}\isamarkuptrue%
\isacommand{definition}\isamarkupfalse%
\ {\isachardoublequoteopen}node{\isacharunderscore}{\isadigit{1}}{\isacharunderscore}a{\isadigit{9}}{\isacharunderscore}{\isadigit{1}}\ {\isacharequal}\ empty{\isacharunderscore}spec\ {\isasymlparr}\isanewline
\ \ acc{\isacharunderscore}blocks\ {\isacharcolon}{\isacharequal}\ {\isacharbrackleft}{\isacharbrackright}{\isacharcomma}\isanewline
\ \ map{\isacharunderscore}blocks\ {\isacharcolon}{\isacharequal}\ {\isacharbrackleft}one{\isacharunderscore}map\ {\isadigit{0}}\ {\isadigit{0}}\ {\isadigit{0}}{\isacharbrackright}\ \isanewline
{\isasymrparr}{\isachardoublequoteclose}%
\begin{isamarkuptext}%
2: DSP. Cannot create interrupts.%
\end{isamarkuptext}\isamarkuptrue%
\isacommand{definition}\isamarkupfalse%
\ {\isachardoublequoteopen}node{\isacharunderscore}{\isadigit{2}}{\isacharunderscore}dsp\ {\isacharequal}\ empty{\isacharunderscore}spec\ {\isasymlparr}\isanewline
\ \ acc{\isacharunderscore}blocks\ {\isacharcolon}{\isacharequal}\ {\isacharbrackleft}{\isacharbrackright}{\isacharcomma}\isanewline
\ \ map{\isacharunderscore}blocks\ {\isacharcolon}{\isacharequal}\ {\isacharbrackleft}{\isacharbrackright}\isanewline
{\isasymrparr}{\isachardoublequoteclose}%
\begin{isamarkuptext}%
M3 Core 0. Cannot create interrupts.%
\end{isamarkuptext}\isamarkuptrue%
\isacommand{definition}\isamarkupfalse%
\ {\isachardoublequoteopen}node{\isacharunderscore}{\isadigit{3}}{\isacharunderscore}m{\isadigit{3}}{\isacharunderscore}{\isadigit{0}}\ {\isacharequal}\ empty{\isacharunderscore}spec\ {\isasymlparr}\isanewline
\ \ acc{\isacharunderscore}blocks\ {\isacharcolon}{\isacharequal}\ {\isacharbrackleft}{\isacharbrackright}{\isacharcomma}\isanewline
\ \ map{\isacharunderscore}blocks\ {\isacharcolon}{\isacharequal}\ {\isacharbrackleft}{\isacharbrackright}\isanewline
{\isasymrparr}{\isachardoublequoteclose}%
\begin{isamarkuptext}%
M3 Core 1. Cannot create interrupts.%
\end{isamarkuptext}\isamarkuptrue%
\isacommand{definition}\isamarkupfalse%
\ {\isachardoublequoteopen}node{\isacharunderscore}{\isadigit{4}}{\isacharunderscore}m{\isadigit{3}}{\isacharunderscore}{\isadigit{1}}\ {\isacharequal}\ empty{\isacharunderscore}spec\ {\isasymlparr}\isanewline
\ \ acc{\isacharunderscore}blocks\ {\isacharcolon}{\isacharequal}\ {\isacharbrackleft}{\isacharbrackright}{\isacharcomma}\isanewline
\ \ map{\isacharunderscore}blocks\ {\isacharcolon}{\isacharequal}\ {\isacharbrackleft}{\isacharbrackright}\isanewline
{\isasymrparr}{\isachardoublequoteclose}%
\begin{isamarkuptext}%
A9 CPU IF 0%
\end{isamarkuptext}\isamarkuptrue%
\isacommand{definition}\isamarkupfalse%
\ {\isachardoublequoteopen}node{\isacharunderscore}{\isadigit{5}}{\isacharunderscore}if{\isacharunderscore}a{\isadigit{9}}{\isacharunderscore}{\isadigit{0}}\ {\isacharequal}\ empty{\isacharunderscore}spec\ {\isasymlparr}\isanewline
\ \ acc{\isacharunderscore}blocks\ {\isacharcolon}{\isacharequal}\ {\isacharbrackleft}ppi{\isacharunderscore}domain{\isacharcomma}\ sgi{\isacharunderscore}domain{\isacharcomma}\ spi{\isacharunderscore}domain{\isacharbrackright}{\isacharcomma}\isanewline
\ \ map{\isacharunderscore}blocks\ {\isacharcolon}{\isacharequal}\ {\isacharbrackleft}{\isacharbrackright}\isanewline
{\isasymrparr}{\isachardoublequoteclose}%
\begin{isamarkuptext}%
A9 CPU IF 1%
\end{isamarkuptext}\isamarkuptrue%
\isacommand{definition}\isamarkupfalse%
\ {\isachardoublequoteopen}node{\isacharunderscore}{\isadigit{6}}{\isacharunderscore}if{\isacharunderscore}a{\isadigit{9}}{\isacharunderscore}{\isadigit{1}}\ {\isacharequal}\ empty{\isacharunderscore}spec\ {\isasymlparr}\isanewline
\ \ acc{\isacharunderscore}blocks\ {\isacharcolon}{\isacharequal}\ {\isacharbrackleft}ppi{\isacharunderscore}domain{\isacharcomma}\ sgi{\isacharunderscore}domain{\isacharcomma}\ spi{\isacharunderscore}domain{\isacharbrackright}{\isacharcomma}\isanewline
\ \ map{\isacharunderscore}blocks\ {\isacharcolon}{\isacharequal}\ {\isacharbrackleft}{\isacharbrackright}\isanewline
{\isasymrparr}{\isachardoublequoteclose}%
\begin{isamarkuptext}%
GIC Dist: The GIC can't change vector numbers, but destination for SPIs is configurable.%
\end{isamarkuptext}\isamarkuptrue%
\isacommand{definition}\isamarkupfalse%
\ {\isachardoublequoteopen}node{\isacharunderscore}{\isadigit{7}}{\isacharunderscore}gic\ {\isacharequal}\ empty{\isacharunderscore}spec\ {\isasymlparr}\isanewline
\ \ acc{\isacharunderscore}blocks\ {\isacharcolon}{\isacharequal}\ {\isacharbrackleft}{\isacharbrackright}{\isacharcomma}\isanewline
\ \ map{\isacharunderscore}blocks\ {\isacharcolon}{\isacharequal}\ {\isacharbrackleft}\isanewline
\ \ one{\isacharunderscore}map\ {\isadigit{7}}{\isadigit{3}}\ {\isadigit{5}}\ {\isadigit{7}}{\isadigit{3}}{\isacharcomma}\ \ \ \ \ \ \ \ \ \ \ \ \ \ \ \ {\isacharparenleft}{\isacharasterisk}\ GPTIMER{\isadigit{5}}\ {\isadigit{4}}{\isadigit{1}}{\isacharplus}{\isadigit{3}}{\isadigit{2}}\ \ to\ core\ {\isadigit{0}}\ {\isacharasterisk}{\isacharparenright}\ \isanewline
\ \ one{\isacharunderscore}map\ {\isadigit{1}}{\isadigit{3}}{\isadigit{1}}\ {\isadigit{5}}\ {\isadigit{1}}{\isadigit{3}}{\isadigit{1}}{\isacharcomma}\ \ \ \ \ \ \ \ \ \ \ \ \ \ {\isacharparenleft}{\isacharasterisk}\ Audio\ {\isadigit{9}}{\isadigit{9}}{\isacharplus}{\isadigit{3}}{\isadigit{2}}\ to\ core\ {\isadigit{0}}\ {\isacharasterisk}{\isacharparenright}\isanewline
\ \ one{\isacharunderscore}map\ {\isadigit{1}}{\isadigit{3}}{\isadigit{2}}\ {\isadigit{5}}\ {\isadigit{1}}{\isadigit{3}}{\isadigit{2}}{\isacharcomma}\ \ \ \ \ \ \ \ \ \ \ \ \ \ {\isacharparenleft}{\isacharasterisk}\ M{\isadigit{3}}\ MMU{\isadigit{2}}\ {\isadigit{1}}{\isadigit{0}}{\isadigit{0}}{\isacharplus}{\isadigit{3}}{\isadigit{2}}\ {\isacharasterisk}{\isacharparenright}\isanewline
\ \ one{\isacharunderscore}map\ {\isadigit{4}}{\isadigit{4}}\ {\isadigit{5}}\ {\isadigit{4}}{\isadigit{4}}{\isacharcomma}\ \ \ \ \ \ \ \ \ \ \ \ \ \ \ \ {\isacharparenleft}{\isacharasterisk}\ SDMA\ interrupts{\isacharcolon}\ {\isadigit{1}}{\isadigit{2}}{\isacharminus}{\isadigit{1}}{\isadigit{5}}{\isacharplus}{\isadigit{3}}{\isadigit{2}}\ to\ core\ {\isadigit{0}}\ {\isacharasterisk}{\isacharparenright}\isanewline
\ \ one{\isacharunderscore}map\ {\isadigit{4}}{\isadigit{5}}\ {\isadigit{5}}\ {\isadigit{4}}{\isadigit{5}}{\isacharcomma}\ \ \ \ \ \ \ \ \ \ \ \ \ \ \ \ {\isacharparenleft}{\isacharasterisk}\ {\isachardot}{\isachardot}{\isachardot}\ to\ core\ {\isadigit{0}}\ {\isacharasterisk}{\isacharparenright}\isanewline
\ \ one{\isacharunderscore}map\ {\isadigit{4}}{\isadigit{6}}\ {\isadigit{6}}\ {\isadigit{4}}{\isadigit{6}}{\isacharcomma}\ \ \ \ \ \ \ \ \ \ \ \ \ \ \ \ {\isacharparenleft}{\isacharasterisk}\ {\isachardot}{\isachardot}{\isachardot}\ to\ core\ {\isadigit{1}}\ {\isacharasterisk}{\isacharparenright}\isanewline
\ \ one{\isacharunderscore}map\ {\isadigit{4}}{\isadigit{7}}\ {\isadigit{6}}\ {\isadigit{4}}{\isadigit{7}}\ \ \ \ \ \ \ \ \ \ \ \ \ \ \ \ \ {\isacharparenleft}{\isacharasterisk}\ {\isachardot}{\isachardot}{\isachardot}\ to\ core\ {\isadigit{1}}\ {\isacharasterisk}{\isacharparenright}\isanewline
{\isacharbrackright}\isanewline
{\isasymrparr}{\isachardoublequoteclose}%
\begin{isamarkuptext}%
DSP INTC: Since the GIC accepts SDMA/interrupts, it must not be accepted here. In
 fact, we do not accept any interrupts, which corresponds to masking everything%
\end{isamarkuptext}\isamarkuptrue%
\isacommand{definition}\isamarkupfalse%
\ {\isachardoublequoteopen}node{\isacharunderscore}{\isadigit{8}}{\isacharunderscore}dsp{\isacharunderscore}intc\ {\isacharequal}\ empty{\isacharunderscore}spec\ {\isasymlparr}\isanewline
\ \ acc{\isacharunderscore}blocks\ {\isacharcolon}{\isacharequal}\ {\isacharbrackleft}{\isacharbrackright}{\isacharcomma}\isanewline
\ \ map{\isacharunderscore}blocks\ {\isacharcolon}{\isacharequal}\ {\isacharbrackleft}{\isacharbrackright}\isanewline
{\isasymrparr}{\isachardoublequoteclose}%
\begin{isamarkuptext}%
NVIC 0: Since the GIC accepts SDMA/ interrupts, it must not be accepted here.%
\end{isamarkuptext}\isamarkuptrue%
\isacommand{definition}\isamarkupfalse%
\ {\isachardoublequoteopen}node{\isacharunderscore}{\isadigit{9}}{\isacharunderscore}nvic{\isacharunderscore}{\isadigit{0}}\ {\isacharequal}\ empty{\isacharunderscore}spec\ {\isasymlparr}\isanewline
\ \ acc{\isacharunderscore}blocks\ {\isacharcolon}{\isacharequal}\ {\isacharbrackleft}{\isacharbrackright}{\isacharcomma}\isanewline
\ \ map{\isacharunderscore}blocks\ {\isacharcolon}{\isacharequal}\ {\isacharbrackleft}{\isacharbrackright}\ \isanewline
{\isasymrparr}{\isachardoublequoteclose}%
\begin{isamarkuptext}%
NVIC 1%
\end{isamarkuptext}\isamarkuptrue%
\isacommand{definition}\isamarkupfalse%
\ {\isachardoublequoteopen}node{\isacharunderscore}{\isadigit{1}}{\isadigit{0}}{\isacharunderscore}nvic{\isacharunderscore}{\isadigit{1}}\ {\isacharequal}\ empty{\isacharunderscore}spec\ {\isasymlparr}\isanewline
\ \ acc{\isacharunderscore}blocks\ {\isacharcolon}{\isacharequal}\ {\isacharbrackleft}{\isacharbrackright}{\isacharcomma}\isanewline
\ \ map{\isacharunderscore}blocks\ {\isacharcolon}{\isacharequal}\ {\isacharbrackleft}{\isacharbrackright}\isanewline
{\isasymrparr}{\isachardoublequoteclose}%
\begin{isamarkuptext}%
A9 Core 0 Private Timer%
\end{isamarkuptext}\isamarkuptrue%
\isacommand{definition}\isamarkupfalse%
\ {\isachardoublequoteopen}node{\isacharunderscore}{\isadigit{1}}{\isadigit{1}}{\isacharunderscore}pt{\isacharunderscore}{\isadigit{0}}\ {\isacharequal}\ empty{\isacharunderscore}spec\ {\isasymlparr}\isanewline
\ \ acc{\isacharunderscore}blocks\ {\isacharcolon}{\isacharequal}\ {\isacharbrackleft}{\isacharbrackright}{\isacharcomma}\isanewline
\ \ map{\isacharunderscore}blocks\ {\isacharcolon}{\isacharequal}\ {\isacharbrackleft}one{\isacharunderscore}map\ {\isadigit{0}}\ {\isadigit{5}}\ {\isadigit{2}}{\isadigit{9}}{\isacharbrackright}\ \isanewline
{\isasymrparr}{\isachardoublequoteclose}%
\begin{isamarkuptext}%
A9 Core 1 Private Timer.%
\end{isamarkuptext}\isamarkuptrue%
\isacommand{definition}\isamarkupfalse%
\ {\isachardoublequoteopen}node{\isacharunderscore}{\isadigit{1}}{\isadigit{2}}{\isacharunderscore}pt{\isacharunderscore}{\isadigit{1}}\ {\isacharequal}\ empty{\isacharunderscore}spec\ {\isasymlparr}\isanewline
\ \ acc{\isacharunderscore}blocks\ {\isacharcolon}{\isacharequal}\ {\isacharbrackleft}{\isacharbrackright}{\isacharcomma}\isanewline
\ \ map{\isacharunderscore}blocks\ {\isacharcolon}{\isacharequal}\ {\isacharbrackleft}one{\isacharunderscore}map\ {\isadigit{0}}\ {\isadigit{6}}\ {\isadigit{2}}{\isadigit{9}}{\isacharbrackright}\ \ \isanewline
{\isasymrparr}{\isachardoublequoteclose}%
\begin{isamarkuptext}%
GPTIMER5: DSP is under NDA, vec 41 is guessed.%
\end{isamarkuptext}\isamarkuptrue%
\isacommand{definition}\isamarkupfalse%
\ {\isachardoublequoteopen}node{\isacharunderscore}{\isadigit{1}}{\isadigit{3}}{\isacharunderscore}gptimer{\isadigit{5}}\ {\isacharequal}\ empty{\isacharunderscore}spec\ {\isasymlparr}\isanewline
\ \ acc{\isacharunderscore}blocks\ {\isacharcolon}{\isacharequal}\ {\isacharbrackleft}{\isacharbrackright}{\isacharcomma}\isanewline
\ \ map{\isacharunderscore}blocks\ {\isacharcolon}{\isacharequal}\ {\isacharbrackleft}one{\isacharunderscore}map\ {\isadigit{0}}\ {\isadigit{1}}{\isadigit{6}}\ {\isadigit{4}}{\isadigit{1}}{\isacharbrackright}\ {\isacharparenleft}{\isacharasterisk}\ destination\ set\ of\ {\isadigit{3}}\ is\ {\isacharbrackleft}{\isacharparenleft}{\isadigit{1}}{\isadigit{6}}{\isacharcomma}{\isadigit{4}}{\isadigit{1}}{\isacharparenright}{\isacharcomma}\ {\isacharparenleft}{\isadigit{8}}{\isacharcomma}{\isadigit{4}}{\isadigit{1}}{\isacharparenright}{\isacharbrackright}\ {\isacharasterisk}{\isacharparenright}\isanewline
{\isasymrparr}{\isachardoublequoteclose}%
\begin{isamarkuptext}%
Audio%
\end{isamarkuptext}\isamarkuptrue%
\isacommand{definition}\isamarkupfalse%
\ {\isachardoublequoteopen}node{\isacharunderscore}{\isadigit{1}}{\isadigit{4}}{\isacharunderscore}audio\ {\isacharequal}\ empty{\isacharunderscore}spec\ {\isasymlparr}\isanewline
\ \ acc{\isacharunderscore}blocks\ {\isacharcolon}{\isacharequal}\ {\isacharbrackleft}{\isacharbrackright}{\isacharcomma}\isanewline
\ \ map{\isacharunderscore}blocks\ {\isacharcolon}{\isacharequal}\ {\isacharbrackleft}one{\isacharunderscore}map\ {\isadigit{0}}\ {\isadigit{1}}{\isadigit{6}}\ {\isadigit{9}}{\isadigit{9}}{\isacharbrackright}\ \isanewline
{\isasymrparr}{\isachardoublequoteclose}%
\begin{isamarkuptext}%
SDMA: Generates four interrupts.%
\end{isamarkuptext}\isamarkuptrue%
\isacommand{definition}\isamarkupfalse%
\ {\isachardoublequoteopen}node{\isacharunderscore}{\isadigit{1}}{\isadigit{5}}{\isacharunderscore}sdma\ {\isacharequal}\ empty{\isacharunderscore}spec\ {\isasymlparr}\isanewline
\ \ acc{\isacharunderscore}blocks\ {\isacharcolon}{\isacharequal}\ {\isacharbrackleft}{\isacharbrackright}{\isacharcomma}\isanewline
\ \ map{\isacharunderscore}blocks\ {\isacharcolon}{\isacharequal}\ {\isacharbrackleft}\isanewline
\ \ \ \ one{\isacharunderscore}map\ {\isadigit{0}}\ {\isadigit{1}}{\isadigit{6}}\ {\isadigit{1}}{\isadigit{2}}{\isacharcomma}\ {\isacharparenleft}{\isacharasterisk}\ destination\ set\ of\ {\isadigit{0}}\ is\ {\isacharbrackleft}{\isacharparenleft}{\isadigit{1}}{\isadigit{6}}{\isacharcomma}{\isadigit{1}}{\isadigit{2}}{\isacharparenright}{\isacharcomma}{\isacharparenleft}{\isadigit{8}}{\isacharcomma}{\isadigit{1}}{\isadigit{8}}{\isacharparenright}{\isacharcomma}{\isacharparenleft}{\isadigit{9}}{\isacharcomma}{\isadigit{1}}{\isadigit{8}}{\isacharparenright}{\isacharcomma}{\isacharparenleft}{\isadigit{1}}{\isadigit{0}}{\isacharcomma}{\isadigit{1}}{\isadigit{8}}{\isacharparenright}{\isacharbrackright}\ {\isacharasterisk}{\isacharparenright}\isanewline
\ \ \ \ one{\isacharunderscore}map\ {\isadigit{1}}\ {\isadigit{1}}{\isadigit{6}}\ {\isadigit{1}}{\isadigit{3}}{\isacharcomma}\ {\isacharparenleft}{\isacharasterisk}\ destination\ set\ of\ {\isadigit{1}}\ is\ {\isacharbrackleft}{\isacharparenleft}{\isadigit{1}}{\isadigit{6}}{\isacharcomma}{\isadigit{1}}{\isadigit{3}}{\isacharparenright}{\isacharcomma}{\isacharparenleft}{\isadigit{8}}{\isacharcomma}{\isadigit{1}}{\isadigit{9}}{\isacharparenright}{\isacharcomma}{\isacharparenleft}{\isadigit{9}}{\isacharcomma}{\isadigit{1}}{\isadigit{9}}{\isacharparenright}{\isacharcomma}{\isacharparenleft}{\isadigit{1}}{\isadigit{0}}{\isacharcomma}{\isadigit{1}}{\isadigit{9}}{\isacharparenright}{\isacharbrackright}\ {\isacharasterisk}{\isacharparenright}\isanewline
\ \ \ \ one{\isacharunderscore}map\ {\isadigit{2}}\ {\isadigit{1}}{\isadigit{6}}\ {\isadigit{1}}{\isadigit{4}}{\isacharcomma}\ {\isacharparenleft}{\isacharasterisk}\ destination\ set\ of\ {\isadigit{2}}\ is\ {\isacharbrackleft}{\isacharparenleft}{\isadigit{1}}{\isadigit{6}}{\isacharcomma}{\isadigit{1}}{\isadigit{4}}{\isacharparenright}{\isacharcomma}{\isacharparenleft}{\isadigit{9}}{\isacharcomma}{\isadigit{2}}{\isadigit{0}}{\isacharparenright}{\isacharcomma}{\isacharparenleft}{\isadigit{1}}{\isadigit{0}}{\isacharcomma}{\isadigit{2}}{\isadigit{0}}{\isacharparenright}{\isacharbrackright}\ {\isacharasterisk}{\isacharparenright}\isanewline
\ \ \ \ one{\isacharunderscore}map\ {\isadigit{3}}\ {\isadigit{1}}{\isadigit{6}}\ {\isadigit{1}}{\isadigit{5}}\ \ {\isacharparenleft}{\isacharasterisk}\ destination\ set\ of\ {\isadigit{3}}\ is\ {\isacharbrackleft}{\isacharparenleft}{\isadigit{1}}{\isadigit{6}}{\isacharcomma}{\isadigit{1}}{\isadigit{5}}{\isacharparenright}{\isacharcomma}{\isacharparenleft}{\isadigit{9}}{\isacharcomma}{\isadigit{2}}{\isadigit{1}}{\isacharparenright}{\isacharcomma}{\isacharparenleft}{\isadigit{1}}{\isadigit{0}}{\isacharcomma}{\isadigit{2}}{\isadigit{1}}{\isacharparenright}{\isacharbrackright}\ {\isacharasterisk}{\isacharparenright}\isanewline
\ \ {\isacharbrackright}\ \ \isanewline
{\isasymrparr}{\isachardoublequoteclose}%
\begin{isamarkuptext}%
SPI: map from datasheet space into GIC space. Adds 32 to each vector number.%
\end{isamarkuptext}\isamarkuptrue%
\isacommand{definition}\isamarkupfalse%
\ {\isachardoublequoteopen}node{\isacharunderscore}{\isadigit{1}}{\isadigit{6}}{\isacharunderscore}spimap\ {\isacharequal}\ empty{\isacharunderscore}spec\ {\isasymlparr}\isanewline
\ \ acc{\isacharunderscore}blocks\ {\isacharcolon}{\isacharequal}\ {\isacharbrackleft}{\isacharbrackright}{\isacharcomma}\isanewline
\ \ map{\isacharunderscore}blocks\ {\isacharcolon}{\isacharequal}\ {\isacharbrackleft}\isanewline
\ \ block{\isacharunderscore}map\ {\isacharparenleft}{\isadigit{0}}{\isacharcomma}\ {\isadigit{9}}{\isadigit{8}}{\isadigit{7}}{\isacharparenright}\ {\isadigit{7}}\ {\isadigit{3}}{\isadigit{2}}\ \ {\isacharparenleft}{\isacharasterisk}\ GIC\ accepts\ at\ most\ {\isadigit{1}}{\isadigit{0}}{\isadigit{1}}{\isadigit{9}}{\isacharminus}{\isadigit{3}}{\isadigit{2}}\ interrupts\ {\isacharasterisk}{\isacharparenright}\isanewline
{\isacharbrackright}\isanewline
{\isasymrparr}{\isachardoublequoteclose}%
\begin{isamarkuptext}%
GPTIMER5%
\end{isamarkuptext}\isamarkuptrue%
\isacommand{definition}\isamarkupfalse%
\ {\isachardoublequoteopen}node{\isacharunderscore}{\isadigit{1}}{\isadigit{7}}{\isacharunderscore}m{\isadigit{3}}mmu\ {\isacharequal}\ empty{\isacharunderscore}spec\ {\isasymlparr}\isanewline
\ \ acc{\isacharunderscore}blocks\ {\isacharcolon}{\isacharequal}\ {\isacharbrackleft}{\isacharbrackright}{\isacharcomma}\isanewline
\ \ map{\isacharunderscore}blocks\ {\isacharcolon}{\isacharequal}\ {\isacharbrackleft}one{\isacharunderscore}map\ {\isadigit{0}}\ {\isadigit{1}}{\isadigit{6}}\ {\isadigit{1}}{\isadigit{0}}{\isadigit{0}}{\isacharbrackright}\ \isanewline
{\isasymrparr}{\isachardoublequoteclose}\isanewline
\isanewline
\isacommand{definition}\isamarkupfalse%
\ {\isachardoublequoteopen}sys\ {\isacharequal}\ {\isacharbrackleft}\isanewline
\ \ {\isacharparenleft}{\isadigit{0}}{\isacharcomma}node{\isacharunderscore}{\isadigit{0}}{\isacharunderscore}a{\isadigit{9}}{\isacharunderscore}{\isadigit{0}}{\isacharparenright}{\isacharcomma}\isanewline
\ \ {\isacharparenleft}{\isadigit{1}}{\isacharcomma}node{\isacharunderscore}{\isadigit{1}}{\isacharunderscore}a{\isadigit{9}}{\isacharunderscore}{\isadigit{1}}{\isacharparenright}{\isacharcomma}\isanewline
\ \ {\isacharparenleft}{\isadigit{2}}{\isacharcomma}node{\isacharunderscore}{\isadigit{2}}{\isacharunderscore}dsp{\isacharparenright}{\isacharcomma}\isanewline
\ \ {\isacharparenleft}{\isadigit{3}}{\isacharcomma}node{\isacharunderscore}{\isadigit{3}}{\isacharunderscore}m{\isadigit{3}}{\isacharunderscore}{\isadigit{0}}{\isacharparenright}{\isacharcomma}\isanewline
\ \ {\isacharparenleft}{\isadigit{4}}{\isacharcomma}node{\isacharunderscore}{\isadigit{4}}{\isacharunderscore}m{\isadigit{3}}{\isacharunderscore}{\isadigit{1}}{\isacharparenright}{\isacharcomma}\isanewline
\ \ {\isacharparenleft}{\isadigit{5}}{\isacharcomma}node{\isacharunderscore}{\isadigit{5}}{\isacharunderscore}if{\isacharunderscore}a{\isadigit{9}}{\isacharunderscore}{\isadigit{0}}{\isacharparenright}{\isacharcomma}\isanewline
\ \ {\isacharparenleft}{\isadigit{6}}{\isacharcomma}node{\isacharunderscore}{\isadigit{6}}{\isacharunderscore}if{\isacharunderscore}a{\isadigit{9}}{\isacharunderscore}{\isadigit{1}}{\isacharparenright}{\isacharcomma}\isanewline
\ \ {\isacharparenleft}{\isadigit{7}}{\isacharcomma}node{\isacharunderscore}{\isadigit{7}}{\isacharunderscore}gic{\isacharparenright}{\isacharcomma}\isanewline
\ \ {\isacharparenleft}{\isadigit{8}}{\isacharcomma}node{\isacharunderscore}{\isadigit{8}}{\isacharunderscore}dsp{\isacharunderscore}intc{\isacharparenright}{\isacharcomma}\isanewline
\ \ {\isacharparenleft}{\isadigit{9}}{\isacharcomma}node{\isacharunderscore}{\isadigit{9}}{\isacharunderscore}nvic{\isacharunderscore}{\isadigit{0}}{\isacharparenright}{\isacharcomma}\isanewline
\ \ {\isacharparenleft}{\isadigit{1}}{\isadigit{0}}{\isacharcomma}node{\isacharunderscore}{\isadigit{1}}{\isadigit{0}}{\isacharunderscore}nvic{\isacharunderscore}{\isadigit{1}}{\isacharparenright}{\isacharcomma}\isanewline
\ \ {\isacharparenleft}{\isadigit{1}}{\isadigit{1}}{\isacharcomma}node{\isacharunderscore}{\isadigit{1}}{\isadigit{1}}{\isacharunderscore}pt{\isacharunderscore}{\isadigit{0}}{\isacharparenright}{\isacharcomma}\isanewline
\ \ {\isacharparenleft}{\isadigit{1}}{\isadigit{2}}{\isacharcomma}node{\isacharunderscore}{\isadigit{1}}{\isadigit{2}}{\isacharunderscore}pt{\isacharunderscore}{\isadigit{1}}{\isacharparenright}{\isacharcomma}\isanewline
\ \ {\isacharparenleft}{\isadigit{1}}{\isadigit{3}}{\isacharcomma}node{\isacharunderscore}{\isadigit{1}}{\isadigit{3}}{\isacharunderscore}gptimer{\isadigit{5}}{\isacharparenright}{\isacharcomma}\isanewline
\ \ {\isacharparenleft}{\isadigit{1}}{\isadigit{4}}{\isacharcomma}node{\isacharunderscore}{\isadigit{1}}{\isadigit{4}}{\isacharunderscore}audio{\isacharparenright}{\isacharcomma}\isanewline
\ \ {\isacharparenleft}{\isadigit{1}}{\isadigit{5}}{\isacharcomma}node{\isacharunderscore}{\isadigit{1}}{\isadigit{5}}{\isacharunderscore}sdma{\isacharparenright}{\isacharcomma}\isanewline
\ \ {\isacharparenleft}{\isadigit{1}}{\isadigit{6}}{\isacharcomma}node{\isacharunderscore}{\isadigit{1}}{\isadigit{6}}{\isacharunderscore}spimap{\isacharparenright}{\isacharcomma}\isanewline
\ \ {\isacharparenleft}{\isadigit{1}}{\isadigit{7}}{\isacharcomma}node{\isacharunderscore}{\isadigit{1}}{\isadigit{7}}{\isacharunderscore}m{\isadigit{3}}mmu{\isacharparenright}\isanewline
{\isacharbrackright}{\isachardoublequoteclose}\ \ \isanewline
\isadelimtheory
\isanewline
\endisadelimtheory
\isatagtheory
\isacommand{end}\isamarkupfalse%
\endisatagtheory
{\isafoldtheory}%
\isadelimtheory
\endisadelimtheory
\end{isabellebody}%

\subsection{A desktop PC}
\label{isabelle:desktop}
\camready{This is the full model representation of the x86 desktop computer 
that we already introduced in ~\autoref{sec:realsystems:desktop}.}
\begin{isabellebody}%
\setisabellecontext{Desktop}%
\isadelimtheory
\endisadelimtheory
\isatagtheory
\endisatagtheory
{\isafoldtheory}%
\isadelimtheory
\endisadelimtheory
\isamarkupsubsubsection{Address spaces%
}
\isamarkuptrue%
\begin{isamarkuptext}%
The Interconnect%
\end{isamarkuptext}\isamarkuptrue%
\isacommand{definition}\isamarkupfalse%
\ {\isachardoublequoteopen}dram{\isacharunderscore}sys{\isadigit{0}}\ {\isacharequal}\ {\isacharparenleft}{\isadigit{0}}x{\isadigit{0}}{\isadigit{0}}{\isadigit{0}}{\isadigit{0}}{\isadigit{0}}{\isadigit{0}}{\isadigit{0}}{\isadigit{0}}{\isadigit{0}}{\isadigit{0}}{\isadigit{1}}{\isadigit{0}}{\isadigit{0}}{\isadigit{0}}{\isadigit{0}}{\isadigit{0}}{\isacharcomma}\ {\isadigit{0}}x{\isadigit{0}}{\isadigit{0}}{\isadigit{0}}{\isadigit{0}}{\isadigit{0}}{\isadigit{0}}{\isadigit{0}}{\isadigit{0}}C{\isadigit{0}}{\isadigit{0}}FFFFF{\isacharparenright}{\isachardoublequoteclose}\ \isanewline
\isacommand{definition}\isamarkupfalse%
\ {\isachardoublequoteopen}dram{\isacharunderscore}sys{\isadigit{1}}\ {\isacharequal}\ {\isacharparenleft}{\isadigit{0}}x{\isadigit{0}}{\isadigit{0}}{\isadigit{0}}{\isadigit{0}}{\isadigit{0}}{\isadigit{0}}{\isadigit{0}}{\isadigit{1}}{\isadigit{0}}{\isadigit{0}}{\isadigit{0}}{\isadigit{0}}{\isadigit{0}}{\isadigit{0}}{\isadigit{0}}{\isadigit{0}}{\isacharcomma}\ {\isadigit{0}}x{\isadigit{0}}{\isadigit{0}}{\isadigit{0}}{\isadigit{0}}{\isadigit{0}}{\isadigit{0}}{\isadigit{0}}{\isadigit{8}}{\isadigit{3}}FFFFFFF{\isacharparenright}{\isachardoublequoteclose}\isanewline
\isacommand{definition}\isamarkupfalse%
\ {\isachardoublequoteopen}pci{\isacharunderscore}devs\ {\isacharequal}\ \ {\isacharparenleft}{\isadigit{0}}x{\isadigit{0}}{\isadigit{0}}{\isadigit{0}}{\isadigit{0}}{\isadigit{0}}{\isadigit{0}}{\isadigit{0}}{\isadigit{0}}C{\isadigit{0}}{\isadigit{8}}{\isadigit{0}}{\isadigit{0}}{\isadigit{0}}{\isadigit{0}}{\isadigit{0}}{\isacharcomma}\ {\isadigit{0}}x{\isadigit{0}}{\isadigit{0}}{\isadigit{0}}{\isadigit{0}}{\isadigit{0}}{\isadigit{0}}{\isadigit{0}}{\isadigit{0}}C{\isadigit{2}}FFFFFF{\isacharparenright}{\isachardoublequoteclose}\isanewline
\isacommand{definition}\isamarkupfalse%
\ {\isachardoublequoteopen}node{\isacharunderscore}{\isadigit{0}}{\isacharunderscore}interconnect\ {\isacharequal}\ empty{\isacharunderscore}spec\ {\isasymlparr}\isanewline
\ \ acc{\isacharunderscore}blocks\ {\isacharcolon}{\isacharequal}\ {\isacharbrackleft}{\isacharbrackright}{\isacharcomma}\isanewline
\ \ map{\isacharunderscore}blocks\ {\isacharcolon}{\isacharequal}\ {\isacharbrackleft}block{\isacharunderscore}map\ dram{\isacharunderscore}sys{\isadigit{0}}\ {\isadigit{1}}\ {\isadigit{0}}x{\isadigit{0}}{\isacharcomma}\ block{\isacharunderscore}map\ dram{\isacharunderscore}sys{\isadigit{1}}\ {\isadigit{1}}\ {\isadigit{0}}xC{\isadigit{0}}{\isadigit{0}}{\isadigit{0}}{\isadigit{0}}{\isadigit{0}}{\isadigit{0}}{\isadigit{0}}{\isacharcomma}\isanewline
\ \ \ \ \ \ \ \ \ \ \ \ \ \ \ \ \ direct{\isacharunderscore}map\ pci{\isacharunderscore}devs\ {\isadigit{2}}{\isacharbrackright}\isanewline
{\isasymrparr}{\isachardoublequoteclose}%
\begin{isamarkuptext}%
The DRAM Controller: has 2 channels with 16G each.%
\end{isamarkuptext}\isamarkuptrue%
\isacommand{definition}\isamarkupfalse%
\ {\isachardoublequoteopen}dram{\isadigit{0}}\ {\isacharequal}\ {\isacharparenleft}{\isadigit{0}}x{\isadigit{0}}{\isadigit{0}}{\isadigit{0}}{\isadigit{0}}{\isadigit{0}}{\isadigit{0}}{\isadigit{0}}{\isadigit{0}}{\isadigit{0}}{\isacharcomma}\ {\isadigit{0}}x{\isadigit{3}}FFFFFFFF{\isacharparenright}{\isachardoublequoteclose}\isanewline
\isacommand{definition}\isamarkupfalse%
\ {\isachardoublequoteopen}dram{\isadigit{1}}\ {\isacharequal}\ {\isacharparenleft}{\isadigit{0}}x{\isadigit{4}}{\isadigit{0}}{\isadigit{0}}{\isadigit{0}}{\isadigit{0}}{\isadigit{0}}{\isadigit{0}}{\isadigit{0}}{\isadigit{0}}{\isacharcomma}\ {\isadigit{0}}x{\isadigit{7}}FFFFFFFF{\isacharparenright}{\isachardoublequoteclose}\isanewline
\isacommand{definition}\isamarkupfalse%
\ {\isachardoublequoteopen}node{\isacharunderscore}{\isadigit{1}}{\isacharunderscore}dram\ {\isacharequal}\ empty{\isacharunderscore}spec\ {\isasymlparr}\isanewline
\ \ acc{\isacharunderscore}blocks\ {\isacharcolon}{\isacharequal}\ {\isacharbrackleft}dram{\isadigit{0}}{\isacharcomma}\ dram{\isadigit{1}}{\isacharbrackright}{\isacharcomma}\isanewline
\ \ map{\isacharunderscore}blocks\ {\isacharcolon}{\isacharequal}\ {\isacharbrackleft}{\isacharbrackright}\isanewline
{\isasymrparr}{\isachardoublequoteclose}%
\begin{isamarkuptext}%
The PCI Root Complex%
\end{isamarkuptext}\isamarkuptrue%
\isacommand{definition}\isamarkupfalse%
\ {\isachardoublequoteopen}xhci\ \ {\isacharequal}\ {\isacharparenleft}{\isadigit{0}}xC{\isadigit{1}}{\isadigit{5}}{\isadigit{8}}{\isadigit{0}}{\isadigit{0}}{\isadigit{0}}{\isadigit{0}}{\isacharcomma}\ {\isadigit{0}}xC{\isadigit{1}}{\isadigit{5}}{\isadigit{8}}FFFF{\isacharparenright}{\isachardoublequoteclose}\isanewline
\isacommand{definition}\isamarkupfalse%
\ {\isachardoublequoteopen}e{\isadigit{1}}{\isadigit{0}}{\isadigit{0}}{\isadigit{0}}\ {\isacharequal}\ {\isacharparenleft}{\isadigit{0}}xC{\isadigit{1}}{\isadigit{3}}{\isadigit{0}}{\isadigit{0}}{\isadigit{0}}{\isadigit{0}}{\isadigit{0}}{\isacharcomma}\ {\isadigit{0}}xC{\isadigit{1}}{\isadigit{3}}FFFFF{\isacharparenright}{\isachardoublequoteclose}\isanewline
\isacommand{definition}\isamarkupfalse%
\ {\isachardoublequoteopen}ahci\ \ {\isacharequal}\ {\isacharparenleft}{\isadigit{0}}xC{\isadigit{1}}{\isadigit{5}}{\isadigit{2}}{\isadigit{0}}{\isadigit{0}}{\isadigit{0}}{\isadigit{0}}{\isacharcomma}\ {\isadigit{0}}xC{\isadigit{1}}{\isadigit{5}}{\isadigit{2}}{\isadigit{0}}{\isadigit{7}}FF{\isacharparenright}{\isachardoublequoteclose}\isanewline
\isacommand{definition}\isamarkupfalse%
\ {\isachardoublequoteopen}vga{\isadigit{2}}\ \ {\isacharequal}\ {\isacharparenleft}{\isadigit{0}}xC{\isadigit{1}}{\isadigit{0}}{\isadigit{1}}{\isadigit{0}}{\isadigit{0}}{\isadigit{0}}{\isadigit{0}}{\isacharcomma}\ {\isadigit{0}}xC{\isadigit{1}}{\isadigit{0}}{\isadigit{1}}{\isadigit{3}}FFF{\isacharparenright}{\isachardoublequoteclose}\isanewline
\isacommand{definition}\isamarkupfalse%
\ {\isachardoublequoteopen}vga{\isadigit{1}}\ \ {\isacharequal}\ {\isacharparenleft}{\isadigit{0}}xC{\isadigit{2}}{\isadigit{0}}{\isadigit{0}}{\isadigit{0}}{\isadigit{0}}{\isadigit{0}}{\isadigit{0}}{\isacharcomma}\ {\isadigit{0}}xC{\isadigit{2}}FFFFFF{\isacharparenright}{\isachardoublequoteclose}\isanewline
\ \ \isanewline
\isacommand{definition}\isamarkupfalse%
\ {\isachardoublequoteopen}vga{\isadigit{4}}\ \ {\isacharequal}\ {\isacharparenleft}{\isadigit{0}}xC{\isadigit{1}}{\isadigit{0}}{\isadigit{0}}{\isadigit{0}}{\isadigit{0}}{\isadigit{0}}{\isadigit{0}}{\isacharcomma}\ {\isadigit{0}}xC{\isadigit{1}}{\isadigit{0}}{\isadigit{0}}FFFF{\isacharparenright}{\isachardoublequoteclose}\isanewline
\isacommand{definition}\isamarkupfalse%
\ {\isachardoublequoteopen}vga{\isadigit{3}}\ \ {\isacharequal}\ {\isacharparenleft}{\isadigit{0}}xC{\isadigit{0}}{\isadigit{8}}{\isadigit{0}}{\isadigit{0}}{\isadigit{0}}{\isadigit{0}}{\isadigit{0}}{\isacharcomma}\ {\isadigit{0}}xC{\isadigit{0}}FFFFFF{\isacharparenright}{\isachardoublequoteclose}\isanewline
\ \ \isanewline
\isacommand{definition}\isamarkupfalse%
\ {\isachardoublequoteopen}node{\isacharunderscore}{\isadigit{2}}{\isacharunderscore}pci\ {\isacharequal}\ empty{\isacharunderscore}spec\ {\isasymlparr}\isanewline
\ \ acc{\isacharunderscore}blocks\ {\isacharcolon}{\isacharequal}\ {\isacharbrackleft}{\isacharbrackright}{\isacharcomma}\isanewline
\ \ map{\isacharunderscore}blocks\ {\isacharcolon}{\isacharequal}\ {\isacharbrackleft}direct{\isacharunderscore}map\ dram{\isacharunderscore}sys{\isadigit{0}}\ {\isadigit{0}}{\isacharcomma}\ direct{\isacharunderscore}map\ dram{\isacharunderscore}sys{\isadigit{1}}\ {\isadigit{0}}{\isacharcomma}\ \isanewline
\ \ \ \ \ \ \ \ \ \ \ \ \ \ \ \ \ direct{\isacharunderscore}map\ xhci\ {\isadigit{3}}{\isacharcomma}\ direct{\isacharunderscore}map\ e{\isadigit{1}}{\isadigit{0}}{\isadigit{0}}{\isadigit{0}}\ {\isadigit{4}}{\isacharcomma}\ \isanewline
\ \ \ \ \ \ \ \ \ \ \ \ \ \ \ \ \ direct{\isacharunderscore}map\ ahci\ {\isadigit{5}}{\isacharcomma}\ direct{\isacharunderscore}map\ vga{\isadigit{1}}\ {\isadigit{6}}{\isacharcomma}\ \isanewline
\ \ \ \ \ \ \ \ \ \ \ \ \ \ \ \ \ direct{\isacharunderscore}map\ vga{\isadigit{2}}\ {\isadigit{6}}{\isacharbrackright}\isanewline
{\isasymrparr}{\isachardoublequoteclose}%
\begin{isamarkuptext}%
XHCI USB Host Controller%
\end{isamarkuptext}\isamarkuptrue%
\isacommand{definition}\isamarkupfalse%
\ {\isachardoublequoteopen}node{\isacharunderscore}{\isadigit{3}}{\isacharunderscore}xhci\ {\isacharequal}\ empty{\isacharunderscore}spec\ {\isasymlparr}\isanewline
\ \ acc{\isacharunderscore}blocks\ {\isacharcolon}{\isacharequal}\ {\isacharbrackleft}xhci{\isacharbrackright}{\isacharcomma}\isanewline
\ \ map{\isacharunderscore}blocks\ {\isacharcolon}{\isacharequal}\ {\isacharbrackleft}direct{\isacharunderscore}map\ dram{\isacharunderscore}sys{\isadigit{0}}\ {\isadigit{2}}{\isacharcomma}\ direct{\isacharunderscore}map\ dram{\isacharunderscore}sys{\isadigit{1}}\ {\isadigit{2}}{\isacharbrackright}\isanewline
{\isasymrparr}{\isachardoublequoteclose}%
\begin{isamarkuptext}%
PCI Network Card%
\end{isamarkuptext}\isamarkuptrue%
\isacommand{definition}\isamarkupfalse%
\ {\isachardoublequoteopen}node{\isacharunderscore}{\isadigit{4}}{\isacharunderscore}e{\isadigit{1}}{\isadigit{0}}{\isadigit{0}}{\isadigit{0}}\ {\isacharequal}\ empty{\isacharunderscore}spec\ {\isasymlparr}\isanewline
\ \ acc{\isacharunderscore}blocks\ {\isacharcolon}{\isacharequal}\ {\isacharbrackleft}e{\isadigit{1}}{\isadigit{0}}{\isadigit{0}}{\isadigit{0}}{\isacharbrackright}{\isacharcomma}\isanewline
\ \ map{\isacharunderscore}blocks\ {\isacharcolon}{\isacharequal}\ {\isacharbrackleft}direct{\isacharunderscore}map\ dram{\isacharunderscore}sys{\isadigit{0}}\ {\isadigit{2}}{\isacharcomma}\ direct{\isacharunderscore}map\ dram{\isacharunderscore}sys{\isadigit{1}}\ {\isadigit{2}}{\isacharbrackright}\isanewline
{\isasymrparr}{\isachardoublequoteclose}%
\begin{isamarkuptext}%
AHCI Disk Controller%
\end{isamarkuptext}\isamarkuptrue%
\isacommand{definition}\isamarkupfalse%
\ {\isachardoublequoteopen}node{\isacharunderscore}{\isadigit{5}}{\isacharunderscore}ahci\ {\isacharequal}\ empty{\isacharunderscore}spec\ {\isasymlparr}\isanewline
\ \ acc{\isacharunderscore}blocks\ {\isacharcolon}{\isacharequal}\ {\isacharbrackleft}ahci{\isacharbrackright}{\isacharcomma}\isanewline
\ \ map{\isacharunderscore}blocks\ {\isacharcolon}{\isacharequal}\ {\isacharbrackleft}direct{\isacharunderscore}map\ dram{\isacharunderscore}sys{\isadigit{0}}\ {\isadigit{2}}{\isacharcomma}\ direct{\isacharunderscore}map\ dram{\isacharunderscore}sys{\isadigit{1}}\ {\isadigit{2}}{\isacharbrackright}\isanewline
{\isasymrparr}{\isachardoublequoteclose}%
\begin{isamarkuptext}%
Graphics Card%
\end{isamarkuptext}\isamarkuptrue%
\isacommand{definition}\isamarkupfalse%
\ {\isachardoublequoteopen}node{\isacharunderscore}{\isadigit{6}}{\isacharunderscore}vga\ {\isacharequal}\ empty{\isacharunderscore}spec\ {\isasymlparr}\isanewline
\ \ acc{\isacharunderscore}blocks\ {\isacharcolon}{\isacharequal}\ {\isacharbrackleft}vga{\isadigit{1}}{\isacharcomma}\ vga{\isadigit{2}}{\isacharbrackright}{\isacharcomma}\isanewline
\ \ map{\isacharunderscore}blocks\ {\isacharcolon}{\isacharequal}\ {\isacharbrackleft}direct{\isacharunderscore}map\ dram{\isacharunderscore}sys{\isadigit{0}}\ {\isadigit{2}}{\isacharcomma}\ direct{\isacharunderscore}map\ dram{\isacharunderscore}sys{\isadigit{1}}\ {\isadigit{2}}{\isacharbrackright}\isanewline
{\isasymrparr}{\isachardoublequoteclose}%
\begin{isamarkuptext}%
CPU Cores%
\end{isamarkuptext}\isamarkuptrue%
\isacommand{definition}\isamarkupfalse%
\ {\isachardoublequoteopen}lapic\ {\isacharequal}\ {\isacharparenleft}{\isadigit{0}}xFEE{\isadigit{0}}{\isadigit{0}}{\isadigit{0}}{\isadigit{0}}{\isadigit{0}}{\isacharcomma}\ {\isadigit{0}}xFEE{\isadigit{0}}FFFF{\isacharparenright}{\isachardoublequoteclose}\isanewline
\isacommand{definition}\isamarkupfalse%
\ {\isachardoublequoteopen}cpu{\isacharunderscore}phys\ {\isacharequal}\ empty{\isacharunderscore}spec\ {\isasymlparr}\isanewline
\ \ acc{\isacharunderscore}blocks\ {\isacharcolon}{\isacharequal}\ {\isacharbrackleft}lapic{\isacharbrackright}{\isacharcomma}\isanewline
\ \ overlay\ {\isacharcolon}{\isacharequal}\ Some\ {\isadigit{0}}\isanewline
{\isasymrparr}{\isachardoublequoteclose}\isanewline
\ \ \isanewline
\isacommand{definition}\isamarkupfalse%
\ {\isachardoublequoteopen}cpu{\isacharunderscore}virt{\isadigit{0}}\ {\isacharequal}\ empty{\isacharunderscore}spec\ {\isasymlparr}\isanewline
\ \ acc{\isacharunderscore}blocks\ {\isacharcolon}{\isacharequal}\ {\isacharbrackleft}{\isacharbrackright}{\isacharcomma}\isanewline
\ \ overlay\ {\isacharcolon}{\isacharequal}\ Some\ {\isadigit{7}}\isanewline
{\isasymrparr}{\isachardoublequoteclose}\isanewline
\isacommand{definition}\isamarkupfalse%
\ {\isachardoublequoteopen}cpu{\isacharunderscore}virt{\isadigit{1}}\ {\isacharequal}\ empty{\isacharunderscore}spec\ {\isasymlparr}\isanewline
\ \ acc{\isacharunderscore}blocks\ {\isacharcolon}{\isacharequal}\ {\isacharbrackleft}{\isacharbrackright}{\isacharcomma}\isanewline
\ \ overlay\ {\isacharcolon}{\isacharequal}\ Some\ {\isadigit{8}}\isanewline
{\isasymrparr}{\isachardoublequoteclose}\isanewline
\isacommand{definition}\isamarkupfalse%
\ {\isachardoublequoteopen}cpu{\isacharunderscore}virt{\isadigit{2}}\ {\isacharequal}\ empty{\isacharunderscore}spec\ {\isasymlparr}\isanewline
\ \ acc{\isacharunderscore}blocks\ {\isacharcolon}{\isacharequal}\ {\isacharbrackleft}{\isacharbrackright}{\isacharcomma}\isanewline
\ \ overlay\ {\isacharcolon}{\isacharequal}\ Some\ {\isadigit{9}}\isanewline
{\isasymrparr}{\isachardoublequoteclose}\isanewline
\isacommand{definition}\isamarkupfalse%
\ {\isachardoublequoteopen}cpu{\isacharunderscore}virt{\isadigit{3}}\ {\isacharequal}\ empty{\isacharunderscore}spec\ {\isasymlparr}\isanewline
\ \ acc{\isacharunderscore}blocks\ {\isacharcolon}{\isacharequal}\ {\isacharbrackleft}{\isacharbrackright}{\isacharcomma}\isanewline
\ \ overlay\ {\isacharcolon}{\isacharequal}\ Some\ {\isadigit{1}}{\isadigit{0}}\isanewline
{\isasymrparr}{\isachardoublequoteclose}\isanewline
\ \ \isanewline
\isacommand{definition}\isamarkupfalse%
\ {\isachardoublequoteopen}sys\ {\isacharequal}\ {\isacharbrackleft}{\isacharparenleft}{\isadigit{0}}{\isacharcomma}\ node{\isacharunderscore}{\isadigit{0}}{\isacharunderscore}interconnect{\isacharparenright}{\isacharcomma}\ \isanewline
\ \ \ \ \ \ \ \ \ \ \ \ \ \ \ \ \ \ \ {\isacharparenleft}{\isadigit{1}}{\isacharcomma}\ node{\isacharunderscore}{\isadigit{1}}{\isacharunderscore}dram{\isacharparenright}{\isacharcomma}\ \isanewline
\ \ \ \ \ \ \ \ \ \ \ \ \ \ \ \ \ \ \ {\isacharparenleft}{\isadigit{2}}{\isacharcomma}\ node{\isacharunderscore}{\isadigit{2}}{\isacharunderscore}pci{\isacharparenright}{\isacharcomma}\ \isanewline
\ \ \ \ \ \ \ \ \ \ \ \ \ \ \ \ \ \ \ {\isacharparenleft}{\isadigit{3}}{\isacharcomma}\ node{\isacharunderscore}{\isadigit{3}}{\isacharunderscore}xhci{\isacharparenright}{\isacharcomma}\ \isanewline
\ \ \ \ \ \ \ \ \ \ \ \ \ \ \ \ \ \ \ {\isacharparenleft}{\isadigit{4}}{\isacharcomma}\ node{\isacharunderscore}{\isadigit{4}}{\isacharunderscore}e{\isadigit{1}}{\isadigit{0}}{\isadigit{0}}{\isadigit{0}}{\isacharparenright}{\isacharcomma}\ \isanewline
\ \ \ \ \ \ \ \ \ \ \ \ \ \ \ \ \ \ \ {\isacharparenleft}{\isadigit{5}}{\isacharcomma}node{\isacharunderscore}{\isadigit{5}}{\isacharunderscore}ahci{\isacharparenright}{\isacharcomma}\ \isanewline
\ \ \ \ \ \ \ \ \ \ \ \ \ \ \ \ \ \ \ {\isacharparenleft}{\isadigit{6}}{\isacharcomma}node{\isacharunderscore}{\isadigit{6}}{\isacharunderscore}vga{\isacharparenright}{\isacharcomma}\ \isanewline
\ \ \ \ \ \ \ \ \ \ \ \ \ \ \ \ \ \ \ {\isacharparenleft}{\isadigit{7}}{\isacharcomma}cpu{\isacharunderscore}phys{\isacharparenright}{\isacharcomma}\ \isanewline
\ \ \ \ \ \ \ \ \ \ \ \ \ \ \ \ \ \ \ {\isacharparenleft}{\isadigit{8}}{\isacharcomma}cpu{\isacharunderscore}phys{\isacharparenright}{\isacharcomma}\isanewline
\ \ \ \ \ \ \ \ \ \ \ \ \ \ \ \ \ \ \ {\isacharparenleft}{\isadigit{9}}{\isacharcomma}cpu{\isacharunderscore}phys{\isacharparenright}{\isacharcomma}\ \isanewline
\ \ \ \ \ \ \ \ \ \ \ \ \ \ \ \ \ \ \ {\isacharparenleft}{\isadigit{1}}{\isadigit{0}}{\isacharcomma}cpu{\isacharunderscore}phys{\isacharparenright}{\isacharcomma}\ \isanewline
\ \ \ \ \ \ \ \ \ \ \ \ \ \ \ \ \ \ \ {\isacharparenleft}{\isadigit{1}}{\isadigit{1}}{\isacharcomma}cpu{\isacharunderscore}virt{\isadigit{0}}{\isacharparenright}{\isacharcomma}\ \isanewline
\ \ \ \ \ \ \ \ \ \ \ \ \ \ \ \ \ \ \ {\isacharparenleft}{\isadigit{1}}{\isadigit{2}}{\isacharcomma}cpu{\isacharunderscore}virt{\isadigit{1}}{\isacharparenright}{\isacharcomma}\ \isanewline
\ \ \ \ \ \ \ \ \ \ \ \ \ \ \ \ \ \ \ {\isacharparenleft}{\isadigit{1}}{\isadigit{3}}{\isacharcomma}cpu{\isacharunderscore}virt{\isadigit{2}}{\isacharparenright}{\isacharcomma}\isanewline
\ \ \ \ \ \ \ \ \ \ \ \ \ \ \ \ \ \ \ {\isacharparenleft}{\isadigit{1}}{\isadigit{4}}{\isacharcomma}cpu{\isacharunderscore}virt{\isadigit{3}}{\isacharparenright}\ {\isacharbrackright}{\isachardoublequoteclose}\isanewline
\isadelimtheory
\ \ \ \ \isanewline
\endisadelimtheory
\isatagtheory
\isacommand{end}\isamarkupfalse%
\endisatagtheory
{\isafoldtheory}%
\isadelimtheory
\endisadelimtheory
\end{isabellebody}%

%
\begin{isabellebody}%
\setisabellecontext{DesktopInt}%
\isadelimtheory
\endisadelimtheory
\isatagtheory
\endisatagtheory
{\isafoldtheory}%
\isadelimtheory
\endisadelimtheory
\isamarkupsubsubsection{Interrupts%
}
\isamarkuptrue%
\begin{isamarkuptext}%
Convention: Interrupt issuing devices start at their 
        node with address 0. If it can trigger multiple interrupts,
        it issues contiguous addresses starting from zero.%
\end{isamarkuptext}\isamarkuptrue%
\begin{isamarkuptext}%
Convention: MSI uses memory 32bit memory writes. We encode such a memory write by
        concatenating the 64bit address with the 32bit data word.%
\end{isamarkuptext}\isamarkuptrue%
\isacommand{definition}\isamarkupfalse%
\ {\isachardoublequoteopen}gfx{\isacharunderscore}msi{\isacharunderscore}write\ {\isacharequal}\ {\isadigit{0}}x{\isadigit{0}}{\isadigit{0}}{\isadigit{0}}{\isadigit{0}}{\isadigit{0}}{\isadigit{0}}{\isadigit{0}}{\isadigit{0}}FEE{\isadigit{0}}{\isadigit{0}}{\isadigit{2}}b{\isadigit{8}}{\isadigit{0}}{\isadigit{0}}{\isadigit{0}}{\isadigit{0}}{\isadigit{0}}{\isadigit{0}}{\isadigit{2}}{\isadigit{9}}{\isachardoublequoteclose}\isanewline
\isacommand{definition}\isamarkupfalse%
\ {\isachardoublequoteopen}nic{\isacharunderscore}msi{\isacharunderscore}write{\isadigit{0}}\ {\isacharequal}\ {\isadigit{0}}x{\isadigit{0}}{\isadigit{0}}{\isadigit{0}}{\isadigit{0}}{\isadigit{0}}{\isadigit{0}}{\isadigit{0}}{\isadigit{0}}FEE{\isadigit{0}}{\isadigit{0}}{\isadigit{2}}b{\isadigit{8}}{\isadigit{0}}{\isadigit{0}}{\isadigit{0}}{\isadigit{0}}{\isadigit{0}}{\isadigit{0}}{\isadigit{7}}D{\isachardoublequoteclose}\isanewline
\isacommand{definition}\isamarkupfalse%
\ {\isachardoublequoteopen}nic{\isacharunderscore}msi{\isacharunderscore}write{\isadigit{1}}\ {\isacharequal}\ {\isadigit{0}}x{\isadigit{0}}{\isadigit{0}}{\isadigit{0}}{\isadigit{0}}{\isadigit{0}}{\isadigit{0}}{\isadigit{0}}{\isadigit{0}}FEE{\isadigit{0}}{\isadigit{0}}{\isadigit{2}}b{\isadigit{8}}{\isadigit{0}}{\isadigit{0}}{\isadigit{0}}{\isadigit{0}}{\isadigit{0}}{\isadigit{0}}{\isadigit{7}}E{\isachardoublequoteclose}\isanewline
\isacommand{definition}\isamarkupfalse%
\ {\isachardoublequoteopen}nic{\isacharunderscore}msi{\isacharunderscore}write{\isadigit{2}}\ {\isacharequal}\ {\isadigit{0}}x{\isadigit{0}}{\isadigit{0}}{\isadigit{0}}{\isadigit{0}}{\isadigit{0}}{\isadigit{0}}{\isadigit{0}}{\isadigit{0}}FEE{\isadigit{0}}{\isadigit{0}}{\isadigit{2}}b{\isadigit{8}}{\isadigit{0}}{\isadigit{0}}{\isadigit{0}}{\isadigit{0}}{\isadigit{0}}{\isadigit{0}}{\isadigit{7}}F{\isachardoublequoteclose}\isanewline
\isacommand{definition}\isamarkupfalse%
\ {\isachardoublequoteopen}nic{\isacharunderscore}msi{\isacharunderscore}write{\isadigit{3}}\ {\isacharequal}\ {\isadigit{0}}x{\isadigit{0}}{\isadigit{0}}{\isadigit{0}}{\isadigit{0}}{\isadigit{0}}{\isadigit{0}}{\isadigit{0}}{\isadigit{0}}FEE{\isadigit{0}}{\isadigit{0}}{\isadigit{2}}b{\isadigit{8}}{\isadigit{0}}{\isadigit{0}}{\isadigit{0}}{\isadigit{0}}{\isadigit{0}}{\isadigit{0}}{\isadigit{8}}{\isadigit{0}}{\isachardoublequoteclose}\isanewline
\isacommand{definition}\isamarkupfalse%
\ {\isachardoublequoteopen}nic{\isacharunderscore}msi{\isacharunderscore}write{\isadigit{4}}\ {\isacharequal}\ {\isadigit{0}}x{\isadigit{0}}{\isadigit{0}}{\isadigit{0}}{\isadigit{0}}{\isadigit{0}}{\isadigit{0}}{\isadigit{0}}{\isadigit{0}}FEE{\isadigit{0}}{\isadigit{0}}{\isadigit{2}}b{\isadigit{8}}{\isadigit{0}}{\isadigit{0}}{\isadigit{0}}{\isadigit{0}}{\isadigit{0}}{\isadigit{0}}{\isadigit{8}}{\isadigit{1}}{\isachardoublequoteclose}\isanewline
\ \ \isanewline
\isacommand{definition}\isamarkupfalse%
\ {\isachardoublequoteopen}x{\isadigit{8}}{\isadigit{6}}{\isacharunderscore}vec{\isacharunderscore}domain\ {\isacharequal}\ {\isacharparenleft}{\isadigit{3}}{\isadigit{2}}{\isacharcomma}{\isadigit{2}}{\isadigit{5}}{\isadigit{5}}{\isacharparenright}{\isachardoublequoteclose}%
\begin{isamarkuptext}%
LAPIC 0%
\end{isamarkuptext}\isamarkuptrue%
\isacommand{definition}\isamarkupfalse%
\ {\isachardoublequoteopen}node{\isacharunderscore}{\isadigit{0}}{\isacharunderscore}lapic{\isacharunderscore}{\isadigit{0}}\ {\isacharequal}\ empty{\isacharunderscore}spec\ {\isasymlparr}\isanewline
\ \ acc{\isacharunderscore}blocks\ {\isacharcolon}{\isacharequal}\ {\isacharbrackleft}x{\isadigit{8}}{\isadigit{6}}{\isacharunderscore}vec{\isacharunderscore}domain{\isacharbrackright}{\isacharcomma}\isanewline
\ \ map{\isacharunderscore}blocks\ {\isacharcolon}{\isacharequal}\ {\isacharbrackleft}{\isacharbrackright}\isanewline
{\isasymrparr}{\isachardoublequoteclose}%
\begin{isamarkuptext}%
LAPIC 1%
\end{isamarkuptext}\isamarkuptrue%
\isacommand{definition}\isamarkupfalse%
\ {\isachardoublequoteopen}node{\isacharunderscore}{\isadigit{1}}{\isacharunderscore}lapic{\isacharunderscore}{\isadigit{1}}\ {\isacharequal}\ empty{\isacharunderscore}spec\ {\isasymlparr}\isanewline
\ \ acc{\isacharunderscore}blocks\ {\isacharcolon}{\isacharequal}\ {\isacharbrackleft}x{\isadigit{8}}{\isadigit{6}}{\isacharunderscore}vec{\isacharunderscore}domain{\isacharbrackright}{\isacharcomma}\isanewline
\ \ map{\isacharunderscore}blocks\ {\isacharcolon}{\isacharequal}\ {\isacharbrackleft}{\isacharbrackright}\isanewline
{\isasymrparr}{\isachardoublequoteclose}%
\begin{isamarkuptext}%
LAPIC 2%
\end{isamarkuptext}\isamarkuptrue%
\isacommand{definition}\isamarkupfalse%
\ {\isachardoublequoteopen}node{\isacharunderscore}{\isadigit{2}}{\isacharunderscore}lapic{\isacharunderscore}{\isadigit{2}}\ {\isacharequal}\ empty{\isacharunderscore}spec\ {\isasymlparr}\isanewline
\ \ acc{\isacharunderscore}blocks\ {\isacharcolon}{\isacharequal}\ {\isacharbrackleft}x{\isadigit{8}}{\isadigit{6}}{\isacharunderscore}vec{\isacharunderscore}domain{\isacharbrackright}{\isacharcomma}\isanewline
\ \ map{\isacharunderscore}blocks\ {\isacharcolon}{\isacharequal}\ {\isacharbrackleft}{\isacharbrackright}\isanewline
{\isasymrparr}{\isachardoublequoteclose}%
\begin{isamarkuptext}%
LAPIC 3%
\end{isamarkuptext}\isamarkuptrue%
\isacommand{definition}\isamarkupfalse%
\ {\isachardoublequoteopen}node{\isacharunderscore}{\isadigit{3}}{\isacharunderscore}lapic{\isacharunderscore}{\isadigit{3}}\ {\isacharequal}\ empty{\isacharunderscore}spec\ {\isasymlparr}\isanewline
\ \ acc{\isacharunderscore}blocks\ {\isacharcolon}{\isacharequal}\ {\isacharbrackleft}x{\isadigit{8}}{\isadigit{6}}{\isacharunderscore}vec{\isacharunderscore}domain{\isacharbrackright}{\isacharcomma}\isanewline
\ \ map{\isacharunderscore}blocks\ {\isacharcolon}{\isacharequal}\ {\isacharbrackleft}{\isacharbrackright}\isanewline
{\isasymrparr}{\isachardoublequoteclose}%
\begin{isamarkuptext}%
IOAPIC 0 to LAPIC%
\end{isamarkuptext}\isamarkuptrue%
\isacommand{definition}\isamarkupfalse%
\ {\isachardoublequoteopen}node{\isacharunderscore}{\isadigit{4}}{\isacharunderscore}ioapic\ {\isacharequal}\ empty{\isacharunderscore}spec\ {\isasymlparr}\isanewline
\ \ acc{\isacharunderscore}blocks\ {\isacharcolon}{\isacharequal}\ {\isacharbrackleft}{\isacharbrackright}{\isacharcomma}\isanewline
\ \ map{\isacharunderscore}blocks\ {\isacharcolon}{\isacharequal}\ {\isacharbrackleft}one{\isacharunderscore}map\ {\isadigit{4}}\ {\isadigit{0}}\ {\isadigit{4}}{\isadigit{8}}{\isacharcomma}\ one{\isacharunderscore}map\ {\isadigit{8}}\ {\isadigit{0}}\ {\isadigit{4}}{\isadigit{0}}{\isacharbrackright}\isanewline
{\isasymrparr}{\isachardoublequoteclose}%
\begin{isamarkuptext}%
LNKA to IOAPIC%
\end{isamarkuptext}\isamarkuptrue%
\isacommand{definition}\isamarkupfalse%
\ {\isachardoublequoteopen}node{\isacharunderscore}{\isadigit{5}}{\isacharunderscore}lnka\ {\isacharequal}\ empty{\isacharunderscore}spec\ {\isasymlparr}\isanewline
\ \ acc{\isacharunderscore}blocks\ {\isacharcolon}{\isacharequal}\ {\isacharbrackleft}{\isacharbrackright}{\isacharcomma}\isanewline
\ \ map{\isacharunderscore}blocks\ {\isacharcolon}{\isacharequal}\ {\isacharbrackleft}one{\isacharunderscore}map\ {\isadigit{0}}\ {\isadigit{4}}\ {\isadigit{4}}{\isacharbrackright}\isanewline
{\isasymrparr}{\isachardoublequoteclose}%
\begin{isamarkuptext}%
USB EHCI to LNKA%
\end{isamarkuptext}\isamarkuptrue%
\isacommand{definition}\isamarkupfalse%
\ {\isachardoublequoteopen}node{\isacharunderscore}{\isadigit{6}}{\isacharunderscore}usb\ {\isacharequal}\ empty{\isacharunderscore}spec\ {\isasymlparr}\isanewline
\ \ acc{\isacharunderscore}blocks\ {\isacharcolon}{\isacharequal}\ {\isacharbrackleft}{\isacharbrackright}{\isacharcomma}\isanewline
\ \ map{\isacharunderscore}blocks\ {\isacharcolon}{\isacharequal}\ {\isacharbrackleft}one{\isacharunderscore}map\ {\isadigit{0}}\ {\isadigit{5}}\ {\isadigit{0}}{\isacharbrackright}\isanewline
{\isasymrparr}{\isachardoublequoteclose}%
\begin{isamarkuptext}%
RTC to IOAPIC%
\end{isamarkuptext}\isamarkuptrue%
\isacommand{definition}\isamarkupfalse%
\ {\isachardoublequoteopen}node{\isacharunderscore}{\isadigit{7}}{\isacharunderscore}rtc\ {\isacharequal}\ empty{\isacharunderscore}spec\ {\isasymlparr}\isanewline
\ \ acc{\isacharunderscore}blocks\ {\isacharcolon}{\isacharequal}\ {\isacharbrackleft}{\isacharbrackright}{\isacharcomma}\isanewline
\ \ map{\isacharunderscore}blocks\ {\isacharcolon}{\isacharequal}\ {\isacharbrackleft}one{\isacharunderscore}map\ {\isadigit{0}}\ {\isadigit{4}}\ {\isadigit{8}}{\isacharbrackright}\isanewline
{\isasymrparr}{\isachardoublequoteclose}%
\begin{isamarkuptext}%
PCH to LAPICs.%
\end{isamarkuptext}\isamarkuptrue%
\isacommand{definition}\isamarkupfalse%
\ {\isachardoublequoteopen}node{\isacharunderscore}{\isadigit{8}}{\isacharunderscore}pch\ {\isacharequal}\ empty{\isacharunderscore}spec\ {\isasymlparr}\isanewline
\ \ acc{\isacharunderscore}blocks\ {\isacharcolon}{\isacharequal}\ {\isacharbrackleft}{\isacharbrackright}{\isacharcomma}\isanewline
\ \ map{\isacharunderscore}blocks\ {\isacharcolon}{\isacharequal}\ {\isacharbrackleft}\isanewline
\ \ \ \ one{\isacharunderscore}map\ nic{\isacharunderscore}msi{\isacharunderscore}write{\isadigit{0}}\ {\isadigit{0}}\ {\isadigit{1}}{\isadigit{2}}{\isadigit{5}}{\isacharcomma}\ \isanewline
\ \ \ \ one{\isacharunderscore}map\ nic{\isacharunderscore}msi{\isacharunderscore}write{\isadigit{1}}\ {\isadigit{0}}\ {\isadigit{1}}{\isadigit{2}}{\isadigit{6}}{\isacharcomma}\ \isanewline
\ \ \ \ one{\isacharunderscore}map\ nic{\isacharunderscore}msi{\isacharunderscore}write{\isadigit{2}}\ {\isadigit{0}}\ {\isadigit{1}}{\isadigit{2}}{\isadigit{7}}{\isacharcomma}\ \isanewline
\ \ \ \ one{\isacharunderscore}map\ nic{\isacharunderscore}msi{\isacharunderscore}write{\isadigit{3}}\ {\isadigit{0}}\ {\isadigit{1}}{\isadigit{2}}{\isadigit{8}}{\isacharcomma}\ \isanewline
\ \ \ \ one{\isacharunderscore}map\ nic{\isacharunderscore}msi{\isacharunderscore}write{\isadigit{4}}\ {\isadigit{0}}\ {\isadigit{1}}{\isadigit{2}}{\isadigit{9}}{\isacharcomma}\ \isanewline
\ \ \ \ one{\isacharunderscore}map\ gfx{\isacharunderscore}msi{\isacharunderscore}write\ {\isadigit{0}}\ {\isadigit{4}}{\isadigit{1}}\ \ \ \isanewline
{\isacharbrackright}\isanewline
{\isasymrparr}{\isachardoublequoteclose}%
\begin{isamarkuptext}%
NIC to PCH. Uses 5 interrupts.%
\end{isamarkuptext}\isamarkuptrue%
\isacommand{definition}\isamarkupfalse%
\ {\isachardoublequoteopen}node{\isacharunderscore}{\isadigit{9}}{\isacharunderscore}nic\ {\isacharequal}\ empty{\isacharunderscore}spec\ {\isasymlparr}\isanewline
\ \ acc{\isacharunderscore}blocks\ {\isacharcolon}{\isacharequal}\ {\isacharbrackleft}{\isacharbrackright}{\isacharcomma}\isanewline
\ \ map{\isacharunderscore}blocks\ {\isacharcolon}{\isacharequal}\ {\isacharbrackleft}\isanewline
\ \ \ \ one{\isacharunderscore}map\ {\isadigit{0}}\ {\isadigit{8}}\ nic{\isacharunderscore}msi{\isacharunderscore}write{\isadigit{0}}{\isacharcomma}\isanewline
\ \ \ \ one{\isacharunderscore}map\ {\isadigit{1}}\ {\isadigit{8}}\ nic{\isacharunderscore}msi{\isacharunderscore}write{\isadigit{1}}{\isacharcomma}\isanewline
\ \ \ \ one{\isacharunderscore}map\ {\isadigit{2}}\ {\isadigit{8}}\ nic{\isacharunderscore}msi{\isacharunderscore}write{\isadigit{2}}{\isacharcomma}\isanewline
\ \ \ \ one{\isacharunderscore}map\ {\isadigit{3}}\ {\isadigit{8}}\ nic{\isacharunderscore}msi{\isacharunderscore}write{\isadigit{3}}{\isacharcomma}\isanewline
\ \ \ \ one{\isacharunderscore}map\ {\isadigit{4}}\ {\isadigit{8}}\ nic{\isacharunderscore}msi{\isacharunderscore}write{\isadigit{4}}\isanewline
{\isacharbrackright}\isanewline
{\isasymrparr}{\isachardoublequoteclose}%
\begin{isamarkuptext}%
GFX to PCH. Uses 1 interrupt.%
\end{isamarkuptext}\isamarkuptrue%
\isacommand{definition}\isamarkupfalse%
\ {\isachardoublequoteopen}node{\isacharunderscore}{\isadigit{1}}{\isadigit{8}}{\isacharunderscore}gfx\ {\isacharequal}\ empty{\isacharunderscore}spec\ {\isasymlparr}\isanewline
\ \ acc{\isacharunderscore}blocks\ {\isacharcolon}{\isacharequal}\ {\isacharbrackleft}{\isacharbrackright}{\isacharcomma}\isanewline
\ \ map{\isacharunderscore}blocks\ {\isacharcolon}{\isacharequal}\ {\isacharbrackleft}\isanewline
\ \ \ \ one{\isacharunderscore}map\ {\isadigit{0}}\ {\isadigit{8}}\ gfx{\isacharunderscore}msi{\isacharunderscore}write\isanewline
{\isacharbrackright}\isanewline
{\isasymrparr}{\isachardoublequoteclose}%
\begin{isamarkuptext}%
Timer0 to LAPIC0%
\end{isamarkuptext}\isamarkuptrue%
\isacommand{definition}\isamarkupfalse%
\ {\isachardoublequoteopen}node{\isacharunderscore}{\isadigit{1}}{\isadigit{0}}{\isacharunderscore}timer{\isadigit{0}}\ {\isacharequal}\ empty{\isacharunderscore}spec\ {\isasymlparr}\isanewline
\ \ acc{\isacharunderscore}blocks\ {\isacharcolon}{\isacharequal}\ {\isacharbrackleft}{\isacharbrackright}{\isacharcomma}\isanewline
\ \ map{\isacharunderscore}blocks\ {\isacharcolon}{\isacharequal}\ {\isacharbrackleft}one{\isacharunderscore}map\ {\isadigit{0}}\ {\isadigit{0}}\ {\isadigit{3}}{\isadigit{2}}{\isacharbrackright}\isanewline
{\isasymrparr}{\isachardoublequoteclose}%
\begin{isamarkuptext}%
Timer1 to LAPIC1%
\end{isamarkuptext}\isamarkuptrue%
\isacommand{definition}\isamarkupfalse%
\ {\isachardoublequoteopen}node{\isacharunderscore}{\isadigit{1}}{\isadigit{1}}{\isacharunderscore}timer{\isadigit{1}}\ {\isacharequal}\ empty{\isacharunderscore}spec\ {\isasymlparr}\isanewline
\ \ acc{\isacharunderscore}blocks\ {\isacharcolon}{\isacharequal}\ {\isacharbrackleft}{\isacharbrackright}{\isacharcomma}\isanewline
\ \ map{\isacharunderscore}blocks\ {\isacharcolon}{\isacharequal}\ {\isacharbrackleft}one{\isacharunderscore}map\ {\isadigit{0}}\ {\isadigit{1}}\ {\isadigit{3}}{\isadigit{2}}{\isacharbrackright}\isanewline
{\isasymrparr}{\isachardoublequoteclose}%
\begin{isamarkuptext}%
Timer2 to LAPIC2%
\end{isamarkuptext}\isamarkuptrue%
\isacommand{definition}\isamarkupfalse%
\ {\isachardoublequoteopen}node{\isacharunderscore}{\isadigit{1}}{\isadigit{2}}{\isacharunderscore}timer{\isadigit{2}}\ {\isacharequal}\ empty{\isacharunderscore}spec\ {\isasymlparr}\isanewline
\ \ acc{\isacharunderscore}blocks\ {\isacharcolon}{\isacharequal}\ {\isacharbrackleft}{\isacharbrackright}{\isacharcomma}\isanewline
\ \ map{\isacharunderscore}blocks\ {\isacharcolon}{\isacharequal}\ {\isacharbrackleft}one{\isacharunderscore}map\ {\isadigit{0}}\ {\isadigit{2}}\ {\isadigit{3}}{\isadigit{2}}{\isacharbrackright}\isanewline
{\isasymrparr}{\isachardoublequoteclose}%
\begin{isamarkuptext}%
Timer3 to LAPIC3%
\end{isamarkuptext}\isamarkuptrue%
\isacommand{definition}\isamarkupfalse%
\ {\isachardoublequoteopen}node{\isacharunderscore}{\isadigit{1}}{\isadigit{3}}{\isacharunderscore}timer{\isadigit{3}}\ {\isacharequal}\ empty{\isacharunderscore}spec\ {\isasymlparr}\isanewline
\ \ acc{\isacharunderscore}blocks\ {\isacharcolon}{\isacharequal}\ {\isacharbrackleft}{\isacharbrackright}{\isacharcomma}\isanewline
\ \ map{\isacharunderscore}blocks\ {\isacharcolon}{\isacharequal}\ {\isacharbrackleft}one{\isacharunderscore}map\ {\isadigit{0}}\ {\isadigit{3}}\ {\isadigit{3}}{\isadigit{2}}{\isacharbrackright}\isanewline
{\isasymrparr}{\isachardoublequoteclose}%
\begin{isamarkuptext}%
TODO:  needs the set destination to make sense. We use the TLB shootdown IPI
as example%
\end{isamarkuptext}\isamarkuptrue%
\begin{isamarkuptext}%
Core 0 to Other APICs.%
\end{isamarkuptext}\isamarkuptrue%
\isacommand{definition}\isamarkupfalse%
\ {\isachardoublequoteopen}node{\isacharunderscore}{\isadigit{1}}{\isadigit{4}}{\isacharunderscore}core{\isadigit{0}}\ {\isacharequal}\ empty{\isacharunderscore}spec\ {\isasymlparr}\isanewline
\ \ acc{\isacharunderscore}blocks\ {\isacharcolon}{\isacharequal}\ {\isacharbrackleft}{\isacharbrackright}{\isacharcomma}\isanewline
\ \ map{\isacharunderscore}blocks\ {\isacharcolon}{\isacharequal}\ {\isacharbrackleft}\isanewline
\ \ \ \ one{\isacharunderscore}map\ {\isadigit{0}}\ {\isadigit{3}}\ {\isadigit{2}}{\isadigit{5}}{\isadigit{1}}\ {\isacharparenleft}{\isacharasterisk}\ {\isacharbrackleft}{\isacharparenleft}{\isadigit{1}}{\isacharcomma}{\isadigit{2}}{\isadigit{5}}{\isadigit{1}}{\isacharparenright}{\isacharcomma}{\isacharparenleft}{\isadigit{2}}{\isacharcomma}{\isadigit{2}}{\isadigit{5}}{\isadigit{1}}{\isacharparenright}{\isacharcomma}{\isacharparenleft}{\isadigit{3}}{\isacharcomma}{\isadigit{2}}{\isadigit{5}}{\isadigit{1}}{\isacharparenright}{\isacharbrackright}\ {\isacharasterisk}{\isacharparenright}\isanewline
\ \ \ \ {\isacharbrackright}\ \isanewline
{\isasymrparr}{\isachardoublequoteclose}%
\begin{isamarkuptext}%
Core 1 to Other APICs.%
\end{isamarkuptext}\isamarkuptrue%
\isacommand{definition}\isamarkupfalse%
\ {\isachardoublequoteopen}node{\isacharunderscore}{\isadigit{1}}{\isadigit{5}}{\isacharunderscore}core{\isadigit{1}}\ {\isacharequal}\ empty{\isacharunderscore}spec\ {\isasymlparr}\isanewline
\ \ acc{\isacharunderscore}blocks\ {\isacharcolon}{\isacharequal}\ {\isacharbrackleft}{\isacharbrackright}{\isacharcomma}\isanewline
\ \ map{\isacharunderscore}blocks\ {\isacharcolon}{\isacharequal}\ {\isacharbrackleft}\isanewline
\ \ \ \ one{\isacharunderscore}map\ {\isadigit{0}}\ {\isadigit{3}}\ {\isadigit{2}}{\isadigit{5}}{\isadigit{1}}{\isacharcomma}\ {\isacharparenleft}{\isacharasterisk}\ {\isacharbrackleft}{\isacharparenleft}{\isadigit{0}}{\isacharcomma}{\isadigit{2}}{\isadigit{5}}{\isadigit{1}}{\isacharparenright}{\isacharcomma}{\isacharparenleft}{\isadigit{2}}{\isacharcomma}{\isadigit{2}}{\isadigit{5}}{\isadigit{1}}{\isacharparenright}{\isacharcomma}{\isacharparenleft}{\isadigit{3}}{\isacharcomma}{\isadigit{2}}{\isadigit{5}}{\isadigit{1}}{\isacharparenright}{\isacharbrackright}\ {\isacharasterisk}{\isacharparenright}\isanewline
\ \ \ \ one{\isacharunderscore}map\ {\isadigit{1}}\ {\isadigit{0}}\ {\isadigit{4}}{\isadigit{8}}\isanewline
\ \ \ \ {\isacharbrackright}\isanewline
{\isasymrparr}{\isachardoublequoteclose}%
\begin{isamarkuptext}%
Core 2 to Other APICs.%
\end{isamarkuptext}\isamarkuptrue%
\isacommand{definition}\isamarkupfalse%
\ {\isachardoublequoteopen}node{\isacharunderscore}{\isadigit{1}}{\isadigit{6}}{\isacharunderscore}core{\isadigit{2}}\ {\isacharequal}\ empty{\isacharunderscore}spec\ {\isasymlparr}\isanewline
\ \ acc{\isacharunderscore}blocks\ {\isacharcolon}{\isacharequal}\ {\isacharbrackleft}{\isacharbrackright}{\isacharcomma}\isanewline
\ \ map{\isacharunderscore}blocks\ {\isacharcolon}{\isacharequal}\ {\isacharbrackleft}\isanewline
\ \ \ \ one{\isacharunderscore}map\ {\isadigit{0}}\ {\isadigit{3}}\ {\isadigit{2}}{\isadigit{5}}{\isadigit{1}}\ {\isacharparenleft}{\isacharasterisk}\ {\isacharbrackleft}{\isacharparenleft}{\isadigit{0}}{\isacharcomma}{\isadigit{2}}{\isadigit{5}}{\isadigit{1}}{\isacharparenright}{\isacharcomma}{\isacharparenleft}{\isadigit{1}}{\isacharcomma}{\isadigit{2}}{\isadigit{5}}{\isadigit{1}}{\isacharparenright}{\isacharcomma}{\isacharparenleft}{\isadigit{3}}{\isacharcomma}{\isadigit{2}}{\isadigit{5}}{\isadigit{1}}{\isacharparenright}{\isacharbrackright}\ {\isacharasterisk}{\isacharparenright}\isanewline
\ \ \ \ {\isacharbrackright}\isanewline
{\isasymrparr}{\isachardoublequoteclose}%
\begin{isamarkuptext}%
Core 3 to Other APICs.%
\end{isamarkuptext}\isamarkuptrue%
\isacommand{definition}\isamarkupfalse%
\ {\isachardoublequoteopen}node{\isacharunderscore}{\isadigit{1}}{\isadigit{7}}{\isacharunderscore}core{\isadigit{3}}\ {\isacharequal}\ empty{\isacharunderscore}spec\ {\isasymlparr}\isanewline
\ \ acc{\isacharunderscore}blocks\ {\isacharcolon}{\isacharequal}\ {\isacharbrackleft}{\isacharbrackright}{\isacharcomma}\isanewline
\ \ map{\isacharunderscore}blocks\ {\isacharcolon}{\isacharequal}\ {\isacharbrackleft}\isanewline
\ \ \ \ one{\isacharunderscore}map\ {\isadigit{0}}\ {\isadigit{3}}\ {\isadigit{2}}{\isadigit{5}}{\isadigit{1}}\ {\isacharparenleft}{\isacharasterisk}\ {\isacharbrackleft}{\isacharparenleft}{\isadigit{0}}{\isacharcomma}{\isadigit{2}}{\isadigit{5}}{\isadigit{1}}{\isacharparenright}{\isacharcomma}{\isacharparenleft}{\isadigit{1}}{\isacharcomma}{\isadigit{2}}{\isadigit{5}}{\isadigit{1}}{\isacharparenright}{\isacharcomma}{\isacharparenleft}{\isadigit{3}}{\isacharcomma}{\isadigit{2}}{\isadigit{5}}{\isadigit{1}}{\isacharparenright}{\isacharbrackright}\ {\isacharasterisk}{\isacharparenright}\isanewline
\ \ \ \ {\isacharbrackright}\isanewline
{\isasymrparr}{\isachardoublequoteclose}\isanewline
\isanewline
\isacommand{definition}\isamarkupfalse%
\ {\isachardoublequoteopen}sys\ {\isacharequal}\ {\isacharbrackleft}\isanewline
\ \ {\isacharparenleft}{\isadigit{0}}{\isacharcomma}node{\isacharunderscore}{\isadigit{0}}{\isacharunderscore}lapic{\isacharunderscore}{\isadigit{0}}{\isacharparenright}{\isacharcomma}\isanewline
\ \ {\isacharparenleft}{\isadigit{1}}{\isacharcomma}node{\isacharunderscore}{\isadigit{1}}{\isacharunderscore}lapic{\isacharunderscore}{\isadigit{1}}{\isacharparenright}{\isacharcomma}\isanewline
\ \ {\isacharparenleft}{\isadigit{2}}{\isacharcomma}node{\isacharunderscore}{\isadigit{2}}{\isacharunderscore}lapic{\isacharunderscore}{\isadigit{2}}{\isacharparenright}{\isacharcomma}\isanewline
\ \ {\isacharparenleft}{\isadigit{3}}{\isacharcomma}node{\isacharunderscore}{\isadigit{3}}{\isacharunderscore}lapic{\isacharunderscore}{\isadigit{3}}{\isacharparenright}{\isacharcomma}\isanewline
\ \ {\isacharparenleft}{\isadigit{4}}{\isacharcomma}node{\isacharunderscore}{\isadigit{4}}{\isacharunderscore}ioapic{\isacharparenright}{\isacharcomma}\isanewline
\ \ {\isacharparenleft}{\isadigit{5}}{\isacharcomma}node{\isacharunderscore}{\isadigit{5}}{\isacharunderscore}lnka{\isacharparenright}{\isacharcomma}\isanewline
\ \ {\isacharparenleft}{\isadigit{6}}{\isacharcomma}node{\isacharunderscore}{\isadigit{6}}{\isacharunderscore}usb{\isacharparenright}{\isacharcomma}\isanewline
\ \ {\isacharparenleft}{\isadigit{7}}{\isacharcomma}node{\isacharunderscore}{\isadigit{7}}{\isacharunderscore}rtc{\isacharparenright}{\isacharcomma}\isanewline
\ \ {\isacharparenleft}{\isadigit{8}}{\isacharcomma}node{\isacharunderscore}{\isadigit{8}}{\isacharunderscore}pch{\isacharparenright}{\isacharcomma}\isanewline
\ \ {\isacharparenleft}{\isadigit{9}}{\isacharcomma}node{\isacharunderscore}{\isadigit{9}}{\isacharunderscore}nic{\isacharparenright}{\isacharcomma}\isanewline
\ \ {\isacharparenleft}{\isadigit{1}}{\isadigit{0}}{\isacharcomma}node{\isacharunderscore}{\isadigit{1}}{\isadigit{0}}{\isacharunderscore}timer{\isadigit{0}}{\isacharparenright}{\isacharcomma}\isanewline
\ \ {\isacharparenleft}{\isadigit{1}}{\isadigit{1}}{\isacharcomma}node{\isacharunderscore}{\isadigit{1}}{\isadigit{1}}{\isacharunderscore}timer{\isadigit{1}}{\isacharparenright}{\isacharcomma}\isanewline
\ \ {\isacharparenleft}{\isadigit{1}}{\isadigit{2}}{\isacharcomma}node{\isacharunderscore}{\isadigit{1}}{\isadigit{2}}{\isacharunderscore}timer{\isadigit{2}}{\isacharparenright}{\isacharcomma}\isanewline
\ \ {\isacharparenleft}{\isadigit{1}}{\isadigit{3}}{\isacharcomma}node{\isacharunderscore}{\isadigit{1}}{\isadigit{3}}{\isacharunderscore}timer{\isadigit{3}}{\isacharparenright}{\isacharcomma}\isanewline
\ \ {\isacharparenleft}{\isadigit{1}}{\isadigit{4}}{\isacharcomma}node{\isacharunderscore}{\isadigit{1}}{\isadigit{4}}{\isacharunderscore}core{\isadigit{0}}{\isacharparenright}{\isacharcomma}\isanewline
\ \ {\isacharparenleft}{\isadigit{1}}{\isadigit{5}}{\isacharcomma}node{\isacharunderscore}{\isadigit{1}}{\isadigit{5}}{\isacharunderscore}core{\isadigit{1}}{\isacharparenright}{\isacharcomma}\isanewline
\ \ {\isacharparenleft}{\isadigit{1}}{\isadigit{6}}{\isacharcomma}node{\isacharunderscore}{\isadigit{1}}{\isadigit{6}}{\isacharunderscore}core{\isadigit{2}}{\isacharparenright}{\isacharcomma}\isanewline
\ \ {\isacharparenleft}{\isadigit{1}}{\isadigit{7}}{\isacharcomma}node{\isacharunderscore}{\isadigit{1}}{\isadigit{7}}{\isacharunderscore}core{\isadigit{3}}{\isacharparenright}{\isacharcomma}\isanewline
\ \ {\isacharparenleft}{\isadigit{1}}{\isadigit{8}}{\isacharcomma}node{\isacharunderscore}{\isadigit{1}}{\isadigit{8}}{\isacharunderscore}gfx{\isacharparenright}\isanewline
{\isacharbrackright}{\isachardoublequoteclose}\isanewline
\isadelimtheory
\isanewline
\endisadelimtheory
\isatagtheory
\isacommand{end}\isamarkupfalse%
\endisatagtheory
{\isafoldtheory}%
\isadelimtheory
\endisadelimtheory
\end{isabellebody}%

\subsection{A heterogeneous x86 Server}
\label{isabelle:server}
\camready{This is the full model representation of the heterogeneous server 
system that we already introduced in ~\autoref{sec:realsystems:server}.}
\begin{isabellebody}%
\setisabellecontext{Server}%
\isadelimtheory
\endisadelimtheory
\isatagtheory
\endisatagtheory
{\isafoldtheory}%
\isadelimtheory
\endisadelimtheory
\isamarkupsubsubsection{Address spaces%
}
\isamarkuptrue%
\begin{isamarkuptext}%
DRAM%
\end{isamarkuptext}\isamarkuptrue%
\isacommand{definition}\isamarkupfalse%
\ {\isachardoublequoteopen}dram{\isacharunderscore}sys{\isadigit{0}}\ {\isacharequal}\ {\isacharparenleft}{\isadigit{0}}x{\isadigit{0}}{\isadigit{0}}{\isadigit{0}}{\isadigit{0}}{\isadigit{0}}{\isadigit{0}}{\isadigit{0}}{\isadigit{0}}{\isadigit{0}}{\isadigit{0}}{\isadigit{0}}{\isadigit{0}}{\isadigit{0}}{\isadigit{0}}{\isadigit{0}}{\isadigit{0}}{\isacharcomma}{\isadigit{0}}x{\isadigit{0}}{\isadigit{0}}{\isadigit{0}}{\isadigit{0}}{\isadigit{0}}{\isadigit{0}}{\isadigit{0}}{\isadigit{0}}{\isadigit{0}}{\isadigit{0}}{\isadigit{0}}{\isadigit{9}}BFFF{\isacharparenright}{\isachardoublequoteclose}\ \isanewline
\isacommand{definition}\isamarkupfalse%
\ {\isachardoublequoteopen}dram{\isacharunderscore}sys{\isadigit{1}}\ {\isacharequal}\ {\isacharparenleft}{\isadigit{0}}x{\isadigit{0}}{\isadigit{0}}{\isadigit{0}}{\isadigit{0}}{\isadigit{0}}{\isadigit{0}}{\isadigit{0}}{\isadigit{0}}{\isadigit{0}}{\isadigit{0}}{\isadigit{1}}{\isadigit{0}}{\isadigit{0}}{\isadigit{0}}{\isadigit{0}}{\isadigit{0}}{\isacharcomma}{\isadigit{0}}x{\isadigit{0}}{\isadigit{0}}{\isadigit{0}}{\isadigit{0}}{\isadigit{0}}{\isadigit{0}}{\isadigit{0}}{\isadigit{0}}BAD{\isadigit{2}}{\isadigit{7}}FFF{\isacharparenright}{\isachardoublequoteclose}\isanewline
\isacommand{definition}\isamarkupfalse%
\ {\isachardoublequoteopen}dram{\isacharunderscore}sys{\isadigit{2}}\ {\isacharequal}\ {\isacharparenleft}{\isadigit{0}}x{\isadigit{0}}{\isadigit{0}}{\isadigit{0}}{\isadigit{0}}{\isadigit{0}}{\isadigit{0}}{\isadigit{0}}{\isadigit{0}}BAF{\isadigit{9}}{\isadigit{0}}{\isadigit{0}}{\isadigit{0}}{\isadigit{0}}{\isacharcomma}{\isadigit{0}}x{\isadigit{0}}{\isadigit{0}}{\isadigit{0}}{\isadigit{0}}{\isadigit{0}}{\isadigit{0}}{\isadigit{0}}{\isadigit{0}}BAFC{\isadigit{4}}FFF{\isacharparenright}{\isachardoublequoteclose}\isanewline
\isacommand{definition}\isamarkupfalse%
\ {\isachardoublequoteopen}dram{\isacharunderscore}sys{\isadigit{3}}\ {\isacharequal}\ {\isacharparenleft}{\isadigit{0}}x{\isadigit{0}}{\isadigit{0}}{\isadigit{0}}{\isadigit{0}}{\isadigit{0}}{\isadigit{0}}{\isadigit{0}}{\isadigit{0}}BAFDA{\isadigit{0}}{\isadigit{0}}{\isadigit{0}}{\isacharcomma}{\isadigit{0}}x{\isadigit{0}}{\isadigit{0}}{\isadigit{0}}{\isadigit{0}}{\isadigit{0}}{\isadigit{0}}{\isadigit{0}}{\isadigit{0}}BB{\isadigit{3}}D{\isadigit{3}}FFF{\isacharparenright}{\isachardoublequoteclose}\isanewline
\isacommand{definition}\isamarkupfalse%
\ {\isachardoublequoteopen}dram{\isacharunderscore}sys{\isadigit{4}}\ {\isacharequal}\ {\isacharparenleft}{\isadigit{0}}x{\isadigit{0}}{\isadigit{0}}{\isadigit{0}}{\isadigit{0}}{\isadigit{0}}{\isadigit{0}}{\isadigit{0}}{\isadigit{0}}BDFAC{\isadigit{0}}{\isadigit{0}}{\isadigit{0}}{\isacharcomma}{\isadigit{0}}x{\isadigit{0}}{\isadigit{0}}{\isadigit{0}}{\isadigit{0}}{\isadigit{0}}{\isadigit{0}}{\isadigit{0}}{\isadigit{0}}BDFFFFFF{\isacharparenright}{\isachardoublequoteclose}\isanewline
\isacommand{definition}\isamarkupfalse%
\ {\isachardoublequoteopen}dram{\isacharunderscore}sys{\isadigit{5}}\ {\isacharequal}\ {\isacharparenleft}{\isadigit{0}}x{\isadigit{0}}{\isadigit{0}}{\isadigit{0}}{\isadigit{0}}{\isadigit{0}}{\isadigit{0}}{\isadigit{0}}{\isadigit{1}}{\isadigit{0}}{\isadigit{0}}{\isadigit{0}}{\isadigit{0}}{\isadigit{0}}{\isadigit{0}}{\isadigit{0}}{\isadigit{0}}{\isacharcomma}{\isadigit{0}}x{\isadigit{0}}{\isadigit{0}}{\isadigit{0}}{\isadigit{0}}{\isadigit{0}}{\isadigit{0}}{\isadigit{2}}{\isadigit{0}}{\isadigit{3}}FFFFFFF{\isacharparenright}{\isachardoublequoteclose}\isanewline
\isacommand{definition}\isamarkupfalse%
\ {\isachardoublequoteopen}dram{\isacharunderscore}sys{\isadigit{6}}\ {\isacharequal}\ {\isacharparenleft}{\isadigit{0}}x{\isadigit{0}}{\isadigit{0}}{\isadigit{0}}{\isadigit{0}}{\isadigit{0}}{\isadigit{0}}{\isadigit{2}}{\isadigit{0}}{\isadigit{4}}{\isadigit{0}}{\isadigit{0}}{\isadigit{0}}{\isadigit{0}}{\isadigit{0}}{\isadigit{0}}{\isadigit{0}}{\isacharcomma}{\isadigit{0}}x{\isadigit{0}}{\isadigit{0}}{\isadigit{0}}{\isadigit{0}}{\isadigit{0}}{\isadigit{0}}{\isadigit{4}}{\isadigit{0}}{\isadigit{3}}FFFFFFF{\isacharparenright}{\isachardoublequoteclose}%
\begin{isamarkuptext}%
PCI Express Ranges%
\end{isamarkuptext}\isamarkuptrue%
\isacommand{definition}\isamarkupfalse%
\ {\isachardoublequoteopen}pci{\isacharunderscore}lo{\isacharunderscore}{\isadigit{0}}\ {\isacharequal}\ {\isacharparenleft}{\isadigit{0}}xD{\isadigit{0}}{\isadigit{0}}{\isadigit{0}}{\isadigit{0}}{\isadigit{0}}{\isadigit{0}}{\isadigit{0}}{\isacharcomma}{\isadigit{0}}xD{\isadigit{0}}EFFFFF{\isacharparenright}{\isachardoublequoteclose}\isanewline
\isacommand{definition}\isamarkupfalse%
\ {\isachardoublequoteopen}pci{\isacharunderscore}lo{\isacharunderscore}{\isadigit{1}}\ {\isacharequal}\ {\isacharparenleft}{\isadigit{0}}xEC{\isadigit{0}}{\isadigit{0}}{\isadigit{0}}{\isadigit{0}}{\isadigit{0}}{\isadigit{0}}{\isacharcomma}{\isadigit{0}}xEC{\isadigit{1}}FFFFF{\isacharparenright}{\isachardoublequoteclose}\ \isanewline
\isacommand{definition}\isamarkupfalse%
\ {\isachardoublequoteopen}pci{\isacharunderscore}hi{\isacharunderscore}{\isadigit{0}}\ {\isacharequal}\ {\isacharparenleft}{\isadigit{0}}x{\isadigit{0}}{\isadigit{0}}{\isadigit{0}}{\isadigit{0}}{\isadigit{3}}{\isadigit{8}}{\isadigit{0}}{\isadigit{0}}{\isadigit{0}}{\isadigit{0}}{\isadigit{0}}{\isadigit{0}}{\isadigit{0}}{\isadigit{0}}{\isadigit{0}}{\isadigit{0}}{\isacharcomma}{\isadigit{0}}x{\isadigit{0}}{\isadigit{0}}{\isadigit{0}}{\isadigit{0}}{\isadigit{3}}{\isadigit{8}}{\isadigit{0}}{\isadigit{2}}{\isadigit{0}}{\isadigit{0}}{\isadigit{9}}FFFFF{\isacharparenright}{\isachardoublequoteclose}\isanewline
\isacommand{definition}\isamarkupfalse%
\ {\isachardoublequoteopen}pci{\isacharunderscore}hi{\isacharunderscore}{\isadigit{1}}\ {\isacharequal}\ {\isacharparenleft}{\isadigit{0}}x{\isadigit{0}}{\isadigit{0}}{\isadigit{0}}{\isadigit{0}}{\isadigit{3}}{\isadigit{8}}{\isadigit{0}}{\isadigit{4}}{\isadigit{0}}{\isadigit{0}}{\isadigit{0}}{\isadigit{0}}{\isadigit{0}}{\isadigit{0}}{\isadigit{0}}{\isadigit{0}}{\isacharcomma}{\isadigit{0}}x{\isadigit{0}}{\isadigit{0}}{\isadigit{0}}{\isadigit{0}}{\isadigit{3}}{\isadigit{8}}{\isadigit{0}}{\isadigit{6}}{\isadigit{0}}{\isadigit{0}}{\isadigit{7}}FFFFF{\isacharparenright}{\isachardoublequoteclose}%
\begin{isamarkuptext}%
The Node 0 Interconnect%
\end{isamarkuptext}\isamarkuptrue%
\isacommand{definition}\isamarkupfalse%
\ {\isachardoublequoteopen}node{\isacharunderscore}{\isadigit{0}}{\isacharunderscore}interconnect\ {\isacharequal}\ empty{\isacharunderscore}spec\ {\isasymlparr}\isanewline
\ \ acc{\isacharunderscore}blocks\ {\isacharcolon}{\isacharequal}\ {\isacharbrackleft}{\isacharbrackright}{\isacharcomma}\isanewline
\ \ map{\isacharunderscore}blocks\ {\isacharcolon}{\isacharequal}\ {\isacharbrackleft}block{\isacharunderscore}map\ dram{\isacharunderscore}sys{\isadigit{0}}\ {\isadigit{2}}\ {\isadigit{0}}x{\isadigit{0}}{\isadigit{0}}{\isadigit{0}}{\isadigit{0}}{\isadigit{0}}{\isadigit{0}}{\isadigit{0}}{\isadigit{0}}{\isadigit{0}}{\isacharcomma}\ block{\isacharunderscore}map\ dram{\isacharunderscore}sys{\isadigit{1}}\ {\isadigit{2}}\ {\isadigit{0}}x{\isadigit{9}}C{\isadigit{0}}{\isadigit{0}}{\isadigit{0}}{\isacharcomma}\isanewline
\ \ \ \ \ \ \ \ \ \ \ \ \ \ \ \ \ block{\isacharunderscore}map\ dram{\isacharunderscore}sys{\isadigit{2}}\ {\isadigit{2}}\ {\isadigit{0}}xBACC{\isadigit{4}}{\isadigit{0}}{\isadigit{0}}{\isadigit{0}}{\isacharcomma}\ block{\isacharunderscore}map\ dram{\isacharunderscore}sys{\isadigit{3}}\ {\isadigit{2}}\ {\isadigit{0}}xBACF{\isadigit{9}}{\isadigit{0}}{\isadigit{0}}{\isadigit{0}}{\isacharcomma}\ \isanewline
\ \ \ \ \ \ \ \ \ \ \ \ \ \ \ \ \ block{\isacharunderscore}map\ dram{\isacharunderscore}sys{\isadigit{4}}\ {\isadigit{2}}\ {\isadigit{0}}xBB{\isadigit{0}}F{\isadigit{3}}{\isadigit{0}}{\isadigit{0}}{\isadigit{0}}{\isacharcomma}\ block{\isacharunderscore}map\ dram{\isacharunderscore}sys{\isadigit{5}}\ {\isadigit{2}}\ {\isadigit{0}}xBB{\isadigit{1}}{\isadigit{4}}{\isadigit{7}}{\isadigit{0}}{\isadigit{0}}{\isadigit{0}}{\isacharcomma}\isanewline
\ \ \ \ \ \ \ \ \ \ \ \ \ \ \ \ \ direct{\isacharunderscore}map\ dram{\isacharunderscore}sys{\isadigit{6}}\ {\isadigit{1}}{\isacharcomma}\ direct{\isacharunderscore}map\ pci{\isacharunderscore}lo{\isacharunderscore}{\isadigit{0}}\ {\isadigit{4}}{\isacharcomma}\ \isanewline
\ \ \ \ \ \ \ \ \ \ \ \ \ \ \ \ \ direct{\isacharunderscore}map\ pci{\isacharunderscore}hi{\isacharunderscore}{\isadigit{0}}\ {\isadigit{4}}{\isacharcomma}\ direct{\isacharunderscore}map\ pci{\isacharunderscore}lo{\isacharunderscore}{\isadigit{1}}\ {\isadigit{1}}{\isacharcomma}\ \isanewline
\ \ \ \ \ \ \ \ \ \ \ \ \ \ \ \ \ direct{\isacharunderscore}map\ pci{\isacharunderscore}hi{\isacharunderscore}{\isadigit{1}}\ {\isadigit{1}}{\isacharbrackright}\isanewline
{\isasymrparr}{\isachardoublequoteclose}%
\begin{isamarkuptext}%
The Node 1 Interconnect%
\end{isamarkuptext}\isamarkuptrue%
\isacommand{definition}\isamarkupfalse%
\ {\isachardoublequoteopen}node{\isacharunderscore}{\isadigit{1}}{\isacharunderscore}interconnect\ {\isacharequal}\ empty{\isacharunderscore}spec\ {\isasymlparr}\isanewline
\ \ acc{\isacharunderscore}blocks\ {\isacharcolon}{\isacharequal}\ {\isacharbrackleft}{\isacharbrackright}{\isacharcomma}\isanewline
\ \ map{\isacharunderscore}blocks\ {\isacharcolon}{\isacharequal}\ {\isacharbrackleft}direct{\isacharunderscore}map\ dram{\isacharunderscore}sys{\isadigit{0}}\ {\isadigit{0}}{\isacharcomma}\ direct{\isacharunderscore}map\ dram{\isacharunderscore}sys{\isadigit{1}}\ {\isadigit{0}}{\isacharcomma}\ \ \isanewline
\ \ \ \ \ \ \ \ \ \ \ \ \ \ \ \ \ direct{\isacharunderscore}map\ dram{\isacharunderscore}sys{\isadigit{2}}\ {\isadigit{0}}{\isacharcomma}\ direct{\isacharunderscore}map\ dram{\isacharunderscore}sys{\isadigit{3}}\ {\isadigit{0}}{\isacharcomma}\ \isanewline
\ \ \ \ \ \ \ \ \ \ \ \ \ \ \ \ \ direct{\isacharunderscore}map\ dram{\isacharunderscore}sys{\isadigit{4}}\ {\isadigit{0}}{\isacharcomma}\ direct{\isacharunderscore}map\ dram{\isacharunderscore}sys{\isadigit{5}}\ {\isadigit{0}}{\isacharcomma}\ \isanewline
\ \ \ \ \ \ \ \ \ \ \ \ \ \ \ \ \ block{\isacharunderscore}map\ dram{\isacharunderscore}sys{\isadigit{6}}\ {\isadigit{3}}\ {\isadigit{0}}x{\isadigit{0}}{\isadigit{2}}{\isadigit{0}}{\isadigit{0}}{\isadigit{0}}{\isadigit{0}}{\isadigit{0}}{\isadigit{0}}{\isadigit{0}}{\isadigit{0}}{\isadigit{0}}{\isacharcomma}\ \ \isanewline
\ \ \ \ \ \ \ \ \ \ \ \ \ \ \ \ \ direct{\isacharunderscore}map\ pci{\isacharunderscore}lo{\isacharunderscore}{\isadigit{0}}\ {\isadigit{0}}{\isacharcomma}\ direct{\isacharunderscore}map\ pci{\isacharunderscore}hi{\isacharunderscore}{\isadigit{0}}\ {\isadigit{0}}{\isacharcomma}\ \isanewline
\ \ \ \ \ \ \ \ \ \ \ \ \ \ \ \ \ direct{\isacharunderscore}map\ pci{\isacharunderscore}lo{\isacharunderscore}{\isadigit{1}}\ {\isadigit{5}}{\isacharcomma}\ direct{\isacharunderscore}map\ pci{\isacharunderscore}hi{\isacharunderscore}{\isadigit{1}}\ {\isadigit{5}}{\isacharbrackright}\isanewline
{\isasymrparr}{\isachardoublequoteclose}%
\begin{isamarkuptext}%
DRAM controller 0: 4 channels with 32G each.%
\end{isamarkuptext}\isamarkuptrue%
\isacommand{definition}\isamarkupfalse%
\ {\isachardoublequoteopen}dram{\isadigit{0}}{\isacharunderscore}{\isadigit{0}}\ {\isacharequal}\ {\isacharparenleft}{\isadigit{0}}x{\isadigit{0}}{\isadigit{0}}{\isadigit{0}}{\isadigit{0}}{\isadigit{0}}{\isadigit{0}}{\isadigit{0}}{\isadigit{0}}{\isadigit{0}}{\isadigit{0}}{\isadigit{0}}{\isacharcomma}\ {\isadigit{0}}x{\isadigit{0}}{\isadigit{0}}{\isadigit{7}}FFFFFFFF{\isacharparenright}{\isachardoublequoteclose}\isanewline
\isacommand{definition}\isamarkupfalse%
\ {\isachardoublequoteopen}dram{\isadigit{0}}{\isacharunderscore}{\isadigit{1}}\ {\isacharequal}\ {\isacharparenleft}{\isadigit{0}}x{\isadigit{0}}{\isadigit{0}}{\isadigit{8}}{\isadigit{0}}{\isadigit{0}}{\isadigit{0}}{\isadigit{0}}{\isadigit{0}}{\isadigit{0}}{\isadigit{0}}{\isadigit{0}}{\isacharcomma}\ {\isadigit{0}}x{\isadigit{0}}{\isadigit{0}}FFFFFFFFF{\isacharparenright}{\isachardoublequoteclose}\isanewline
\isacommand{definition}\isamarkupfalse%
\ {\isachardoublequoteopen}dram{\isadigit{0}}{\isacharunderscore}{\isadigit{2}}\ {\isacharequal}\ {\isacharparenleft}{\isadigit{0}}x{\isadigit{0}}{\isadigit{1}}{\isadigit{0}}{\isadigit{0}}{\isadigit{0}}{\isadigit{0}}{\isadigit{0}}{\isadigit{0}}{\isadigit{0}}{\isadigit{0}}{\isadigit{0}}{\isacharcomma}\ {\isadigit{0}}x{\isadigit{0}}{\isadigit{1}}{\isadigit{7}}FFFFFFFF{\isacharparenright}{\isachardoublequoteclose}\ \ \isanewline
\isacommand{definition}\isamarkupfalse%
\ {\isachardoublequoteopen}dram{\isadigit{0}}{\isacharunderscore}{\isadigit{3}}\ {\isacharequal}\ {\isacharparenleft}{\isadigit{0}}x{\isadigit{0}}{\isadigit{1}}{\isadigit{8}}{\isadigit{0}}{\isadigit{0}}{\isadigit{0}}{\isadigit{0}}{\isadigit{0}}{\isadigit{0}}{\isadigit{0}}{\isadigit{0}}{\isacharcomma}\ {\isadigit{0}}x{\isadigit{0}}{\isadigit{1}}FFFFFFFFF{\isacharparenright}{\isachardoublequoteclose}\isanewline
\isacommand{definition}\isamarkupfalse%
\ {\isachardoublequoteopen}node{\isacharunderscore}{\isadigit{2}}{\isacharunderscore}dram\ {\isacharequal}\ empty{\isacharunderscore}spec\ {\isasymlparr}\isanewline
\ \ acc{\isacharunderscore}blocks\ {\isacharcolon}{\isacharequal}\ {\isacharbrackleft}dram{\isadigit{0}}{\isacharunderscore}{\isadigit{0}}{\isacharcomma}\ dram{\isadigit{0}}{\isacharunderscore}{\isadigit{1}}{\isacharcomma}\ dram{\isadigit{0}}{\isacharunderscore}{\isadigit{2}}{\isacharcomma}\ dram{\isadigit{0}}{\isacharunderscore}{\isadigit{3}}{\isacharbrackright}{\isacharcomma}\isanewline
\ \ map{\isacharunderscore}blocks\ {\isacharcolon}{\isacharequal}\ {\isacharbrackleft}{\isacharbrackright}\isanewline
{\isasymrparr}{\isachardoublequoteclose}%
\begin{isamarkuptext}%
DRAM controller 1: 4 channels with 32G each%
\end{isamarkuptext}\isamarkuptrue%
\isacommand{definition}\isamarkupfalse%
\ {\isachardoublequoteopen}dram{\isadigit{1}}{\isacharunderscore}{\isadigit{0}}\ {\isacharequal}\ {\isacharparenleft}{\isadigit{0}}x{\isadigit{0}}{\isadigit{2}}{\isadigit{0}}{\isadigit{0}}{\isadigit{0}}{\isadigit{0}}{\isadigit{0}}{\isadigit{0}}{\isadigit{0}}{\isadigit{0}}{\isadigit{0}}{\isacharcomma}\ {\isadigit{0}}x{\isadigit{0}}{\isadigit{2}}{\isadigit{7}}ffffffff{\isacharparenright}{\isachardoublequoteclose}\isanewline
\isacommand{definition}\isamarkupfalse%
\ {\isachardoublequoteopen}dram{\isadigit{1}}{\isacharunderscore}{\isadigit{1}}\ {\isacharequal}\ {\isacharparenleft}{\isadigit{0}}x{\isadigit{0}}{\isadigit{2}}{\isadigit{8}}{\isadigit{0}}{\isadigit{0}}{\isadigit{0}}{\isadigit{0}}{\isadigit{0}}{\isadigit{0}}{\isadigit{0}}{\isadigit{0}}{\isacharcomma}\ {\isadigit{0}}x{\isadigit{0}}{\isadigit{2}}FFFFFFFFF{\isacharparenright}{\isachardoublequoteclose}\isanewline
\isacommand{definition}\isamarkupfalse%
\ {\isachardoublequoteopen}dram{\isadigit{1}}{\isacharunderscore}{\isadigit{2}}\ {\isacharequal}\ {\isacharparenleft}{\isadigit{0}}x{\isadigit{0}}{\isadigit{3}}{\isadigit{0}}{\isadigit{0}}{\isadigit{0}}{\isadigit{0}}{\isadigit{0}}{\isadigit{0}}{\isadigit{0}}{\isadigit{0}}{\isadigit{0}}{\isacharcomma}\ {\isadigit{0}}x{\isadigit{0}}{\isadigit{3}}{\isadigit{7}}FFFFFFFF{\isacharparenright}{\isachardoublequoteclose}\ \ \isanewline
\isacommand{definition}\isamarkupfalse%
\ {\isachardoublequoteopen}dram{\isadigit{1}}{\isacharunderscore}{\isadigit{3}}\ {\isacharequal}\ {\isacharparenleft}{\isadigit{0}}x{\isadigit{0}}{\isadigit{3}}{\isadigit{8}}{\isadigit{0}}{\isadigit{0}}{\isadigit{0}}{\isadigit{0}}{\isadigit{0}}{\isadigit{0}}{\isadigit{0}}{\isadigit{0}}{\isacharcomma}\ {\isadigit{0}}x{\isadigit{0}}{\isadigit{3}}FFFFFFFFF{\isacharparenright}{\isachardoublequoteclose}\isanewline
\isacommand{definition}\isamarkupfalse%
\ {\isachardoublequoteopen}node{\isacharunderscore}{\isadigit{3}}{\isacharunderscore}dram\ {\isacharequal}\ empty{\isacharunderscore}spec\ {\isasymlparr}\isanewline
\ \ acc{\isacharunderscore}blocks\ {\isacharcolon}{\isacharequal}\ {\isacharbrackleft}dram{\isadigit{1}}{\isacharunderscore}{\isadigit{0}}{\isacharcomma}\ dram{\isadigit{1}}{\isacharunderscore}{\isadigit{1}}{\isacharcomma}\ dram{\isadigit{1}}{\isacharunderscore}{\isadigit{2}}{\isacharcomma}\ dram{\isadigit{1}}{\isacharunderscore}{\isadigit{3}}{\isacharbrackright}{\isacharcomma}\isanewline
\ \ map{\isacharunderscore}blocks\ {\isacharcolon}{\isacharequal}\ {\isacharbrackleft}{\isacharbrackright}\isanewline
{\isasymrparr}{\isachardoublequoteclose}%
\begin{isamarkuptext}%
PCI Express Devices%
\end{isamarkuptext}\isamarkuptrue%
\isacommand{definition}\isamarkupfalse%
\ {\isachardoublequoteopen}phi{\isadigit{0}}{\isacharunderscore}gddr\ {\isacharequal}\ {\isacharparenleft}{\isadigit{0}}x{\isadigit{3}}{\isadigit{8}}{\isadigit{0}}{\isadigit{0}}{\isadigit{0}}{\isadigit{0}}{\isadigit{0}}{\isadigit{0}}{\isadigit{0}}{\isadigit{0}}{\isadigit{0}}{\isadigit{0}}{\isacharcomma}\ {\isadigit{0}}x{\isadigit{3}}{\isadigit{8}}{\isadigit{0}}{\isadigit{1}}FFFFFFFF{\isacharparenright}{\isachardoublequoteclose}\isanewline
\isacommand{definition}\isamarkupfalse%
\ {\isachardoublequoteopen}phi{\isadigit{0}}{\isacharunderscore}mmio\ {\isacharequal}\ {\isacharparenleft}{\isadigit{0}}xD{\isadigit{0}}C{\isadigit{0}}{\isadigit{0}}{\isadigit{0}}{\isadigit{0}}{\isadigit{0}}{\isacharcomma}\ {\isadigit{0}}xd{\isadigit{0}}C{\isadigit{1}}FFFF{\isacharparenright}{\isachardoublequoteclose}\isanewline
\isacommand{definition}\isamarkupfalse%
\ {\isachardoublequoteopen}phi{\isadigit{1}}{\isacharunderscore}gddr\ {\isacharequal}\ {\isacharparenleft}{\isadigit{0}}x{\isadigit{3}}{\isadigit{8}}{\isadigit{0}}{\isadigit{4}}{\isadigit{0}}{\isadigit{0}}{\isadigit{0}}{\isadigit{0}}{\isadigit{0}}{\isadigit{0}}{\isadigit{0}}{\isadigit{0}}{\isacharcomma}\ {\isadigit{0}}x{\isadigit{3}}{\isadigit{8}}{\isadigit{0}}{\isadigit{5}}FFFFFFFF{\isacharparenright}{\isachardoublequoteclose}\isanewline
\isacommand{definition}\isamarkupfalse%
\ {\isachardoublequoteopen}phi{\isadigit{1}}{\isacharunderscore}mmio\ {\isacharequal}\ {\isacharparenleft}{\isadigit{0}}xDC{\isadigit{2}}{\isadigit{0}}{\isadigit{0}}{\isadigit{0}}{\isadigit{0}}{\isadigit{0}}{\isacharcomma}\ {\isadigit{0}}xEC{\isadigit{2}}{\isadigit{1}}FFFF{\isacharparenright}{\isachardoublequoteclose}\ \ \isanewline
\isacommand{definition}\isamarkupfalse%
\ {\isachardoublequoteopen}dma{\isadigit{0}}\ {\isacharequal}\ {\isacharparenleft}{\isadigit{0}}x{\isadigit{3}}{\isadigit{8}}{\isadigit{0}}{\isadigit{3}}FFF{\isadigit{9}}{\isadigit{0}}{\isadigit{0}}{\isadigit{0}}{\isadigit{0}}{\isacharcomma}\ {\isadigit{0}}x{\isadigit{3}}{\isadigit{8}}{\isadigit{0}}{\isadigit{3}}FFF{\isadigit{2}}{\isadigit{3}}FFF{\isacharparenright}{\isachardoublequoteclose}\isanewline
\isacommand{definition}\isamarkupfalse%
\ {\isachardoublequoteopen}dma{\isadigit{1}}\ {\isacharequal}\ {\isacharparenleft}{\isadigit{0}}x{\isadigit{3}}{\isadigit{8}}{\isadigit{0}}{\isadigit{7}}FFF{\isadigit{6}}{\isadigit{0}}{\isadigit{0}}{\isadigit{0}}{\isadigit{0}}{\isacharcomma}\ {\isadigit{0}}x{\isadigit{3}}{\isadigit{8}}{\isadigit{0}}{\isadigit{7}}FFF{\isadigit{0}}{\isadigit{3}}FFF{\isacharparenright}{\isachardoublequoteclose}\isanewline
\isacommand{definition}\isamarkupfalse%
\ {\isachardoublequoteopen}ahci\ {\isacharequal}\ {\isacharparenleft}{\isadigit{0}}xD{\isadigit{0}}F{\isadigit{0}}{\isadigit{0}}{\isadigit{0}}{\isadigit{0}}{\isadigit{0}}{\isacharcomma}\ {\isadigit{0}}xD{\isadigit{0}}F{\isadigit{0}}{\isadigit{0}}{\isadigit{7}}FF{\isacharparenright}{\isachardoublequoteclose}\isanewline
\isacommand{definition}\isamarkupfalse%
\ {\isachardoublequoteopen}ehci\ {\isacharequal}\ {\isacharparenleft}{\isadigit{0}}xD{\isadigit{0}}F{\isadigit{1}}{\isadigit{0}}{\isadigit{0}}{\isadigit{0}}{\isadigit{0}}{\isacharcomma}\ {\isadigit{0}}xD{\isadigit{0}}F{\isadigit{1}}{\isadigit{0}}{\isadigit{3}}FF{\isacharparenright}{\isachardoublequoteclose}\isanewline
\isacommand{definition}\isamarkupfalse%
\ {\isachardoublequoteopen}ioapic{\isadigit{0}}\ {\isacharequal}\ {\isacharparenleft}{\isadigit{0}}xD{\isadigit{0}}F{\isadigit{6}}{\isadigit{0}}{\isadigit{0}}{\isadigit{0}}{\isadigit{0}}{\isacharcomma}\ {\isadigit{0}}xD{\isadigit{0}}F{\isadigit{6}}{\isadigit{0}}FFF{\isacharparenright}{\isachardoublequoteclose}\isanewline
\isacommand{definition}\isamarkupfalse%
\ {\isachardoublequoteopen}ioapic{\isadigit{1}}\ {\isacharequal}\ {\isacharparenleft}{\isadigit{0}}xEC{\isadigit{3}}{\isadigit{0}}{\isadigit{0}}{\isadigit{0}}{\isadigit{0}}{\isadigit{0}}{\isacharcomma}\ {\isadigit{0}}xEC{\isadigit{3}}{\isadigit{0}}{\isadigit{0}}FFF{\isacharparenright}{\isachardoublequoteclose}\isanewline
\isacommand{definition}\isamarkupfalse%
\ {\isachardoublequoteopen}e{\isadigit{1}}{\isadigit{0}}{\isadigit{0}}{\isadigit{0}}\ {\isacharequal}\ {\isacharparenleft}{\isadigit{0}}xD{\isadigit{0}}{\isadigit{9}}{\isadigit{6}}{\isadigit{0}}{\isadigit{0}}{\isadigit{0}}{\isadigit{0}}{\isacharcomma}\ {\isadigit{0}}xD{\isadigit{0}}{\isadigit{9}}{\isadigit{7}}FFFFF{\isacharparenright}{\isachardoublequoteclose}%
\begin{isamarkuptext}%
PCI Root Complex 0%
\end{isamarkuptext}\isamarkuptrue%
\isacommand{definition}\isamarkupfalse%
\ {\isachardoublequoteopen}node{\isacharunderscore}{\isadigit{4}}{\isacharunderscore}pci\ {\isacharequal}\ empty{\isacharunderscore}spec\ {\isasymlparr}\isanewline
\ \ acc{\isacharunderscore}blocks\ {\isacharcolon}{\isacharequal}\ {\isacharbrackleft}{\isacharbrackright}{\isacharcomma}\isanewline
\ \ map{\isacharunderscore}blocks\ {\isacharcolon}{\isacharequal}\ {\isacharbrackleft}direct{\isacharunderscore}map\ dram{\isacharunderscore}sys{\isadigit{0}}\ {\isadigit{0}}{\isacharcomma}\ direct{\isacharunderscore}map\ dram{\isacharunderscore}sys{\isadigit{1}}\ {\isadigit{0}}{\isacharcomma}\ \isanewline
\ \ \ \ \ \ \ \ \ \ \ \ \ \ \ \ \ direct{\isacharunderscore}map\ dram{\isacharunderscore}sys{\isadigit{2}}\ {\isadigit{0}}{\isacharcomma}\ direct{\isacharunderscore}map\ dram{\isacharunderscore}sys{\isadigit{3}}\ {\isadigit{0}}{\isacharcomma}\ \isanewline
\ \ \ \ \ \ \ \ \ \ \ \ \ \ \ \ \ direct{\isacharunderscore}map\ dram{\isacharunderscore}sys{\isadigit{4}}\ {\isadigit{0}}{\isacharcomma}\ direct{\isacharunderscore}map\ dram{\isacharunderscore}sys{\isadigit{5}}\ {\isadigit{0}}{\isacharcomma}\ \isanewline
\ \ \ \ \ \ \ \ \ \ \ \ \ \ \ \ \ direct{\isacharunderscore}map\ dram{\isacharunderscore}sys{\isadigit{6}}\ {\isadigit{0}}{\isacharcomma}\ direct{\isacharunderscore}map\ dma{\isadigit{0}}\ {\isadigit{9}}{\isacharcomma}\ \isanewline
\ \ \ \ \ \ \ \ \ \ \ \ \ \ \ \ \ direct{\isacharunderscore}map\ ahci\ {\isadigit{1}}{\isadigit{2}}{\isacharcomma}\ direct{\isacharunderscore}map\ ehci\ {\isadigit{1}}{\isadigit{1}}{\isacharcomma}\ \isanewline
\ \ \ \ \ \ \ \ \ \ \ \ \ \ \ \ \ direct{\isacharunderscore}map\ ioapic{\isadigit{0}}\ {\isadigit{7}}{\isacharcomma}\ direct{\isacharunderscore}map\ e{\isadigit{1}}{\isadigit{0}}{\isadigit{0}}{\isadigit{0}}\ {\isadigit{6}}{\isacharcomma}\isanewline
\ \ \ \ \ \ \ \ \ \ \ \ \ \ \ \ \ block{\isacharunderscore}map\ phi{\isadigit{0}}{\isacharunderscore}gddr\ {\isadigit{1}}{\isadigit{3}}\ {\isadigit{0}}x{\isadigit{0}}{\isacharcomma}\ \isanewline
\ \ \ \ \ \ \ \ \ \ \ \ \ \ \ \ \ block{\isacharunderscore}map\ phi{\isadigit{0}}{\isacharunderscore}mmio\ {\isadigit{1}}{\isadigit{3}}\ {\isadigit{0}}x{\isadigit{0}}{\isadigit{8}}{\isadigit{0}}{\isadigit{0}}{\isadigit{7}}D{\isadigit{0}}{\isadigit{0}}{\isadigit{0}}{\isadigit{0}}{\isacharbrackright}\isanewline
{\isasymrparr}{\isachardoublequoteclose}%
\begin{isamarkuptext}%
PCI Root Complex 1%
\end{isamarkuptext}\isamarkuptrue%
\isacommand{definition}\isamarkupfalse%
\ {\isachardoublequoteopen}node{\isacharunderscore}{\isadigit{5}}{\isacharunderscore}pci\ {\isacharequal}\ empty{\isacharunderscore}spec\ {\isasymlparr}\isanewline
\ \ acc{\isacharunderscore}blocks\ {\isacharcolon}{\isacharequal}\ {\isacharbrackleft}{\isacharbrackright}{\isacharcomma}\isanewline
\ \ map{\isacharunderscore}blocks\ {\isacharcolon}{\isacharequal}\ {\isacharbrackleft}direct{\isacharunderscore}map\ dram{\isacharunderscore}sys{\isadigit{0}}\ {\isadigit{0}}{\isacharcomma}\ direct{\isacharunderscore}map\ dram{\isacharunderscore}sys{\isadigit{1}}\ {\isadigit{0}}{\isacharcomma}\ \isanewline
\ \ \ \ \ \ \ \ \ \ \ \ \ \ \ \ \ direct{\isacharunderscore}map\ dram{\isacharunderscore}sys{\isadigit{2}}\ {\isadigit{0}}{\isacharcomma}\ direct{\isacharunderscore}map\ dram{\isacharunderscore}sys{\isadigit{3}}\ {\isadigit{0}}{\isacharcomma}\ \isanewline
\ \ \ \ \ \ \ \ \ \ \ \ \ \ \ \ \ direct{\isacharunderscore}map\ dram{\isacharunderscore}sys{\isadigit{4}}\ {\isadigit{0}}{\isacharcomma}\ direct{\isacharunderscore}map\ dram{\isacharunderscore}sys{\isadigit{5}}\ {\isadigit{0}}{\isacharcomma}\ \isanewline
\ \ \ \ \ \ \ \ \ \ \ \ \ \ \ \ \ direct{\isacharunderscore}map\ dram{\isacharunderscore}sys{\isadigit{6}}\ {\isadigit{0}}{\isacharcomma}\ direct{\isacharunderscore}map\ dma{\isadigit{1}}\ {\isadigit{1}}{\isadigit{0}}{\isacharcomma}\isanewline
\ \ \ \ \ \ \ \ \ \ \ \ \ \ \ \ \ direct{\isacharunderscore}map\ ioapic{\isadigit{1}}\ {\isadigit{8}}{\isacharcomma}\ direct{\isacharunderscore}map\ dma{\isadigit{1}}\ {\isadigit{1}}{\isadigit{0}}{\isacharcomma}\isanewline
\ \ \ \ \ \ \ \ \ \ \ \ \ \ \ \ \ block{\isacharunderscore}map\ phi{\isadigit{1}}{\isacharunderscore}gddr\ {\isadigit{1}}{\isadigit{3}}\ {\isadigit{0}}{\isacharcomma}\ \isanewline
\ \ \ \ \ \ \ \ \ \ \ \ \ \ \ \ \ block{\isacharunderscore}map\ phi{\isadigit{1}}{\isacharunderscore}mmio\ {\isadigit{1}}{\isadigit{3}}\ {\isadigit{0}}x{\isadigit{0}}{\isadigit{8}}{\isadigit{0}}{\isadigit{0}}{\isadigit{7}}D{\isadigit{0}}{\isadigit{0}}{\isadigit{0}}{\isadigit{0}}{\isacharbrackright}\isanewline
{\isasymrparr}{\isachardoublequoteclose}%
\begin{isamarkuptext}%
e1000 Network Card%
\end{isamarkuptext}\isamarkuptrue%
\isacommand{definition}\isamarkupfalse%
\ {\isachardoublequoteopen}node{\isacharunderscore}{\isadigit{6}}{\isacharunderscore}e{\isadigit{1}}{\isadigit{0}}{\isadigit{0}}{\isadigit{0}}\ {\isacharequal}\ empty{\isacharunderscore}spec\ {\isasymlparr}\isanewline
\ \ acc{\isacharunderscore}blocks\ {\isacharcolon}{\isacharequal}\ {\isacharbrackleft}e{\isadigit{1}}{\isadigit{0}}{\isadigit{0}}{\isadigit{0}}{\isacharbrackright}{\isacharcomma}\isanewline
\ \ overlay\ {\isacharcolon}{\isacharequal}\ Some\ {\isadigit{4}}\isanewline
{\isasymrparr}{\isachardoublequoteclose}%
\begin{isamarkuptext}%
IO APICs%
\end{isamarkuptext}\isamarkuptrue%
\isacommand{definition}\isamarkupfalse%
\ {\isachardoublequoteopen}node{\isacharunderscore}{\isadigit{7}}{\isacharunderscore}ioapic\ {\isacharequal}\ empty{\isacharunderscore}spec\ {\isasymlparr}\isanewline
\ \ acc{\isacharunderscore}blocks\ {\isacharcolon}{\isacharequal}\ {\isacharbrackleft}ioapic{\isadigit{0}}{\isacharbrackright}{\isacharcomma}\isanewline
\ \ overlay\ {\isacharcolon}{\isacharequal}\ Some\ {\isadigit{4}}\isanewline
{\isasymrparr}{\isachardoublequoteclose}\isanewline
\isanewline
\isacommand{definition}\isamarkupfalse%
\ {\isachardoublequoteopen}node{\isacharunderscore}{\isadigit{8}}{\isacharunderscore}ioapic\ {\isacharequal}\ empty{\isacharunderscore}spec\ {\isasymlparr}\isanewline
\ \ acc{\isacharunderscore}blocks\ {\isacharcolon}{\isacharequal}\ {\isacharbrackleft}ioapic{\isadigit{1}}{\isacharbrackright}{\isacharcomma}\isanewline
\ \ overlay\ {\isacharcolon}{\isacharequal}\ Some\ {\isadigit{5}}\isanewline
{\isasymrparr}{\isachardoublequoteclose}%
\begin{isamarkuptext}%
DMA Engines%
\end{isamarkuptext}\isamarkuptrue%
\isacommand{definition}\isamarkupfalse%
\ {\isachardoublequoteopen}node{\isacharunderscore}{\isadigit{9}}{\isacharunderscore}dma\ {\isacharequal}\ empty{\isacharunderscore}spec\ {\isasymlparr}\isanewline
\ \ acc{\isacharunderscore}blocks\ {\isacharcolon}{\isacharequal}\ {\isacharbrackleft}dma{\isadigit{0}}{\isacharbrackright}{\isacharcomma}\isanewline
\ \ overlay\ {\isacharcolon}{\isacharequal}\ Some\ {\isadigit{4}}\isanewline
{\isasymrparr}{\isachardoublequoteclose}\isanewline
\isanewline
\isacommand{definition}\isamarkupfalse%
\ {\isachardoublequoteopen}node{\isacharunderscore}{\isadigit{1}}{\isadigit{0}}{\isacharunderscore}dma\ {\isacharequal}\ empty{\isacharunderscore}spec\ {\isasymlparr}\isanewline
\ \ acc{\isacharunderscore}blocks\ {\isacharcolon}{\isacharequal}\ {\isacharbrackleft}dma{\isadigit{1}}{\isacharbrackright}{\isacharcomma}\isanewline
\ \ overlay\ {\isacharcolon}{\isacharequal}\ Some\ {\isadigit{5}}\isanewline
{\isasymrparr}{\isachardoublequoteclose}%
\begin{isamarkuptext}%
USB Host Controller%
\end{isamarkuptext}\isamarkuptrue%
\isacommand{definition}\isamarkupfalse%
\ {\isachardoublequoteopen}node{\isacharunderscore}{\isadigit{1}}{\isadigit{1}}{\isacharunderscore}ehci\ {\isacharequal}\ empty{\isacharunderscore}spec\ {\isasymlparr}\isanewline
\ \ acc{\isacharunderscore}blocks\ {\isacharcolon}{\isacharequal}\ {\isacharbrackleft}ehci{\isacharbrackright}{\isacharcomma}\isanewline
\ \ overlay\ {\isacharcolon}{\isacharequal}\ Some\ {\isadigit{4}}\isanewline
{\isasymrparr}{\isachardoublequoteclose}%
\begin{isamarkuptext}%
AHCI Disk controller%
\end{isamarkuptext}\isamarkuptrue%
\isacommand{definition}\isamarkupfalse%
\ {\isachardoublequoteopen}node{\isacharunderscore}{\isadigit{1}}{\isadigit{2}}{\isacharunderscore}ahci\ {\isacharequal}\ empty{\isacharunderscore}spec\ {\isasymlparr}\isanewline
\ \ acc{\isacharunderscore}blocks\ {\isacharcolon}{\isacharequal}\ {\isacharbrackleft}ahci{\isacharbrackright}{\isacharcomma}\isanewline
\ \ overlay\ {\isacharcolon}{\isacharequal}\ Some\ {\isadigit{4}}\isanewline
{\isasymrparr}{\isachardoublequoteclose}%
\begin{isamarkuptext}%
Xeon Phi 0%
\end{isamarkuptext}\isamarkuptrue%
\isacommand{definition}\isamarkupfalse%
\ {\isachardoublequoteopen}phi{\isacharunderscore}gddr\ {\isacharequal}\ {\isacharparenleft}{\isadigit{0}}{\isacharcomma}\ {\isadigit{0}}x{\isadigit{1}}{\isadigit{8}}{\isadigit{0}}{\isadigit{0}}{\isadigit{0}}{\isadigit{0}}{\isadigit{0}}{\isadigit{0}}{\isadigit{0}}{\isacharparenright}{\isachardoublequoteclose}\isanewline
\isacommand{definition}\isamarkupfalse%
\ {\isachardoublequoteopen}phi{\isacharunderscore}sbox\ {\isacharequal}\ {\isacharparenleft}{\isadigit{0}}x{\isadigit{0}}{\isadigit{8}}{\isadigit{0}}{\isadigit{0}}{\isadigit{7}}D{\isadigit{0}}{\isadigit{0}}{\isadigit{0}}{\isadigit{0}}{\isacharcomma}{\isadigit{0}}x{\isadigit{8}}{\isadigit{0}}{\isadigit{0}}{\isadigit{7}}E{\isadigit{0}}{\isadigit{0}}{\isadigit{0}}{\isadigit{0}}{\isacharparenright}{\isachardoublequoteclose}\isanewline
\isacommand{definition}\isamarkupfalse%
\ {\isachardoublequoteopen}phi{\isacharunderscore}sysmem\ {\isacharequal}\ {\isacharparenleft}{\isadigit{0}}x{\isadigit{8}}{\isadigit{0}}{\isadigit{0}}{\isadigit{0}}{\isadigit{0}}{\isadigit{0}}{\isadigit{0}}{\isadigit{0}}{\isadigit{0}}{\isadigit{0}}{\isacharcomma}\ {\isadigit{0}}xFFFFFFFFFF{\isacharparenright}{\isachardoublequoteclose}\isanewline
\isacommand{definition}\isamarkupfalse%
\ {\isachardoublequoteopen}phi{\isacharunderscore}lut{\isacharunderscore}e{\isadigit{0}}{\isadigit{0}}\ {\isacharequal}\ {\isacharparenleft}{\isadigit{0}}x{\isadigit{0}}{\isadigit{0}}{\isadigit{0}}{\isadigit{0}}{\isadigit{0}}{\isadigit{0}}{\isadigit{0}}{\isadigit{0}}{\isadigit{0}}{\isadigit{0}}{\isacharcomma}\ {\isadigit{0}}x{\isadigit{0}}{\isadigit{3}}FFFFFFFF{\isacharparenright}{\isachardoublequoteclose}\ \ \isanewline
\isanewline
\isacommand{definition}\isamarkupfalse%
\ {\isachardoublequoteopen}node{\isacharunderscore}{\isadigit{1}}{\isadigit{3}}{\isacharunderscore}phi\ {\isacharequal}\ empty{\isacharunderscore}spec\ {\isasymlparr}\isanewline
\ \ acc{\isacharunderscore}blocks\ {\isacharcolon}{\isacharequal}\ {\isacharbrackleft}phi{\isacharunderscore}gddr{\isacharcomma}\ phi{\isacharunderscore}sbox{\isacharbrackright}{\isacharcomma}\isanewline
\ \ map{\isacharunderscore}blocks\ {\isacharcolon}{\isacharequal}\ {\isacharbrackleft}block{\isacharunderscore}map\ phi{\isacharunderscore}sysmem\ {\isadigit{1}}{\isadigit{4}}\ {\isadigit{0}}x{\isadigit{0}}{\isacharbrackright}\isanewline
{\isasymrparr}{\isachardoublequoteclose}\isanewline
\isanewline
\isacommand{definition}\isamarkupfalse%
\ {\isachardoublequoteopen}node{\isacharunderscore}{\isadigit{1}}{\isadigit{4}}{\isacharunderscore}lut{\isadigit{0}}\ {\isacharequal}\ empty{\isacharunderscore}spec\ {\isasymlparr}\isanewline
\ \ acc{\isacharunderscore}blocks\ {\isacharcolon}{\isacharequal}\ {\isacharbrackleft}{\isacharbrackright}{\isacharcomma}\isanewline
\ \ \ map{\isacharunderscore}blocks\ {\isacharcolon}{\isacharequal}\ {\isacharbrackleft}block{\isacharunderscore}map\ phi{\isacharunderscore}lut{\isacharunderscore}e{\isadigit{0}}{\isadigit{0}}\ {\isadigit{1}}{\isadigit{7}}\ {\isadigit{0}}x{\isadigit{0}}{\isacharbrackright}\isanewline
{\isasymrparr}{\isachardoublequoteclose}%
\begin{isamarkuptext}%
Xeon Phi 1%
\end{isamarkuptext}\isamarkuptrue%
\isacommand{definition}\isamarkupfalse%
\ {\isachardoublequoteopen}node{\isacharunderscore}{\isadigit{1}}{\isadigit{5}}{\isacharunderscore}phi\ {\isacharequal}\ empty{\isacharunderscore}spec\ {\isasymlparr}\isanewline
\ \ acc{\isacharunderscore}blocks\ {\isacharcolon}{\isacharequal}\ {\isacharbrackleft}phi{\isacharunderscore}gddr{\isacharcomma}\ phi{\isacharunderscore}sbox{\isacharbrackright}{\isacharcomma}\isanewline
\ \ map{\isacharunderscore}blocks\ {\isacharcolon}{\isacharequal}\ {\isacharbrackleft}block{\isacharunderscore}map\ phi{\isacharunderscore}sysmem\ {\isadigit{1}}{\isadigit{6}}\ {\isadigit{0}}x{\isadigit{0}}{\isacharbrackright}\isanewline
{\isasymrparr}{\isachardoublequoteclose}\isanewline
\isanewline
\isacommand{definition}\isamarkupfalse%
\ {\isachardoublequoteopen}node{\isacharunderscore}{\isadigit{1}}{\isadigit{6}}{\isacharunderscore}lut{\isadigit{1}}\ {\isacharequal}\ empty{\isacharunderscore}spec\ {\isasymlparr}\isanewline
\ \ acc{\isacharunderscore}blocks\ {\isacharcolon}{\isacharequal}\ {\isacharbrackleft}{\isacharbrackright}{\isacharcomma}\isanewline
\ \ map{\isacharunderscore}blocks\ {\isacharcolon}{\isacharequal}\ {\isacharbrackleft}block{\isacharunderscore}map\ phi{\isacharunderscore}lut{\isacharunderscore}e{\isadigit{0}}{\isadigit{0}}\ {\isadigit{1}}{\isadigit{8}}\ {\isadigit{0}}x{\isadigit{0}}{\isacharbrackright}\isanewline
{\isasymrparr}{\isachardoublequoteclose}%
\begin{isamarkuptext}%
IO-MMU%
\end{isamarkuptext}\isamarkuptrue%
\isacommand{definition}\isamarkupfalse%
\ {\isachardoublequoteopen}iommu{\isacharunderscore}map\ {\isacharequal}\ {\isacharparenleft}{\isadigit{0}}x{\isadigit{0}}{\isadigit{0}}{\isadigit{0}}{\isadigit{0}}{\isadigit{0}}{\isadigit{0}}{\isadigit{0}}{\isadigit{0}}{\isadigit{0}}{\isadigit{0}}{\isacharcomma}\ {\isadigit{0}}x{\isadigit{0}}{\isadigit{3}}FFFFFFFF{\isacharparenright}{\isachardoublequoteclose}\ \ \isanewline
\isanewline
\isacommand{definition}\isamarkupfalse%
\ {\isachardoublequoteopen}node{\isacharunderscore}{\isadigit{1}}{\isadigit{7}}{\isacharunderscore}iommu\ {\isacharequal}\ empty{\isacharunderscore}spec\ {\isasymlparr}\isanewline
\ \ acc{\isacharunderscore}blocks\ {\isacharcolon}{\isacharequal}\ {\isacharbrackleft}{\isacharbrackright}{\isacharcomma}\isanewline
\ \ map{\isacharunderscore}blocks\ {\isacharcolon}{\isacharequal}\ {\isacharbrackleft}direct{\isacharunderscore}map\ iommu{\isacharunderscore}map\ {\isadigit{4}}{\isacharbrackright}\isanewline
{\isasymrparr}{\isachardoublequoteclose}\isanewline
\isanewline
\isacommand{definition}\isamarkupfalse%
\ {\isachardoublequoteopen}node{\isacharunderscore}{\isadigit{1}}{\isadigit{8}}{\isacharunderscore}iommu\ {\isacharequal}\ empty{\isacharunderscore}spec\ {\isasymlparr}\isanewline
\ \ acc{\isacharunderscore}blocks\ {\isacharcolon}{\isacharequal}\ {\isacharbrackleft}{\isacharbrackright}{\isacharcomma}\isanewline
\ \ map{\isacharunderscore}blocks\ {\isacharcolon}{\isacharequal}\ {\isacharbrackleft}direct{\isacharunderscore}map\ iommu{\isacharunderscore}map\ {\isadigit{5}}{\isacharbrackright}\isanewline
{\isasymrparr}{\isachardoublequoteclose}%
\begin{isamarkuptext}%
CPU Cores%
\end{isamarkuptext}\isamarkuptrue%
\isacommand{definition}\isamarkupfalse%
\ {\isachardoublequoteopen}lapic\ {\isacharequal}\ {\isacharparenleft}{\isadigit{0}}xFEE{\isadigit{0}}{\isadigit{0}}{\isadigit{0}}{\isadigit{0}}{\isadigit{0}}{\isacharcomma}\ {\isadigit{0}}xFEE{\isadigit{0}}FFFF{\isacharparenright}{\isachardoublequoteclose}\isanewline
\isacommand{definition}\isamarkupfalse%
\ {\isachardoublequoteopen}cpu{\isacharunderscore}phys{\isadigit{0}}\ {\isacharequal}\ empty{\isacharunderscore}spec\ {\isasymlparr}\isanewline
\ \ acc{\isacharunderscore}blocks\ {\isacharcolon}{\isacharequal}\ {\isacharbrackleft}lapic{\isacharbrackright}{\isacharcomma}\isanewline
\ \ overlay\ {\isacharcolon}{\isacharequal}\ Some\ {\isadigit{0}}\isanewline
{\isasymrparr}{\isachardoublequoteclose}\isanewline
\isanewline
\isacommand{definition}\isamarkupfalse%
\ {\isachardoublequoteopen}cpu{\isacharunderscore}phys{\isadigit{1}}\ {\isacharequal}\ empty{\isacharunderscore}spec\ {\isasymlparr}\isanewline
\ \ acc{\isacharunderscore}blocks\ {\isacharcolon}{\isacharequal}\ {\isacharbrackleft}lapic{\isacharbrackright}{\isacharcomma}\isanewline
\ \ overlay\ {\isacharcolon}{\isacharequal}\ Some\ {\isadigit{1}}\isanewline
{\isasymrparr}{\isachardoublequoteclose}\isanewline
\isanewline
\isacommand{definition}\isamarkupfalse%
\ {\isachardoublequoteopen}cpu{\isacharunderscore}phi{\isadigit{0}}\ {\isacharequal}\ empty{\isacharunderscore}spec\ {\isasymlparr}\isanewline
\ \ acc{\isacharunderscore}blocks\ {\isacharcolon}{\isacharequal}\ {\isacharbrackleft}lapic{\isacharbrackright}{\isacharcomma}\isanewline
\ \ overlay\ {\isacharcolon}{\isacharequal}\ Some\ {\isadigit{1}}{\isadigit{3}}\isanewline
{\isasymrparr}{\isachardoublequoteclose}\isanewline
\isanewline
\isacommand{definition}\isamarkupfalse%
\ {\isachardoublequoteopen}cpu{\isacharunderscore}phi{\isadigit{1}}\ {\isacharequal}\ empty{\isacharunderscore}spec\ {\isasymlparr}\isanewline
\ \ acc{\isacharunderscore}blocks\ {\isacharcolon}{\isacharequal}\ {\isacharbrackleft}lapic{\isacharbrackright}{\isacharcomma}\isanewline
\ \ overlay\ {\isacharcolon}{\isacharequal}\ Some\ {\isadigit{1}}{\isadigit{5}}\isanewline
{\isasymrparr}{\isachardoublequoteclose}\isanewline
\isanewline
\isacommand{definition}\isamarkupfalse%
\ {\isachardoublequoteopen}sys\ {\isacharequal}\ {\isacharbrackleft}{\isacharparenleft}{\isadigit{0}}{\isacharcomma}\ node{\isacharunderscore}{\isadigit{0}}{\isacharunderscore}interconnect{\isacharparenright}{\isacharcomma}\ \isanewline
\ \ \ \ \ \ \ \ \ \ \ \ \ \ \ \ \ \ \ {\isacharparenleft}{\isadigit{1}}{\isacharcomma}\ node{\isacharunderscore}{\isadigit{1}}{\isacharunderscore}interconnect{\isacharparenright}{\isacharcomma}\ \isanewline
\ \ \ \ \ \ \ \ \ \ \ \ \ \ \ \ \ \ \ {\isacharparenleft}{\isadigit{2}}{\isacharcomma}\ node{\isacharunderscore}{\isadigit{2}}{\isacharunderscore}dram{\isacharparenright}{\isacharcomma}\ \isanewline
\ \ \ \ \ \ \ \ \ \ \ \ \ \ \ \ \ \ \ {\isacharparenleft}{\isadigit{3}}{\isacharcomma}\ node{\isacharunderscore}{\isadigit{3}}{\isacharunderscore}dram{\isacharparenright}{\isacharcomma}\ \isanewline
\ \ \ \ \ \ \ \ \ \ \ \ \ \ \ \ \ \ \ {\isacharparenleft}{\isadigit{4}}{\isacharcomma}\ node{\isacharunderscore}{\isadigit{4}}{\isacharunderscore}pci{\isacharparenright}{\isacharcomma}\ \isanewline
\ \ \ \ \ \ \ \ \ \ \ \ \ \ \ \ \ \ \ {\isacharparenleft}{\isadigit{5}}{\isacharcomma}\ node{\isacharunderscore}{\isadigit{5}}{\isacharunderscore}pci{\isacharparenright}{\isacharcomma}\ \isanewline
\ \ \ \ \ \ \ \ \ \ \ \ \ \ \ \ \ \ \ {\isacharparenleft}{\isadigit{6}}{\isacharcomma}\ node{\isacharunderscore}{\isadigit{6}}{\isacharunderscore}e{\isadigit{1}}{\isadigit{0}}{\isadigit{0}}{\isadigit{0}}{\isacharparenright}{\isacharcomma}\ \isanewline
\ \ \ \ \ \ \ \ \ \ \ \ \ \ \ \ \ \ \ {\isacharparenleft}{\isadigit{7}}{\isacharcomma}\ node{\isacharunderscore}{\isadigit{7}}{\isacharunderscore}ioapic{\isacharparenright}{\isacharcomma}\ \isanewline
\ \ \ \ \ \ \ \ \ \ \ \ \ \ \ \ \ \ \ {\isacharparenleft}{\isadigit{8}}{\isacharcomma}\ node{\isacharunderscore}{\isadigit{8}}{\isacharunderscore}ioapic{\isacharparenright}{\isacharcomma}\ \isanewline
\ \ \ \ \ \ \ \ \ \ \ \ \ \ \ \ \ \ \ {\isacharparenleft}{\isadigit{9}}{\isacharcomma}\ node{\isacharunderscore}{\isadigit{9}}{\isacharunderscore}dma{\isacharparenright}{\isacharcomma}\ \isanewline
\ \ \ \ \ \ \ \ \ \ \ \ \ \ \ \ \ \ \ {\isacharparenleft}{\isadigit{1}}{\isadigit{0}}{\isacharcomma}\ node{\isacharunderscore}{\isadigit{1}}{\isadigit{0}}{\isacharunderscore}dma{\isacharparenright}{\isacharcomma}\isanewline
\ \ \ \ \ \ \ \ \ \ \ \ \ \ \ \ \ \ \ {\isacharparenleft}{\isadigit{1}}{\isadigit{1}}{\isacharcomma}\ node{\isacharunderscore}{\isadigit{1}}{\isadigit{1}}{\isacharunderscore}ehci{\isacharparenright}{\isacharcomma}\ \isanewline
\ \ \ \ \ \ \ \ \ \ \ \ \ \ \ \ \ \ \ {\isacharparenleft}{\isadigit{1}}{\isadigit{2}}{\isacharcomma}\ node{\isacharunderscore}{\isadigit{1}}{\isadigit{2}}{\isacharunderscore}ahci{\isacharparenright}{\isacharcomma}\ \isanewline
\ \ \ \ \ \ \ \ \ \ \ \ \ \ \ \ \ \ \ {\isacharparenleft}{\isadigit{1}}{\isadigit{3}}{\isacharcomma}\ node{\isacharunderscore}{\isadigit{1}}{\isadigit{3}}{\isacharunderscore}phi{\isacharparenright}{\isacharcomma}\isanewline
\ \ \ \ \ \ \ \ \ \ \ \ \ \ \ \ \ \ \ {\isacharparenleft}{\isadigit{1}}{\isadigit{4}}{\isacharcomma}\ node{\isacharunderscore}{\isadigit{1}}{\isadigit{4}}{\isacharunderscore}lut{\isadigit{0}}{\isacharparenright}{\isacharcomma}\ \isanewline
\ \ \ \ \ \ \ \ \ \ \ \ \ \ \ \ \ \ \ {\isacharparenleft}{\isadigit{1}}{\isadigit{5}}{\isacharcomma}\ node{\isacharunderscore}{\isadigit{1}}{\isadigit{5}}{\isacharunderscore}phi{\isacharparenright}{\isacharcomma}\ \isanewline
\ \ \ \ \ \ \ \ \ \ \ \ \ \ \ \ \ \ \ {\isacharparenleft}{\isadigit{1}}{\isadigit{6}}{\isacharcomma}\ node{\isacharunderscore}{\isadigit{1}}{\isadigit{6}}{\isacharunderscore}lut{\isadigit{1}}{\isacharparenright}{\isacharcomma}\isanewline
\ \ \ \ \ \ \ \ \ \ \ \ \ \ \ \ \ \ \ {\isacharparenleft}{\isadigit{1}}{\isadigit{7}}{\isacharcomma}\ node{\isacharunderscore}{\isadigit{1}}{\isadigit{7}}{\isacharunderscore}iommu{\isacharparenright}{\isacharcomma}\ \isanewline
\ \ \ \ \ \ \ \ \ \ \ \ \ \ \ \ \ \ \ {\isacharparenleft}{\isadigit{1}}{\isadigit{8}}{\isacharcomma}\ node{\isacharunderscore}{\isadigit{1}}{\isadigit{8}}{\isacharunderscore}iommu{\isacharparenright}{\isacharbrackright}\ {\isacharat}\isanewline
\ \ \ \ \ \ \ \ \ \ \ \ \ \ \ \ \ \ \ repeat{\isacharunderscore}node\ cpu{\isacharunderscore}phys{\isadigit{0}}\ {\isadigit{2}}{\isadigit{0}}\ {\isadigit{1}}{\isadigit{0}}\ {\isacharat}\isanewline
\ \ \ \ \ \ \ \ \ \ \ \ \ \ \ \ \ \ \ repeat{\isacharunderscore}node\ cpu{\isacharunderscore}phys{\isadigit{1}}\ {\isadigit{3}}{\isadigit{0}}\ {\isadigit{1}}{\isadigit{0}}\ {\isacharat}\isanewline
\ \ \ \ \ \ \ \ \ \ \ \ \ \ \ \ \ \ \ repeat{\isacharunderscore}node\ cpu{\isacharunderscore}phi{\isadigit{0}}\ {\isadigit{4}}{\isadigit{0}}\ {\isadigit{6}}{\isadigit{0}}\ {\isacharat}\isanewline
\ \ \ \ \ \ \ \ \ \ \ \ \ \ \ \ \ \ \ repeat{\isacharunderscore}node\ cpu{\isacharunderscore}phi{\isadigit{1}}\ {\isadigit{1}}{\isadigit{0}}{\isadigit{0}}\ {\isadigit{6}}{\isadigit{0}}{\isachardoublequoteclose}\isanewline
\isadelimtheory
\endisadelimtheory
\isatagtheory
\isacommand{end}\isamarkupfalse%
\endisatagtheory
{\isafoldtheory}%
\isadelimtheory
\endisadelimtheory
\end{isabellebody}%

%
\begin{isabellebody}%
\setisabellecontext{ServerInt}%
\isadelimtheory
\endisadelimtheory
\isatagtheory
\endisatagtheory
{\isafoldtheory}%
\isadelimtheory
\endisadelimtheory
\isamarkupsubsubsection{Interrupts%
}
\isamarkuptrue%
\isacommand{definition}\isamarkupfalse%
\ {\isachardoublequoteopen}phi{\isadigit{0}}{\isacharunderscore}elapic{\isacharunderscore}rcv{\isadigit{0}}\ {\isacharequal}\ {\isadigit{0}}x{\isadigit{2}}{\isadigit{9}}{\isachardoublequoteclose}\ \ \isanewline
\isacommand{definition}\isamarkupfalse%
\ {\isachardoublequoteopen}phi{\isadigit{1}}{\isacharunderscore}elapic{\isacharunderscore}rcv{\isadigit{0}}\ {\isacharequal}\ {\isadigit{0}}x{\isadigit{2}}{\isadigit{9}}{\isachardoublequoteclose}\ \ \isanewline
\isacommand{definition}\isamarkupfalse%
\ {\isachardoublequoteopen}phi{\isadigit{0}}{\isacharunderscore}msi{\isacharunderscore}write{\isadigit{0}}\ {\isacharequal}\ {\isadigit{0}}x{\isadigit{0}}{\isadigit{0}}{\isadigit{0}}{\isadigit{0}}{\isadigit{0}}{\isadigit{0}}{\isadigit{0}}{\isadigit{0}}FEE{\isadigit{0}}{\isadigit{0}}{\isadigit{2}}B{\isadigit{8}}{\isadigit{0}}{\isadigit{0}}{\isadigit{0}}{\isadigit{0}}{\isadigit{0}}{\isadigit{0}}{\isadigit{2}}{\isadigit{9}}{\isachardoublequoteclose}\isanewline
\isacommand{definition}\isamarkupfalse%
\ {\isachardoublequoteopen}phi{\isadigit{1}}{\isacharunderscore}msi{\isacharunderscore}write{\isadigit{0}}\ {\isacharequal}\ {\isadigit{0}}x{\isadigit{0}}{\isadigit{0}}{\isadigit{0}}{\isadigit{0}}{\isadigit{0}}{\isadigit{0}}{\isadigit{0}}{\isadigit{0}}FEE{\isadigit{0}}{\isadigit{0}}{\isadigit{2}}B{\isadigit{8}}{\isadigit{0}}{\isadigit{0}}{\isadigit{0}}{\isadigit{0}}{\isadigit{0}}{\isadigit{0}}{\isadigit{7}}D{\isachardoublequoteclose}\isanewline
\isanewline
\isacommand{definition}\isamarkupfalse%
\ {\isachardoublequoteopen}x{\isadigit{8}}{\isadigit{6}}{\isacharunderscore}vec{\isacharunderscore}domain\ {\isacharequal}\ {\isacharparenleft}{\isadigit{3}}{\isadigit{2}}{\isacharcomma}{\isadigit{2}}{\isadigit{5}}{\isadigit{5}}{\isacharparenright}{\isachardoublequoteclose}%
\begin{isamarkuptext}%
Host LAPIC 0%
\end{isamarkuptext}\isamarkuptrue%
\isacommand{definition}\isamarkupfalse%
\ {\isachardoublequoteopen}node{\isacharunderscore}{\isadigit{0}}{\isacharunderscore}lapic{\isadigit{0}}\ {\isacharequal}\ empty{\isacharunderscore}spec\ {\isasymlparr}\isanewline
\ \ acc{\isacharunderscore}blocks\ {\isacharcolon}{\isacharequal}\ {\isacharbrackleft}x{\isadigit{8}}{\isadigit{6}}{\isacharunderscore}vec{\isacharunderscore}domain{\isacharbrackright}{\isacharcomma}\isanewline
\ \ map{\isacharunderscore}blocks\ {\isacharcolon}{\isacharequal}\ {\isacharbrackleft}{\isacharbrackright}\isanewline
{\isasymrparr}{\isachardoublequoteclose}%
\begin{isamarkuptext}%
Host LAPIC 1%
\end{isamarkuptext}\isamarkuptrue%
\isacommand{definition}\isamarkupfalse%
\ {\isachardoublequoteopen}node{\isacharunderscore}{\isadigit{1}}{\isacharunderscore}lapic{\isadigit{1}}\ {\isacharequal}\ empty{\isacharunderscore}spec\ {\isasymlparr}\isanewline
\ \ acc{\isacharunderscore}blocks\ {\isacharcolon}{\isacharequal}\ {\isacharbrackleft}x{\isadigit{8}}{\isadigit{6}}{\isacharunderscore}vec{\isacharunderscore}domain{\isacharbrackright}{\isacharcomma}\isanewline
\ \ map{\isacharunderscore}blocks\ {\isacharcolon}{\isacharequal}\ {\isacharbrackleft}{\isacharbrackright}\isanewline
{\isasymrparr}{\isachardoublequoteclose}%
\begin{isamarkuptext}%
Core 0 to APICs%
\end{isamarkuptext}\isamarkuptrue%
\isacommand{definition}\isamarkupfalse%
\ {\isachardoublequoteopen}node{\isacharunderscore}{\isadigit{2}}{\isacharunderscore}core{\isadigit{0}}\ {\isacharequal}\ empty{\isacharunderscore}spec\ {\isasymlparr}\isanewline
\ \ acc{\isacharunderscore}blocks\ {\isacharcolon}{\isacharequal}\ {\isacharbrackleft}{\isacharbrackright}{\isacharcomma}\isanewline
\ \ map{\isacharunderscore}blocks\ {\isacharcolon}{\isacharequal}\ {\isacharbrackleft}one{\isacharunderscore}map\ {\isadigit{0}}\ {\isadigit{3}}\ {\isadigit{2}}{\isadigit{5}}{\isadigit{1}}{\isacharbrackright}\isanewline
{\isasymrparr}{\isachardoublequoteclose}%
\begin{isamarkuptext}%
Core 1 to APICs%
\end{isamarkuptext}\isamarkuptrue%
\isacommand{definition}\isamarkupfalse%
\ {\isachardoublequoteopen}node{\isacharunderscore}{\isadigit{3}}{\isacharunderscore}core{\isadigit{1}}\ {\isacharequal}\ empty{\isacharunderscore}spec\ {\isasymlparr}\isanewline
\ \ acc{\isacharunderscore}blocks\ {\isacharcolon}{\isacharequal}\ {\isacharbrackleft}{\isacharbrackright}{\isacharcomma}\isanewline
\ \ map{\isacharunderscore}blocks\ {\isacharcolon}{\isacharequal}\ {\isacharbrackleft}one{\isacharunderscore}map\ {\isadigit{0}}\ {\isadigit{3}}\ {\isadigit{2}}{\isadigit{5}}{\isadigit{1}}{\isacharbrackright}\isanewline
{\isasymrparr}{\isachardoublequoteclose}%
\begin{isamarkuptext}%
Timer0 to LAPIC0%
\end{isamarkuptext}\isamarkuptrue%
\isacommand{definition}\isamarkupfalse%
\ {\isachardoublequoteopen}node{\isacharunderscore}{\isadigit{4}}{\isacharunderscore}timer{\isadigit{0}}\ {\isacharequal}\ empty{\isacharunderscore}spec\ {\isasymlparr}\isanewline
\ \ acc{\isacharunderscore}blocks\ {\isacharcolon}{\isacharequal}\ {\isacharbrackleft}{\isacharbrackright}{\isacharcomma}\isanewline
\ \ map{\isacharunderscore}blocks\ {\isacharcolon}{\isacharequal}\ {\isacharbrackleft}one{\isacharunderscore}map\ {\isadigit{0}}\ {\isadigit{0}}\ {\isadigit{3}}{\isadigit{2}}{\isacharbrackright}\isanewline
{\isasymrparr}{\isachardoublequoteclose}%
\begin{isamarkuptext}%
Timer1 to LAPIC1%
\end{isamarkuptext}\isamarkuptrue%
\isacommand{definition}\isamarkupfalse%
\ {\isachardoublequoteopen}node{\isacharunderscore}{\isadigit{5}}{\isacharunderscore}timer{\isadigit{1}}\ {\isacharequal}\ empty{\isacharunderscore}spec\ {\isasymlparr}\isanewline
\ \ acc{\isacharunderscore}blocks\ {\isacharcolon}{\isacharequal}\ {\isacharbrackleft}{\isacharbrackright}{\isacharcomma}\isanewline
\ \ map{\isacharunderscore}blocks\ {\isacharcolon}{\isacharequal}\ {\isacharbrackleft}one{\isacharunderscore}map\ {\isadigit{0}}\ {\isadigit{1}}\ {\isadigit{3}}{\isadigit{2}}{\isacharbrackright}\isanewline
{\isasymrparr}{\isachardoublequoteclose}%
\begin{isamarkuptext}%
IOMMU to LAPICs.%
\end{isamarkuptext}\isamarkuptrue%
\isacommand{definition}\isamarkupfalse%
\ {\isachardoublequoteopen}node{\isacharunderscore}{\isadigit{6}}{\isacharunderscore}iommu\ {\isacharequal}\ empty{\isacharunderscore}spec\ {\isasymlparr}\isanewline
\ \ acc{\isacharunderscore}blocks\ {\isacharcolon}{\isacharequal}\ {\isacharbrackleft}{\isacharbrackright}{\isacharcomma}\isanewline
\ \ map{\isacharunderscore}blocks\ {\isacharcolon}{\isacharequal}\ {\isacharbrackleft}\isanewline
\ \ \ \ one{\isacharunderscore}map\ phi{\isadigit{0}}{\isacharunderscore}msi{\isacharunderscore}write{\isadigit{0}}\ {\isadigit{0}}\ {\isadigit{1}}{\isadigit{2}}{\isadigit{5}}{\isacharcomma}\isanewline
\ \ \ \ one{\isacharunderscore}map\ phi{\isadigit{1}}{\isacharunderscore}msi{\isacharunderscore}write{\isadigit{0}}\ {\isadigit{0}}\ {\isadigit{1}}{\isadigit{2}}{\isadigit{6}}\isanewline
{\isacharbrackright}\isanewline
{\isasymrparr}{\isachardoublequoteclose}%
\begin{isamarkuptext}%
Phi0 LAPIC0%
\end{isamarkuptext}\isamarkuptrue%
\isacommand{definition}\isamarkupfalse%
\ {\isachardoublequoteopen}node{\isacharunderscore}{\isadigit{7}}{\isacharunderscore}phi{\isadigit{0}}{\isacharunderscore}lapic{\isadigit{0}}\ {\isacharequal}\ empty{\isacharunderscore}spec\ {\isasymlparr}\isanewline
\ \ acc{\isacharunderscore}blocks\ {\isacharcolon}{\isacharequal}\ {\isacharbrackleft}x{\isadigit{8}}{\isadigit{6}}{\isacharunderscore}vec{\isacharunderscore}domain{\isacharbrackright}{\isacharcomma}\isanewline
\ \ map{\isacharunderscore}blocks\ {\isacharcolon}{\isacharequal}\ {\isacharbrackleft}{\isacharbrackright}\isanewline
{\isasymrparr}{\isachardoublequoteclose}%
\begin{isamarkuptext}%
Phi0 LAPIC1%
\end{isamarkuptext}\isamarkuptrue%
\isacommand{definition}\isamarkupfalse%
\ {\isachardoublequoteopen}node{\isacharunderscore}{\isadigit{8}}{\isacharunderscore}phi{\isadigit{0}}{\isacharunderscore}lapic{\isadigit{1}}\ {\isacharequal}\ empty{\isacharunderscore}spec\ {\isasymlparr}\isanewline
\ \ acc{\isacharunderscore}blocks\ {\isacharcolon}{\isacharequal}\ {\isacharbrackleft}x{\isadigit{8}}{\isadigit{6}}{\isacharunderscore}vec{\isacharunderscore}domain{\isacharbrackright}{\isacharcomma}\isanewline
\ \ map{\isacharunderscore}blocks\ {\isacharcolon}{\isacharequal}\ {\isacharbrackleft}{\isacharbrackright}\isanewline
{\isasymrparr}{\isachardoublequoteclose}%
\begin{isamarkuptext}%
Phi0 Core 0 to APICs%
\end{isamarkuptext}\isamarkuptrue%
\isacommand{definition}\isamarkupfalse%
\ {\isachardoublequoteopen}node{\isacharunderscore}{\isadigit{9}}{\isacharunderscore}phi{\isadigit{0}}{\isacharunderscore}core{\isadigit{0}}\ {\isacharequal}\ empty{\isacharunderscore}spec\ {\isasymlparr}\isanewline
\ \ acc{\isacharunderscore}blocks\ {\isacharcolon}{\isacharequal}\ {\isacharbrackleft}{\isacharbrackright}{\isacharcomma}\isanewline
\ \ map{\isacharunderscore}blocks\ {\isacharcolon}{\isacharequal}\ {\isacharbrackleft}one{\isacharunderscore}map\ {\isadigit{0}}\ {\isadigit{8}}\ {\isadigit{2}}{\isadigit{5}}{\isadigit{1}}{\isacharbrackright}\isanewline
{\isasymrparr}{\isachardoublequoteclose}%
\begin{isamarkuptext}%
Phi0 Core 1 to APICs%
\end{isamarkuptext}\isamarkuptrue%
\isacommand{definition}\isamarkupfalse%
\ {\isachardoublequoteopen}node{\isacharunderscore}{\isadigit{1}}{\isadigit{0}}{\isacharunderscore}phi{\isadigit{0}}{\isacharunderscore}core{\isadigit{1}}\ {\isacharequal}\ empty{\isacharunderscore}spec\ {\isasymlparr}\isanewline
\ \ acc{\isacharunderscore}blocks\ {\isacharcolon}{\isacharequal}\ {\isacharbrackleft}{\isacharbrackright}{\isacharcomma}\isanewline
\ \ map{\isacharunderscore}blocks\ {\isacharcolon}{\isacharequal}\ {\isacharbrackleft}one{\isacharunderscore}map\ {\isadigit{0}}\ {\isadigit{7}}\ {\isadigit{2}}{\isadigit{5}}{\isadigit{1}}{\isacharbrackright}\isanewline
{\isasymrparr}{\isachardoublequoteclose}%
\begin{isamarkuptext}%
Phi0 Timer0 to Phi0 LAPIC0%
\end{isamarkuptext}\isamarkuptrue%
\isacommand{definition}\isamarkupfalse%
\ {\isachardoublequoteopen}node{\isacharunderscore}{\isadigit{1}}{\isadigit{1}}{\isacharunderscore}phi{\isadigit{0}}{\isacharunderscore}timer{\isadigit{0}}\ {\isacharequal}\ empty{\isacharunderscore}spec\ {\isasymlparr}\isanewline
\ \ acc{\isacharunderscore}blocks\ {\isacharcolon}{\isacharequal}\ {\isacharbrackleft}{\isacharbrackright}{\isacharcomma}\isanewline
\ \ map{\isacharunderscore}blocks\ {\isacharcolon}{\isacharequal}\ {\isacharbrackleft}one{\isacharunderscore}map\ {\isadigit{0}}\ {\isadigit{7}}\ {\isadigit{3}}{\isadigit{2}}{\isacharbrackright}\isanewline
{\isasymrparr}{\isachardoublequoteclose}%
\begin{isamarkuptext}%
Phi0 Timer1 to Phi0 LAPIC1%
\end{isamarkuptext}\isamarkuptrue%
\isacommand{definition}\isamarkupfalse%
\ {\isachardoublequoteopen}node{\isacharunderscore}{\isadigit{1}}{\isadigit{2}}{\isacharunderscore}phi{\isadigit{0}}{\isacharunderscore}timer{\isadigit{1}}\ {\isacharequal}\ empty{\isacharunderscore}spec\ {\isasymlparr}\isanewline
\ \ acc{\isacharunderscore}blocks\ {\isacharcolon}{\isacharequal}\ {\isacharbrackleft}{\isacharbrackright}{\isacharcomma}\isanewline
\ \ map{\isacharunderscore}blocks\ {\isacharcolon}{\isacharequal}\ {\isacharbrackleft}one{\isacharunderscore}map\ {\isadigit{0}}\ {\isadigit{8}}\ {\isadigit{3}}{\isadigit{2}}{\isacharbrackright}\isanewline
{\isasymrparr}{\isachardoublequoteclose}%
\begin{isamarkuptext}%
Phi0 IOAPIC0 -> LAPICs%
\end{isamarkuptext}\isamarkuptrue%
\isacommand{definition}\isamarkupfalse%
\ {\isachardoublequoteopen}node{\isacharunderscore}{\isadigit{1}}{\isadigit{3}}{\isacharunderscore}phi{\isadigit{0}}{\isacharunderscore}ioapic\ {\isacharequal}\ empty{\isacharunderscore}spec\ {\isasymlparr}\isanewline
\ \ acc{\isacharunderscore}blocks\ {\isacharcolon}{\isacharequal}\ {\isacharbrackleft}{\isacharbrackright}{\isacharcomma}\isanewline
\ \ map{\isacharunderscore}blocks\ {\isacharcolon}{\isacharequal}\ {\isacharbrackleft}one{\isacharunderscore}map\ {\isadigit{0}}\ {\isadigit{7}}\ {\isadigit{3}}{\isadigit{3}}{\isacharbrackright}\isanewline
{\isasymrparr}{\isachardoublequoteclose}%
\begin{isamarkuptext}%
Phi0 Thermal to IOAPIC%
\end{isamarkuptext}\isamarkuptrue%
\isacommand{definition}\isamarkupfalse%
\ {\isachardoublequoteopen}node{\isacharunderscore}{\isadigit{1}}{\isadigit{4}}{\isacharunderscore}phi{\isadigit{0}}{\isacharunderscore}rtc\ {\isacharequal}\ empty{\isacharunderscore}spec\ {\isasymlparr}\isanewline
\ \ acc{\isacharunderscore}blocks\ {\isacharcolon}{\isacharequal}\ {\isacharbrackleft}{\isacharbrackright}{\isacharcomma}\isanewline
\ \ map{\isacharunderscore}blocks\ {\isacharcolon}{\isacharequal}\ {\isacharbrackleft}one{\isacharunderscore}map\ {\isadigit{0}}\ {\isadigit{1}}{\isadigit{3}}\ {\isadigit{0}}{\isacharbrackright}\isanewline
{\isasymrparr}{\isachardoublequoteclose}%
\begin{isamarkuptext}%
Phi0 ELAPIC to LAPICs%
\end{isamarkuptext}\isamarkuptrue%
\isacommand{definition}\isamarkupfalse%
\ {\isachardoublequoteopen}node{\isacharunderscore}{\isadigit{1}}{\isadigit{5}}{\isacharunderscore}elapic\ {\isacharequal}\ empty{\isacharunderscore}spec\ {\isasymlparr}\isanewline
\ \ acc{\isacharunderscore}blocks\ {\isacharcolon}{\isacharequal}\ {\isacharbrackleft}{\isacharbrackright}{\isacharcomma}\isanewline
\ \ map{\isacharunderscore}blocks\ {\isacharcolon}{\isacharequal}\ {\isacharbrackleft}one{\isacharunderscore}map\ phi{\isadigit{0}}{\isacharunderscore}elapic{\isacharunderscore}rcv{\isadigit{0}}\ {\isadigit{7}}\ {\isadigit{0}}x{\isadigit{2}}{\isadigit{9}}{\isacharbrackright}\isanewline
{\isasymrparr}{\isachardoublequoteclose}%
\begin{isamarkuptext}%
Phi0 SBOX to IOMMU%
\end{isamarkuptext}\isamarkuptrue%
\isacommand{definition}\isamarkupfalse%
\ {\isachardoublequoteopen}node{\isacharunderscore}{\isadigit{1}}{\isadigit{6}}{\isacharunderscore}sbox\ {\isacharequal}\ empty{\isacharunderscore}spec\ {\isasymlparr}\isanewline
\ \ acc{\isacharunderscore}blocks\ {\isacharcolon}{\isacharequal}\ {\isacharbrackleft}{\isacharbrackright}{\isacharcomma}\isanewline
\ {\isacharparenleft}{\isacharasterisk}\ There\ should\ be\ a\ more\ expressive\ way\ of\ modeling\ the\ input\ {\isacharasterisk}{\isacharparenright}\isanewline
\ \ map{\isacharunderscore}blocks\ {\isacharcolon}{\isacharequal}\ {\isacharbrackleft}one{\isacharunderscore}map\ {\isadigit{0}}\ {\isadigit{5}}\ phi{\isadigit{0}}{\isacharunderscore}msi{\isacharunderscore}write{\isadigit{0}}{\isacharbrackright}\isanewline
{\isasymrparr}{\isachardoublequoteclose}%
\begin{isamarkuptext}%
Phi1 LAPIC0%
\end{isamarkuptext}\isamarkuptrue%
\isacommand{definition}\isamarkupfalse%
\ {\isachardoublequoteopen}node{\isacharunderscore}{\isadigit{1}}{\isadigit{7}}{\isacharunderscore}phi{\isadigit{1}}{\isacharunderscore}lapic{\isadigit{0}}\ {\isacharequal}\ empty{\isacharunderscore}spec\ {\isasymlparr}\isanewline
\ \ acc{\isacharunderscore}blocks\ {\isacharcolon}{\isacharequal}\ {\isacharbrackleft}x{\isadigit{8}}{\isadigit{6}}{\isacharunderscore}vec{\isacharunderscore}domain{\isacharbrackright}{\isacharcomma}\isanewline
\ \ map{\isacharunderscore}blocks\ {\isacharcolon}{\isacharequal}\ {\isacharbrackleft}{\isacharbrackright}\isanewline
{\isasymrparr}{\isachardoublequoteclose}%
\begin{isamarkuptext}%
Phi1 LAPIC1%
\end{isamarkuptext}\isamarkuptrue%
\isacommand{definition}\isamarkupfalse%
\ {\isachardoublequoteopen}node{\isacharunderscore}{\isadigit{1}}{\isadigit{8}}{\isacharunderscore}phi{\isadigit{1}}{\isacharunderscore}lapic{\isadigit{1}}\ {\isacharequal}\ empty{\isacharunderscore}spec\ {\isasymlparr}\isanewline
\ \ acc{\isacharunderscore}blocks\ {\isacharcolon}{\isacharequal}\ {\isacharbrackleft}x{\isadigit{8}}{\isadigit{6}}{\isacharunderscore}vec{\isacharunderscore}domain{\isacharbrackright}{\isacharcomma}\isanewline
\ \ map{\isacharunderscore}blocks\ {\isacharcolon}{\isacharequal}\ {\isacharbrackleft}{\isacharbrackright}\isanewline
{\isasymrparr}{\isachardoublequoteclose}%
\begin{isamarkuptext}%
Phi1 Core 0 to APICs%
\end{isamarkuptext}\isamarkuptrue%
\isacommand{definition}\isamarkupfalse%
\ {\isachardoublequoteopen}node{\isacharunderscore}{\isadigit{1}}{\isadigit{9}}{\isacharunderscore}phi{\isadigit{1}}{\isacharunderscore}core{\isadigit{0}}\ {\isacharequal}\ empty{\isacharunderscore}spec\ {\isasymlparr}\isanewline
\ \ acc{\isacharunderscore}blocks\ {\isacharcolon}{\isacharequal}\ {\isacharbrackleft}{\isacharbrackright}{\isacharcomma}\isanewline
\ \ map{\isacharunderscore}blocks\ {\isacharcolon}{\isacharequal}\ {\isacharbrackleft}one{\isacharunderscore}map\ {\isadigit{0}}\ {\isadigit{1}}{\isadigit{8}}\ {\isadigit{2}}{\isadigit{5}}{\isadigit{1}}{\isacharbrackright}\isanewline
{\isasymrparr}{\isachardoublequoteclose}%
\begin{isamarkuptext}%
Phi1 Core 1 to APICs%
\end{isamarkuptext}\isamarkuptrue%
\isacommand{definition}\isamarkupfalse%
\ {\isachardoublequoteopen}node{\isacharunderscore}{\isadigit{2}}{\isadigit{0}}{\isacharunderscore}phi{\isadigit{1}}{\isacharunderscore}core{\isadigit{1}}\ {\isacharequal}\ empty{\isacharunderscore}spec\ {\isasymlparr}\isanewline
\ \ acc{\isacharunderscore}blocks\ {\isacharcolon}{\isacharequal}\ {\isacharbrackleft}{\isacharbrackright}{\isacharcomma}\isanewline
\ \ map{\isacharunderscore}blocks\ {\isacharcolon}{\isacharequal}\ {\isacharbrackleft}one{\isacharunderscore}map\ {\isadigit{0}}\ {\isadigit{1}}{\isadigit{7}}\ {\isadigit{2}}{\isadigit{5}}{\isadigit{1}}{\isacharbrackright}\isanewline
{\isasymrparr}{\isachardoublequoteclose}%
\begin{isamarkuptext}%
Phi1 Timer0 to Phi1 LAPIC0%
\end{isamarkuptext}\isamarkuptrue%
\isacommand{definition}\isamarkupfalse%
\ {\isachardoublequoteopen}node{\isacharunderscore}{\isadigit{2}}{\isadigit{1}}{\isacharunderscore}phi{\isadigit{1}}{\isacharunderscore}timer{\isadigit{0}}\ {\isacharequal}\ empty{\isacharunderscore}spec\ {\isasymlparr}\isanewline
\ \ acc{\isacharunderscore}blocks\ {\isacharcolon}{\isacharequal}\ {\isacharbrackleft}{\isacharbrackright}{\isacharcomma}\isanewline
\ \ map{\isacharunderscore}blocks\ {\isacharcolon}{\isacharequal}\ {\isacharbrackleft}one{\isacharunderscore}map\ {\isadigit{0}}\ {\isadigit{1}}{\isadigit{7}}\ {\isadigit{3}}{\isadigit{2}}{\isacharbrackright}\isanewline
{\isasymrparr}{\isachardoublequoteclose}%
\begin{isamarkuptext}%
Phi1 Timer1 to Phi1 LAPIC1%
\end{isamarkuptext}\isamarkuptrue%
\isacommand{definition}\isamarkupfalse%
\ {\isachardoublequoteopen}node{\isacharunderscore}{\isadigit{2}}{\isadigit{2}}{\isacharunderscore}phi{\isadigit{1}}{\isacharunderscore}timer{\isadigit{1}}\ {\isacharequal}\ empty{\isacharunderscore}spec\ {\isasymlparr}\isanewline
\ \ acc{\isacharunderscore}blocks\ {\isacharcolon}{\isacharequal}\ {\isacharbrackleft}{\isacharbrackright}{\isacharcomma}\isanewline
\ \ map{\isacharunderscore}blocks\ {\isacharcolon}{\isacharequal}\ {\isacharbrackleft}one{\isacharunderscore}map\ {\isadigit{0}}\ {\isadigit{1}}{\isadigit{8}}\ {\isadigit{3}}{\isadigit{2}}{\isacharbrackright}\isanewline
{\isasymrparr}{\isachardoublequoteclose}%
\begin{isamarkuptext}%
Phi1 IOAPIC0 to LAPICs%
\end{isamarkuptext}\isamarkuptrue%
\isacommand{definition}\isamarkupfalse%
\ {\isachardoublequoteopen}node{\isacharunderscore}{\isadigit{2}}{\isadigit{3}}{\isacharunderscore}phi{\isadigit{1}}{\isacharunderscore}ioapic\ {\isacharequal}\ empty{\isacharunderscore}spec\ {\isasymlparr}\isanewline
\ \ acc{\isacharunderscore}blocks\ {\isacharcolon}{\isacharequal}\ {\isacharbrackleft}{\isacharbrackright}{\isacharcomma}\isanewline
\ \ map{\isacharunderscore}blocks\ {\isacharcolon}{\isacharequal}\ {\isacharbrackleft}one{\isacharunderscore}map\ {\isadigit{0}}\ {\isadigit{1}}{\isadigit{7}}\ {\isadigit{3}}{\isadigit{3}}{\isacharbrackright}\isanewline
{\isasymrparr}{\isachardoublequoteclose}%
\begin{isamarkuptext}%
Phi1 Thermal to IOAPIC%
\end{isamarkuptext}\isamarkuptrue%
\isacommand{definition}\isamarkupfalse%
\ {\isachardoublequoteopen}node{\isacharunderscore}{\isadigit{2}}{\isadigit{4}}{\isacharunderscore}phi{\isadigit{1}}{\isacharunderscore}rtc\ {\isacharequal}\ empty{\isacharunderscore}spec\ {\isasymlparr}\isanewline
\ \ acc{\isacharunderscore}blocks\ {\isacharcolon}{\isacharequal}\ {\isacharbrackleft}{\isacharbrackright}{\isacharcomma}\isanewline
\ \ map{\isacharunderscore}blocks\ {\isacharcolon}{\isacharequal}\ {\isacharbrackleft}one{\isacharunderscore}map\ {\isadigit{0}}\ {\isadigit{2}}{\isadigit{3}}\ {\isadigit{0}}{\isacharbrackright}\isanewline
{\isasymrparr}{\isachardoublequoteclose}%
\begin{isamarkuptext}%
Phi1 ELAPIC to LAPICs%
\end{isamarkuptext}\isamarkuptrue%
\isacommand{definition}\isamarkupfalse%
\ {\isachardoublequoteopen}node{\isacharunderscore}{\isadigit{2}}{\isadigit{5}}{\isacharunderscore}elapic\ {\isacharequal}\ empty{\isacharunderscore}spec\ {\isasymlparr}\isanewline
\ \ acc{\isacharunderscore}blocks\ {\isacharcolon}{\isacharequal}\ {\isacharbrackleft}{\isacharbrackright}{\isacharcomma}\isanewline
\ \ map{\isacharunderscore}blocks\ {\isacharcolon}{\isacharequal}\ {\isacharbrackleft}one{\isacharunderscore}map\ phi{\isadigit{1}}{\isacharunderscore}elapic{\isacharunderscore}rcv{\isadigit{0}}\ {\isadigit{1}}{\isadigit{7}}\ {\isadigit{0}}x{\isadigit{2}}{\isadigit{9}}{\isacharbrackright}\isanewline
{\isasymrparr}{\isachardoublequoteclose}%
\begin{isamarkuptext}%
Phi1 SBOX to IOMMU%
\end{isamarkuptext}\isamarkuptrue%
\isacommand{definition}\isamarkupfalse%
\ {\isachardoublequoteopen}node{\isacharunderscore}{\isadigit{2}}{\isadigit{6}}{\isacharunderscore}sbox\ {\isacharequal}\ empty{\isacharunderscore}spec\ {\isasymlparr}\isanewline
\ \ acc{\isacharunderscore}blocks\ {\isacharcolon}{\isacharequal}\ {\isacharbrackleft}{\isacharbrackright}{\isacharcomma}\isanewline
\ {\isacharparenleft}{\isacharasterisk}\ There\ should\ be\ a\ more\ expressive\ way\ of\ modeling\ the\ input\ {\isacharasterisk}{\isacharparenright}\isanewline
\ \ map{\isacharunderscore}blocks\ {\isacharcolon}{\isacharequal}\ {\isacharbrackleft}one{\isacharunderscore}map\ {\isadigit{0}}\ {\isadigit{5}}\ phi{\isadigit{1}}{\isacharunderscore}msi{\isacharunderscore}write{\isadigit{0}}{\isacharbrackright}\isanewline
{\isasymrparr}{\isachardoublequoteclose}\isanewline
\ \ \isanewline
\isacommand{definition}\isamarkupfalse%
\ {\isachardoublequoteopen}sys\ {\isacharequal}\ {\isacharbrackleft}\isanewline
\ \ {\isacharparenleft}{\isadigit{0}}{\isacharcomma}node{\isacharunderscore}{\isadigit{0}}{\isacharunderscore}lapic{\isadigit{0}}{\isacharparenright}{\isacharcomma}\isanewline
\ \ {\isacharparenleft}{\isadigit{1}}{\isacharcomma}node{\isacharunderscore}{\isadigit{1}}{\isacharunderscore}lapic{\isadigit{1}}{\isacharparenright}{\isacharcomma}\isanewline
\ \ {\isacharparenleft}{\isadigit{2}}{\isacharcomma}node{\isacharunderscore}{\isadigit{2}}{\isacharunderscore}core{\isadigit{0}}{\isacharparenright}{\isacharcomma}\isanewline
\ \ {\isacharparenleft}{\isadigit{3}}{\isacharcomma}node{\isacharunderscore}{\isadigit{3}}{\isacharunderscore}core{\isadigit{1}}{\isacharparenright}{\isacharcomma}\isanewline
\ \ {\isacharparenleft}{\isadigit{4}}{\isacharcomma}node{\isacharunderscore}{\isadigit{4}}{\isacharunderscore}timer{\isadigit{0}}{\isacharparenright}{\isacharcomma}\isanewline
\ \ {\isacharparenleft}{\isadigit{5}}{\isacharcomma}node{\isacharunderscore}{\isadigit{5}}{\isacharunderscore}timer{\isadigit{1}}{\isacharparenright}{\isacharcomma}\isanewline
\ \ {\isacharparenleft}{\isadigit{6}}{\isacharcomma}node{\isacharunderscore}{\isadigit{6}}{\isacharunderscore}iommu{\isacharparenright}{\isacharcomma}\isanewline
\ \ {\isacharparenleft}{\isadigit{7}}{\isacharcomma}node{\isacharunderscore}{\isadigit{7}}{\isacharunderscore}phi{\isadigit{0}}{\isacharunderscore}lapic{\isadigit{0}}{\isacharparenright}{\isacharcomma}\isanewline
\ \ {\isacharparenleft}{\isadigit{8}}{\isacharcomma}node{\isacharunderscore}{\isadigit{8}}{\isacharunderscore}phi{\isadigit{0}}{\isacharunderscore}lapic{\isadigit{1}}{\isacharparenright}{\isacharcomma}\isanewline
\ \ {\isacharparenleft}{\isadigit{9}}{\isacharcomma}node{\isacharunderscore}{\isadigit{9}}{\isacharunderscore}phi{\isadigit{0}}{\isacharunderscore}core{\isadigit{0}}{\isacharparenright}{\isacharcomma}\isanewline
\ \ {\isacharparenleft}{\isadigit{1}}{\isadigit{0}}{\isacharcomma}node{\isacharunderscore}{\isadigit{1}}{\isadigit{0}}{\isacharunderscore}phi{\isadigit{0}}{\isacharunderscore}core{\isadigit{1}}{\isacharparenright}{\isacharcomma}\isanewline
\ \ {\isacharparenleft}{\isadigit{1}}{\isadigit{1}}{\isacharcomma}node{\isacharunderscore}{\isadigit{1}}{\isadigit{1}}{\isacharunderscore}phi{\isadigit{0}}{\isacharunderscore}timer{\isadigit{0}}{\isacharparenright}{\isacharcomma}\isanewline
\ \ {\isacharparenleft}{\isadigit{1}}{\isadigit{2}}{\isacharcomma}node{\isacharunderscore}{\isadigit{1}}{\isadigit{2}}{\isacharunderscore}phi{\isadigit{0}}{\isacharunderscore}timer{\isadigit{1}}{\isacharparenright}{\isacharcomma}\isanewline
\ \ {\isacharparenleft}{\isadigit{1}}{\isadigit{3}}{\isacharcomma}node{\isacharunderscore}{\isadigit{1}}{\isadigit{3}}{\isacharunderscore}phi{\isadigit{0}}{\isacharunderscore}ioapic{\isacharparenright}{\isacharcomma}\isanewline
\ \ {\isacharparenleft}{\isadigit{1}}{\isadigit{4}}{\isacharcomma}node{\isacharunderscore}{\isadigit{1}}{\isadigit{4}}{\isacharunderscore}phi{\isadigit{0}}{\isacharunderscore}rtc{\isacharparenright}{\isacharcomma}\isanewline
\ \ {\isacharparenleft}{\isadigit{1}}{\isadigit{5}}{\isacharcomma}node{\isacharunderscore}{\isadigit{1}}{\isadigit{5}}{\isacharunderscore}elapic{\isacharparenright}{\isacharcomma}\isanewline
\ \ {\isacharparenleft}{\isadigit{1}}{\isadigit{6}}{\isacharcomma}node{\isacharunderscore}{\isadigit{1}}{\isadigit{6}}{\isacharunderscore}sbox{\isacharparenright}{\isacharcomma}\isanewline
\ \ {\isacharparenleft}{\isadigit{1}}{\isadigit{7}}{\isacharcomma}node{\isacharunderscore}{\isadigit{1}}{\isadigit{7}}{\isacharunderscore}phi{\isadigit{1}}{\isacharunderscore}lapic{\isadigit{0}}{\isacharparenright}{\isacharcomma}\isanewline
\ \ {\isacharparenleft}{\isadigit{1}}{\isadigit{8}}{\isacharcomma}node{\isacharunderscore}{\isadigit{1}}{\isadigit{8}}{\isacharunderscore}phi{\isadigit{1}}{\isacharunderscore}lapic{\isadigit{1}}{\isacharparenright}{\isacharcomma}\isanewline
\ \ {\isacharparenleft}{\isadigit{1}}{\isadigit{9}}{\isacharcomma}node{\isacharunderscore}{\isadigit{1}}{\isadigit{9}}{\isacharunderscore}phi{\isadigit{1}}{\isacharunderscore}core{\isadigit{0}}{\isacharparenright}{\isacharcomma}\isanewline
\ \ {\isacharparenleft}{\isadigit{2}}{\isadigit{0}}{\isacharcomma}node{\isacharunderscore}{\isadigit{2}}{\isadigit{0}}{\isacharunderscore}phi{\isadigit{1}}{\isacharunderscore}core{\isadigit{1}}{\isacharparenright}{\isacharcomma}\isanewline
\ \ {\isacharparenleft}{\isadigit{2}}{\isadigit{1}}{\isacharcomma}node{\isacharunderscore}{\isadigit{2}}{\isadigit{1}}{\isacharunderscore}phi{\isadigit{1}}{\isacharunderscore}timer{\isadigit{0}}{\isacharparenright}{\isacharcomma}\isanewline
\ \ {\isacharparenleft}{\isadigit{2}}{\isadigit{2}}{\isacharcomma}node{\isacharunderscore}{\isadigit{2}}{\isadigit{2}}{\isacharunderscore}phi{\isadigit{1}}{\isacharunderscore}timer{\isadigit{1}}{\isacharparenright}{\isacharcomma}\isanewline
\ \ {\isacharparenleft}{\isadigit{2}}{\isadigit{3}}{\isacharcomma}node{\isacharunderscore}{\isadigit{2}}{\isadigit{3}}{\isacharunderscore}phi{\isadigit{1}}{\isacharunderscore}ioapic{\isacharparenright}{\isacharcomma}\isanewline
\ \ {\isacharparenleft}{\isadigit{2}}{\isadigit{4}}{\isacharcomma}node{\isacharunderscore}{\isadigit{2}}{\isadigit{4}}{\isacharunderscore}phi{\isadigit{1}}{\isacharunderscore}rtc{\isacharparenright}{\isacharcomma}\isanewline
\ \ {\isacharparenleft}{\isadigit{2}}{\isadigit{5}}{\isacharcomma}node{\isacharunderscore}{\isadigit{2}}{\isadigit{5}}{\isacharunderscore}elapic{\isacharparenright}{\isacharcomma}\isanewline
\ \ {\isacharparenleft}{\isadigit{2}}{\isadigit{6}}{\isacharcomma}node{\isacharunderscore}{\isadigit{2}}{\isadigit{6}}{\isacharunderscore}sbox{\isacharparenright}\isanewline
{\isacharbrackright}{\isachardoublequoteclose}\isanewline
\isadelimtheory
\isanewline
\endisadelimtheory
\isatagtheory
\isacommand{end}\isamarkupfalse%
\endisatagtheory
{\isafoldtheory}%
\isadelimtheory
\endisadelimtheory
\end{isabellebody}%

%
%
\subsection{Cluster Server}
\label{isabelle:cluster}
\begin{figure}[!ht]
    \centering 
    \includegraphics[width=\textwidth]{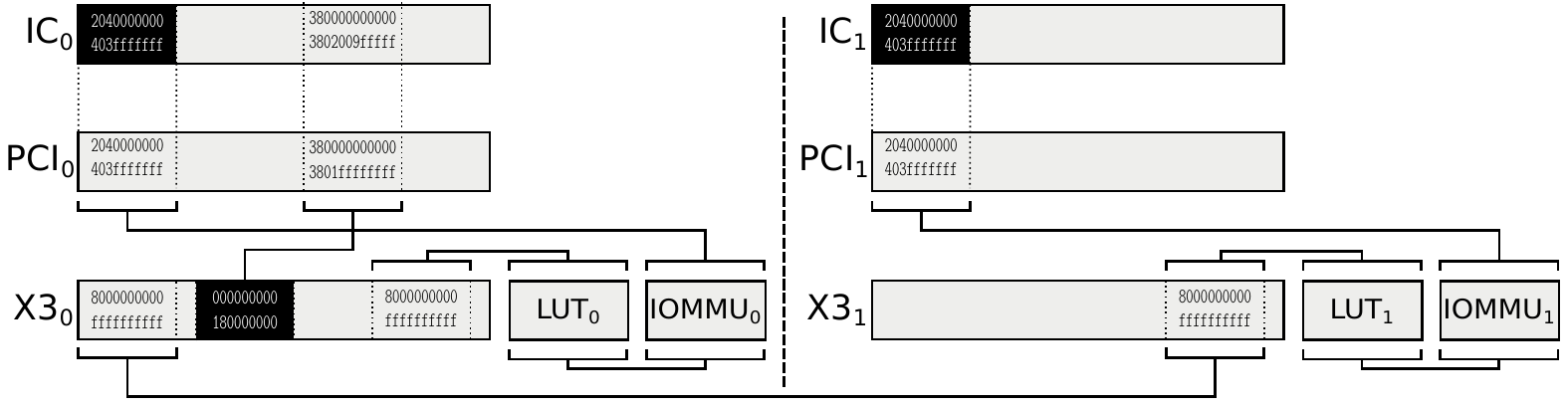}
    \caption{Schemantic overview of a heterogeneous server cluster with Xeon Phi
        co-processors and InfiniBand interconnect}
    \label{fig:system:cluster}
\end{figure}

We now consider four of the servers with two Xeon Phi co-processors from 
~\autoref{sec:realsystems:server} above. In addition, we install one Mellanox 
ConnectX3 InfiniBand card into each of those servers. This additional hardware 
allows direct read/write access to remote memory using the one-sided RDMA 
operations. This means that one machine has direct access to the RAM of another 
machine. We show the set up in ~\autoref{fig:system:cluster}. Note, we 
omitted the other hardware devices from the figure for better readability.
\camtodo{Explain}
Conceptually, the ConnectX3 card be seen as a computer on its own: it features 
translation and protection mechanisms and runs a firmware. Besides accessing 
the host RAM it can issue loads and stores on the Infiniband network which end 
up on a different machine. 

While memory can be accessed without involvement of the CPU using one-sided 
RDMA operations, the use of two sided operations can trigger an interrupt at 
the remote machine. The InfiniBand card will trigger and interrupt which is 
being forwarded to the destination core on the remote machine. Software can 
then figure out the source by polling the completion queue.

\begin{isabellebody}%
\setisabellecontext{Cluster}%
\isadelimtheory
\endisadelimtheory
\isatagtheory
\endisatagtheory
{\isafoldtheory}%
\isadelimtheory
\endisadelimtheory
\isamarkupsubsubsection{Model representation%
}
\isamarkuptrue%
\isacommand{definition}\isamarkupfalse%
\ {\isachardoublequoteopen}dram\ {\isacharequal}\ {\isacharparenleft}{\isadigit{0}}x{\isadigit{0}}{\isadigit{0}}{\isadigit{0}}{\isadigit{0}}{\isadigit{0}}{\isadigit{0}}{\isadigit{0}}{\isadigit{0}}{\isadigit{0}}{\isadigit{0}}{\isadigit{0}}{\isadigit{0}}{\isadigit{0}}{\isadigit{0}}{\isadigit{0}}{\isadigit{0}}{\isacharcomma}{\isadigit{0}}x{\isadigit{0}}{\isadigit{0}}{\isadigit{0}}{\isadigit{0}}{\isadigit{0}}{\isadigit{0}}{\isadigit{2}}{\isadigit{0}}{\isadigit{3}}FFFFFFF{\isacharparenright}{\isachardoublequoteclose}\ \isanewline
\isacommand{definition}\isamarkupfalse%
\ {\isachardoublequoteopen}pci\ {\isacharequal}\ {\isacharparenleft}{\isadigit{0}}x{\isadigit{0}}{\isadigit{0}}{\isadigit{0}}{\isadigit{0}}{\isadigit{3}}{\isadigit{8}}{\isadigit{0}}{\isadigit{0}}{\isadigit{0}}{\isadigit{0}}{\isadigit{0}}{\isadigit{0}}{\isadigit{0}}{\isadigit{0}}{\isadigit{0}}{\isadigit{0}}{\isacharcomma}{\isadigit{0}}x{\isadigit{0}}{\isadigit{0}}{\isadigit{0}}{\isadigit{0}}{\isadigit{3}}{\isadigit{8}}{\isadigit{0}}{\isadigit{2}}{\isadigit{0}}{\isadigit{0}}{\isadigit{9}}FFFFF{\isacharparenright}{\isachardoublequoteclose}%
\begin{isamarkuptext}%
The Interconnect:%
\end{isamarkuptext}\isamarkuptrue%
\isacommand{definition}\isamarkupfalse%
\ {\isachardoublequoteopen}node{\isacharunderscore}{\isadigit{0}}{\isacharunderscore}m{\isadigit{0}}{\isacharunderscore}interconnect\ {\isacharequal}\ empty{\isacharunderscore}spec\ {\isasymlparr}\isanewline
\ \ acc{\isacharunderscore}blocks\ {\isacharcolon}{\isacharequal}\ {\isacharbrackleft}dram{\isacharbrackright}{\isacharcomma}\isanewline
\ \ map{\isacharunderscore}blocks\ {\isacharcolon}{\isacharequal}\ {\isacharbrackleft}direct{\isacharunderscore}map\ pci\ {\isadigit{2}}{\isacharbrackright}\isanewline
{\isasymrparr}{\isachardoublequoteclose}\isanewline
\ \ \isanewline
\isacommand{definition}\isamarkupfalse%
\ {\isachardoublequoteopen}node{\isacharunderscore}{\isadigit{1}}{\isacharunderscore}m{\isadigit{1}}{\isacharunderscore}interconnect\ {\isacharequal}\ empty{\isacharunderscore}spec\ {\isasymlparr}\isanewline
\ \ acc{\isacharunderscore}blocks\ {\isacharcolon}{\isacharequal}\ {\isacharbrackleft}dram{\isacharbrackright}{\isacharcomma}\isanewline
\ \ map{\isacharunderscore}blocks\ {\isacharcolon}{\isacharequal}\ {\isacharbrackleft}direct{\isacharunderscore}map\ pci\ {\isadigit{3}}{\isacharbrackright}\isanewline
{\isasymrparr}{\isachardoublequoteclose}%
\begin{isamarkuptext}%
PCI Root Complexes:%
\end{isamarkuptext}\isamarkuptrue%
\isacommand{definition}\isamarkupfalse%
\ {\isachardoublequoteopen}node{\isacharunderscore}{\isadigit{2}}{\isacharunderscore}m{\isadigit{0}}{\isacharunderscore}pci\ {\isacharequal}\ empty{\isacharunderscore}spec\ {\isasymlparr}\isanewline
\ \ acc{\isacharunderscore}blocks\ {\isacharcolon}{\isacharequal}\ {\isacharbrackleft}{\isacharbrackright}{\isacharcomma}\isanewline
\ \ map{\isacharunderscore}blocks\ {\isacharcolon}{\isacharequal}\ {\isacharbrackleft}direct{\isacharunderscore}map\ dram\ {\isadigit{0}}{\isacharcomma}\ direct{\isacharunderscore}map\ pci\ {\isadigit{2}}{\isacharbrackright}\isanewline
{\isasymrparr}{\isachardoublequoteclose}\isanewline
\ \ \isanewline
\isacommand{definition}\isamarkupfalse%
\ {\isachardoublequoteopen}node{\isacharunderscore}{\isadigit{3}}{\isacharunderscore}m{\isadigit{1}}{\isacharunderscore}pci\ {\isacharequal}\ empty{\isacharunderscore}spec\ {\isasymlparr}\isanewline
\ \ acc{\isacharunderscore}blocks\ {\isacharcolon}{\isacharequal}\ {\isacharbrackleft}{\isacharbrackright}{\isacharcomma}\isanewline
\ \ map{\isacharunderscore}blocks\ {\isacharcolon}{\isacharequal}\ {\isacharbrackleft}direct{\isacharunderscore}map\ dram\ {\isadigit{1}}{\isacharcomma}\ direct{\isacharunderscore}map\ pci\ {\isadigit{3}}{\isacharbrackright}\isanewline
{\isasymrparr}{\isachardoublequoteclose}%
\begin{isamarkuptext}%
Infiniband Cards (RDMA):%
\end{isamarkuptext}\isamarkuptrue%
\isacommand{definition}\isamarkupfalse%
\ {\isachardoublequoteopen}remote{\isacharunderscore}ram\ {\isacharequal}\ {\isacharparenleft}{\isadigit{0}}x{\isadigit{0}}{\isadigit{0}}{\isadigit{0}}{\isadigit{0}}{\isadigit{0}}{\isadigit{0}}{\isadigit{0}}{\isadigit{0}}{\isadigit{0}}{\isadigit{0}}{\isadigit{0}}{\isadigit{0}}{\isadigit{0}}{\isadigit{0}}{\isadigit{0}}{\isadigit{0}}{\isacharcomma}{\isadigit{0}}x{\isadigit{0}}{\isadigit{0}}{\isadigit{0}}{\isadigit{0}}{\isadigit{0}}{\isadigit{0}}{\isadigit{2}}{\isadigit{0}}{\isadigit{3}}FFFFFF{\isacharparenright}{\isachardoublequoteclose}\isanewline
\isacommand{definition}\isamarkupfalse%
\ {\isachardoublequoteopen}host{\isacharunderscore}ram\ {\isacharequal}\ {\isacharparenleft}{\isadigit{0}}x{\isadigit{0}}{\isadigit{0}}{\isadigit{0}}{\isadigit{8}}{\isadigit{0}}{\isadigit{0}}{\isadigit{0}}{\isadigit{0}}{\isadigit{0}}{\isadigit{0}}{\isadigit{0}}{\isadigit{0}}{\isadigit{0}}{\isadigit{0}}{\isadigit{0}}{\isacharcomma}{\isadigit{0}}x{\isadigit{0}}{\isadigit{0}}{\isadigit{0}}{\isadigit{8}}{\isadigit{0}}{\isadigit{0}}{\isadigit{0}}{\isadigit{2}}{\isadigit{0}}{\isadigit{3}}FFFFFF{\isacharparenright}{\isachardoublequoteclose}\isanewline
\isacommand{definition}\isamarkupfalse%
\ {\isachardoublequoteopen}node{\isacharunderscore}{\isadigit{4}}{\isacharunderscore}m{\isadigit{1}}{\isacharunderscore}cx{\isadigit{3}}\ {\isacharequal}\ empty{\isacharunderscore}spec\ {\isasymlparr}\isanewline
\ \ acc{\isacharunderscore}blocks\ {\isacharcolon}{\isacharequal}\ {\isacharbrackleft}pci{\isacharbrackright}{\isacharcomma}\isanewline
\ \ map{\isacharunderscore}blocks\ {\isacharcolon}{\isacharequal}\ {\isacharbrackleft}block{\isacharunderscore}map\ host{\isacharunderscore}ram\ {\isadigit{6}}\ {\isadigit{0}}x{\isadigit{0}}{\isacharcomma}\ block{\isacharunderscore}map\ remote{\isacharunderscore}ram\ {\isadigit{5}}\ {\isadigit{0}}x{\isadigit{0}}{\isadigit{0}}{\isadigit{0}}{\isadigit{8}}{\isadigit{0}}{\isadigit{0}}{\isadigit{0}}{\isadigit{0}}{\isadigit{0}}{\isadigit{0}}{\isadigit{0}}{\isadigit{0}}{\isadigit{0}}{\isadigit{0}}{\isadigit{0}}{\isacharbrackright}\isanewline
{\isasymrparr}{\isachardoublequoteclose}\isanewline
\isacommand{definition}\isamarkupfalse%
\ {\isachardoublequoteopen}node{\isacharunderscore}{\isadigit{5}}{\isacharunderscore}m{\isadigit{1}}{\isacharunderscore}cx{\isadigit{3}}\ {\isacharequal}\ empty{\isacharunderscore}spec\ {\isasymlparr}\isanewline
\ \ acc{\isacharunderscore}blocks\ {\isacharcolon}{\isacharequal}\ {\isacharbrackleft}pci{\isacharbrackright}{\isacharcomma}\isanewline
\ \ map{\isacharunderscore}blocks\ {\isacharcolon}{\isacharequal}\ {\isacharbrackleft}block{\isacharunderscore}map\ host{\isacharunderscore}ram\ {\isadigit{7}}\ {\isadigit{0}}x{\isadigit{0}}{\isacharcomma}\ block{\isacharunderscore}map\ remote{\isacharunderscore}ram\ {\isadigit{4}}\ {\isadigit{0}}x{\isadigit{0}}{\isadigit{0}}{\isadigit{0}}{\isadigit{8}}{\isadigit{0}}{\isadigit{0}}{\isadigit{0}}{\isadigit{0}}{\isadigit{0}}{\isadigit{0}}{\isadigit{0}}{\isadigit{0}}{\isadigit{0}}{\isadigit{0}}{\isacharbrackright}\isanewline
{\isasymrparr}{\isachardoublequoteclose}%
\begin{isamarkuptext}%
MMUs:%
\end{isamarkuptext}\isamarkuptrue%
\isacommand{definition}\isamarkupfalse%
\ {\isachardoublequoteopen}node{\isacharunderscore}{\isadigit{6}}{\isacharunderscore}m{\isadigit{1}}{\isacharunderscore}tab\ {\isacharequal}\ empty{\isacharunderscore}spec\ {\isasymlparr}\isanewline
\ \ acc{\isacharunderscore}blocks\ {\isacharcolon}{\isacharequal}\ {\isacharbrackleft}{\isacharbrackright}{\isacharcomma}\isanewline
\ \ map{\isacharunderscore}blocks\ {\isacharcolon}{\isacharequal}\ {\isacharbrackleft}{\isacharbrackright}{\isacharcomma}\isanewline
\ \ overlay\ {\isacharcolon}{\isacharequal}\ Some\ {\isadigit{8}}\isanewline
{\isasymrparr}{\isachardoublequoteclose}\isanewline
\isacommand{definition}\isamarkupfalse%
\ {\isachardoublequoteopen}node{\isacharunderscore}{\isadigit{7}}{\isacharunderscore}m{\isadigit{1}}{\isacharunderscore}tab\ {\isacharequal}\ empty{\isacharunderscore}spec\ {\isasymlparr}\isanewline
\ \ acc{\isacharunderscore}blocks\ {\isacharcolon}{\isacharequal}\ {\isacharbrackleft}{\isacharbrackright}{\isacharcomma}\isanewline
\ \ map{\isacharunderscore}blocks\ {\isacharcolon}{\isacharequal}\ {\isacharbrackleft}{\isacharbrackright}{\isacharcomma}\isanewline
\ \ overlay\ {\isacharcolon}{\isacharequal}\ Some\ {\isadigit{9}}\isanewline
{\isasymrparr}{\isachardoublequoteclose}%
\begin{isamarkuptext}%
IO-MMUs:%
\end{isamarkuptext}\isamarkuptrue%
\isacommand{definition}\isamarkupfalse%
\ {\isachardoublequoteopen}node{\isacharunderscore}{\isadigit{8}}{\isacharunderscore}m{\isadigit{1}}{\isacharunderscore}iommu\ {\isacharequal}\ empty{\isacharunderscore}spec\ {\isasymlparr}\isanewline
\ \ acc{\isacharunderscore}blocks\ {\isacharcolon}{\isacharequal}\ {\isacharbrackleft}{\isacharbrackright}{\isacharcomma}\isanewline
\ \ map{\isacharunderscore}blocks\ {\isacharcolon}{\isacharequal}\ {\isacharbrackleft}direct{\isacharunderscore}map\ dram\ {\isadigit{2}}{\isacharbrackright}\isanewline
{\isasymrparr}{\isachardoublequoteclose}\isanewline
\isacommand{definition}\isamarkupfalse%
\ {\isachardoublequoteopen}node{\isacharunderscore}{\isadigit{9}}{\isacharunderscore}m{\isadigit{1}}{\isacharunderscore}iommu\ {\isacharequal}\ empty{\isacharunderscore}spec\ {\isasymlparr}\isanewline
\ \ acc{\isacharunderscore}blocks\ {\isacharcolon}{\isacharequal}\ {\isacharbrackleft}{\isacharbrackright}{\isacharcomma}\isanewline
\ \ map{\isacharunderscore}blocks\ {\isacharcolon}{\isacharequal}\ {\isacharbrackleft}direct{\isacharunderscore}map\ dram\ {\isadigit{3}}{\isacharbrackright}\isanewline
{\isasymrparr}{\isachardoublequoteclose}\isanewline
\isanewline
\isacommand{definition}\isamarkupfalse%
\ {\isachardoublequoteopen}sys\ {\isacharequal}\ {\isacharbrackleft}{\isacharparenleft}{\isadigit{0}}{\isacharcomma}\ node{\isacharunderscore}{\isadigit{0}}{\isacharunderscore}m{\isadigit{0}}{\isacharunderscore}interconnect{\isacharparenright}{\isacharcomma}\ \isanewline
\ \ \ \ \ \ \ \ \ \ \ \ \ \ \ \ \ \ \ {\isacharparenleft}{\isadigit{1}}{\isacharcomma}\ node{\isacharunderscore}{\isadigit{1}}{\isacharunderscore}m{\isadigit{1}}{\isacharunderscore}interconnect{\isacharparenright}{\isacharcomma}\ \isanewline
\ \ \ \ \ \ \ \ \ \ \ \ \ \ \ \ \ \ \ {\isacharparenleft}{\isadigit{2}}{\isacharcomma}\ node{\isacharunderscore}{\isadigit{2}}{\isacharunderscore}m{\isadigit{0}}{\isacharunderscore}pci{\isacharparenright}{\isacharcomma}\ \isanewline
\ \ \ \ \ \ \ \ \ \ \ \ \ \ \ \ \ \ \ {\isacharparenleft}{\isadigit{3}}{\isacharcomma}\ node{\isacharunderscore}{\isadigit{3}}{\isacharunderscore}m{\isadigit{1}}{\isacharunderscore}pci{\isacharparenright}{\isacharcomma}\ \isanewline
\ \ \ \ \ \ \ \ \ \ \ \ \ \ \ \ \ \ \ {\isacharparenleft}{\isadigit{4}}{\isacharcomma}\ node{\isacharunderscore}{\isadigit{4}}{\isacharunderscore}m{\isadigit{1}}{\isacharunderscore}cx{\isadigit{3}}{\isacharparenright}{\isacharcomma}\ \isanewline
\ \ \ \ \ \ \ \ \ \ \ \ \ \ \ \ \ \ \ {\isacharparenleft}{\isadigit{5}}{\isacharcomma}\ node{\isacharunderscore}{\isadigit{5}}{\isacharunderscore}m{\isadigit{1}}{\isacharunderscore}cx{\isadigit{3}}{\isacharparenright}{\isacharcomma}\ \isanewline
\ \ \ \ \ \ \ \ \ \ \ \ \ \ \ \ \ \ \ {\isacharparenleft}{\isadigit{6}}{\isacharcomma}\ node{\isacharunderscore}{\isadigit{6}}{\isacharunderscore}m{\isadigit{1}}{\isacharunderscore}tab{\isacharparenright}{\isacharcomma}\ \isanewline
\ \ \ \ \ \ \ \ \ \ \ \ \ \ \ \ \ \ \ {\isacharparenleft}{\isadigit{7}}{\isacharcomma}\ node{\isacharunderscore}{\isadigit{7}}{\isacharunderscore}m{\isadigit{1}}{\isacharunderscore}tab{\isacharparenright}{\isacharcomma}\ \isanewline
\ \ \ \ \ \ \ \ \ \ \ \ \ \ \ \ \ \ \ {\isacharparenleft}{\isadigit{8}}{\isacharcomma}\ node{\isacharunderscore}{\isadigit{8}}{\isacharunderscore}m{\isadigit{1}}{\isacharunderscore}iommu{\isacharparenright}{\isacharcomma}\ \isanewline
\ \ \ \ \ \ \ \ \ \ \ \ \ \ \ \ \ \ \ {\isacharparenleft}{\isadigit{9}}{\isacharcomma}\ node{\isacharunderscore}{\isadigit{9}}{\isacharunderscore}m{\isadigit{1}}{\isacharunderscore}iommu{\isacharparenright}{\isacharbrackright}{\isachardoublequoteclose}\isanewline
\isadelimtheory
\isanewline
\endisadelimtheory
\isatagtheory
\isacommand{end}\isamarkupfalse%
\endisatagtheory
{\isafoldtheory}%
\isadelimtheory
\endisadelimtheory
\end{isabellebody}%

%
%
\subsection{The Single Chip Cloud Computer}
\label{sec:system:scc}
\begin{figure}[!ht]
    \centering 
    \includegraphics[width=\textwidth]{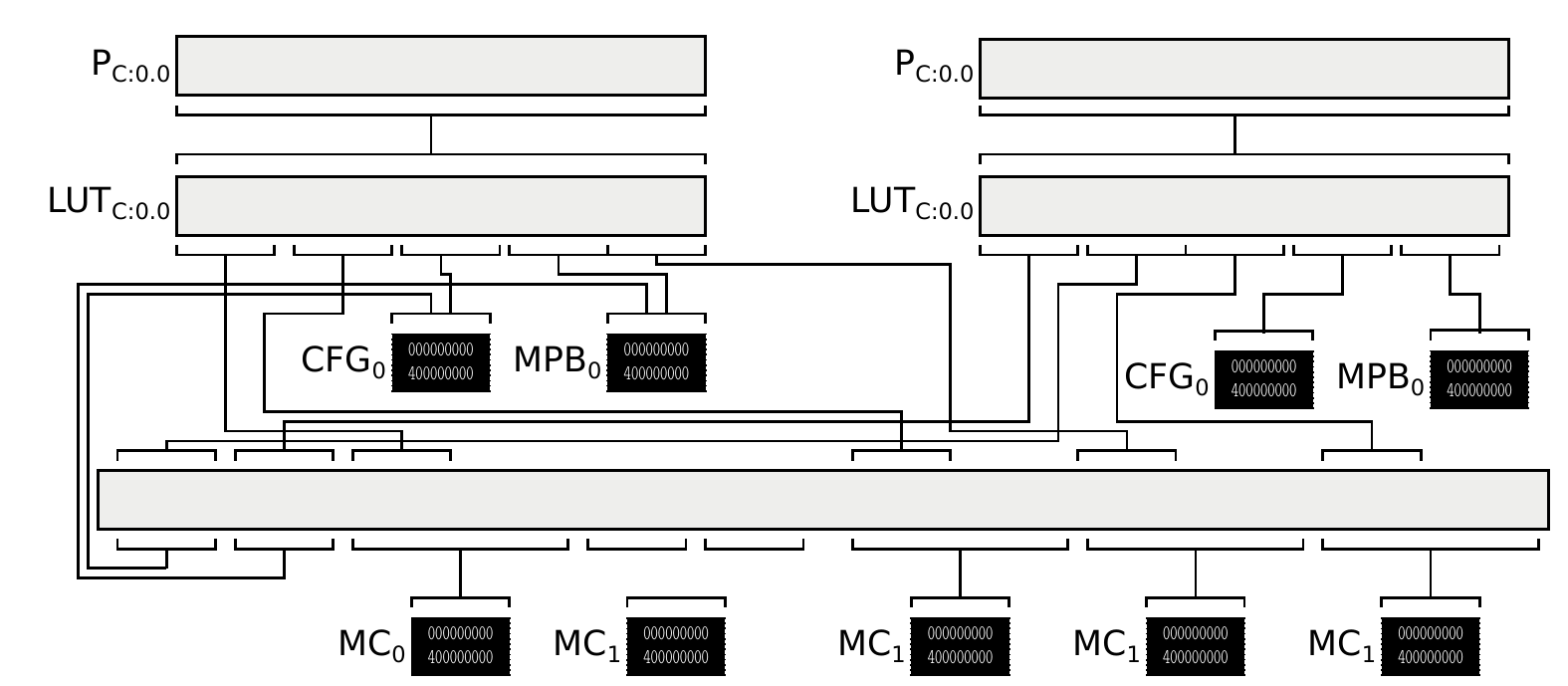}
    \caption{Schematic overview of the single chip cloud computer}
    \label{fig:system:scc}
\end{figure}

Intel's single chip cloud computer (SCC~\cite{scc}) is a 48-core research 
microprocessor consisting of 24 two-core tiles interconnected in a (x,y)-mesh 
with a 24 router network and hardware support for message-passing. We show the 
schema on~\autoref{fig:system:scc}.
\camtodo{Explain}
On the chip there are four DDR3 memory controllers (MC) each of which is able 
to address 16GB of memory using physical address extensions (PAE) -- 34-bit 
addressing mode. Therefore the SCC has a maximum memory of 64GB which exists on 
the SCC board. However, each core is able to access at most 4GB of at the same 
time.

Access to memory from each core is configured using a per-core system address 
lookup table (LUT). The LUT maps the core's 32-bit physical addresses to the 
extended 46-bit system address. There are 256 entries in the table each of 
which translating a 16MB segment to any memory location e.g.\ DRAM, local 
message passing buffer (MPB), configuration registers of any tile. The LUT can 
be reprogrammed given a core has a valid mapping to the configuration space. 

The lookup table is indexed by the bits \texttt{31..24} of the address, while 
bits \texttt{23..0} is the offset into the 16MB region. The lookup table will 
produce a 46-bit output address. Where bits \texttt{33..24} together with the 
offset are the 34-bit extended physical address. In addition, a tuple $(bypass, 
destination, port)$ is out pot, that defines whether or not the router on the 
tile is bypassed, the destination tile ID and the port which should be used 
(identifies config registers, message passing buffer or memory controller)

The resources from the SCC are accessible from the host for initial bootstrap 
and communication later on. The interface supports IO and DMA 
transfers~\cite{Peter:202011:SCC}.

The single chip has a very simple interrupt system. Each core has a local
APIC with the standard private timer. In addition, the LAPIC exhibits a memory
mapped register that can be used to trigger one of its vectors by any core.
There are no card specific interrupt sources that can be routed to the cores,
neither exists there are any way of broadcasting an interrupt.

\begin{isabellebody}%
\setisabellecontext{SCC}%
\isadelimtheory
\endisadelimtheory
\isatagtheory
\endisatagtheory
{\isafoldtheory}%
\isadelimtheory
\endisadelimtheory
\isamarkupsubsubsection{Model representation%
}
\isamarkuptrue%
\label{isabelle:scc}
\begin{isamarkuptext}%
Memory Controllers: 16GB each%
\end{isamarkuptext}\isamarkuptrue%
\isacommand{definition}\isamarkupfalse%
\ {\isachardoublequoteopen}dram\ {\isacharequal}\ {\isacharparenleft}{\isadigit{0}}x{\isadigit{0}}{\isadigit{0}}{\isadigit{0}}{\isadigit{0}}{\isadigit{0}}{\isadigit{0}}{\isadigit{0}}{\isadigit{0}}{\isadigit{0}}{\isacharcomma}{\isadigit{0}}x{\isadigit{3}}FFFFFFFF{\isacharparenright}{\isachardoublequoteclose}\ \isanewline
\isacommand{definition}\isamarkupfalse%
\ {\isachardoublequoteopen}node{\isacharunderscore}mc\ {\isacharequal}\ empty{\isacharunderscore}spec\ {\isasymlparr}\isanewline
\ \ acc{\isacharunderscore}blocks\ {\isacharcolon}{\isacharequal}\ {\isacharbrackleft}dram{\isacharbrackright}{\isacharcomma}\isanewline
\ \ map{\isacharunderscore}blocks\ {\isacharcolon}{\isacharequal}\ {\isacharbrackleft}{\isacharbrackright}\isanewline
{\isasymrparr}{\isachardoublequoteclose}%
\begin{isamarkuptext}%
Local Memory / Message Passing buffer of 16kB%
\end{isamarkuptext}\isamarkuptrue%
\isacommand{definition}\isamarkupfalse%
\ {\isachardoublequoteopen}lmb\ {\isacharequal}\ {\isacharparenleft}{\isadigit{0}}x{\isadigit{0}}{\isadigit{0}}{\isadigit{0}}{\isacharcomma}\ {\isadigit{0}}xFFFF{\isacharparenright}{\isachardoublequoteclose}\isanewline
\isacommand{definition}\isamarkupfalse%
\ {\isachardoublequoteopen}node{\isacharunderscore}lmb\ {\isacharequal}\ empty{\isacharunderscore}spec\ {\isasymlparr}\isanewline
\ \ acc{\isacharunderscore}blocks\ {\isacharcolon}{\isacharequal}\ {\isacharbrackleft}lmb{\isacharbrackright}{\isacharcomma}\isanewline
\ \ map{\isacharunderscore}blocks\ {\isacharcolon}{\isacharequal}\ {\isacharbrackleft}{\isacharbrackright}\isanewline
{\isasymrparr}{\isachardoublequoteclose}%
\begin{isamarkuptext}%
Local Configuration Registers%
\end{isamarkuptext}\isamarkuptrue%
\isacommand{definition}\isamarkupfalse%
\ {\isachardoublequoteopen}conf\ {\isacharequal}\ {\isacharparenleft}{\isadigit{0}}x{\isadigit{0}}{\isadigit{0}}{\isadigit{0}}{\isadigit{0}}{\isadigit{0}}{\isadigit{0}}{\isacharcomma}\ {\isadigit{0}}x{\isadigit{7}}FFFFF{\isacharparenright}{\isachardoublequoteclose}\isanewline
\isacommand{definition}\isamarkupfalse%
\ {\isachardoublequoteopen}node{\isacharunderscore}conf\ {\isacharequal}\ empty{\isacharunderscore}spec\ {\isasymlparr}\isanewline
\ \ acc{\isacharunderscore}blocks\ {\isacharcolon}{\isacharequal}\ {\isacharbrackleft}conf{\isacharbrackright}{\isacharcomma}\isanewline
\ \ map{\isacharunderscore}blocks\ {\isacharcolon}{\isacharequal}\ {\isacharbrackleft}{\isacharbrackright}\isanewline
{\isasymrparr}{\isachardoublequoteclose}%
\begin{isamarkuptext}%
2D Mesh Network%
\end{isamarkuptext}\isamarkuptrue%
\isacommand{definition}\isamarkupfalse%
\ {\isachardoublequoteopen}dram{\isadigit{0}}\ {\isacharequal}\ {\isacharparenleft}{\isadigit{0}}x{\isadigit{0}}{\isadigit{0}}{\isadigit{1}}{\isadigit{8}}{\isadigit{0}}{\isadigit{0}}{\isadigit{0}}{\isadigit{0}}{\isadigit{0}}{\isadigit{0}}{\isadigit{0}}{\isadigit{0}}{\isacharcomma}{\isadigit{0}}x{\isadigit{0}}{\isadigit{0}}{\isadigit{1}}BFFFFFFFF{\isacharparenright}{\isachardoublequoteclose}\isanewline
\isacommand{definition}\isamarkupfalse%
\ {\isachardoublequoteopen}dram{\isadigit{1}}\ {\isacharequal}\ {\isacharparenleft}{\isadigit{0}}x{\isadigit{0}}{\isadigit{0}}F{\isadigit{0}}{\isadigit{0}}{\isadigit{0}}{\isadigit{0}}{\isadigit{0}}{\isadigit{0}}{\isadigit{0}}{\isadigit{0}}{\isadigit{0}}{\isacharcomma}{\isadigit{0}}x{\isadigit{0}}{\isadigit{0}}{\isadigit{1}}{\isadigit{3}}FFFFFFFF{\isacharparenright}{\isachardoublequoteclose}\isanewline
\isacommand{definition}\isamarkupfalse%
\ {\isachardoublequoteopen}dram{\isadigit{2}}\ {\isacharequal}\ {\isacharparenleft}{\isadigit{0}}x{\isadigit{0}}{\isadigit{4}}{\isadigit{1}}{\isadigit{8}}{\isadigit{0}}{\isadigit{0}}{\isadigit{0}}{\isadigit{0}}{\isadigit{0}}{\isadigit{0}}{\isadigit{0}}{\isadigit{0}}{\isacharcomma}{\isadigit{0}}x{\isadigit{0}}{\isadigit{3}}{\isadigit{1}}BFFFFFFFF{\isacharparenright}{\isachardoublequoteclose}\isanewline
\isacommand{definition}\isamarkupfalse%
\ {\isachardoublequoteopen}dram{\isadigit{3}}\ {\isacharequal}\ {\isacharparenleft}{\isadigit{0}}x{\isadigit{0}}{\isadigit{4}}F{\isadigit{0}}{\isadigit{0}}{\isadigit{0}}{\isadigit{0}}{\isadigit{0}}{\isadigit{0}}{\isadigit{0}}{\isadigit{0}}{\isadigit{0}}{\isacharcomma}{\isadigit{0}}x{\isadigit{0}}{\isadigit{3}}F{\isadigit{3}}FFFFFFFF{\isacharparenright}{\isachardoublequoteclose}\isanewline
\isanewline
\isacommand{definition}\isamarkupfalse%
\ {\isachardoublequoteopen}conf{\isacharunderscore}{\isadigit{0}}{\isadigit{0}}\ {\isacharequal}\ {\isacharparenleft}{\isadigit{0}}x{\isadigit{0}}{\isadigit{0}}{\isadigit{0}}{\isadigit{8}}{\isadigit{0}}{\isadigit{0}}{\isadigit{0}}{\isadigit{0}}{\isadigit{0}}{\isadigit{0}}{\isadigit{0}}{\isadigit{0}}{\isacharcomma}\ {\isadigit{0}}x{\isadigit{0}}{\isadigit{0}}{\isadigit{0}}{\isadigit{8}}{\isadigit{0}}{\isadigit{0}}{\isadigit{7}}FFFFF{\isacharparenright}{\isachardoublequoteclose}\isanewline
\isacommand{definition}\isamarkupfalse%
\ {\isachardoublequoteopen}conf{\isacharunderscore}{\isadigit{0}}{\isadigit{1}}\ {\isacharequal}\ {\isacharparenleft}{\isadigit{0}}x{\isadigit{0}}{\isadigit{0}}{\isadigit{2}}{\isadigit{8}}{\isadigit{0}}{\isadigit{0}}{\isadigit{0}}{\isadigit{0}}{\isadigit{0}}{\isadigit{0}}{\isadigit{0}}{\isadigit{0}}{\isacharcomma}\ {\isadigit{0}}x{\isadigit{0}}{\isadigit{0}}{\isadigit{2}}{\isadigit{8}}{\isadigit{0}}{\isadigit{0}}{\isadigit{7}}FFFFF{\isacharparenright}{\isachardoublequoteclose}\isanewline
\isacommand{definition}\isamarkupfalse%
\ {\isachardoublequoteopen}mpb{\isacharunderscore}{\isadigit{0}}{\isadigit{0}}\ \ {\isacharequal}\ {\isacharparenleft}{\isadigit{0}}x{\isadigit{0}}{\isadigit{0}}{\isadigit{0}}C{\isadigit{0}}{\isadigit{0}}{\isadigit{0}}{\isadigit{0}}{\isadigit{0}}{\isadigit{0}}{\isadigit{0}}{\isadigit{0}}{\isacharcomma}\ {\isadigit{0}}x{\isadigit{0}}{\isadigit{0}}{\isadigit{0}}c{\isadigit{0}}{\isadigit{0}}{\isadigit{7}}FFFFF{\isacharparenright}{\isachardoublequoteclose}\isanewline
\isacommand{definition}\isamarkupfalse%
\ {\isachardoublequoteopen}mbp{\isacharunderscore}{\isadigit{0}}{\isadigit{1}}\ \ {\isacharequal}\ {\isacharparenleft}{\isadigit{0}}x{\isadigit{0}}{\isadigit{0}}{\isadigit{2}}C{\isadigit{0}}{\isadigit{0}}{\isadigit{0}}{\isadigit{0}}{\isadigit{0}}{\isadigit{0}}{\isadigit{0}}{\isadigit{0}}{\isacharcomma}\ {\isadigit{0}}x{\isadigit{0}}{\isadigit{0}}{\isadigit{2}}c{\isadigit{0}}{\isadigit{0}}{\isadigit{7}}FFFFF{\isacharparenright}{\isachardoublequoteclose}\ \ \isanewline
\isacommand{definition}\isamarkupfalse%
\ {\isachardoublequoteopen}sif\ \ \ \ \ {\isacharequal}\ {\isacharparenleft}{\isadigit{0}}x{\isadigit{0}}{\isadigit{0}}F{\isadigit{4}}{\isadigit{0}}{\isadigit{0}}{\isadigit{0}}{\isadigit{0}}{\isadigit{0}}{\isadigit{0}}{\isadigit{0}}{\isadigit{0}}{\isacharcomma}\ {\isadigit{0}}x{\isadigit{0}}{\isadigit{0}}F{\isadigit{7}}FFFFFFFF{\isacharparenright}{\isachardoublequoteclose}\ \isanewline
\isacommand{definition}\isamarkupfalse%
\ {\isachardoublequoteopen}node{\isacharunderscore}{\isadigit{0}}{\isacharunderscore}interconnect\ {\isacharequal}\ empty{\isacharunderscore}spec\ {\isasymlparr}\isanewline
\ \ acc{\isacharunderscore}blocks\ {\isacharcolon}{\isacharequal}\ {\isacharbrackleft}{\isacharbrackright}{\isacharcomma}\isanewline
\ \ map{\isacharunderscore}blocks\ {\isacharcolon}{\isacharequal}\ {\isacharbrackleft}block{\isacharunderscore}map\ dram{\isadigit{0}}\ {\isadigit{1}}\ {\isadigit{0}}x{\isadigit{0}}{\isacharcomma}\ block{\isacharunderscore}map\ dram{\isadigit{1}}\ {\isadigit{2}}\ {\isadigit{0}}x{\isadigit{0}}{\isacharcomma}\isanewline
\ \ \ \ \ \ \ \ \ \ \ \ \ \ \ \ \ block{\isacharunderscore}map\ dram{\isadigit{2}}\ {\isadigit{2}}\ {\isadigit{0}}x{\isadigit{0}}{\isacharcomma}\ block{\isacharunderscore}map\ dram{\isadigit{3}}\ {\isadigit{3}}\ {\isadigit{0}}x{\isadigit{0}}{\isacharcomma}\isanewline
\ \ \ \ \ \ \ \ \ \ \ \ \ \ \ \ \ block{\isacharunderscore}map\ mpb{\isacharunderscore}{\isadigit{0}}{\isadigit{0}}\ {\isadigit{4}}\ {\isadigit{0}}x{\isadigit{0}}{\isacharcomma}\ block{\isacharunderscore}map\ mbp{\isacharunderscore}{\isadigit{0}}{\isadigit{1}}\ {\isadigit{5}}\ {\isadigit{0}}x{\isadigit{0}}{\isacharcomma}\isanewline
\ \ \ \ \ \ \ \ \ \ \ \ \ \ \ \ \ block{\isacharunderscore}map\ conf{\isacharunderscore}{\isadigit{0}}{\isadigit{0}}\ {\isadigit{6}}\ {\isadigit{0}}x{\isadigit{0}}{\isacharcomma}\ block{\isacharunderscore}map\ conf{\isacharunderscore}{\isadigit{0}}{\isadigit{1}}\ {\isadigit{7}}\ {\isadigit{0}}x{\isadigit{0}}{\isacharcomma}\isanewline
\ \ \ \ \ \ \ \ \ \ \ \ \ \ \ \ \ block{\isacharunderscore}map\ sif\ {\isadigit{4}}{\isadigit{4}}\ {\isadigit{0}}x{\isadigit{0}}{\isacharbrackright}\isanewline
{\isasymrparr}{\isachardoublequoteclose}%
\begin{isamarkuptext}%
Core 0.0%
\end{isamarkuptext}\isamarkuptrue%
\isacommand{definition}\isamarkupfalse%
\ {\isachardoublequoteopen}vram\ {\isacharequal}\ {\isacharparenleft}{\isadigit{0}}x{\isadigit{0}}{\isadigit{0}}{\isadigit{0}}{\isadigit{0}}{\isadigit{0}}{\isadigit{0}}{\isadigit{0}}{\isacharcomma}\ {\isadigit{0}}x{\isadigit{0}}FFFFFF{\isacharparenright}{\isachardoublequoteclose}\isanewline
\isacommand{definition}\isamarkupfalse%
\ {\isachardoublequoteopen}node{\isacharunderscore}{\isadigit{9}}{\isacharunderscore}vas{\isadigit{0}}{\isadigit{0}}\ {\isacharequal}\ \ empty{\isacharunderscore}spec\ {\isasymlparr}\isanewline
\ \ acc{\isacharunderscore}blocks\ {\isacharcolon}{\isacharequal}\ {\isacharbrackleft}{\isacharbrackright}{\isacharcomma}\isanewline
\ \ map{\isacharunderscore}blocks\ {\isacharcolon}{\isacharequal}\ {\isacharbrackleft}block{\isacharunderscore}map\ vram\ {\isadigit{1}}{\isadigit{0}}\ {\isadigit{0}}x{\isadigit{0}}\ {\isacharbrackright}\isanewline
{\isasymrparr}{\isachardoublequoteclose}\isanewline
\ \ \isanewline
\isacommand{definition}\isamarkupfalse%
\ {\isachardoublequoteopen}cpu{\isacharunderscore}phys\ {\isacharequal}\ \ empty{\isacharunderscore}spec\ {\isasymlparr}\isanewline
\ \ acc{\isacharunderscore}blocks\ {\isacharcolon}{\isacharequal}\ {\isacharbrackleft}{\isacharbrackright}{\isacharcomma}\isanewline
\ \ overlay\ {\isacharcolon}{\isacharequal}\ Some\ {\isadigit{1}}{\isadigit{1}}\isanewline
{\isasymrparr}{\isachardoublequoteclose}\isanewline
\isanewline
\isacommand{definition}\isamarkupfalse%
\ {\isachardoublequoteopen}lut{\isadigit{0}}{\isadigit{0}}{\isacharunderscore}cfg\ {\isacharequal}\ {\isacharparenleft}{\isadigit{0}}x{\isadigit{3}}{\isadigit{0}}{\isadigit{0}}{\isadigit{0}}{\isadigit{0}}{\isadigit{0}}{\isadigit{0}}{\isacharcomma}\ {\isadigit{0}}x{\isadigit{3}}FFFFFF{\isacharparenright}{\isachardoublequoteclose}\isanewline
\isacommand{definition}\isamarkupfalse%
\ {\isachardoublequoteopen}lut{\isadigit{0}}{\isadigit{0}}{\isacharunderscore}mpb\ {\isacharequal}\ {\isacharparenleft}{\isadigit{0}}x{\isadigit{1}}{\isadigit{0}}{\isadigit{0}}{\isadigit{0}}{\isadigit{0}}{\isadigit{0}}{\isadigit{0}}{\isacharcomma}\ {\isadigit{0}}x{\isadigit{1}}FFFFFF{\isacharparenright}{\isachardoublequoteclose}\isanewline
\isacommand{definition}\isamarkupfalse%
\ {\isachardoublequoteopen}node{\isacharunderscore}{\isadigit{1}}{\isadigit{1}}{\isacharunderscore}lut{\isadigit{0}}{\isadigit{0}}\ \ {\isacharequal}\ \ empty{\isacharunderscore}spec\ {\isasymlparr}\isanewline
\ \ acc{\isacharunderscore}blocks\ {\isacharcolon}{\isacharequal}\ {\isacharbrackleft}{\isacharbrackright}{\isacharcomma}\isanewline
\ \ map{\isacharunderscore}blocks\ {\isacharcolon}{\isacharequal}\ {\isacharbrackleft}block{\isacharunderscore}map\ lut{\isadigit{0}}{\isadigit{0}}{\isacharunderscore}cfg\ {\isadigit{5}}\ {\isadigit{0}}x{\isadigit{0}}{\isacharcomma}\ block{\isacharunderscore}map\ lut{\isadigit{0}}{\isadigit{0}}{\isacharunderscore}mpb\ {\isadigit{6}}\ {\isadigit{0}}x{\isadigit{0}}{\isacharbrackright}{\isacharcomma}\isanewline
\ \ overlay\ {\isacharcolon}{\isacharequal}\ Some\ {\isadigit{0}}\isanewline
{\isasymrparr}{\isachardoublequoteclose}%
\begin{isamarkuptext}%
Core 1.0%
\end{isamarkuptext}\isamarkuptrue%
\isacommand{definition}\isamarkupfalse%
\ {\isachardoublequoteopen}node{\isacharunderscore}{\isadigit{1}}{\isadigit{2}}{\isacharunderscore}vas{\isadigit{0}}{\isadigit{1}}\ {\isacharequal}\ \ empty{\isacharunderscore}spec\ {\isasymlparr}\isanewline
\ \ acc{\isacharunderscore}blocks\ {\isacharcolon}{\isacharequal}\ {\isacharbrackleft}{\isacharbrackright}{\isacharcomma}\isanewline
\ \ map{\isacharunderscore}blocks\ {\isacharcolon}{\isacharequal}\ {\isacharbrackleft}block{\isacharunderscore}map\ vram\ {\isadigit{1}}{\isadigit{3}}\ {\isadigit{0}}x{\isadigit{0}}\ {\isacharbrackright}\isanewline
{\isasymrparr}{\isachardoublequoteclose}\isanewline
\isanewline
\isacommand{definition}\isamarkupfalse%
\ {\isachardoublequoteopen}lut{\isadigit{0}}{\isadigit{1}}{\isacharunderscore}cfg\ {\isacharequal}\ {\isacharparenleft}{\isadigit{0}}x{\isadigit{3}}{\isadigit{0}}{\isadigit{0}}{\isadigit{0}}{\isadigit{0}}{\isadigit{0}}{\isadigit{0}}{\isacharcomma}\ {\isadigit{0}}x{\isadigit{3}}FFFFFF{\isacharparenright}{\isachardoublequoteclose}\isanewline
\isacommand{definition}\isamarkupfalse%
\ {\isachardoublequoteopen}lut{\isadigit{0}}{\isadigit{1}}{\isacharunderscore}mpb\ {\isacharequal}\ {\isacharparenleft}{\isadigit{0}}x{\isadigit{1}}{\isadigit{0}}{\isadigit{0}}{\isadigit{0}}{\isadigit{0}}{\isadigit{0}}{\isadigit{0}}{\isacharcomma}\ {\isadigit{0}}x{\isadigit{1}}FFFFFF{\isacharparenright}{\isachardoublequoteclose}\isanewline
\isacommand{definition}\isamarkupfalse%
\ {\isachardoublequoteopen}node{\isacharunderscore}{\isadigit{1}}{\isadigit{4}}{\isacharunderscore}lut{\isadigit{0}}{\isadigit{1}}\ \ {\isacharequal}\ \ empty{\isacharunderscore}spec\ {\isasymlparr}\isanewline
\ \ acc{\isacharunderscore}blocks\ {\isacharcolon}{\isacharequal}\ {\isacharbrackleft}{\isacharbrackright}{\isacharcomma}\isanewline
\ \ map{\isacharunderscore}blocks\ {\isacharcolon}{\isacharequal}\ {\isacharbrackleft}block{\isacharunderscore}map\ lut{\isadigit{0}}{\isadigit{1}}{\isacharunderscore}cfg\ {\isadigit{6}}\ {\isadigit{0}}x{\isadigit{0}}{\isacharcomma}\ block{\isacharunderscore}map\ lut{\isadigit{0}}{\isadigit{1}}{\isacharunderscore}mpb\ {\isadigit{7}}\ {\isadigit{0}}x{\isadigit{0}}{\isacharbrackright}{\isacharcomma}\isanewline
\ \ overlay\ {\isacharcolon}{\isacharequal}\ Some\ {\isadigit{0}}\isanewline
{\isasymrparr}{\isachardoublequoteclose}\isanewline
\ \ \isanewline
\isacommand{definition}\isamarkupfalse%
\ {\isachardoublequoteopen}sys\ {\isacharequal}\ {\isacharbrackleft}{\isacharparenleft}{\isadigit{0}}{\isacharcomma}\ node{\isacharunderscore}{\isadigit{0}}{\isacharunderscore}interconnect{\isacharparenright}{\isacharcomma}\ \isanewline
\ \ \ \ \ \ \ \ \ \ \ \ \ \ \ \ \ \ \ {\isacharparenleft}{\isadigit{1}}{\isacharcomma}node{\isacharunderscore}mc{\isacharparenright}{\isacharcomma}\ \ \isanewline
\ \ \ \ \ \ \ \ \ \ \ \ \ \ \ \ \ \ \ {\isacharparenleft}{\isadigit{2}}{\isacharcomma}node{\isacharunderscore}mc{\isacharparenright}{\isacharcomma}\ \ \isanewline
\ \ \ \ \ \ \ \ \ \ \ \ \ \ \ \ \ \ \ {\isacharparenleft}{\isadigit{3}}{\isacharcomma}node{\isacharunderscore}mc{\isacharparenright}{\isacharcomma}\ \ \isanewline
\ \ \ \ \ \ \ \ \ \ \ \ \ \ \ \ \ \ \ {\isacharparenleft}{\isadigit{4}}{\isacharcomma}node{\isacharunderscore}mc{\isacharparenright}{\isacharcomma}\isanewline
\ \ \ \ \ \ \ \ \ \ \ \ \ \ \ \ \ \ \ {\isacharparenleft}{\isadigit{5}}{\isacharcomma}node{\isacharunderscore}lmb{\isacharparenright}{\isacharcomma}\ \ \isanewline
\ \ \ \ \ \ \ \ \ \ \ \ \ \ \ \ \ \ \ {\isacharparenleft}{\isadigit{6}}{\isacharcomma}node{\isacharunderscore}lmb{\isacharparenright}{\isacharcomma}\ \ \isanewline
\ \ \ \ \ \ \ \ \ \ \ \ \ \ \ \ \ \ \ {\isacharparenleft}{\isadigit{7}}{\isacharcomma}node{\isacharunderscore}conf{\isacharparenright}{\isacharcomma}\ \ \isanewline
\ \ \ \ \ \ \ \ \ \ \ \ \ \ \ \ \ \ \ {\isacharparenleft}{\isadigit{8}}{\isacharcomma}node{\isacharunderscore}conf{\isacharparenright}{\isacharcomma}\ \isanewline
\ \ \ \ \ \ \ \ \ \ \ \ \ \ \ \ \ \ \ {\isacharparenleft}{\isadigit{9}}{\isacharcomma}\ node{\isacharunderscore}{\isadigit{9}}{\isacharunderscore}vas{\isadigit{0}}{\isadigit{0}}{\isacharparenright}{\isacharcomma}\isanewline
\ \ \ \ \ \ \ \ \ \ \ \ \ \ \ \ \ \ \ {\isacharparenleft}{\isadigit{1}}{\isadigit{0}}{\isacharcomma}\ cpu{\isacharunderscore}phys{\isacharparenright}{\isacharcomma}\ \ \isanewline
\ \ \ \ \ \ \ \ \ \ \ \ \ \ \ \ \ \ \ {\isacharparenleft}{\isadigit{1}}{\isadigit{1}}{\isacharcomma}\ node{\isacharunderscore}{\isadigit{1}}{\isadigit{1}}{\isacharunderscore}lut{\isadigit{0}}{\isadigit{0}}{\isacharparenright}{\isacharcomma}\ \isanewline
\ \ \ \ \ \ \ \ \ \ \ \ \ \ \ \ \ \ \ {\isacharparenleft}{\isadigit{1}}{\isadigit{2}}{\isacharcomma}\ node{\isacharunderscore}{\isadigit{1}}{\isadigit{2}}{\isacharunderscore}vas{\isadigit{0}}{\isadigit{1}}{\isacharparenright}{\isacharcomma}\ \isanewline
\ \ \ \ \ \ \ \ \ \ \ \ \ \ \ \ \ \ \ {\isacharparenleft}{\isadigit{1}}{\isadigit{3}}{\isacharcomma}\ cpu{\isacharunderscore}phys{\isacharparenright}{\isacharcomma}\ \isanewline
\ \ \ \ \ \ \ \ \ \ \ \ \ \ \ \ \ \ \ {\isacharparenleft}{\isadigit{1}}{\isadigit{4}}{\isacharcomma}\ node{\isacharunderscore}{\isadigit{1}}{\isadigit{4}}{\isacharunderscore}lut{\isadigit{0}}{\isadigit{1}}{\isacharparenright}\isanewline
{\isacharbrackright}{\isachardoublequoteclose}\isanewline
\isadelimtheory
\isanewline
\endisadelimtheory
\isatagtheory
\isacommand{end}\isamarkupfalse%
\endisatagtheory
{\isafoldtheory}%
\isadelimtheory
\endisadelimtheory
\end{isabellebody}%


\end{document}